\RequirePackage{fix-cm}
\documentclass[smallextended]{svjour3}       
\smartqed  
\usepackage{graphicx}%
\usepackage{amsmath}%
\usepackage{newtxmath}

\usepackage{multirow}%
\usepackage[title]{appendix}%

\usepackage{booktabs}%

\usepackage{xspace}

\usepackage{tcolorbox} 
\usepackage{colortbl} 
\usepackage{subcaption} 
\usepackage{pifont}

\usepackage{threeparttable}
\usepackage{tabularx}
\usepackage{placeins}
\usepackage{needspace}
\usepackage{makecell}
\usepackage{fontawesome}

\usepackage{float} 
\newfloat{listing}{htbp}{lop}
\floatname{listing}{Listing}
\usepackage{adjustbox}

\usepackage{xcolor}
\usepackage{listings}
\lstdefinestyle{promptstyle}{
    basicstyle=\ttfamily\small,          
    frame=none,                          
    backgroundcolor=\color{gray!10},     
    breaklines=true,                     
    escapeinside={||},                   
}

\lstdefinestyle{pythonstyle}{
    language=Python,                     
    basicstyle=\ttfamily\small,          
    frame=none,                          
    keywordstyle=\color{blue},           
    stringstyle=\color{red},             
    commentstyle=\color{gray},           
    breaklines=true,                     
    escapeinside={||},                   
}

\lstdefinelanguage{json}{
    basicstyle=\ttfamily\small,      
    frame=none,                      
    showstringspaces=false,          
    breaklines=true,                 
    tabsize=2,                       
    stringstyle=\color{red},         
    keywordstyle=\color{blue},       
    commentstyle=\color{gray},       
    morekeywords={true, false, null} 
}

\lstdefinestyle{javabox}{
    language=Java,
    basicstyle=\ttfamily\small,
    keywordstyle=\color{blue}\bfseries,
    commentstyle=\color{gray}\itshape,
    stringstyle=\color{brown},
    frame=single,                  
    rulecolor=\color{black},       
    numberstyle=\tiny\color{gray},
    breaklines=true,               
    showstringspaces=false,
    tabsize=4,
    escapeinside={(*@}{@*)} 
}

\newcommand{\e}[1]{\times10^{#1}}

\usepackage{natbib}
\usepackage[colorlinks = true,
            linkcolor = blue,
            urlcolor  = blue,
            citecolor = blue,
            anchorcolor = blue]{hyperref}

\newcommand{\Instruction}{\textit{Instruction}\xspace}
\newcommand{\InstructionLabel}{\textit{InstructionLabel}\xspace}
\newcommand{\InstructionInfill}{\textit{InstructionMask}\xspace}

\newcommand{\subfig}[2]{\ref{#1}\subref{#2}}

\begin{document}

\title{HAFix: History-Augmented Large Language Models for Bug Fixing
}


\author{Yu Shi
\and
Abdul Ali Bangash
\and
Emad Fallahzadeh
\and
Bram Adams
\and
Ahmed E. Hassan
}


\institute{Yu Shi \at
              School of Computing, Queen's University, Kingston, Ontario, Canada. \\
              \email{y.shi@queensu.ca}\\           %
              \url{https://orcid.org/0009-0005-6083-0932}
           \and
           Abdul Ali Bangash \at
              SBASSE, Lahore University of Management Sciences, Lahore, Pakistan. \\
              \email{abdulali@lums.edu.pk}\\           %
              \url{https://orcid.org/0000-0002-5311-6061}
           \and
           Emad Fallahzadeh \at
              School of Computing, Queen's University, Kingston, Ontario, Canada. \\
              \email{emad.fallahzadeh@queensu.ca}\\           %
              \url{https://orcid.org/0009-0005-5024-4868}
           \and
           Bram Adams \at
              School of Computing, Queen's University, Kingston, Ontario, Canada. \\
              \email{bram.adams@queensu.ca}\\
              \url{https://orcid.org/0000-0001-7213-4006}
           \and
           Ahmed E. Hassan \at
              School of Computing, Queen's University, Kingston, Ontario, Canada. \\
              \email{hassan@queensu.ca}\\
              \url{https://orcid.org/0000-0001-7749-5513}
}

\date{Received: date / Accepted: date}

\maketitle

\begin{abstract}
Recent studies have explored the performance of Large Language Models (LLMs) on various Software Engineering (SE) tasks, such as code generation and bug fixing. However, these approaches typically rely on the context data from the current snapshot of the project, overlooking the potential of rich historical data residing in real-world software repositories. Additionally, the impact of prompt styles on LLM performance for SE tasks within a historical context remains underexplored. To address these gaps, we propose \textbf{HAFix}, which stands for \underline{H}istory-\underline{A}ugmented LLMs on Bug \underline{Fix}ing, a novel approach that leverages seven individual historical heuristics associated with bugs and aggregates the results of these heuristics (HAFix-Agg) to enhance LLMs' bug-fixing capabilities. To empirically evaluate HAFix, we employ three Code LLMs (i.e., Code Llama, DeepSeek-Coder and DeepSeek-Coder-V2-Lite models) on 51 single-line Python bugs from BugsInPy and 116 single-line Java bugs from Defects4J. Our evaluation demonstrates that multiple HAFix heuristics (e.g., FN-modified and FN-all on Defects4J) achieve statistically significant improvements with large effect sizes compared to a non-historical baseline inspired by GitHub Copilot. Furthermore, the aggregated HAFix variant HAFix-Agg achieves substantial improvements with large effect sizes by combining the complementary strengths of individual heuristics, increasing bug-fixing rates relatively by an average of 45.05\% on BugsInPy and 49.92\% on Defects4J relative to the corresponding baseline. Moreover, within the context of historical heuristics, we identify the \Instruction prompt style as the most effective template compared to the \InstructionLabel and \InstructionInfill for LLMs in bug fixing. Finally, we evaluate the cost of HAFix in terms of inference time and token usage, and provide a pragmatic trade-off analysis of the cost and bug-fixing performance, offering valuable insights for the practical deployment of our approach in real-world scenarios.


\end{abstract}

\section{Introduction}\label{intro}

Large language models (LLMs) have emerged as transformative tools in software engineering (SE), with applications spanning code generation and completion \citep{lu2021codexglue,chen2021evaluating,du2023classeval,yu2024codereval,li2024deveval,zhang2024codeagent}, and bug fixing \citep{jiang2023impact,zhang2024diversity,fan2023automated,xia2023revisiting,li2024hybrid,jimenez2023swe,aleithan2024swe}. These models, such as CodeBERT \citep{feng2020codebert}, CodeT5 \citep{wang2021codet5}, Codex \citep{chen2021evaluating}, Code Llama \citep{roziere2023code} and DeepSeek-Coder \citep{guo2024deepseek}, have demonstrated significant capabilities in understanding and generating code based on various contextual inputs. The main key to unlocking the potential of LLMs is finding ways to more effectively leverage the full spectrum of context available in software development, particularly when it comes to understanding and resolving bugs.


While recent advances have focused on evaluating LLM performance by utilizing contextual information such as buggy-line-surrounded function code snippets to guide LLMs in bug fixing \citep{ahmad2021unified,wang2021codet5,niu2022spt,xia2022less,jiang2023impact,fan2023automated,xia2023revisiting,jimenez2023swe,xia2024automated,zirak2024improving},
the potential of incorporating historical context remains underexplored, except for \citet{le2016history}'s use of previously-appearing fix patterns from different projects to guide the current bug fixing. Historical data, such as information from blame commits, encapsulates the incremental evolution of a project, reflecting the developers' intent, bugs origins, and the rationale behind previous fixes. This historical information draws inspiration from early MSR (Mining Software Repositories) works, which emphasized the importance of historical data in understanding the evolution of software bugs \citep{sliwerski2005changes,hassan2006mining}. These insights can be instrumental in understanding the context of bugs and guiding LLM in bug-fixing strategies. However, there is still a significant gap in leveraging rich history data to help LLMs with bug fixing.

Moreover, prompt design is another critical factor influencing LLM performance in bug fixing, as the way in which information is presented to LLMs can significantly influence the relevance and quality of the generated code \citep{sclar2023quantifying}. While recent work \citep{jiang2023impact} studied if the LLMs can make good use of the buggy line, they did not systematically evaluate how different prompt styles perform when incorporating the historical context of a bug alongside natural language instructions. This highlights the need to explore which prompt style works best in leveraging historical heuristics to enhance LLM bug-fixing performance.

While historical heuristics may enhance bug-fixing performance, they also come with increased prompt size, monetary cost, and longer inference time. This raises the need to explore how historical data can be effectively utilized in a cost-efficient manner with LLMs. Although prior work, such as \citet{xia2024automated}, has examined the price cost of ChatGPT for fixing a single bug, it did not consider the role of historical data in LLM-based bug fixing, i.e., the increased prompt because of the rich historical context information. Similarly, \citet{jiang2023impact} provided the analysis of the model size and its relationship with performance, without delving into the broader implications of incorporating historical heuristics for LLMs on bug fixing. In other words, a comprehensive exploration of the trade-offs between bug-fixing performance, inference cost (e.g., in terms of price or the number of tokens), and time efficiency when leveraging historical heuristics remains missing.

Based on these research gaps, we aim to investigate how historical context, particularly blame commit data, can enhance LLM performance in bug fixing. Additionally, we aim to evaluate three distinct prompt styles: Instruction (\Instruction), Instruction with the buggy line labeled (\InstructionLabel), and Infill (\InstructionInfill), to assess their impact on bug-fixing effectiveness when used with historical context. Furthermore, our objective is to analyze the trade-offs between computational cost, time efficiency, and bug-fixing performance to provide actionable insights into the practical use of historical data in LLM-based bug fixing.

Based on these objectives, we determine the following research questions to explore:
\begin{itemize}
    \item \textbf{RQ1:} How much do history-augmented LLMs improve bug fixing compared to models without historical context?
    
    \item \textbf{RQ2:} How do different prompt styles impact the bug-fixing performance of history-augmented LLMs?

    \item \textbf{RQ3:} What is the cost of history-augmented LLMs on bug fixing?
\end{itemize}

To address our research questions, we propose HAFix, which stands for History-Augmented LLMs on Bug Fixing, a novel approach that leverages individual historical heuristics associated with bugs and aggregates the results of these heuristics (HAFix-Agg) to enhance LLMs’ bug-fixing capabilities. To empirically evaluate HAFix, we used two existing datasets: 51 single-line Python bugs from BugsInPy \citep{widyasari2020bugsinpy} and 116 Java bugs from Defects4J \citep{just2014defects4j}, both focusing on real-world projects with rich development histories. For RQ1, we explored the integration of historical data by mining and categorizing historical data from the blame commit and designing different historical heuristics to augment our baseline. The baseline was inspired by how GitHub Copilot \citep{copilot} constructs prompts in practice \citep{copilotprompt}. We also compare our approach with two state-of-the-art (SOTA) LLM-based bug-fixing methods, ChatRepair \citep{xia2024automated} and ITER \citep{ye2024iter}, to further contextualize its effectiveness. We employed CodeLlama-Instruct-7B\citep{roziere2023code}, DeepSeek-Coder-Instruct-6.7B \citep{guo2024deepseek}, and DeepSeek-Coder-V2-Lite-Instruct-16B \citep{zhu2024deepseek} as our subject models. The pass@k metric was used to assess the likelihood of generating correct results across multiple model outputs.

For RQ2, we investigated the influence of prompt styles on bug-fixing effectiveness. Using the baseline and the top-performing approaches from RQ1 HAFix-Agg, we systematically tested three distinct prompt styles: \Instruction, \InstructionLabel, and \InstructionInfill. The results identified the optimal combination of prompt style and historical context.

For RQ3, we analyzed the cost and efficiency of inference using HAFix-Agg with the \Instruction prompt style, as it was the best-performing configuration from RQ2. We measured inference time by recording the duration required to generate 10 outputs via nucleus sampling, ensuring consistency by running all experiments on identical infrastructure. To estimate the inference monetary cost, we calculate the number of input and output tokens. Additionally, we defined four execution scenarios: Exhaustive, EarlyStop (ES), ES-AccSorted, and ES-UniSorted, to explore the trade-offs between bug-fixing performance and computational cost, including inference time and inference token usage.

This study makes the following contributions:
\begin{itemize}
    \item \textbf{Leveraging Historical Context for LLM-based Bug Fixing}: This work investigates the impact of integrating various history-augmented heuristics derived from blame commits into LLM prompts to evaluate bug-fixing performance. It introduces an innovative approach called HAFix, demonstrating how leveraging historical context enhances bug-fixing performance and expands the model's capability to resolve complex bugs. Our findings show that multiple HAFix heuristics achieve statistically significant improvements over the baseline with large effect sizes, and that HAFix-Agg improves bug-fixing performance by an average of 45.05\% on BugsInPy and 49.92\% on Defects4J relative to the corresponding baseline.
        
    \item \textbf{Comprehensive Prompt Style Evaluation}: We comprehensively analyze three distinct prompt styles, including \Instruction, \InstructionLabel, and \InstructionInfill, and reveal their respective impacts on LLM bug-fixing performance. This study identifies the \Instruction prompt as the most effective style in maximizing the performance of HAFix for leveraging historical context.
    
    \item \textbf{Pragmatic Performance-Cost-Efficiency Analysis}: We provide an in-depth evaluation of the trade-offs between bug-fixing performance, inference token usage, and time efficiency across various historical heuristics and their execution sequences. Strategies such as the early stop of heuristic execution can reduce inference time and token consumption by an average of 69\% and 73\%, respectively, while maintaining competitive performance.

\end{itemize}
Our contributions lay the groundwork for leveraging historical data and optimal prompt design to improve LLM-based generated code and bug fixing, providing actionable guidance for balancing performance and cost in real-world applications.

\section{Related Work}\label{Related Work}

\subsection{Usage of LLMs in the Context of Software Engineering}\label{Usage of LLMs in the Context of Software Engineering}

Large Language Models (LLMs) have rapidly become a valuable tool in software engineering (SE), enabling various tasks such as code generation, code completion, and automated program repair (APR). Recent advancements in LLMs, including models like CodeBERT \citep{feng2020codebert}, CodeT5 \citep{wang2021codet5}, Codex \citep{chen2021evaluating}, Code Llama \citep{roziere2023code} and DeepSeek-Coder \citep{guo2024deepseek}, have shown strong performance in understanding and generating code based on local and repository-level contexts.

\subsubsection{Code Generation}\label{Code Generation}

In the context of code generation, recent works have explored various benchmarks and methods to enhance and evaluate LLM performance. RepoBench \citep{{liu2023repobench}} retrieves the most relevant code snippets from other files for code completion, but all retrieved context information is from the current snapshot of the project. Similarly, ClassEval \citep{du2023classeval} focuses on class-level code generation, identifying limitations in how LLMs handle class structures and dependencies. Additionally, CoderEval \citep{yu2024codereval} introduces a benchmark for pragmatic code generation, highlighting areas where LLMs need improvement to generate functional, maintainable code for real-world applications. The DevEval \citep{li2024deveval} benchmark assesses models like GPT-4 and Code Llama on real-world software projects, revealing challenges in generating practical code. RepoHyper \citep{phan2024repohyper} improves code completion by constructing semantic graphs, allowing LLMs to prioritize relevant code snippets but still without considering history context. CodePlan \citep{bairi2024codeplan} introduces a planning-based approach, where LLMs generate sequences of code edits based on context from the current project snapshot, showing promise in large-scale code modifications.

\subsubsection{Automated Bug Fixing}\label{Automated Bug Fixing}

In the LLM-based bug fixing field, most works focus on single-line bugs and providing models with the buggy-line surrounded code snippets \citep{lu2021codexglue,guo2020graphcodebert,ahmad2021unified,chakraborty2021multi,wang2021codet5,zhang2022coditt5,niu2022spt,chakraborty2022natgen,xia2022less}. Recent works have investigated various methods to enhance the effectiveness of LLMs in fixing software bugs. \citet{fan2023automated} have shown that given proper instructions such as information from fault localization, LLMs show promising results and can outperform traditional bug-fixing tools. \citet{xia2023revisiting} highlighted the importance of leveraging fine-tuning and prompting to harness the power of LLM with identifiers extracted from lines that are very similar to the
buggy line. The SWE-bench benchmark \citep{jimenez2023swe} evaluates LLMs on real-world GitHub issues, identifying that while LLMs can resolve straightforward bugs, they often falter on complex, context-dependent issues. \citet{jiang2023impact} demonstrated that while LLMs show promising results, they struggle to effectively utilize the buggy line, but when fine-tuned, they exhibit enhanced bug-fixing capabilities, although they may potentially over-rely on the buggy line. Furthermore, \citet{hossain2024deep} conducted a deep dive into bug localization and repair, localizing and fixing bugs at the token granularity rather than the traditional line granularity, resulting in substantial improvements in bug-fixing performance.

More recent works start considering the repository-level context information but only from the current snapshot of the project. RepoBugs \citep{chen2024large} introduces repository-level benchmarks, revealing that LLMs perform better when provided with extensive repository-level context. In another work, \citet{prenner2024out} examine the impact of local context from the current snapshot of the project in neural program repair, revealing that increasing context size significantly improves performance while emphasizing the need for clear context documentation and adequate datasets. Furthermore, domain adaptation techniques have been proposed to align models with specific codebases \citep{zirak2024improving}, enhancing repair success rates, while hybrid approaches \citep{li2024hybrid} combining LLMs with program analysis provide promising results in generating more reliable fixes. Finally, \citet{zhang2024diversity} highlight the integration of diverse software engineering agents to enhance the effectiveness of LLMs in solving real-world GitHub issues, utilizing the same contextual input across different agents.

One of the state-of-the-art bug-fixing tools in practice is GitHub Copilot \citep{copilot}. According to the official prompt engineering guidelines of GitHub Copilot \citep{copilotprompt}, when generating code suggestions, it uses the lines immediately before and after the user's current cursor position, as well as information from other files open in the editor and the URLs or file paths to provide relevant context. This information is derived solely from the current project snapshot, without incorporating historical data. 

Unlike previous research that primarily focuses on leveraging context information from the current snapshot of the project for code generation and automated bug fixing, our work explores the largely untapped potential of historical context from previous snapshots of the project in enhancing LLM performance for software engineering tasks. While recent studies have expanded the input context window of LLMs, utilized repository-level information, or employed domain adaptation techniques, they have not incorporated the code's evolutionary history to inform bug fixing or code generation. Our study addresses this gap by systematically evaluating the impact of historical context on LLM-based automated bug fixing, providing insights that could generalize to other software engineering tasks where the history of code evolution is a crucial factor. To evaluate the impact of historical context, we design a prompt inspired by GitHub Copilot’s practices as a baseline, reflecting traditional bug-fixing methods. The details of the baseline prompt and its implementation are discussed in Subsection \ref{Baseline Data Collection}.

\subsection{LLM Prompt Style vs. Bug Fixing}\label{LLM Prompt Style in Bug Fixing}

There is limited research on prompt styles for large language models (LLMs) in bug fixing, yet LLMs demonstrate sensitivity to prompt template choices \citep{sclar2023quantifying}. The work by \citet{xia2022less} introduces a cloze-style APR approach that directly leverages LLMs without requiring any fine-tuning or retraining on bug-fix datasets, framing the repair process as a cloze task to predict masked code snippets. Additionally, the study by \citet{jiang2023impact} evaluates various code language models (CLMs) for APR, determining whether to explicitly label the buggy line or mask it in the prompt template based on the corresponding pre-training task style of the LLM. Furthermore, \citet{xia2024automated} present a conversation-driven approach that employs cloze-style prompts, interspersing patch generation with immediate feedback to enhance the interaction between the model and the repair task. Recent work by \citet{sclar2023quantifying} emphasizes the critical nature of prompt formatting, demonstrating that even minor variation can lead to drastic performance changes.

In contrast to these studies, our research evaluates both the usage of instruction and infill prompts, to measure the impact of various prompt styles (instruction, instruction with buggy lines labeled, and infilling) on LLM performance in bug fixing. Our analysis focuses on how these prompt styles affect the model's effectiveness when applied in conjunction with various history heuristics. This investigation offers a unique perspective on the relationship between prompt design and bug repair performance, contributing to a more nuanced understanding of how customized prompts can enhance LLM capabilities in bug-fixing.

\subsection{LLM Cost vs. Bug Fixing}\label{LLM Cost in Bug Fixing}

Recent research has increasingly focused on analyzing the cost implications of using large language models (LLMs) for bug fixing. \citet{jiang2023impact}'s study on the impact of code language models on APR examines the trade-off between model size and bug-fixing capability, showing that while larger models offer higher success rates, they also incur greater computational costs. Additionally, \citet{xia2024automated} introduce a conversation-driven APR approach using ChatGPT, achieving an average repair cost of \$0.42 per bug, emphasizing the cost-effectiveness of using conversational LLMs in automated bug fixing. \citet{hidvegi2024cigar} propose a cost-efficient program repair method that minimizes token costs by optimizing prompts and leveraging strategies like summarizing responses and patch multiplication while maintaining high bug-fixing performance. Similarly, \citet{nayab2024concise} explore how the length of LLM-generated outputs influences both inference cost and model performance, offering strategies to minimize unnecessary token generation for more cost-effective results. \citet{shekhar2024towards} optimize LLM usage costs by predicting output quality and selecting models to balance quality, cost, and latency, showing significant improvements in cost-efficiency and quality.

In contrast to prior studies that primarily analyze monetary inference costs or focus on performance in isolation, our work adopts a more comprehensive perspective by jointly examining inference time, token-based cost, and their trade-offs with bug-fixing performance. We investigate how our history-augmented bug-fixing approach achieves a balance between high effectiveness and cost-efficiency. In particular, we investigate the cost of using different historical heuristics for LLM-based bug fixing, providing a pragmatic cost estimation by calculating the number of input and output tokens. This holistic approach addresses a gap in the current literature, offering insights into the large-scale, practical application of LLMs in real-world software engineering scenarios.

\section{HAFix: History-Augmented LLMs for Bug Fixing}\label{HAFix: History-Augmented LLMs for Bug Fixing}

We introduce HAFix (History-Augmented LLMs on Bug Fixing), a novel approach that enhances bug-fixing capabilities by incorporating historical heuristics extracted from blame commit data. By integrating historical data into the prompts, HAFix provides the model with additional context, aiding in identifying the root cause of the bug and generating a possible solution to solve it. Yet, what is the most relevant historical context data for a bug? We explore this question from the perspectives of temporal and spatial analysis.

The spatial aspect of bug fixing involves understanding the structural and positional context within the codebase. This approach, commonly utilized by tools like GitHub Copilot \citep{copilot}, emphasizes using information from the current snapshot of the code, such as the surrounding function code and file structure \citep{copilotprompt}. By focusing on the specific function and narrowing it down to the buggy line, the model is provided with the most relevant spatial context, minimizing irrelevant data. This approach prevents the model from being distracted by unrelated sections of the code and ensures that the model's attention is directed at the precise location of the bug. For example, file names and function-level code surrounding the buggy line provide context to pinpoint where the bug exists and how it might be fixed.

From the temporal perspective of a bug, the commit that last touches the buggy line (blame commit) will give the most closely related information about how this buggy code is modified \citep{sliwerski2005changes,hassan2006mining}. This information includes details of the changes made, the reasoning behind these changes, and the broader context of other modifications within the same commit. This temporal analysis draws inspiration from early MSR (Mining Software Repositories) techniques, which emphasized the importance of historical data in understanding the evolution of software bugs. By integrating both spatial and temporal data, HAFix combines established practices with novel insights to improve bug-fixing capability.

\begin{figure}[!t]
     \begin{center}
         \includegraphics[scale=0.271]{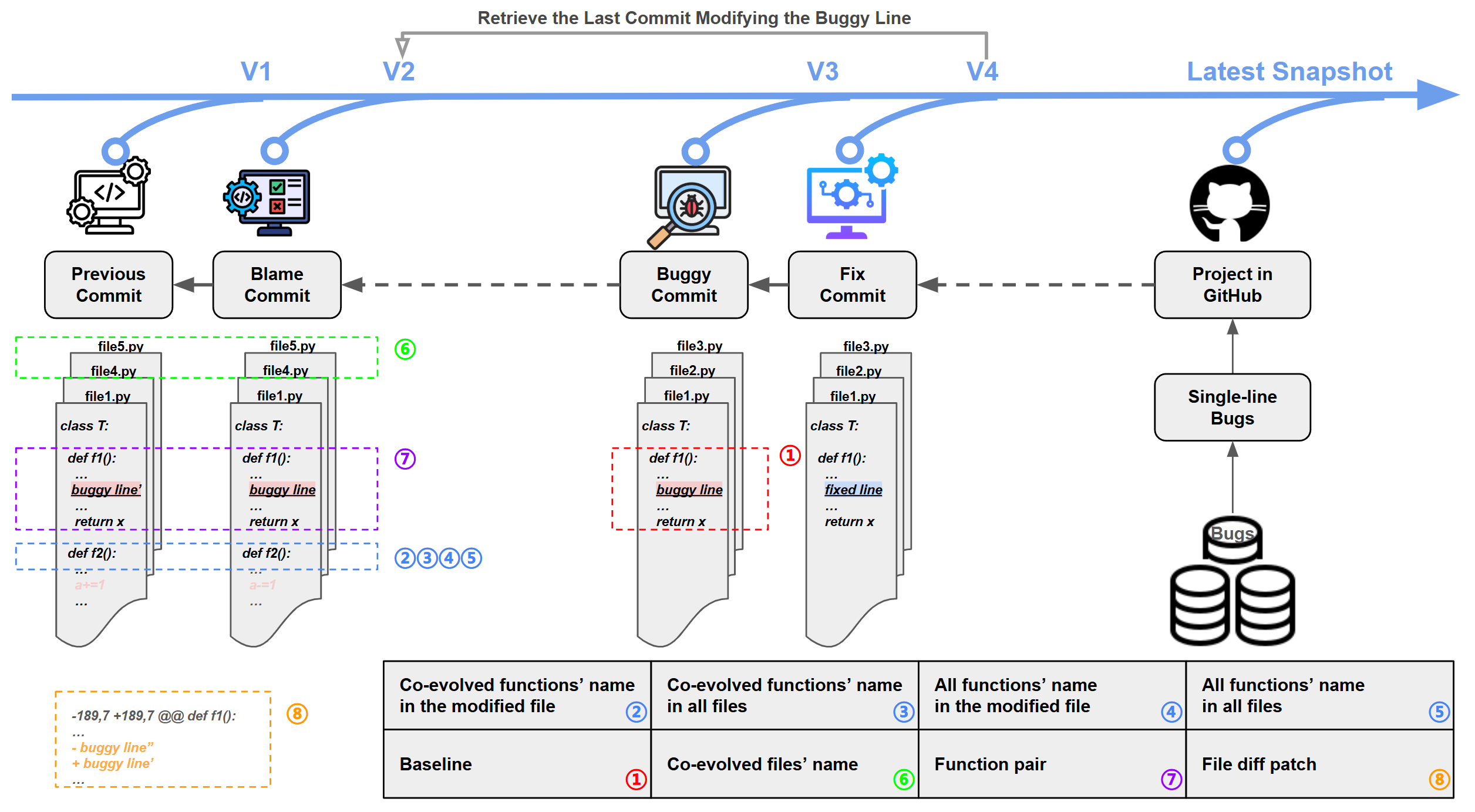}
        \end{center}
        \caption{Dataset collection for HAFix: \ding{172} represents the data used for the baseline, while \ding{173} to \ding{179} represent the data for various historical heuristics. V4 refers to the snapshot of the project version where the bug fix was committed, and V3 is the snapshot of the previous version containing the bug. V2 is the snapshot of the last commit modifying the buggy line in the V4 snapshot, while V1 is the snapshot of the commit preceding V2. The rationale for selecting the blame commit and these historical heuristics are detailed in Section \ref{HAFix: History-Augmented LLMs for Bug Fixing}.}
        \label{fig:data_collection}
\end{figure}

Figure \ref{fig:data_collection} illustrates the data collection process for HAFix, which is structured across multiple stages corresponding to different project versions (V1, V2, V3, V4). The data collection process begins at V4, which is the snapshot of the project version where the fixed code was committed (fix commit). Next, we trace back to V3, which is the version preceding V4 that still contains the buggy code (buggy commit). Using PyDriller \citep{Spadini2018}, we then identify V2 as the version that last modifies the buggy code line (blame commit). Lastly, we trace back to V1, which is the snapshot right before V2, to pair it with V2 as the historical data of the blame commit. V4 and V3 capture the direct and actual changes made to resolve the bug, which will be used for the baseline detailed in the following Subsection \ref{Baseline Data Collection}. V2 and V1 contain the historical data of the bug, we will use these different heuristics to enhance the baseline, which will be detailed in Subsection \ref{Historical Heuristics Prompt}.

\subsection{History-Augmented Bug Fixing: Historical Heuristics Prompt}\label{Historical Heuristics Prompt}

The key innovation in HAFix lies in the augmentation of bug-fixing prompts with historical context. We extract this context by mining the blame commit, which represents the last modification of the buggy line. This process provides temporal insights that highlight how the code evolved, shedding light on potential root causes of the bug. As shown in Figure \ref{fig:data_collection}, to construct history-augmented prompts, we extract the following seven heuristics from the blame commit (V2) and the previous commit (V1). We focus on these seven heuristics because they represent well-established \citep{kamei2012large,adams2010identifying,hassan2008road}, blame-commit-derived signals that balance contextual richness and inference cost. While other historical signals, such as dependency or test case changes, may also be informative, we leave their integration to future work to maintain a tractable scope. 

For clarity, we group these heuristics into three types: (1) name-based heuristics that summarize which functions or files are involved (CFN-modified, CFN-all, FN-modified, FN-all, FLN-all), (2) code-evolution heuristics that capture before-and-after function changes (FN-pair), and (3) patch-level heuristics that expose concrete edits (FL-diff).
These heuristics include:

\begin{itemize}
    \item \textbf{Co-evolved Functions' Names in the Modified Buggy File (CFN-modified)}: The names of functions modified within the buggy file in commit V2. These are important because they provide context about which specific functions were altered and may directly influence the buggy line.
    \item \textbf{Co-evolved Functions’ Names in All Modified Files (CFN-all)}: The names of functions modified across all modified files in commit V2. This information helps in understanding broader structural changes within the codebase that may indirectly impact the buggy line.
    \item \textbf{All Functions’ Names in the Modified Buggy File (FN-modified)}: The names of all functions (whether changed or not) in the modified buggy file in the commit of V2. This allows the model to understand the function structure in the file, offering context to better locate the bug within its function.
    \item \textbf{All Functions' Names in All Modified Files (FN-all)}: The names of all functions (whether changed or not) in all modified files in the commit of V2. This information helps capture a wider scope of the code changes and potential interactions between different functions across the codebase.
    \item \textbf{Co-evolved Files' Names (FLN-all)}: The names of changed files of the commit in V2. This provides a broad context of which files were modified, potentially indicating areas in the code that might affect the buggy line or the system’s behavior.
    \item \textbf{Function Code Pairs (FN-pair)}: The function code before and after the blame commit in V1 and V2. This temporal context helps identify how the buggy function evolved and provides clues about what changes directly contributed to the bug’s introduction.
    \item \textbf{File Diff Patch (FL-diff)}: The diff patch from the git diff command in the commit of V2. This allows us to see the exact code changes made, providing precise details on what was modified, which can aid in pinpointing the cause of the bug.
\end{itemize}

These heuristics were chosen to offer a comprehensive yet focused historical snapshot, essential for understanding both the bug’s cause and the code structure surrounding it.

\subsection{HAFix-Agg: Aggregated HAFix Variant}\label{HAFix-Agg}

Embedding all heuristic data directly into a single prompt would result in excessively large inputs, increasing both computational cost and latency. To address this, we introduce HAFix-Agg, a variant that aggregates the LLM results of individual heuristics, which helps reduce the prompt size and computational cost of model inference. While HAFix-Agg aims to improve performance by leveraging insights from multiple heuristics, it comes at the expense of higher inference costs, as each heuristic requires separate inference runs. This trade-off allows for broader coverage of potential fixes, making HAFix-Agg particularly useful for complex bug-fixing scenarios where a single heuristic may be insufficient. We assess this variant's prediction performance (RQ1/2) and further explore its cost-effectiveness in RQ3, providing insights into its feasibility and scalability for real-world deployment.

\section{Empirical Evaluation of HAFix}\label{Empirical Evaluation of HAFix}

To empirically validate the effectiveness of HAFix, we conduct a comprehensive evaluation using real-world Python and Java bugs. This section details our dataset selection, history-agnostic baseline, data collection process, model selection, prompt construction, experimental pipeline, and inference infrastructure.

\subsection{Dataset Selection}\label{Dataset selection}

We evaluate HAFix using two real-world bug datasets: BugsInPy \citep{widyasari2020bugsinpy} and Defects4J \citep{just2014defects4j}. This selection allows us to assess the generalizability of HAFix across Python and Java, which rank among the top four popular programming languages \citep{tiobe_index}. They differ in typing paradigms (dynamic for Python, static for Java) and are widely used in the APR domain. In addition to language diversity, we require datasets that include real-world projects with test cases, allowing us to verify the functional correctness of model-generated code against the developer's original fixed code. Lastly, we prioritize datasets with rich development history, enabling us to mine the historical context data for each bug.

We select BugsInPy collected by \citet{widyasari2020bugsinpy} as our subject benchmark dataset for several reasons. BugsInPy is a comprehensive, hand-curated dataset with 493 real-world bugs from 17 large, non-trivial Python projects. The bugs in BugsInPy are carefully selected to meet specific criteria: they must involve changes in the source code, excluding modifications like configurations or build scripts. Additionally, the bugs should be reproducible, with at least one test case failing on the faulty version, and they must be isolated from unrelated changes, such as refactoring or feature additions. These criteria ensure the quality of the bugs for our study. The projects included in BugsInPy span various domains such as machine learning, developer tools, scientific computing, and web frameworks.

Based on the similar selection criteria above, we include Defects4J \citep{just2014defects4j} as a second benchmark dataset, which is widely used in the program repair community for evaluating APR techniques on real-world Java bugs \citep{xia2022less,jiang2023impact,hossain2024deep,lutellier2020coconut,jiang2021cure,bouzenia2024repairagent,ye2024iter}. At the time of writing, Defects4J contains 854 Java bugs collected from 17 different open-source projects, spanning various domains such as data visualization (Chart), compiler construction (Closure), and date/time utilities (Time). The inclusion of both Python and Java datasets with diverse, widely used projects improves the external validity of our evaluation and enables assessment of HAFix’s effectiveness across real-world scenarios.

\begin{table}[!htbp]
    \centering
    \caption{Summary of 51 subject bugs from BugsInPy and 116 subject bugs from Defects4J, along with associated project source information (as of the time of writing). The columns \# of Bugs, \# of LOC, and \# of GitHub Stars donate the number of bugs, lines of code, and GitHub stars, respectively.}
  \begin{tabular}{l r r r r}
    \toprule
    \textbf{Dataset} & \textbf{Project} & \textbf{\# of Bugs} & \textbf{\# of LOC} & \textbf{\# of GH Stars} \\
    \midrule
    \multirow{11}{*}{BugsInPy} 
      & sanic    & 1 & 77k & 18.1k \\
      & luigi    & 9 & 44k & 17.9k \\
      & youtube-dl  & 7 & 139k & 132k \\
      & ansible  & 2 & 237k & 63k \\
      & scrapy   & 4 & 479k & 53.1k \\
      & pandas   & 15 & 457k & 43.8k \\
      & thefuck  & 6 & 11k & 85.4k \\
      & tornado  & 2 & 29k & 21.7k \\
      & fastapi  & 1 & 165k & 77.6k \\
      & black    & 1 & 118k & 39.2k \\
      & tqdm     & 3 & 7k & 28.7k \\
    \midrule
    \multirow{16}{*}{Defects4J}
      & Chart 			&  8 & 133k & 1.3k   \\
      & Cli 				&  5 & 14k & 371   \\
      & Closure 			& 19 & 618k & 7.5k   \\
      & Codec 			&  8 & 26k & 469   \\
      & Collections 		&  3 & 77k & 706   \\
      & Compress 		&  3 & 80k & 369   \\
      & Csv 				&  3 & 12k & 392   \\
      & Gson 			&  2 & 38k & 23.9k   \\
      & JacksonCore 		&  3 & 89k & 2.3k   \\
      & JacksonDatabind  & 12 & 21k & 3.6k   \\
      & Jsoup 			& 14 & 34k & 11.2k   \\
      & JxPath 			&  1 & 27k & 34   \\
      & Lang  			&  9 & 102k & 2.8k   \\
      & Math  			& 20 & 153k & 616   \\
      & Mockito  		&  3 & 65k & 15.2k   \\
      & Time  			&  3 & 97k & 5k   \\
   \bottomrule
  \end{tabular}
  \label{datasets_summary}
\end{table}

Following previous work \citep{xia2022less,ye2022neural,jiang2023impact,prenner2024out, xia2022less,jesse2023large}, we focus on single-line bugs, i.e., bugs whose fixes are focused on one line of code. This design choice aligns with recent large-scale studies on real-world bug fixes. For example, the ManySStuBs4J dataset \citep{karampatsis2020often} includes over 153,000 single-statement bug fixes from 1,000 Java projects and reports that such bugs occur roughly once per 1,600-2,500 lines of code. In Python, the TSSB-3M dataset \citep{richter2022tssb} collects over 3 million single-line bug fixes, with 72\% fixable using just a few AST-level edits. These findings confirm that single-line bugs are both common and suitable for controlled, interpretable evaluation. Thus, we follow this setting to clearly assess the impact of historical context in bug fixing.

To the best of our knowledge, limited research has evaluated LLM performance on bug fixing across both single-line Python and Java bugs. Furthermore, by starting with single-line bugs, we can establish a solid foundation before tackling more complex cases, such as multi-line or even multi-hunk bugs (i.e., bugs where the faulty code lines are not contiguous) in future work.

To identify single-line bugs, we examined the code changes in the commit that fixed the bug (the ``fix commit"). Specifically, we used the open-source tool PyDriller \citep{Spadini2018} along with the fix commit ID provided by BugsInPy and Defects4J to locate the fix commit. We then verified whether the commit contained only one change in a single Python or Java file, excluding test files. Further, we checked if the intersection of added and deleted lines involved a single line of code change while excluding no-code lines such as blank lines or comments. Through this process, we identified 68 single-line Python bugs from BugsInPy and 118 single-line Java bugs from Defects4J.

We manually validated each single-line bug to ensure it met the specified criteria above. With the location and isolation of bugs confirmed during the initial identification above, our primary focus was verifying reproducibility. This step involved running test cases to confirm that they can pass in the fix commit and fail in the buggy commit (the immediate predecessor). For example, if the test cases of a bug pass in both the fixed and buggy commits or if they fail in both, we filter out such cases.

Ultimately, out of the original 68 and 118 bugs, we obtained a subject dataset of 51 high-quality single-line bugs from BugsInPy and 116 high-quality single-line bugs from Defects4J, which is a similar dataset size as prior works \citep{prenner2022can,kolak2022patch,fan2023automated,peng2024domain,chen2024large}. Table \ref{datasets_summary} lists the project sources of these subject bugs, including their number of lines of code and GitHub star counts. We believe that the popularity of these projects highlights the representativeness of these bugs. For each bug, we rely on the fix commit ID to locate the corrected code and its corresponding test cases. We also use this commit ID to trace back to the buggy commit (the immediate predecessor) and mine the necessary data for our study, which will be detailed below.

To provide a deeper understanding of the dataset and demonstrate the effort involved in its curation, we selected one representative example from the BugsInPy dataset. Note that both BugsInPy and Defects4J follow this schema. This example was chosen to highlight the diversity and complexity of the bugs included in the dataset. We provide detailed information in the example, including the commit description, heuristic values, and relevant metadata. Full details of the selected example can be found in the Appendix \ref{appendix:RepresentativeExamples}.

\subsection{History-Agnostic Baseline}\label{History-Agnostic Baseline}
To establish a baseline for comparison, we need to design a prompt that reflects traditional bug-fixing practices without incorporating historical data. As mentioned in Subsection \ref{Automated Bug Fixing}, this baseline prompt serves as a reference point for evaluating the effectiveness of history-augmented approaches. We design our baseline prompt inspired by how GitHub Copilot \citep{copilot} processes user prompts, as it is one of the most widely used coding assistant tools for bug fixing in practice. While the specific prompt template used by GitHub Copilot is not publicly available, the official prompt engineering guidelines of GitHub Copilot \citep{copilotprompt}, mention that when generating code suggestions, it uses the lines immediately before and after the user's current cursor position, as well as information from other files open in the editor and the URLs or file paths to provide relevant context. This information is derived solely from the current project snapshot, without incorporating historical data. The detailed baseline data collection process is described in Section \ref{Baseline Data Collection}. In addition to this baseline, we also compare our approach with two state-of-the-art (SOTA) LLM-based bug-fixing methods, ChatRepair \citep{xia2024automated} and ITER \citep{ye2024iter}, to further contextualize its effectiveness.

\subsection{Data Collection}\label{Data Mining}

The data collection process follows the staged approach depicted in Figure \ref{fig:data_collection}, progressing from the most recent fix commit (V4) to the earliest commit (V1).

\subsubsection{Baseline Data Collection}\label{Baseline Data Collection}

As introduced in Section \ref{History-Agnostic Baseline}, our history-agnostic baseline reflects traditional bug-fixing practices without using historical context. For each bug in our subject dataset, we mined the non-history data for constructing the baseline prompt from the fix commit (V4) and buggy commit (V3), providing spatial context such as function-level code and the buggy line without temporal information. The fields we mined include:

\begin{itemize}
    \item \textbf{Project Name}: This field provides the LLM with the repository name associated with the bug.
    
    \item \textbf{Buggy File Name and Path}: Specifies the name and path of the buggy file that was modified to fix the bug, crucial for locating the bug within the project's codebase. This field remains consistent between V4 and V3, as no file renamings were observed in the studied bugs.
    
    \item \textbf{Buggy Line Location}: Since our focus is on bug repair rather than fault localization, we explicitly provide the LLM with the precise buggy line code. The buggy line code should remain consistent across all commit snapshots from V2 to V3, we extract it only from V3 for simplicity.
    
    \item \textbf{Buggy Function Name}: Indicates the specific function where the bug was located, providing more precise localization within the file. The field should be the same in V4 and V3.

    \item \textbf{Function Code Before and After the Fix Commit}: It provides LLM with the whole buggy line surrounded with the function-level code before and after the fix, allowing for a detailed examination of the changes at the function level. The function code before the fix commit is from V3 and after is from V4.

    \item \textbf{Bug Description}: This field provides the essential bug context, which will be detailed in Subsection \ref{Data Mining}, using the cleaned-up version mined from GitHub issue pages or commit messages, ensuring no post-fix details were included to prevent data leakage.
\end{itemize}

We also provide an example of the baseline prompt built based on data collected above in Listing \ref{BaselinePromptExample} in Appendix \ref{appendix:PromptExample}.

Note that we do not mine the entire buggy or fixed file code, but instead narrow the scope to the function-level code snippet surrounding the buggy line, since this provides sufficient context to understand the single-line bug while avoiding noise from unrelated parts of the file. During the data collection stage, before LLM inference, we employ AST (Abstract Syntax Tree) matching to accurately locate and extract the modified function from both the buggy and blame commits. We use existing libraries (ast module in Python and PyDriller) to resolve name collisions by parameter signatures and parent nodes (e.g., class or file). This ensures consistent extraction for constructing the \textit{Function Code Before and After the Fix Commit} field described in this section and the \textit{Function Code Pairs} field described in Section \ref{Historical Heuristics Prompt}. Moreover, for the bug description field, we consider mining both the commit message and the corresponding GitHub issue page as detailed below.

\vspace{0.3cm} 
\noindent \textbf{Bug Description Mining}

\begin{figure}[!htbp]
     \begin{center}
         \includegraphics[scale=0.4]{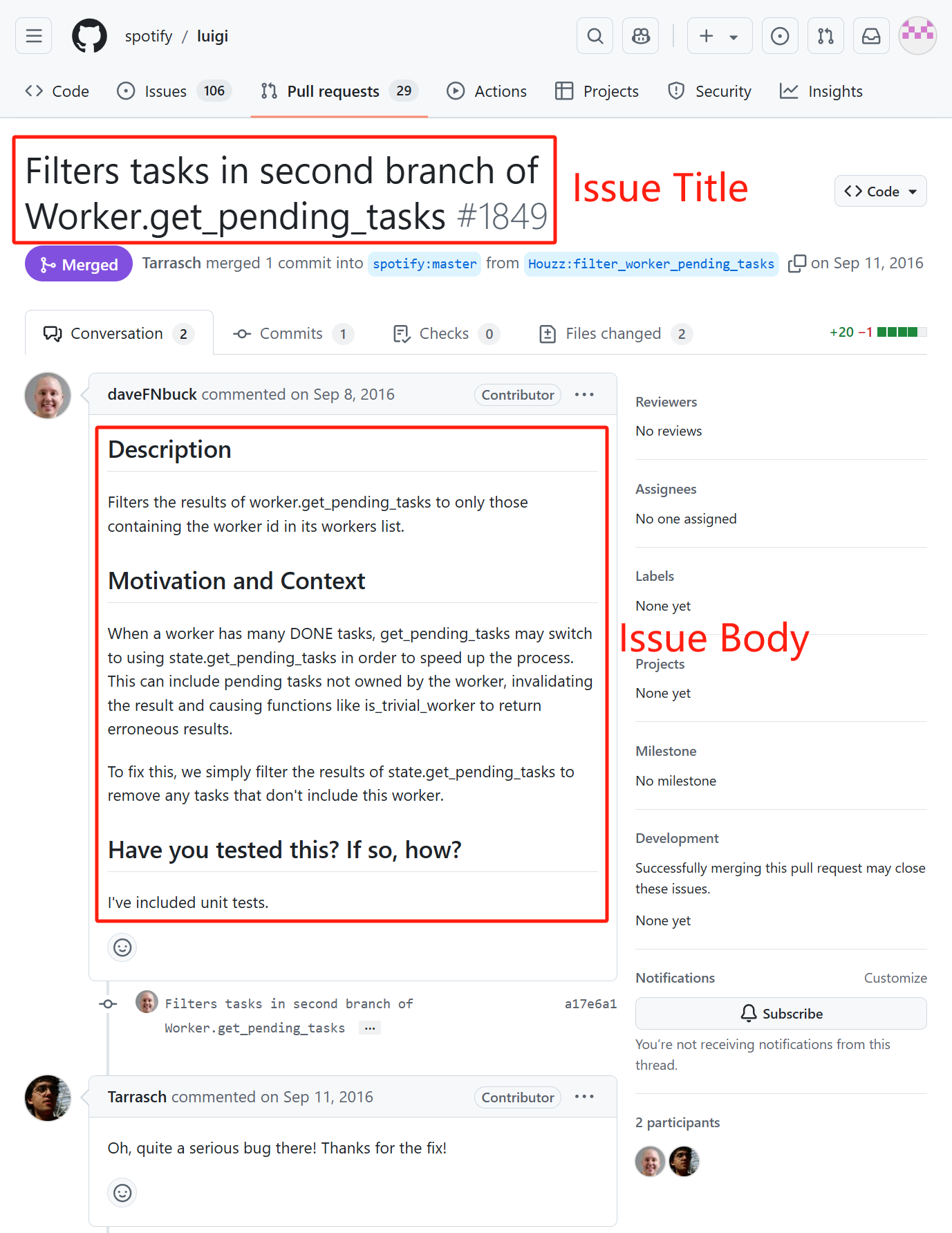}
        \end{center}
        \caption{An example of the bug description we mined from the GitHub issue page.}
        \label{fig:bug_description_example}
\end{figure}

In practice, before fixing a bug, developers often have access to contextual information such as how the bug manifests, its consequences, and any error output. Therefore, it is crucial to incorporate this context into our approach to ensure a comprehensive evaluation. However, fix commit messages typically lack this level of detail. To supplement this, we mined the corresponding bug description information for each bug. For BugsInPy, we extract the bug description from the GitHub issue page. We began by manually identifying whether the commit message contained a link to the relevant GitHub issue page. If a link is present, we use it to retrieve the issue; otherwise, we fall back to the commit message itself. GitHub issue pages generally provide detailed descriptions, discussions, and the steps taken to resolve the bug. For Defects4J, we use the official bug report links provided by the dataset. For both datasets, we employ either the open-source tool GHApi \citep{ghapi} to mine GitHub issues or the Python \texttt{requests} API to extract content from bug report pages. Specifically, we extract the title and body (the initial comment block) of the issue page or the bug report, which provides a concise yet informative bug description. Figure \ref{fig:bug_description_example} provides an example of the bug description information that we mined from the GitHub issue page. We combine the issue title and body, highlighted within the two red-circled boxes, to create the bug description.

To avoid data leakage in our empirical evaluation, we have to ensure that the bug description that we mined from GitHub issue pages, bug reports, or fix commit messages does not contain post-hoc details about how a bug was fixed. To avoid such an issue, we manually checked each bug description and filtered the information that was too closely related to the fix for the bug. Our goal was to ensure that the model focused on understanding the general context and nature of the bug, without exposure to the exact fix, which could otherwise compromise our evaluation of the model's ability to independently generate a correct solution.

\subsubsection{Historical Data Collection (HAFix Heuristics)}\label{Historical Data Collection (HAFix Heuristics)}

While non-history data provides valuable spatial context, it lacks insights into the evolution of the bug. To bridge this gap, we incorporate historical data collection to enhance the prompts with temporal context. For each bug in our subject dataset, we mined the history data for constructing the prompt of each HAFix heuristic from the blame commit (V2) and previous commit (V1), providing buggy line-related temporal context such as co-evolved files, functions, and diffs. The fields we mined follow the Subsection \ref{Historical Heuristics Prompt}.

The combination of non-history and historical data collection forms the foundation of HAFix, enriching bug-fixing prompts with both spatial and temporal perspectives to improve LLM performance in bug fixing.

\subsection{Model Selection}\label{Model selection}
We evaluate HAFix using three state-of-the-art and open-source code LLMs: (1) \textbf{CodeLlama-Instruct-7B}, (2) \textbf{DeepSeek-Coder-Instruct-6.7B}, and (3) \textbf{DeepSeek-Coder-V2-Lite-Instruct-16B}. These models are selected based on instruction-following capability, support for code infilling, and compatibility with our computational constraints.

One of the primary considerations was the model size. The CodeLlama-Instruct-7B and DeepSeek-Coder-Instruct-6.7B models require approximately 13GB in FP16 precision, making them suitable for single-GPU execution while maintaining strong generation performance. To assess HAFix on larger models, we additionally include the DeepSeek-Coder-V2-Lite-Instruct-16B model, which requires approximately 31GB in FP16. In contrast, larger variants with FP16 precision, such as CodeLlama-34B ($\sim$63GB) or DeepSeek-Coder-V2-236B ($\sim$472GB), are excluded due to their significantly higher hardware demands.

All selected models are instruction-tuned, enabling them to understand structured prompts and follow task-specific instructions more effectively. They also support code infilling, which allows the models to generate appropriate replacements for masked faulty code segments based on the surrounding context. We omit base or domain-specific variants (e.g., Python-tuned models) as they either lack instruction-following capabilities or are not optimized for structured prompt understanding, potentially leading to degraded performance in our evaluation.

\subsection{Prompt Construction}\label{Prompt styles}
\begin{figure}[t]
     \begin{center}
         \includegraphics[scale=0.25]{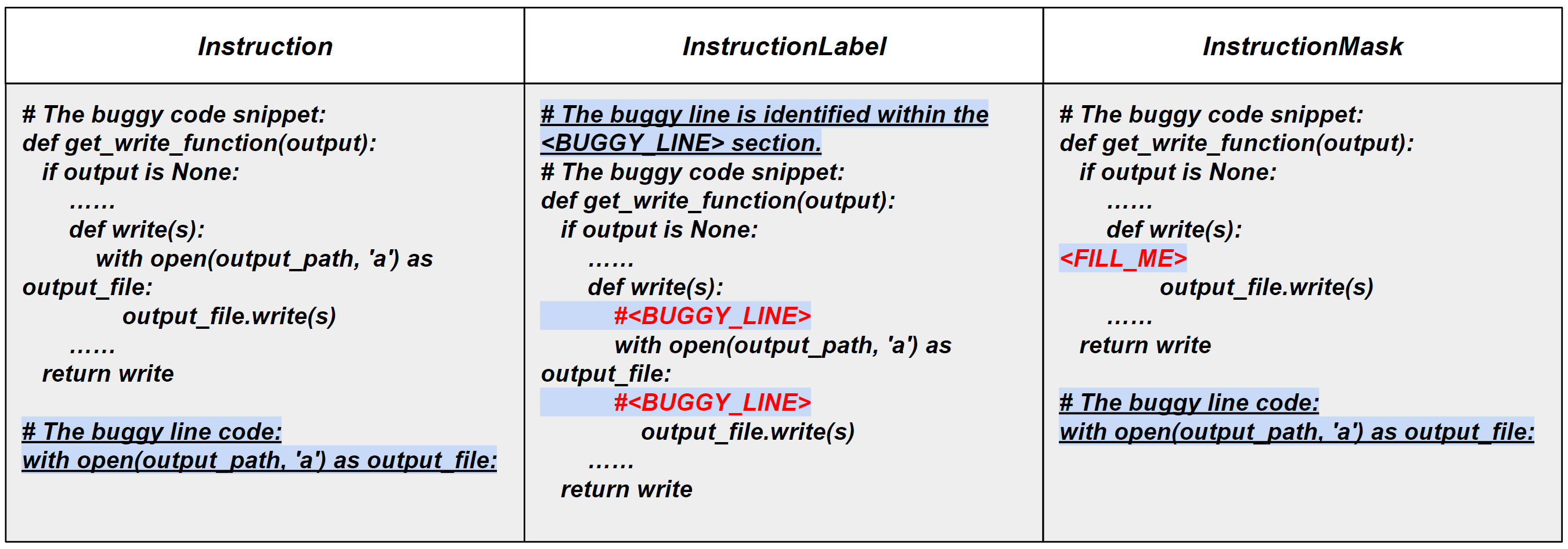}
        \end{center}
        \caption{Example of three prompt styles.}
        \label{fig:prompt_examples}
\end{figure}

We designed and evaluated three different prompt styles to provide input to the LLM to fix bugs, as different prompting styles can influence the outputs generated by LLMs \citep{sclar2023quantifying}. Each prompt style is applicable to the baseline and each heuristic. Figure \ref{fig:prompt_examples} provides an example for three prompt styles. These styles are categorized as follows:

\begin{itemize}
    \item \Instruction: This prompt style presents the entire buggy code snippet and highlights the buggy line in the instruction text. It includes the function with the bug, along with the full implementation context. In Figure \ref{fig:prompt_examples}, the first box demonstrates this style, with the buggy line highlighted in the instruction text at the bottom.
    
    \item \InstructionLabel: This prompt style labels the buggy line within the function code to provide more precise guidance. The buggy line is tagged with \textless BUGGY\_LINE\textgreater, directing the LLM's focus to the specific part of the code that needs fixing. The second box in Figure \ref{fig:prompt_examples} showcases this style, where the buggy line is labeled and tagged in both the function and the instruction.
    
    \item \InstructionInfill: This prompt style masks the buggy line with a placeholder, \textless FILL\_ME\textgreater, and highlights it in the instruction text. The LLM then generates the correct code to replace the masked line. The third box in Figure \ref{fig:prompt_examples} illustrates this style, with the masked line in the function and the corresponding instruction guiding the model to generate a fix.
\end{itemize}

The rationale for using the \Instruction is that previous LLM-based bug-fixing approaches only provide the model with buggy code \citep{lu2021codexglue}. However, most LLMs, including Code Llama and the DeepSeek models, are trained to understand both natural and programming languages. For \InstructionLabel, we were inspired by previous work \citep{jiang2023impact} that explicitly labeled the buggy line within the functional code. Finally, we evaluated \InstructionInfill based on the infilling capabilities of our subject models and their original design for code completion. To the best of our knowledge, no prior work has compared prompt styles on these models, such as Code Llama. Inspired by this, we explored whether masking and regenerating the buggy line could yield better performance than direct fixes.

\subsection{Experimental Pipeline}\label{Experimental pipeline}
\begin{figure}[t]
     \begin{center}
         \includegraphics[scale=0.35]{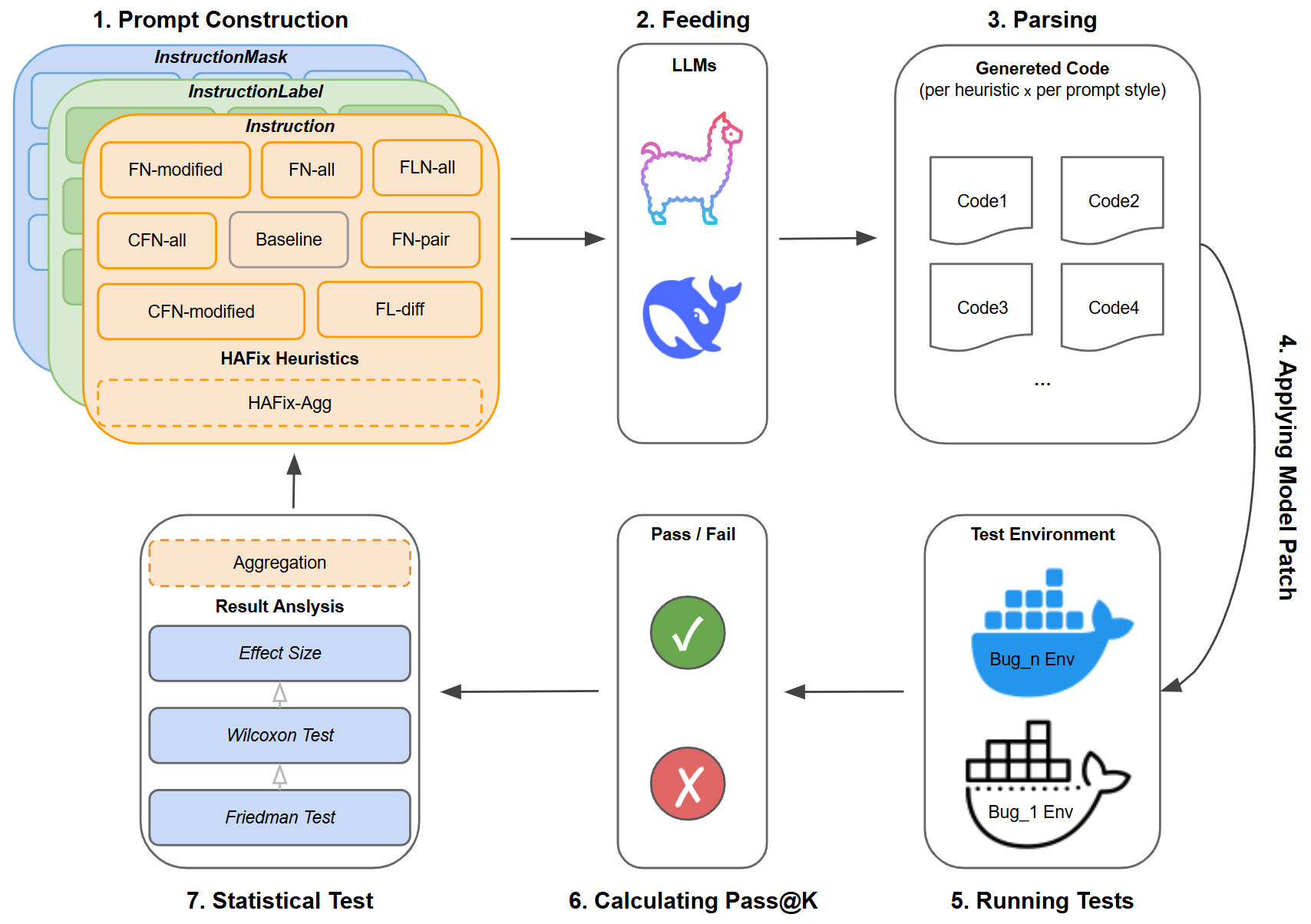}
        \end{center}
        \caption{HAFix architecture and evaluation pipeline.}
        \label{fig:approach_pipeline}
\end{figure}

With the selected model and prepared prompt styles, we can now feed the prompts to the models and initiate the experimental pipeline. Figure \ref{fig:approach_pipeline} shows an overview of the HAFix architecture and evaluation pipeline, which we describe step by step.

\begin{enumerate}
    \item Constructing and Categorizing Prompts. The first step in our experimental pipeline is to construct prompts in three distinct styles. As illustrated in Figure \ref{fig:approach_pipeline}, the \Instruction prompt group includes a baseline (see Subsection \ref{Baseline Data Collection}) and seven history-based HAFix heuristics, each variant enriched with varying historical information mined from bug blame commit data (see Subsection \ref{Historical Data Collection (HAFix Heuristics)}). In both the baseline and these historical heuristics variants, the buggy code is presented in the \Instruction style. Additionally, for \InstructionLabel and \InstructionInfill styles, the buggy line and code are presented in the baseline and heuristics by following the corresponding structure as described in Subsection \ref{Prompt styles}.

    \item Feeding Prompts to LLMs. The constructed prompts are then input into our subject LLMs to generate potential fixes for each bug individually. Each model processes the prompt, leveraging the historical and contextual information provided within the prompt, and outputs ten potential bug fixes. This output typically consists of function-level code, which is essential for the next steps in the pipeline.
    
    \item Parsing the Model-Generated Function-Level Code. The outputs generated by our subject LLMs are parsed to extract the specific function-level code snippet. This is because, even though we explicitly instruct the models to generate only the fixed function-level code snippet, it often generates additional or unnecessary text, such as code explanations or unrelated code. To address this, we conducted multiple inferences with our prompt, observed the general output patterns, and developed regular expressions in our implementation to reliably isolate the desired function-level code snippet.
    
    \item Applying the Model-Generated Code to the Original Fixed Code. We begin by using Git commands to track the fixed snapshot of the project (commit V4 in Figure \ref{fig:data_collection}). Next, we locate the fixed file using the file path and name and identify the fixed function code by its start and end line numbers within the buggy function. To ensure a rollback option after test evaluation, we create a temporary backup (File\_Backup) in the same directory. We then replace the original fixed code with the model-generated function-level code, so that we can later verify if this fix passes the test cases in the corresponding commit.
    
    \item Running Test Cases in a Docker Environment. After reintegrating the code into the project, the next step is to validate the effectiveness of the generated fix. We install all dependencies for each bug in each project within a Docker environment, which ensures isolation and reproducibility during testing. The bug’s test cases are then executed, providing a consistent platform for evaluating the correctness of the generated code. After this test evaluation, we restore the original state by deleting the current file and renaming the backup file ((File\_Backup)) to its original file name. To determine whether a bug is successfully fixed, we consider it resolved if at least one of the n samples (where n=10) generated by nucleus sampling passes the test cases, demonstrating functional correctness.
    
    \item Calculating Pass@k as Evaluation Metric. In line with previous studies \citep{chen2021evaluating,du2023classeval,yu2024codereval,parasaram2024fact}, we evaluate the functional correctness of programs by executing test cases to calculate the pass@k as shown in Formula \ref{eq:pass_at_k}. This step measures the success rate of the model-generated code over k attempts. Specifically, we generate $n$ code samples by issuing $n$ separate queries to the LLM ($n \geq k$), then count the number of correct programs $c$ that pass the test cases ($c \leq n$), and calculate the Pass@k.
    \begin{equation}
    \text{Pass@}k := \underset{\text{bugs}}{\mathbb{E}} \left[ 1 - \frac{\binom{n-c}{k}}{\binom{n}{k}} \right]
    \label{eq:pass_at_k}
    \end{equation}

    Where $\mathbb{E}$ denotes the mean Pass@k computed over all bugs in the dataset. When $n-c < k$ (i.e., when the number of failed attempts is less than $k$), Pass@k equals 1, as success is guaranteed within $k$ attempts. For example, if 3 out of 10 generated samples pass the test cases, \( c = 3 \) and \( n = 10 \), allowing us to compute pass@k for \( k = 1, 3, 5, 10 \). We compute pass@k for each bug and aggregate the results across all bugs to derive the overall performance of HAFix and baseline approaches.

    Pass@k is chosen because it reflects the likelihood of the bug being fixed within k attempts, aligning with realistic bug-fixing scenarios where multiple solutions can be attempted. Compared to metrics used by previous works such as the number of bugs fixed or exact match \citep{jiang2023impact,lu2021codexglue,wang2021codet5,xia2024automated}, pass@k considers the distribution of correct fixes across multiple attempts rather than evaluating bug fixes in a binary manner. This provides a more nuanced measure of model effectiveness.
    
    \item Conducting Statistical Test. To assess statistical significance, we apply the Friedman test, followed by Wilcoxon signed-rank tests for post-hoc analysis, comparing pass@k distributions across different heuristics and prompt styles. The Friedman test, a non-parametric paired test, is used to detect differences across more than two related distributions, while the Wilcoxon signed-rank test is well-suited for pairwise comparisons in non-parametric data. To complement statistical significance, we calculate the effect size using the Rank-Biserial Correlation ($r_{\mathrm{rb}}$), which quantifies the magnitude of pairwise differences. The Rank-Biserial Correlation ranges from -1 to 1, where values closer to -1 or 1 indicate stronger effects. Adopting common benchmarks \citep{cohen1992power}, we interpret effect sizes based on the absolute value $|r_{\mathrm{rb}}|$: $|r_{\mathrm{rb}}| < 0.1$ as negligible, $0.1 \le |r_{\mathrm{rb}}| < 0.3$ as small, $0.3 \le |r_{\mathrm{rb}}| < 0.5$ as medium, and $|r_{\mathrm{rb}}| \ge 0.5$ as large. When reporting statistical comparisons, we present both $p$-values and $r_{\mathrm{rb}}$ to distinguish statistically significant differences from negligible effects. These tests are consistently applied across RQ1, RQ2, and RQ3 to ensure uniform and robust statistical evaluation.

\end{enumerate}

\subsection{Inference Infrastructure and Hyper-parameters}\label{Inference Infrastructure and Hyper-parameters}
For the infrastructure supporting our experiments, we selected an Nvidia A100 GPU with 80 GB of memory. It provides ample space to load our subject model (ranging from approximately 13GB to 31GB, as described in Subsection \ref{Model selection}), along with the additional memory needed for processing large batches of data. We employ the Ollama framework \citep{Ollama} to serve all subject models. Following prior works \citep{roziere2023code,li2024deveval}, we adopt nucleus sampling as the decoding strategy with a temperature of 0.4 and a top-p value of 0.95. While Ollama supports customization of sampling parameters such as top-$p$, it does not support generating multiple outputs in a single query. To obtain multiple samples (e.g., for pass@$k$ computation), we perform repeated queries using identical decoding settings. As hyper-parameter tuning was not our primary focus, we adopted these values from prior studies and left further exploration for future work. While the model generally produced stable outputs, it occasionally included unrelated text \citep{macedo2024exploring}, such as code explanations or irrelevant snippets. To address this, we applied the extraction rules described in Subsection \ref{Experimental pipeline}. These rules, developed by analyzing common output patterns, ensured the reliable isolation of function-level code snippets and maintained the stability of the evaluated fixed code samples.

\section{Empirical Results}\label{Research Questions}

\subsection{RQ1: How Much Do History-Augmented LLMs Improve Bug Fixing Compared to Models Without Historical Context?}\label{RQ1}
\subsubsection{Motivation}\label{RQ1Motivation}
Large Language Models (LLMs) have demonstrated remarkable capabilities in SE tasks such as code generation \citep{lu2021codexglue,chen2021evaluating,du2023classeval,yu2024codereval,li2024deveval,zhang2024codeagent} and bug-fixing \citep{jiang2023impact,zhang2024diversity,fan2023automated,xia2023revisiting,li2024hybrid}. However, it remains uncertain whether incorporating historical context from software repositories, such as the blame commit of a bug, can further enhance their effectiveness in bug-fixing. The blame commit identifies the last modification to the buggy code, offering critical context for understanding its root cause. This context has been used in the MSR community for decades as a heuristic to identify the bug-introducing commit (SZZ) \citep{sliwerski2005changes,hassan2006mining}. As discussed in Section \ref{HAFix: History-Augmented LLMs for Bug Fixing}, examining the changes in the blame commit has long been used to understand the evolution of software bugs. This research question explores whether leveraging various history heuristics derived from the blame commit can improve LLMs' bug-fixing performance.

\subsubsection{Approach}\label{RQ1Approach}

To evaluate the impact of history heuristics, we use a baseline prompt inspired by GitHub Copilot’s prompt data \citep{copilotprompt}, given that the latter is one of the most widely adopted coding assistant tools. The detailed prompt design for baseline is presented in Subsection \ref{Baseline Data Collection}. As discussed in Subsection \ref{Historical Heuristics Prompt}, we empirically evaluate seven historical information into several heuristics: co-evolved functions' names in the modified buggy files (CFN-modified), co-evolved functions' names in all modified files (CFN-all), all functions' names in the modified buggy file (FN-modified), all functions' names in all modified files (FN-all), co-evolved files' names (FLN-all), function code pairs (FN-pair), and file diff patches (FL-diff). When testing different heuristics, we always provide the baseline information first, then append the heuristic data to ensure a fair comparison of their impact on top of the baseline.

Additionally, we propose an aggregated approach named HAFix-Agg, as described in Subsection \ref{HAFix-Agg}, which combines the results of all heuristics to assess its potential for improvement over the performance of individual heuristics. Note that in this research question, we used the \Instruction prompt style for all experiments to maintain consistency.

\subsubsection{Results}\label{RQ1Result}

\begin{figure}[!htbp]
  \centering
  \begin{subfigure}[b]{0.48\textwidth}
    \includegraphics[width=\textwidth]{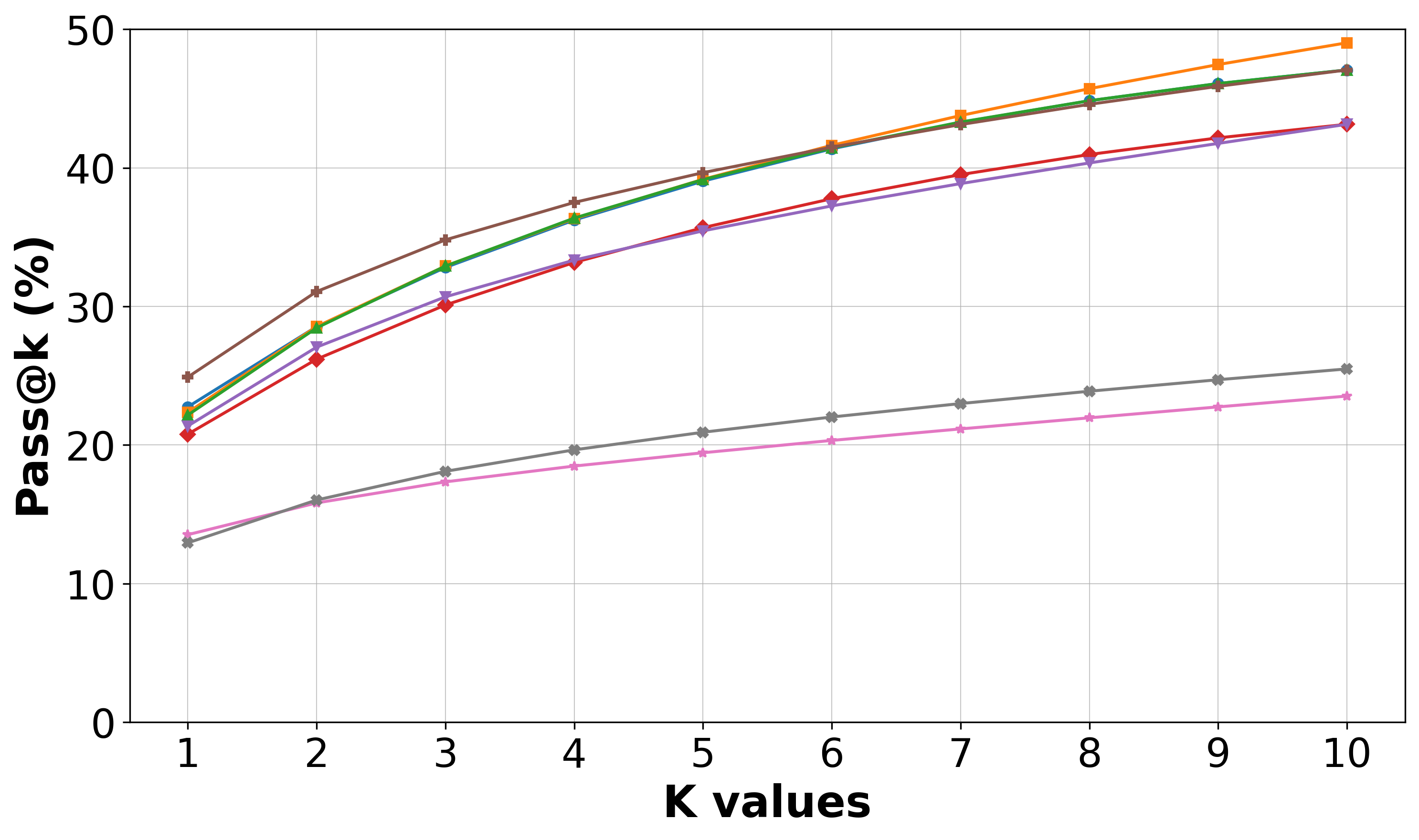}
    \caption{CodeLlama-Instruct-7B on BugsInPy}
    \label{rq1_passk_baseline_heuristics_2_datasets_3_models:a}
  \end{subfigure}
  \hfill
  \begin{subfigure}[b]{0.48\textwidth}
    \includegraphics[width=\textwidth]{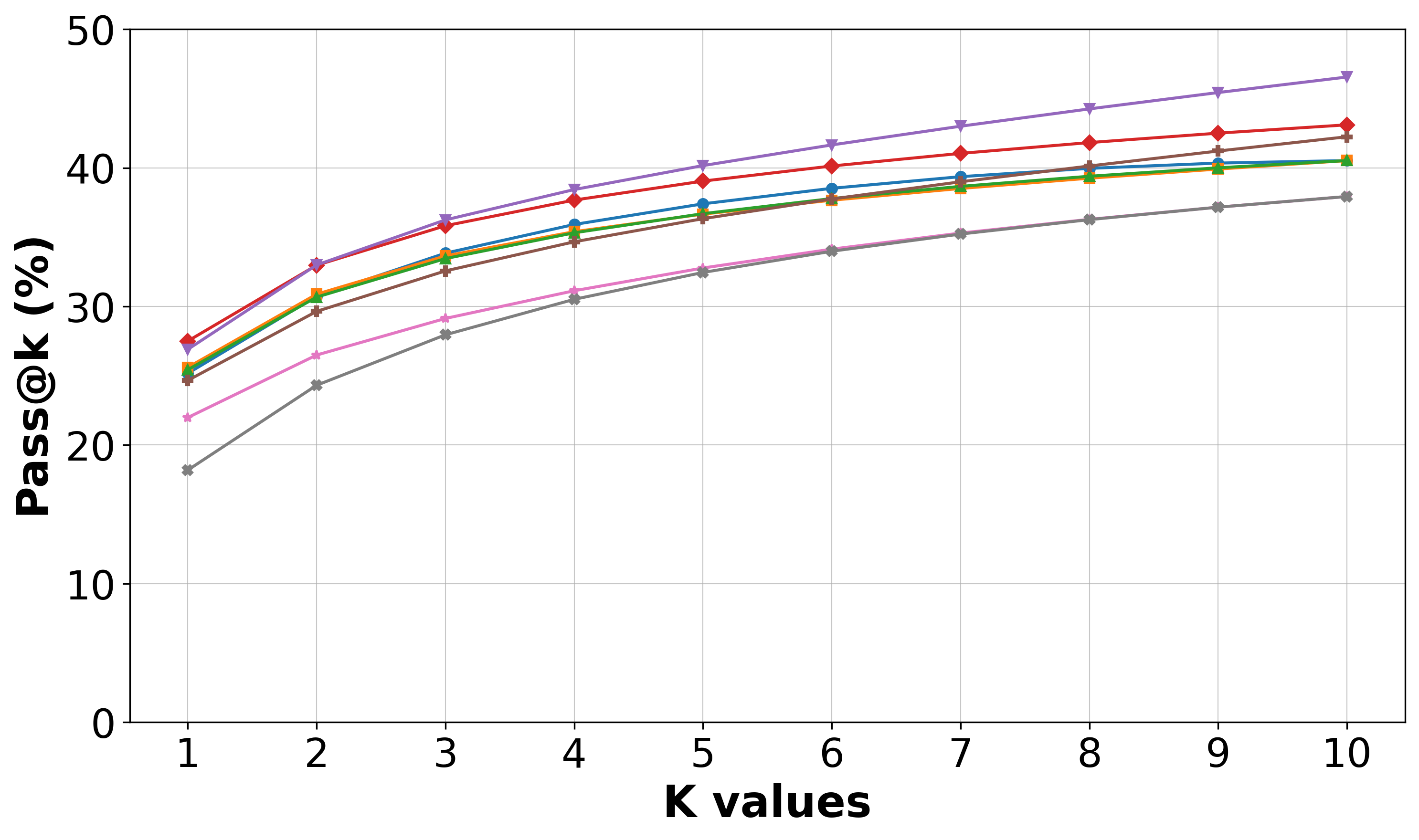}
    \caption{CodeLlama-Instruct-7B on Defects4J}
    \label{rq1_passk_baseline_heuristics_2_datasets_3_models:b}
  \end{subfigure}

  \vspace{1em} 
    
  \begin{subfigure}[b]{0.48\textwidth}
    \includegraphics[width=\textwidth]{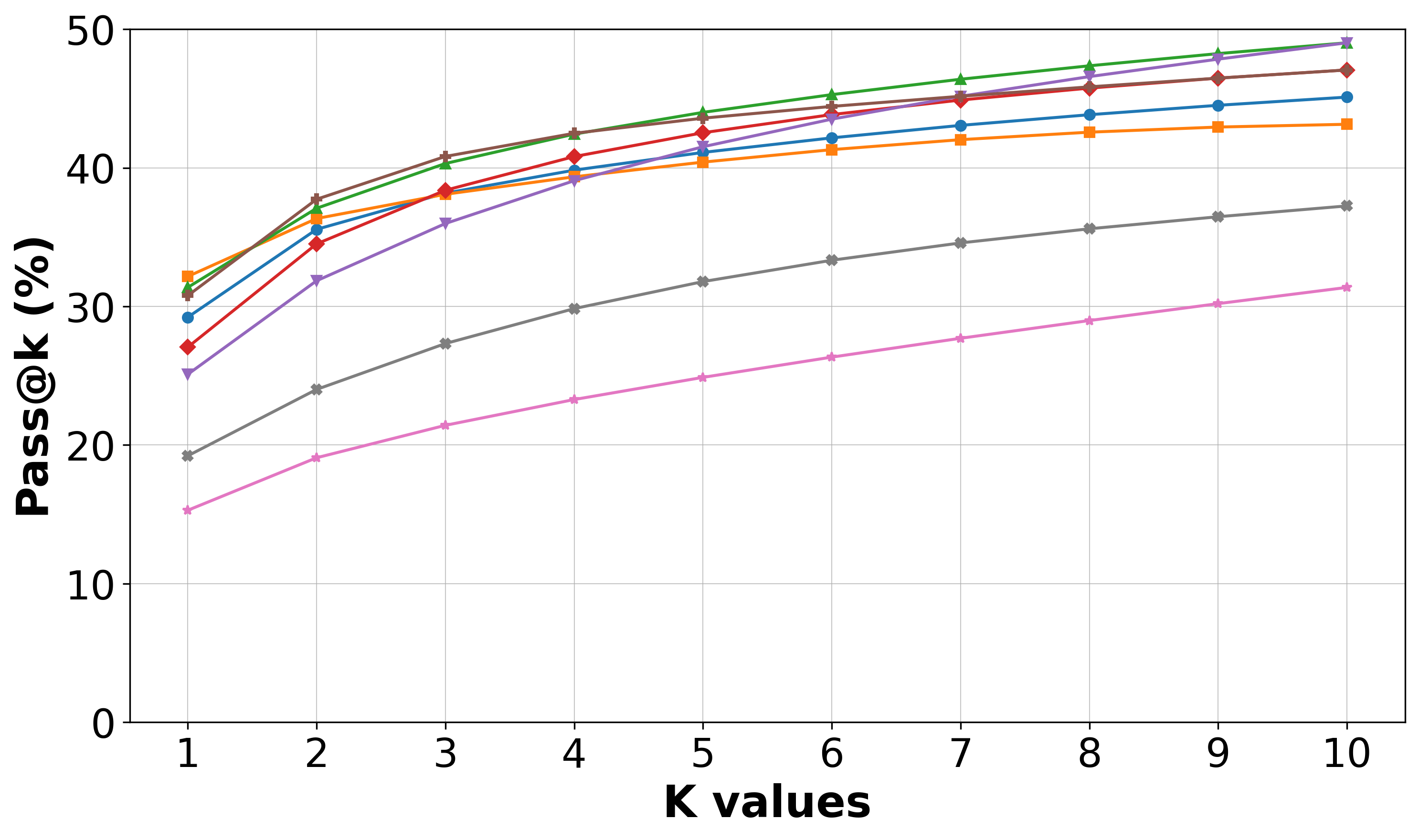}
    \caption{DeepSeek-Coder-Instruct-6.7B on \\BugsInPy}
    \label{rq1_passk_baseline_heuristics_2_datasets_3_models:c}
  \end{subfigure}
  \hfill
  \begin{subfigure}[b]{0.48\textwidth}
    \includegraphics[width=\textwidth]{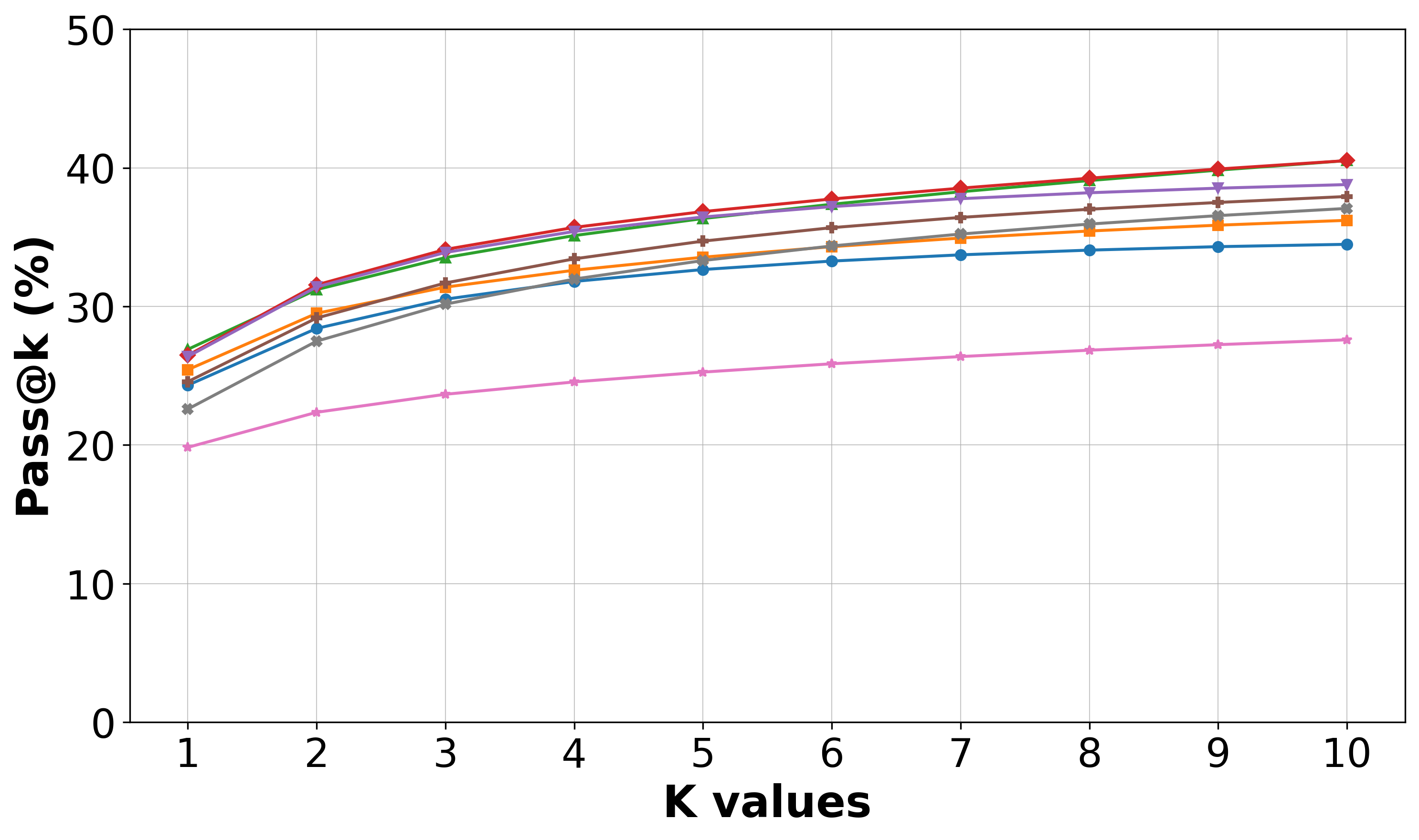}
    \caption{DeepSeek-Coder-Instruct-6.7B on \\Defects4J}
    \label{rq1_passk_baseline_heuristics_2_datasets_3_models:d}
  \end{subfigure}

  \vspace{1em} 
    
  \begin{subfigure}[b]{0.48\textwidth}
    \includegraphics[width=\textwidth]{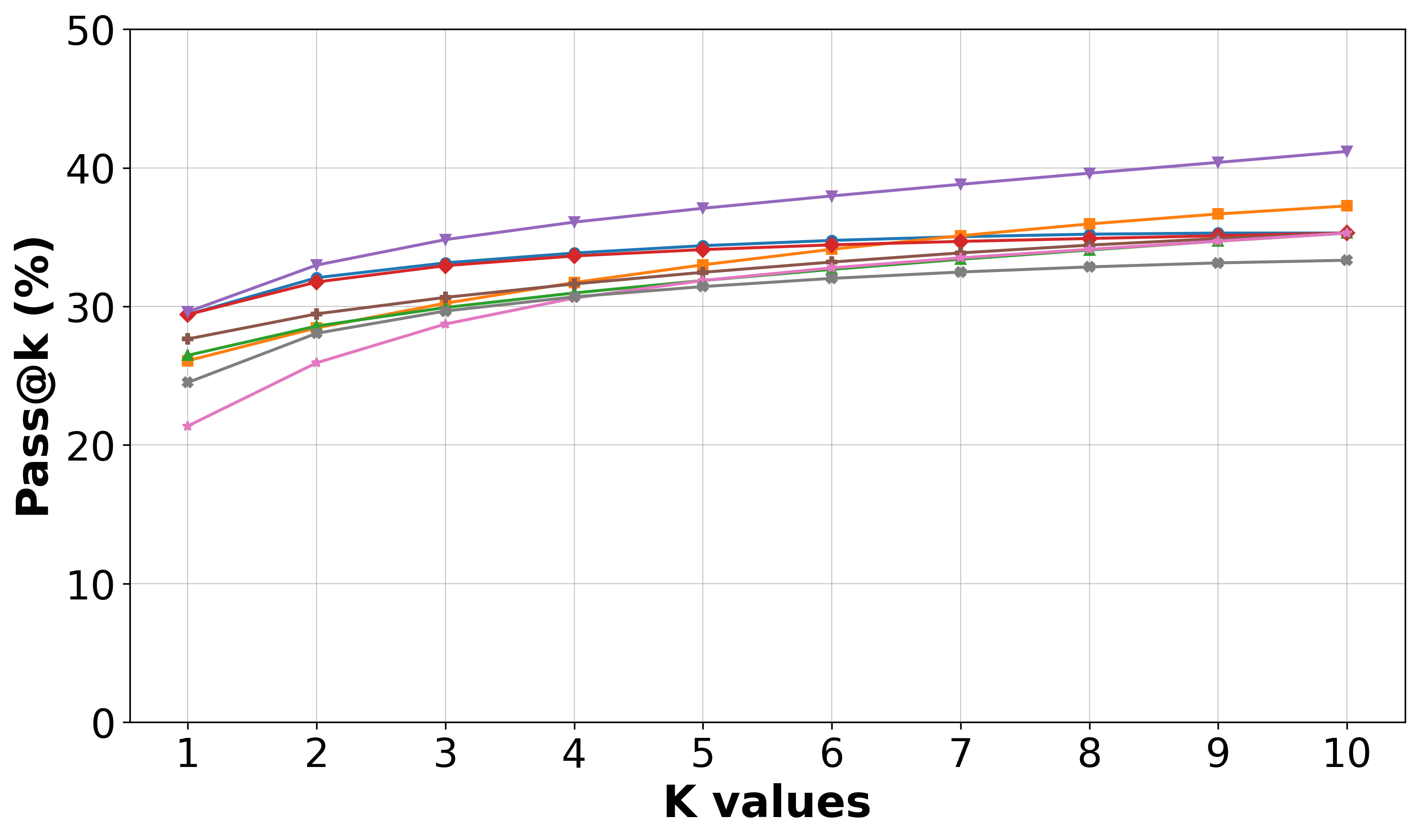}
    \caption{DeepSeek-Coder-V2-Lite-Instruct-16B \\on BugsInPy}
    \label{rq1_passk_baseline_heuristics_2_datasets_3_models:e}
  \end{subfigure}
  \hfill
  \begin{subfigure}[b]{0.48\textwidth}
    \includegraphics[width=\textwidth]{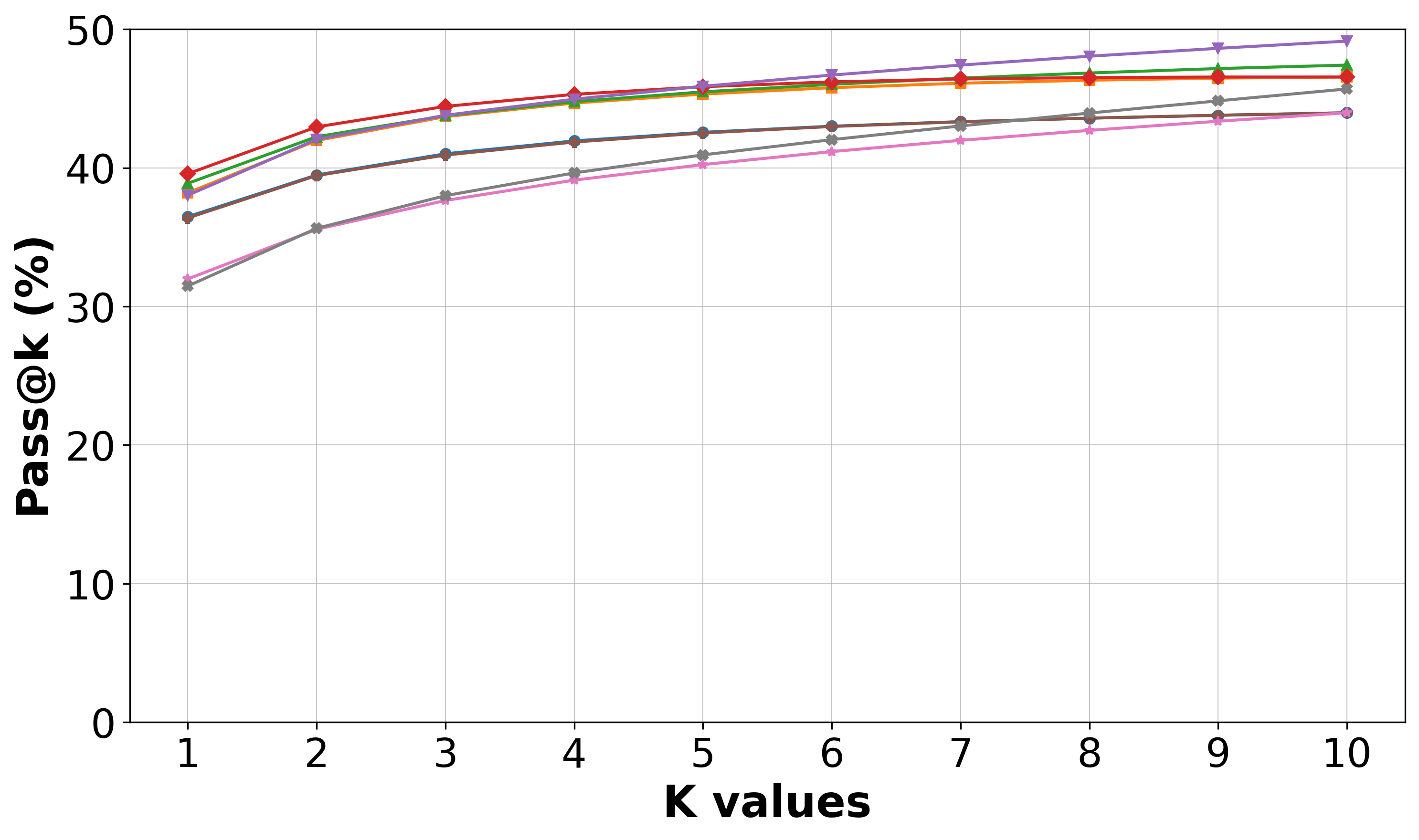}
    \caption{DeepSeek-Coder-V2-Lite-Instruct-16B \\on Defects4J}
    \label{rq1_passk_baseline_heuristics_2_datasets_3_models:f}
  \end{subfigure}

  \vspace{1em}
  \begin{adjustbox}{width=0.3\textwidth}
    \includegraphics{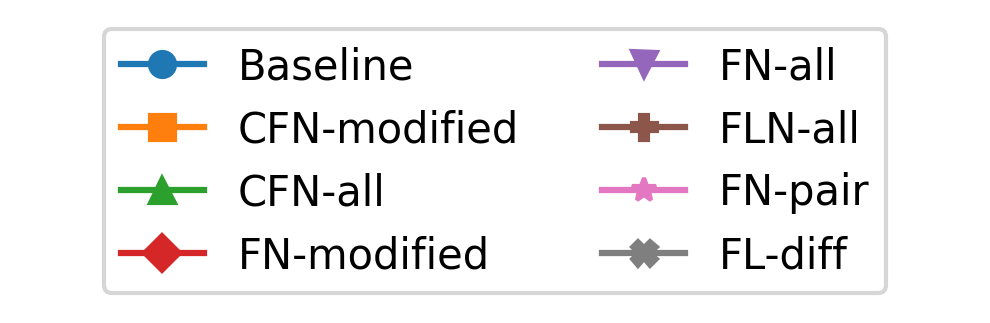}
  \end{adjustbox}

  \caption{Pass@k (\%) comparison of baseline and seven HAFix heuristics for bug-fixing performance across two datasets and three models.}
  \label{fig:rq1_passk_baseline_heuristics_2_datasets_3_models}
\end{figure}

\textbf{Overall, several HAFix heuristics show consistent and significant improvements over the baseline, particularly on Defects4J and with DeepSeek-Coder models.} Figure \ref{fig:rq1_passk_baseline_heuristics_2_datasets_3_models} illustrates the trends in Pass@k rates for the baseline and the seven heuristics across a range of k values from 1 to 10, evaluated on two datasets and three subject models, providing a comprehensive view of their comparative performance on bug fixing. Notably, certain heuristics, such as FN-modified and FN-all in CodeLlama-Instruct-7B on Defects4J (Figure \subfig{fig:rq1_passk_baseline_heuristics_2_datasets_3_models}{rq1_passk_baseline_heuristics_2_datasets_3_models:b}), show consistent improvements over the baseline as k increases. Similar consistent improvements can be observed in other configurations as illustrated in Figures \subfig{fig:rq1_passk_baseline_heuristics_2_datasets_3_models}{rq1_passk_baseline_heuristics_2_datasets_3_models:c} through Figure \subfig{fig:rq1_passk_baseline_heuristics_2_datasets_3_models}{rq1_passk_baseline_heuristics_2_datasets_3_models:f}. Table \ref{RQ1_passk} summarizes Pass@k results for k=1, 5, and 10 (n=10) across all datasets and models. We selected these k values because they are commonly reported and provide a balanced perspective on performance at lower, mid, and higher thresholds \citep{du2023classeval,yu2024codereval}. For example, on CodeLlama-Instruct-7B with Defects4J, compared to the baseline's rates of 25.17\% at Pass@1, 37.40\% at Pass@5 and 40.52\% at Pass@10, FN-modified shows an improvement of approximately 9\% at Pass@1, 4\% at Pass@5 and 6\% at Pass@10, while FN-all improves by about 7\% at Pass@1, 7\% at Pass@5 and 15\% at Pass@10 over the baseline.

\FloatBarrier
\begin{table}[!htbp]
  \centering
  \small
  \caption{Pass@k (k=1, 5 and 10) for baseline and various HAFix heuristics across three models and two datasets. The heuristic achieving the highest Pass@k value for a given dataset in each column is highlighted in bold.}
  \label{RQ1_passk}
  \begin{tabular}{l l l r r r}
    \toprule
    \textbf{Model} & \textbf{Dataset} & \textbf{Heuristic} & \textbf{Pass@1} & \textbf{Pass@5} & \textbf{Pass@10} \\
    \midrule
    \multirow{16}{*}{\shortstack{CodeLlama\\-Instruct-7B}} 
      & \multirow{8}{*}{BugsInPy}
         & Baseline      & 22.75\% & 39.04\% & 47.06\% \\
      &  & CFN-modified  & 22.35\% & 39.17\% & \textbf{49.02\%} \\
      &  & CFN-all       & 22.16\% & 39.15\% & 47.06\% \\
      &  & FN-modified   & 20.78\% & 35.68\% & 43.14\% \\
      &  & FN-all        & 21.37\% & 35.45\% & 43.14\% \\
      &  & FLN-all       & \textbf{24.90\%} & \textbf{39.65\%} & 47.06\% \\
      &  & FN-pair       & 13.53\% & 19.44\% & 23.53\% \\
      &  & FL-diff       & 12.94\% & 20.92\% & 25.49\% \\ 
      \cmidrule{2-6}
      & \multirow{8}{*}{Defects4J}
         & Baseline      & 25.17\% & 37.40\% & 40.52\% \\
      &  & CFN-modified  & 25.60\% & 36.66\% & 40.52\% \\
      &  & CFN-all       & 25.43\% & 36.69\% & 40.52\% \\
      &  & FN-modified   & \textbf{27.50\%} & 39.04\% & 43.10\% \\
      &  & FN-all        & 26.90\% & \textbf{40.16\%} & \textbf{46.55\%} \\
      &  & FLN-all       & 24.66\% & 36.33\% & 42.24\% \\
      &  & FN-pair       & 21.98\% & 32.77\% & 37.93\% \\
      &  & FL-diff       & 18.19\% & 32.45\% & 37.93\% \\
    \midrule
    \multirow{16}{*}{\shortstack{DeepSeek-Coder\\-Instruct-6.7B}}
      & \multirow{8}{*}{BugsInPy}
         & Baseline      & 29.22\% & 41.11\% & 45.10\% \\
      &  & CFN-modified  & \textbf{32.16\%} & 40.41\% & 43.14\% \\
      &  & CFN-all       & 31.37\% & \textbf{44.00\%} & \textbf{49.02\%} \\
      &  & FN-modified   & 27.06\% & 42.53\% & 47.06\% \\
      &  & FN-all        & 25.10\% & 41.52\% & \textbf{49.02\%} \\
      &  & FLN-all       & 30.78\% & 43.58\% & 47.06\% \\
      &  & FN-pair       & 15.29\% & 24.88\% & 31.37\% \\
      &  & FL-diff       & 19.22\% & 31.79\% & 37.25\% \\
      \cmidrule{2-6}
      & \multirow{8}{*}{Defects4J}
         & Baseline       & 24.31\% & 32.66\% & 34.48\% \\
      &  & CFN-modified   & 25.43\% & 33.55\% & 36.21\% \\
      &  & CFN-all        & \textbf{26.90\%} & 36.34\% & 40.52\% \\
      &  & FN-modified    & 26.47\% & \textbf{36.84\%} & 40.52\% \\
      &  & FN-all         & 26.38\% & 36.45\% & \textbf{38.79\%} \\
      &  & FLN-all        & 24.57\% & 34.71\% & 37.93\% \\
      &  & FN-pair        & 19.83\% & 25.26\% & 27.59\% \\
      &  & FL-diff        & 22.59\% & 33.31\% & 37.07\% \\
    \midrule
    \multirow{16}{*}{\shortstack{DeepSeek-Coder-V2\\-Lite-Instruct-16B}}
      & \multirow{8}{*}{BugsInPy}
         & Baseline      & 29.41\% & 34.38\% & 35.29\% \\
      &  & CFN-modified  & 26.08\% & 33.00\% & 37.25\% \\
      &  & CFN-all       & 26.47\% & 31.87\% & 35.29\% \\
      &  & FN-modified   & 29.41\% & 34.10\% & 35.29\% \\
      &  & FN-all        & \textbf{29.61\%} & \textbf{37.08\%} & \textbf{41.18\%} \\
      &  & FLN-all       & 27.65\% & 32.46\% & 35.29\% \\
      &  & FN-pair       & 21.37\% & 31.87\% & 35.29\% \\
      &  & FL-diff       & 24.51\% & 31.43\% & 33.33\% \\
      \cmidrule{2-6}
      & \multirow{8}{*}{Defects4J}
         & Baseline       & 36.47\% & 42.56\% & 43.97\% \\
      &  & CFN-modified   & 38.19\% & 45.32\% & 46.55\% \\
      &  & CFN-all        & 38.88\% & 45.48\% & 47.41\% \\
      &  & FN-modified    & \textbf{39.57\%} & 45.85\% & 46.55\% \\
      &  & FN-all         & 38.02\% & \textbf{45.88\%} & \textbf{49.14\%} \\
      &  & FLN-all        & 36.38\% & 42.51\% & 43.97\% \\
      &  & FN-pair        & 31.98\% & 40.23\% & 43.97\% \\
      &  & FL-diff        & 31.47\% & 40.93\% & 45.69\% \\
    \bottomrule
    \end{tabular}
\end{table}

\begin{table}[!htbp]
  \centering
  \small
  \caption{Statistical comparison between each HAFix heuristic and its corresponding baseline using Pass@k. $p_{\mathrm{F}}$ denotes the $p$-value from the Friedman test, $p$ denotes the $p$-value from the pairwise Wilcoxon signed-rank tests, and $r_{\mathrm{rb}}$ denotes the effect size calculated using the Rank-Biserial Correlation. A Bonferroni-corrected significance threshold of $\alpha = 0.0071$ ($0.05/7$) is applied for pairwise comparisons. Heuristics with significantly better performance compared to the corresponding baseline are highlighted in bold and marked with \textcolor{green}{$\uparrow$}, while those with significantly worse performance are marked with \textcolor{red}{$\downarrow$}. Non‑significant differences are left unmarked. $\dagger$ flags large effects ($|r_{\mathrm{rb}}|\ge 0.5$) that are not significant after Bonferroni correction ($p\ge 0.0071$).}
  \label{tab:wilcox_rq1_passk_baseline_heuristics_2_datasets_3_models}
  \resizebox{\textwidth}{!}{
  \begin{tabular}{lllrrr}
    \toprule
    \textbf{Model} & \textbf{Dataset} & \textbf{Heuristic} & \textbf{$p_{\text{F}}$} & \textbf{$p$} & \textbf{$r_{\mathrm{rb}}$} \\
    \midrule
    \multirow{16}{*}{\shortstack{CodeLlama\\-Instruct-7B}} 
      & \multirow{7}{*}{BugsInPy}
                                  & CFN-modified$\,\dagger$            & \multirow{7}{*}{$3.631 \times 10^{-11}$} & 0.0440 & 0.82 \\
      &                           & CFN-all$\,\dagger$                 & & 0.5541 & 0.64 \\
      &                           & FN-modified\,\textcolor{red}{$\downarrow$}             & & 0.0059 & -1.00 \\
      &                           & FN-all\,\textcolor{red}{$\downarrow$}                  & & 0.0020 & -1.00 \\
      &                           & FLN-all$\,\dagger$                 & & 0.1235 & 0.67 \\
      &                           & FN-pair\,\textcolor{red}{$\downarrow$}                 & & 0.0020 & -1.00 \\
      &                           & FL-diff\,\textcolor{red}{$\downarrow$}                 & & 0.0020 & -1.00 \\
    \cmidrule{2-6}
      & \multirow{7}{*}{Defects4J}
                                  & CFN-modified   & \multirow{7}{*}{$3.879 \times 10^{-11}$} & 0.0330 & -0.49 \\
      &                           & CFN-all$\,\dagger$        & & 0.0178 & -0.56 \\
      &                           & \textbf{FN-modified}\,\textcolor{green}{$\uparrow$}   & & \textbf{0.0020} & 1.00 \\
      &                           & \textbf{FN-all}\,\textcolor{green}{$\uparrow$}        & & \textbf{0.0020} & 1.00 \\
      &                           & FLN-all       & & 0.2754 & -0.42 \\
      &                           & FN-pair\,\textcolor{red}{$\downarrow$}       & & 0.0020 & -1.00 \\
      &                           & FL-diff\,\textcolor{red}{$\downarrow$}       & & 0.0020 & -1.00 \\
    \midrule
    \multirow{14}{*}{\shortstack{DeepSeek-Coder\\-Instruct-6.7B}}
      & \multirow{7}{*}{BugsInPy}
                                  & CFN-modified  & \multirow{7}{*}{$6.08 \times 10^{-10}$} & 0.1934 & -0.49 \\
      &                           & \textbf{CFN-all}\,\textcolor{green}{$\uparrow$}       & & \textbf{0.0020} & 1.00 \\
      &                           & FN-modified$\,\dagger$   & & 0.1533 & 0.53 \\
      &                           & FN-all        & & 0.8457 & 0.09 \\
      &                           & \textbf{FLN-all}\,\textcolor{green}{$\uparrow$}       & & \textbf{0.0059} & 1.00 \\
      &                           & FN-pair\,\textcolor{red}{$\downarrow$}       & & 0.0020 & -1.00 \\
      &                           & FL-diff\,\textcolor{red}{$\downarrow$}       & & 0.0020 & -1.00 \\
    \cmidrule{2-6}
      & \multirow{7}{*}{Defects4J}
                                  & \textbf{CFN-modified}\,\textcolor{green}{$\uparrow$}  & \multirow{7}{*}{$1.061 \times 10^{-11}$} & \textbf{0.0020} & 1.00 \\
      &                           & \textbf{CFN-all}\,\textcolor{green}{$\uparrow$}       & & \textbf{0.0020} & 1.00 \\
      &                           & \textbf{FN-modified}\,\textcolor{green}{$\uparrow$}   & & \textbf{0.0020} & 1.00 \\
      &                           & \textbf{FN-all}\,\textcolor{green}{$\uparrow$}        & & \textbf{0.0020} & 1.00 \\
      &                           & \textbf{FLN-all}\,\textcolor{green}{$\uparrow$}       & & \textbf{0.0020} & 1.00 \\
      &                           & FN-pair\,\textcolor{red}{$\downarrow$}       & & 0.0020 & -1.00 \\
      &                           & FL-diff$\,\dagger$       & & 0.1602 & 0.53 \\
    \midrule
    \multirow{14}{*}{\shortstack{DeepSeek-Coder-V2\\-Lite-Instruct-16B}}
      & \multirow{7}{*}{BugsInPy}
                                  & CFN-modified  & \multirow{7}{*}{$1.632 \times 10^{-10}$} & 0.2023 & -0.47 \\
      &                           & CFN-all$\,\dagger$       & & 0.0092 & -0.64 \\
      &                           & FN-modified   & & 0.0143 & -0.31 \\
      &                           & \textbf{FN-all}\,\textcolor{green}{$\uparrow$}        & & \textbf{0.0020} & 1.00 \\
      &                           & FLN-all$\,\dagger$       & & 0.0092 & -0.64 \\
      &                           & FN-pair$\,\dagger$       & & 0.0092 & -0.64 \\
      &                           & FL-diff\,\textcolor{red}{$\downarrow$}       & & 0.0020 & -1.00 \\
    \cmidrule{2-6}
      & \multirow{7}{*}{Defects4J}
                                  & \textbf{CFN-modified}\,\textcolor{green}{$\uparrow$}  & \multirow{7}{*}{$3.879 \times 10^{-11}$} & \textbf{0.0059} & 1.00 \\
      &                           & \textbf{CFN-all}\,\textcolor{green}{$\uparrow$}       & & \textbf{0.0020} & 1.00 \\
      &                           & \textbf{FN-modified}\,\textcolor{green}{$\uparrow$}   & & \textbf{0.0020} & 1.00 \\
      &                           & \textbf{FN-all}\,\textcolor{green}{$\uparrow$}        & & \textbf{0.0020} & 1.00 \\
      &                           & FLN-all       & & 0.0223 & -0.02 \\
      &                           & FN-pair$\,\dagger$       & & 0.0092 & -0.64 \\
      &                           & FL-diff$\,\dagger$       & & 0.1309 & -0.56 \\
    \bottomrule
  \end{tabular}
  }
\end{table}

To assess which heuristics perform significantly better than the corresponding baseline, we first conducted Friedman tests across the eight data groups of Pass@k (corresponding to each heuristic and the baseline) for each model-dataset configuration. Each group spans k values from 1 to 10. We then performed pairwise Wilcoxon signed-rank tests as post-hoc analysis to compare the baseline with seven heuristics, applying a Bonferroni-corrected significance threshold of 0.0071 ($0.05/7$) for multiple comparisons. Lastly, to quantify the magnitude of these improvements, we calculated the effect size using the Rank-Biserial Correlation. The results are shown in Table \ref{tab:wilcox_rq1_passk_baseline_heuristics_2_datasets_3_models}. All Friedman test p-values ($p_{\text{F}}$) are below 0.001, indicating that significant differences exist among the groups. The Wilcoxon tests ($p$), together with the effect sizes ($r_{\mathrm{rb}}$), provide further insights. For CodeLlama on BugsInPy, no heuristic significantly outperforms the baseline. However, for CodeLlama on Defects4J, both FN-modified and FN-all achieved significant gains over baseline ($p = 0.0020$) with effect sizes of 1.00 (large), indicating perfect dominance. For DeepSeek-Coder-Instruct-6.7B, CFN-all significantly outperforms the baseline on both datasets ($p = 0.0020$), and FLN-all showed significant improvements with $p = 0.0059$ on BugsInPy and $p = 0.0020$ on Defects4J. On Defects4J specifically, CFN-modified, FN-modified, and FN-all also achieve significant improvements over the baseline ($p = 0.0020$), each with a large effect size of 1.00. Similarly, for DeepSeek-Coder-V2-Lite-Instruct-16B, FN-all significantly outperformed the baseline across both datasets ($p = 0.0020$, $r_{\mathrm{rb}} = 1.00$ (large)). On Defects4J, CFN-modified, CFN-all, and FN-modified also showed significant improvements over baseline (all $p = 0.0020$ and $r_{\mathrm{rb}} = 1.00$ (large)).

\begin{table}[!htbp]
  \centering
  \small
  \caption{Number and Percentage of bugs fixed by baseline and individual heuristics of HAFix. Bugs\# and Percent\# represent the number and percentage of bugs being fixed. BugsU\# represents the number of bugs uniquely solved compared to the corresponding baseline. The highest value for a given dataset in each column is highlighted in bold.}
  \resizebox{\textwidth}{!}{
  \begin{tabular}{l l l r r r}
    \toprule
    \textbf{Model} & \textbf{Dataset} & \textbf{Heuristic} & \textbf{Bugs\#} & \textbf{Percent\#} & \textbf{BugsU\#} \\
    \midrule
    \multirow{16}{*}{\shortstack{CodeLlama\\-Instruct-7B}}
      & \multirow{8}{*}{BugsInPy}
         & Baseline        & 24 & 47.06\% & - \\ 
      &  & CFN-modified    & \textbf{25} & \textbf{49.02}\% & \textbf{6} \\ 
      &  & CFN-all         & 24 & 47.06\% & 3 \\ 
      &  & FN-modified     & 22 & 43.14\% & 4 \\ 
      &  & FN-all          & 22 & 43.14\% & 5 \\ 
      &  & FLN-all         & 24 & 47.06\% & 3 \\ 
      &  & FN-pair         & 12 & 23.53\% & 3 \\
      &  & FL-diff         & 13 & 25.49\% & 2 \\
      \cmidrule{2-6}
      & \multirow{8}{*}{Defects4J}
         & Baseline        & 47 & 40.52\% & - \\
      &  & CFN-modified    & 47 & 40.52\% & 8 \\ 
      &  & CFN-all         & 47 & 40.52\% & 7 \\ 
      &  & FN-modified     & 50 & 43.10\% & 9 \\ 
      &  & FN-all          & \textbf{54} & \textbf{46.55\%} & \textbf{14} \\ 
      &  & FLN-all         & 49 & 42.24\% & 7 \\ 
      &  & FN-pair         & 44 & 37.93\% & 9 \\ 
      &  & FL-diff         & 44 & 37.93\% & 10 \\ 
    \midrule
    \multirow{16}{*}{\shortstack{DeepSeek-Coder\\-Instruct-6.7B}}
      & \multirow{8}{*}{BugsInPy}
         & Baseline        & 23 & 45.10\% & - \\
      &  & CFN-modified    & 22 & 43.14\% & 2 \\ 
      &  & CFN-all         & \textbf{25} & \textbf{49.02\%} & 2 \\ 
      &  & FN-modified     & 24 & 47.06\% & 2 \\ 
      &  & FN-all          & \textbf{25} & \textbf{49.02\%} & \textbf{6} \\ 
      &  & FLN-all         & 24 & 47.06\% & 3 \\ 
      &  & FN-pair         & 16 & 31.37\% & 3 \\ 
      &  & FL-diff         & 19 & 37.25\% & 4 \\ 
      \cmidrule{2-6}
      & \multirow{8}{*}{Defects4J}
         & Baseline        & 40 & 34.48\% & - \\
      &  & CFN-modified    & 42 & 36.21\% & 6 \\ 
      &  & CFN-all         & \textbf{47} & \textbf{40.52\%} & 10 \\ 
      &  & FN-modified     & \textbf{47} & \textbf{40.52\%} & 9 \\ 
      &  & FN-all          & 45 & 38.79\% & 11 \\ 
      &  & FLN-all         & 44 & 37.93\% & 8 \\ 
      &  & FN-pair         & 32 & 27.59\% & 6 \\ 
      &  & FL-diff         & 43 & 37.07\% & \textbf{12} \\ 
    \midrule
    \multirow{16}{*}{\shortstack{DeepSeek-Coder-V2\\-Lite-Instruct-16B}}
      & \multirow{8}{*}{BugsInPy}
         & Baseline        & 18 & 35.29\% & - \\
      &  & CFN-modified    & 19 & 37.25\% & 4 \\ 
      &  & CFN-all         & 18 & 35.29\% & 3 \\ 
      &  & FN-modified     & 18 & 35.29\% & 3 \\ 
      &  & FN-all          & \textbf{21} & \textbf{41.18\%} & \textbf{6} \\ 
      &  & FLN-all         & 18 & 35.29\% & 3 \\ 
      &  & FN-pair         & 18 & 35.29\% & 3 \\ 
      &  & FL-diff         & 17 & 33.33\% & 3 \\ 
      \cmidrule{2-6}
      & \multirow{8}{*}{Defects4J}
         & Baseline        & 51 & 43.97\% & - \\
      &  & CFN-modified    & 54 & 46.55\% & 4 \\ 
      &  & CFN-all         & 55 & 47.41\% & 8 \\ 
      &  & FN-modified     & 54 & 46.55\% & 7 \\ 
      &  & FN-all          & \textbf{57} & \textbf{49.14\%} & 10 \\ 
      &  & FLN-all         & 51 & 43.97\% & 5 \\ 
      &  & FN-pair         & 51 & 43.97\% & 11 \\ 
      &  & FL-diff         & 53 & 45.69\% & \textbf{13} \\ 
    \bottomrule
    \end{tabular}
    }
    \label{rq1_acc_percentage}
\end{table}


\textbf{Each HAFix heuristic contributes complementary improvements by uniquely fixing an average of about 3 bugs (+16.30\%) on BugsInPy and 9 bugs (+19.24\%) on Defects4J that the baseline fails to address.} While the statistical analysis confirms significant improvements for several heuristics, a closer examination reveals additional noteworthy trends. Table \ref{rq1_acc_percentage} summarizes the number and percentage of bugs fixed by the baseline and each HAFix heuristic. The columns Bugs\# and Percent\# indicate the number and percentage of bugs fixed, respectively, while BugsU\# represents the number of bugs uniquely fixed compared to the baseline. There are 51 subject bugs in BugsInPy and 116 subject bugs in Defects4J. A bug is considered fixed if at least one of the generated samples passes all test cases, as defined in Subsection \ref{Experimental pipeline}. We found that while individual heuristics generally yield comparable or slightly improved overall fix rates, several of them recover substantial unique bugs, with an average of 3 in BugsInPy (+16.30\%) and 9 in Defects4J (+19.24\%) compared to the corresponding baseline. For example, CFN-all and FN-modified each fix 47 bugs in DeepSeek-Coder-Instruct-6.7B on Defects4J, representing a 17.5\% improvement over the baseline, which fixes 40 bugs. In terms of unique fixes, FN-all is particularly effective: it fixes 14 additional bugs over the baseline in CodeLlama-Instruct-7B on Defects4J, and also contributes 11 and 10 uniquely fixed bugs in DeepSeek-Coder-Instruct-6.7B and DeepSeek-Coder-V2-Lite-Instruct-16B, respectively, on the same dataset. On BugsInPy, FN-all still demonstrates complementary strength, uniquely fixing 5, 6, and 6 bugs across the three models.

\begin{figure}[!htbp]
    \centering
    \begin{subfigure}[b]{0.4\textwidth} 
        \centering
        \includegraphics[width=\textwidth]{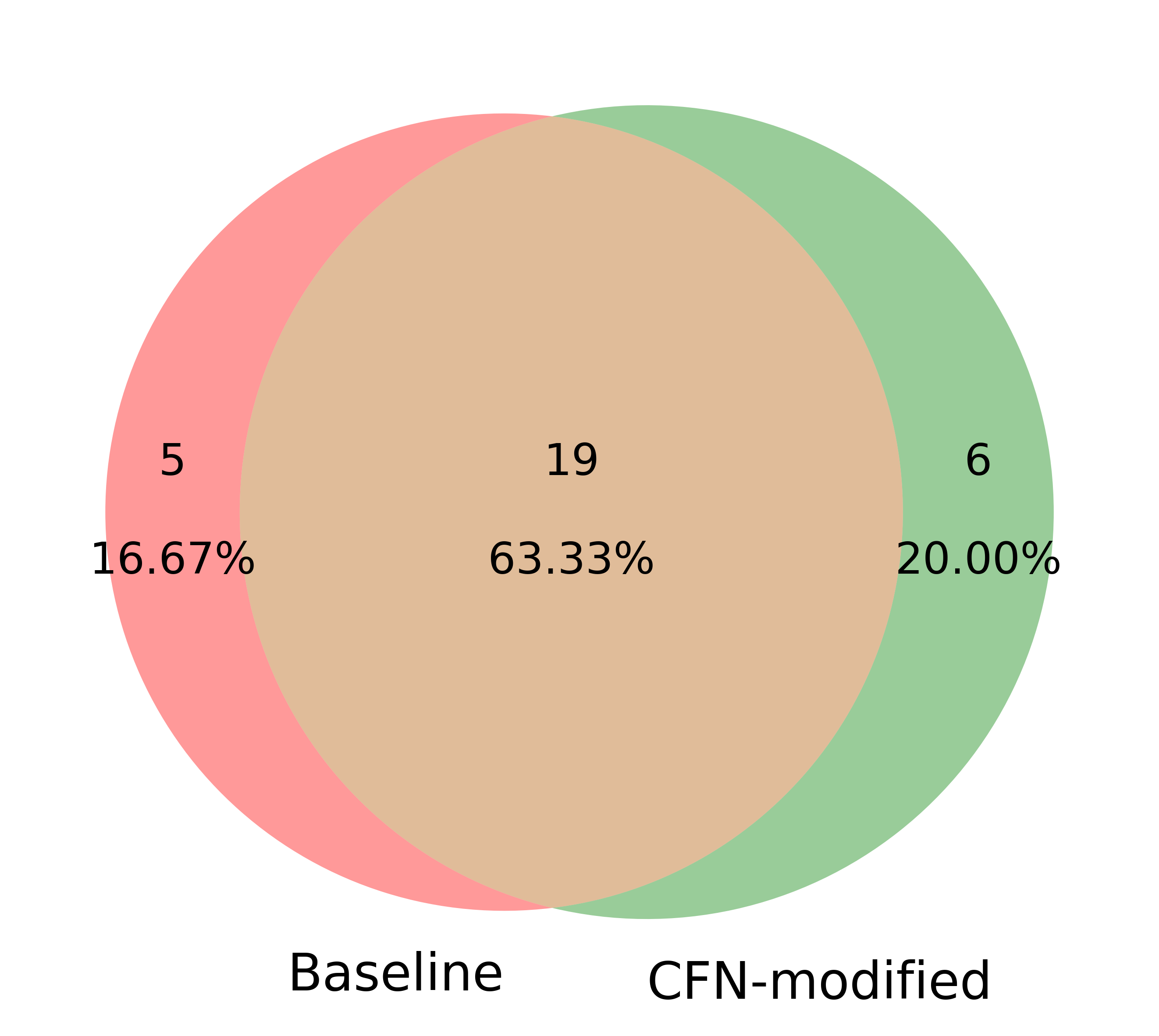} 
        \caption{CFN-modified fixes 6 more bugs compared to the baseline.}
        \label{venn-diagram-CodeLlama-Instruct-7B-BugsInPy:a}
    \end{subfigure}
    \hfill 
    \begin{subfigure}[b]{0.4\textwidth} 
        \centering
        \includegraphics[width=\textwidth]{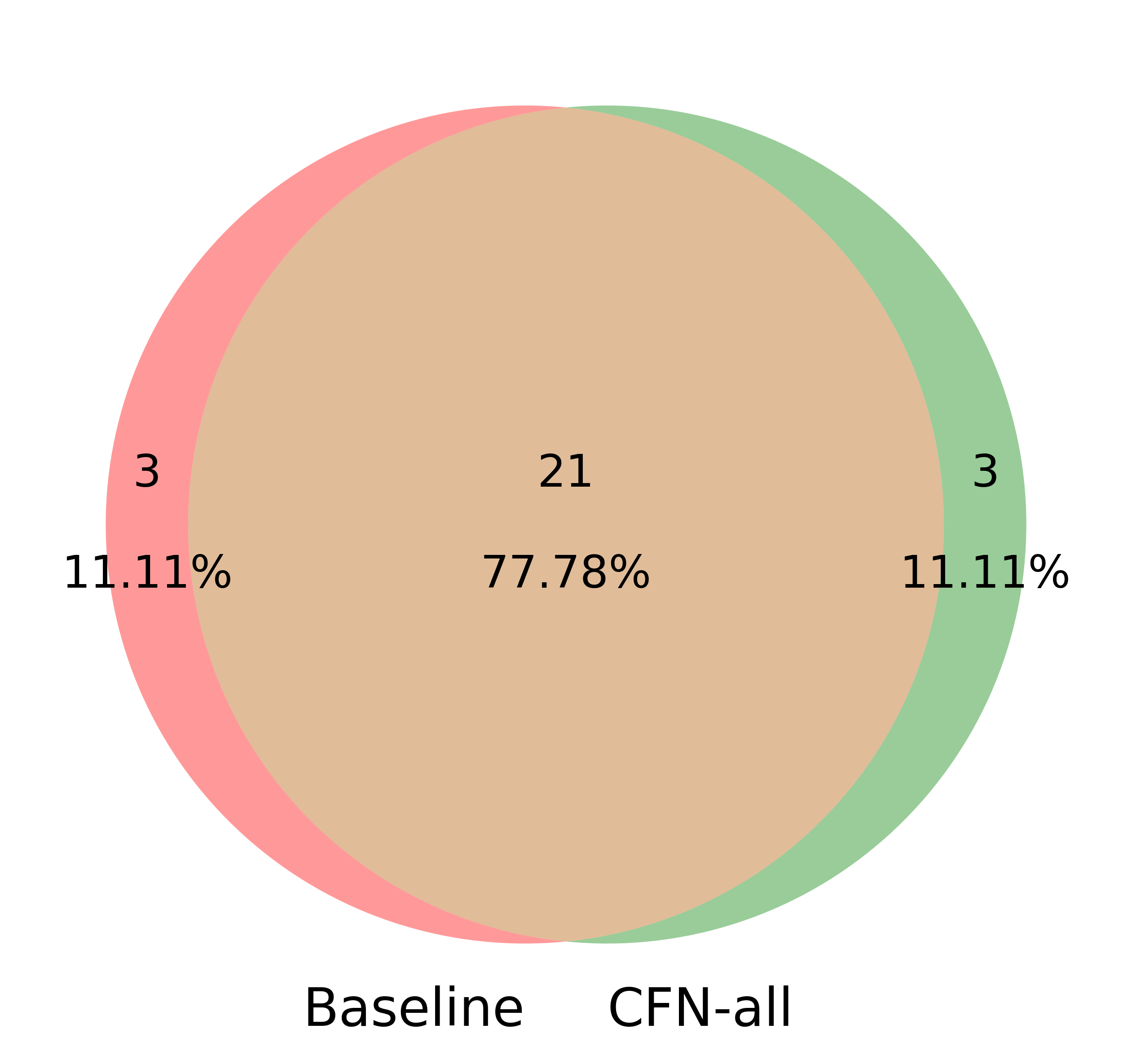} 
        \caption{CFN-all fixes 3 more bugs compared to the baseline.}
        \label{venn-diagram-CodeLlama-Instruct-7B-BugsInPy:b}
    \end{subfigure}
    \vspace{-0.1cm}
    \\
    \begin{subfigure}[b]{0.4\textwidth} 
        \centering
        \includegraphics[width=\textwidth]{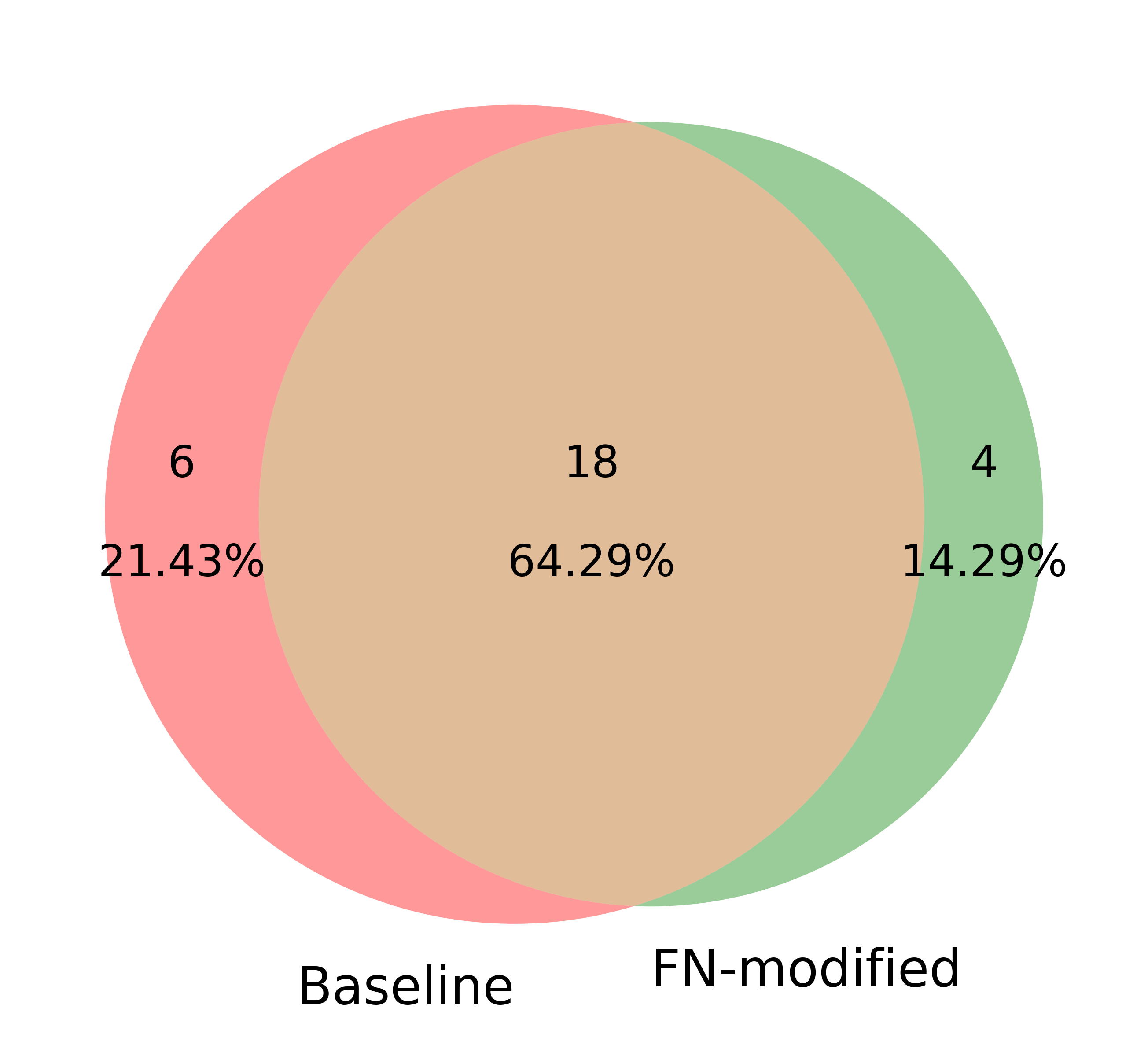} 
        \caption{FN-modified fixes 4 more bugs compared to the baseline.}
        \label{venn-diagram-CodeLlama-Instruct-7B-BugsInPy:c}
    \end{subfigure}
    \hfill
    \begin{subfigure}[b]{0.4\textwidth} 
        \centering
        \includegraphics[width=\textwidth]{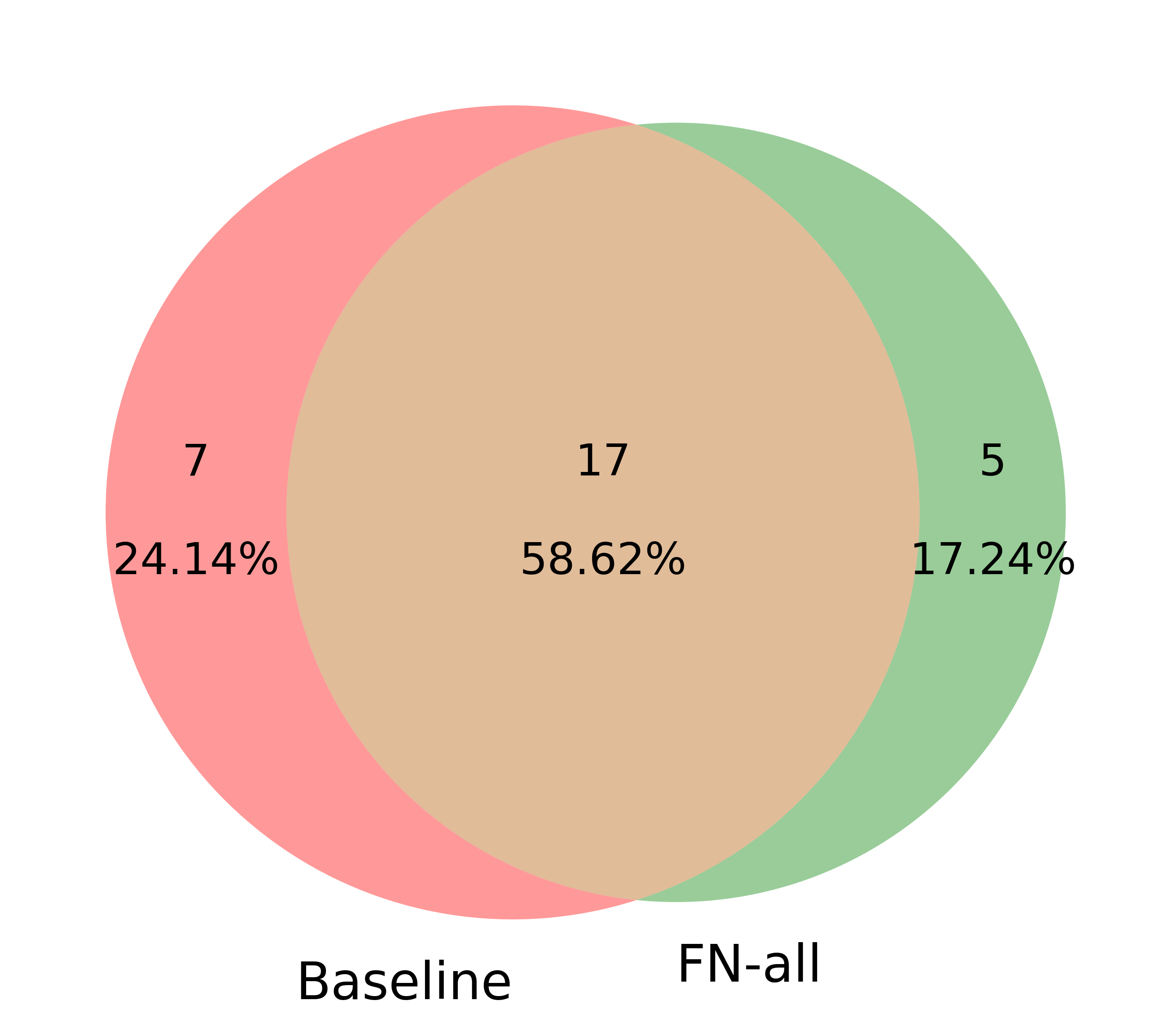} 
        \caption{FN-all fixes 5 more bugs compared to the baseline.}
        \label{venn-diagram-CodeLlama-Instruct-7B-BugsInPy:d}
    \end{subfigure}
    \vspace{-0.1cm}
    \\
    \begin{subfigure}[b]{0.4\textwidth} 
        \centering
        \includegraphics[width=\textwidth]{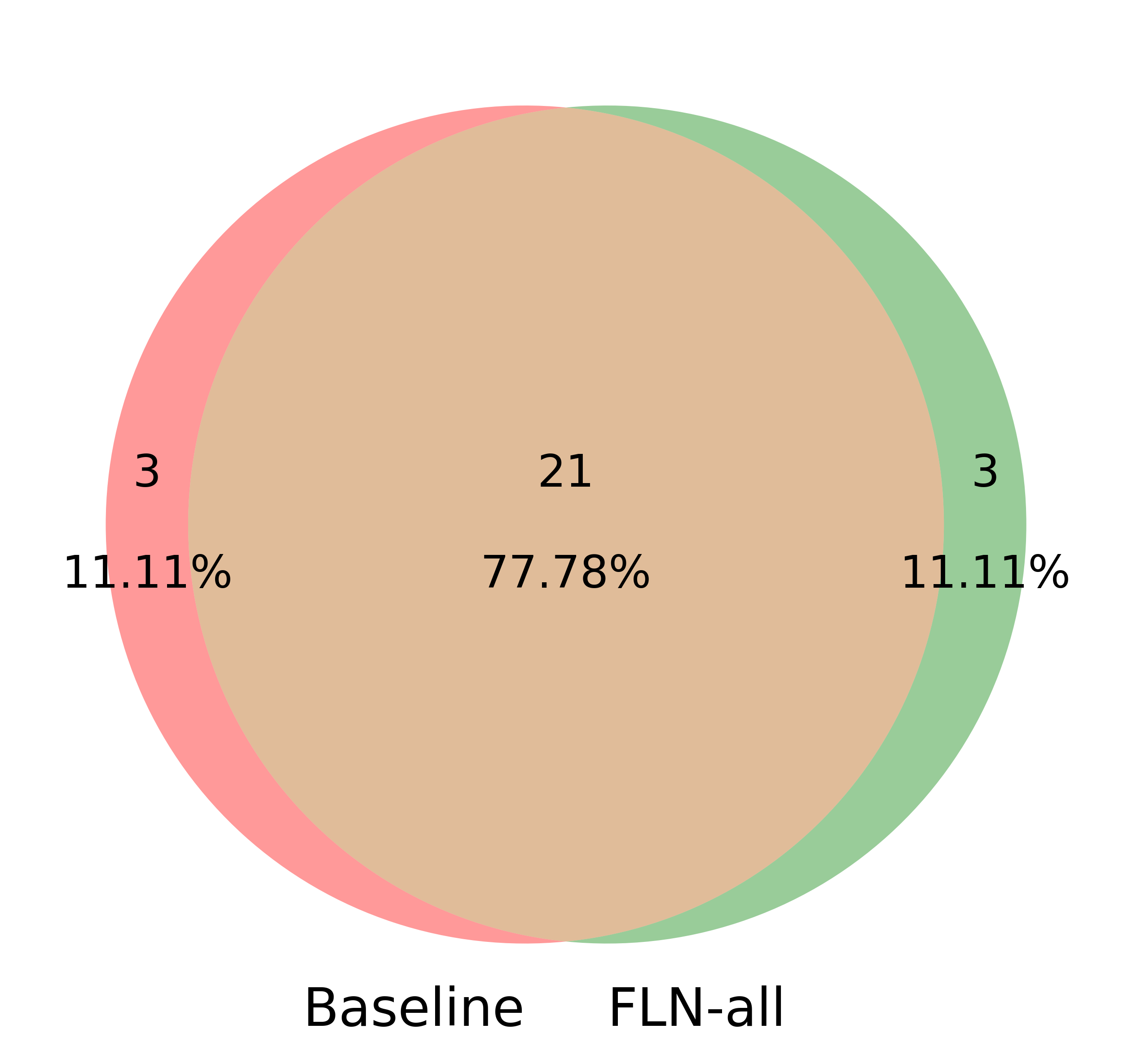} 
        \caption{FLN-all fixes 3 more bugs compared to the baseline.}
        \label{venn-diagram-CodeLlama-Instruct-7B-BugsInPy:e}
    \end{subfigure}
    \hfill
    \begin{subfigure}[b]{0.4\textwidth} 
        \centering
        \includegraphics[width=\textwidth]{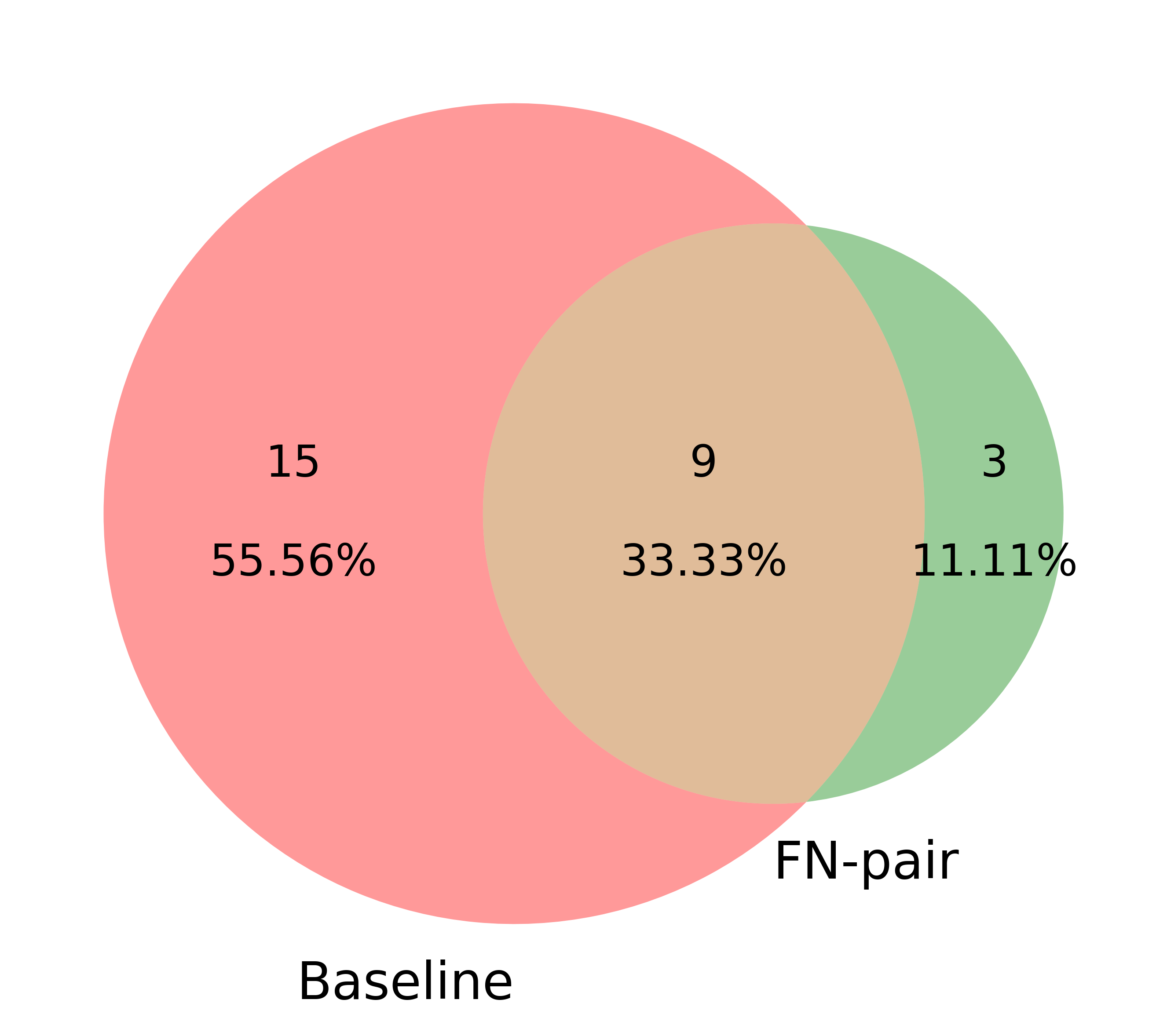} 
        \caption{FN-pair fixes 3 more bugs compared to the baseline.}
        \label{venn-diagram-CodeLlama-Instruct-7B-BugsInPy:f}
    \end{subfigure}
    \vspace{-0.1cm}
    \\
    \begin{subfigure}[b]{0.4\textwidth} 
        \centering
        \includegraphics[width=\textwidth]{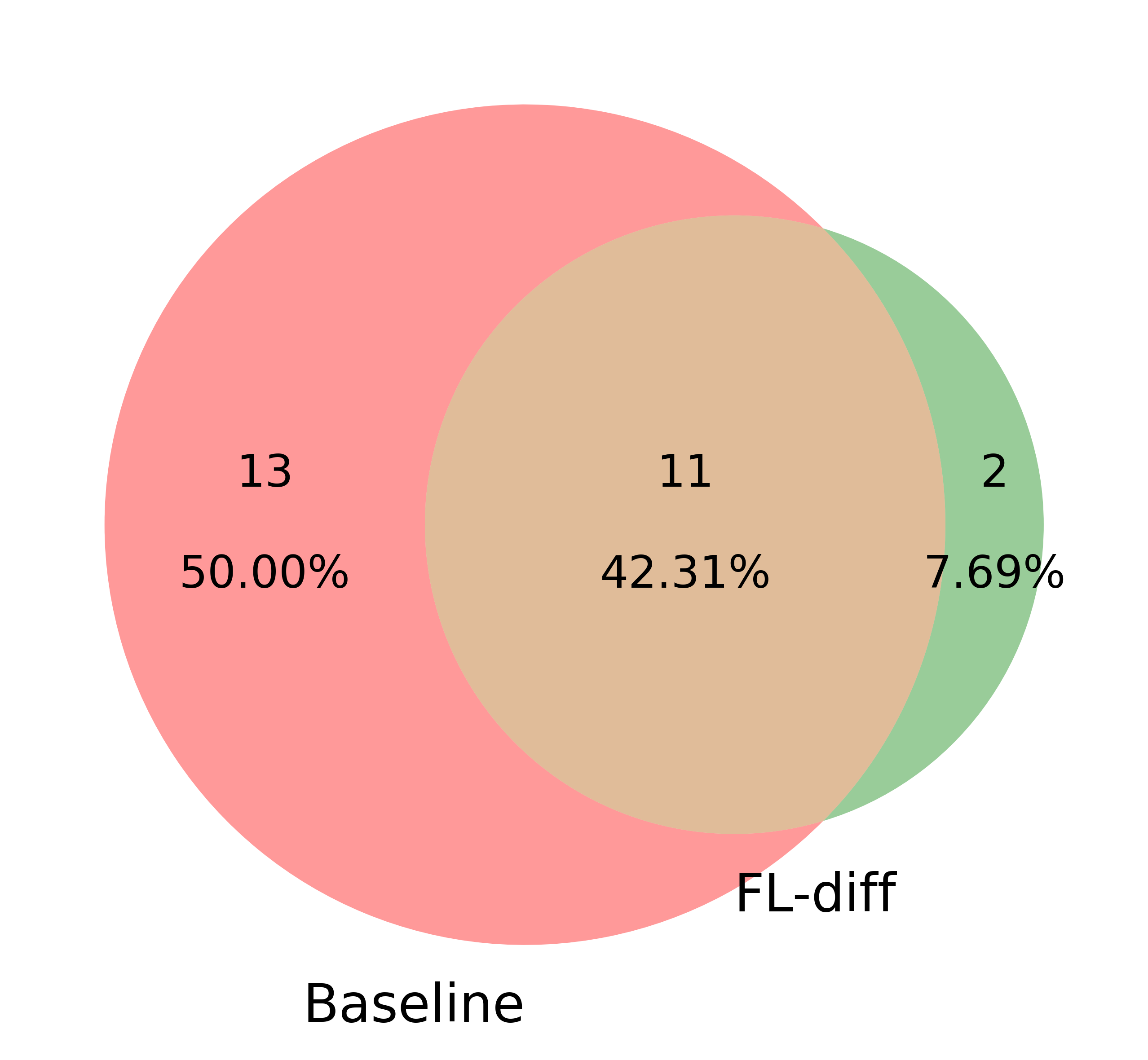} 
        \caption{FL-diff fixes 2 more bugs compared to the baseline.}
        \label{venn-diagram-CodeLlama-Instruct-7B-BugsInPy:g}
    \end{subfigure}
    \caption{CodeLlama-Instruct-7B on BugsInPy. Venn diagrams comparing the number of bugs fixed by the baseline (red) and the seven individual HAFix heuristics (green), with the overlapping region (brown) indicating bugs fixed by both the baseline and the heuristic. Numbers and percentages within each region denote the count and proportion of bugs fixed.}
    \label{fig:venn-diagram-CodeLlama-Instruct-7B-BugsInPy}
\end{figure}

\begin{figure}[!htbp]
    \centering
    \begin{subfigure}[b]{0.4\textwidth} 
        \centering
        \includegraphics[width=\textwidth]{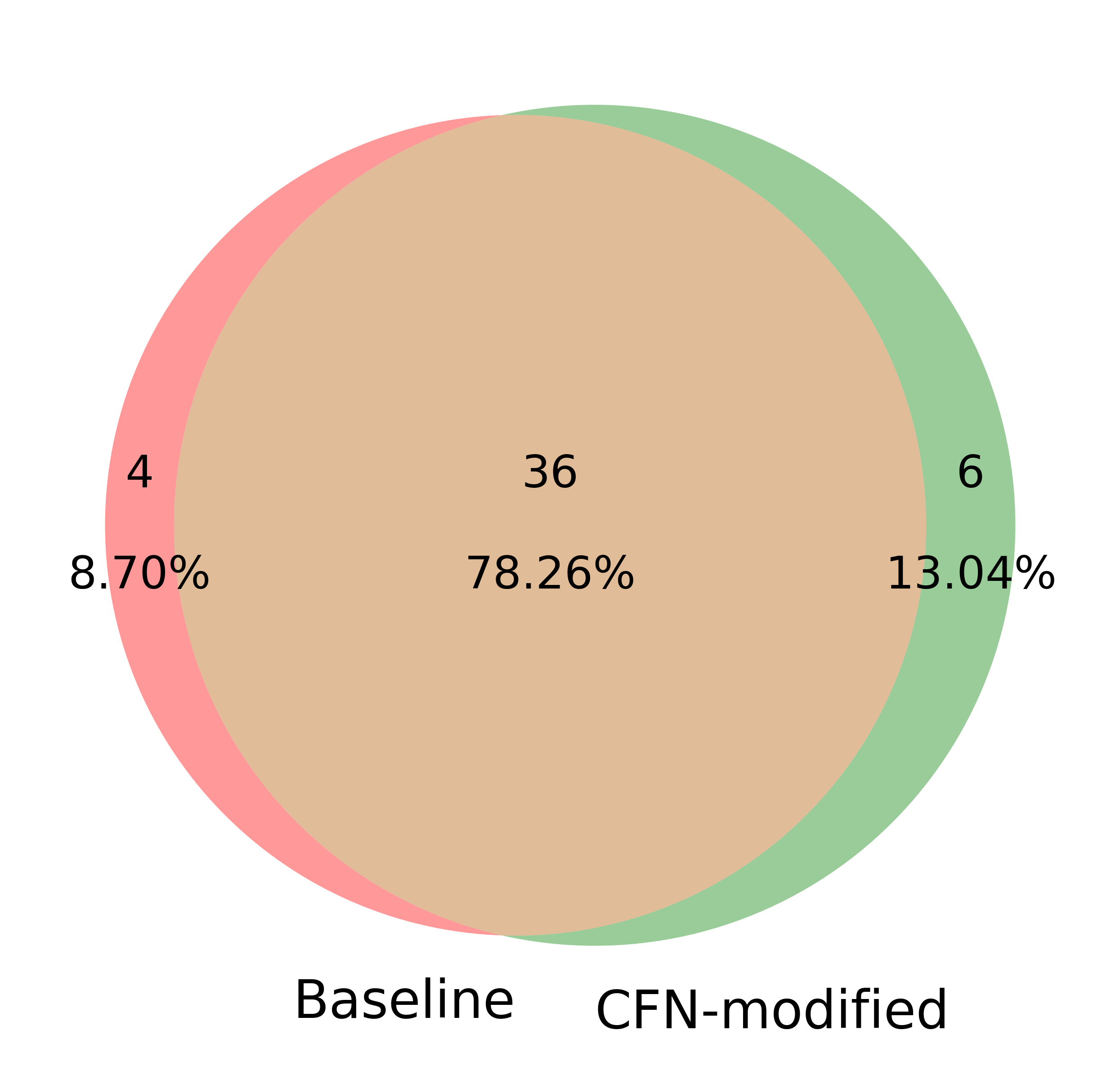} 
        \caption{CFN-modified fixes 6 more bugs compared to the baseline.}
        \label{venn-diagram-DeepSeek-Coder-Instruct-6.7B-Defects4J:a}
    \end{subfigure}
    \hfill 
    \begin{subfigure}[b]{0.4\textwidth} 
        \centering
        \includegraphics[width=\textwidth]{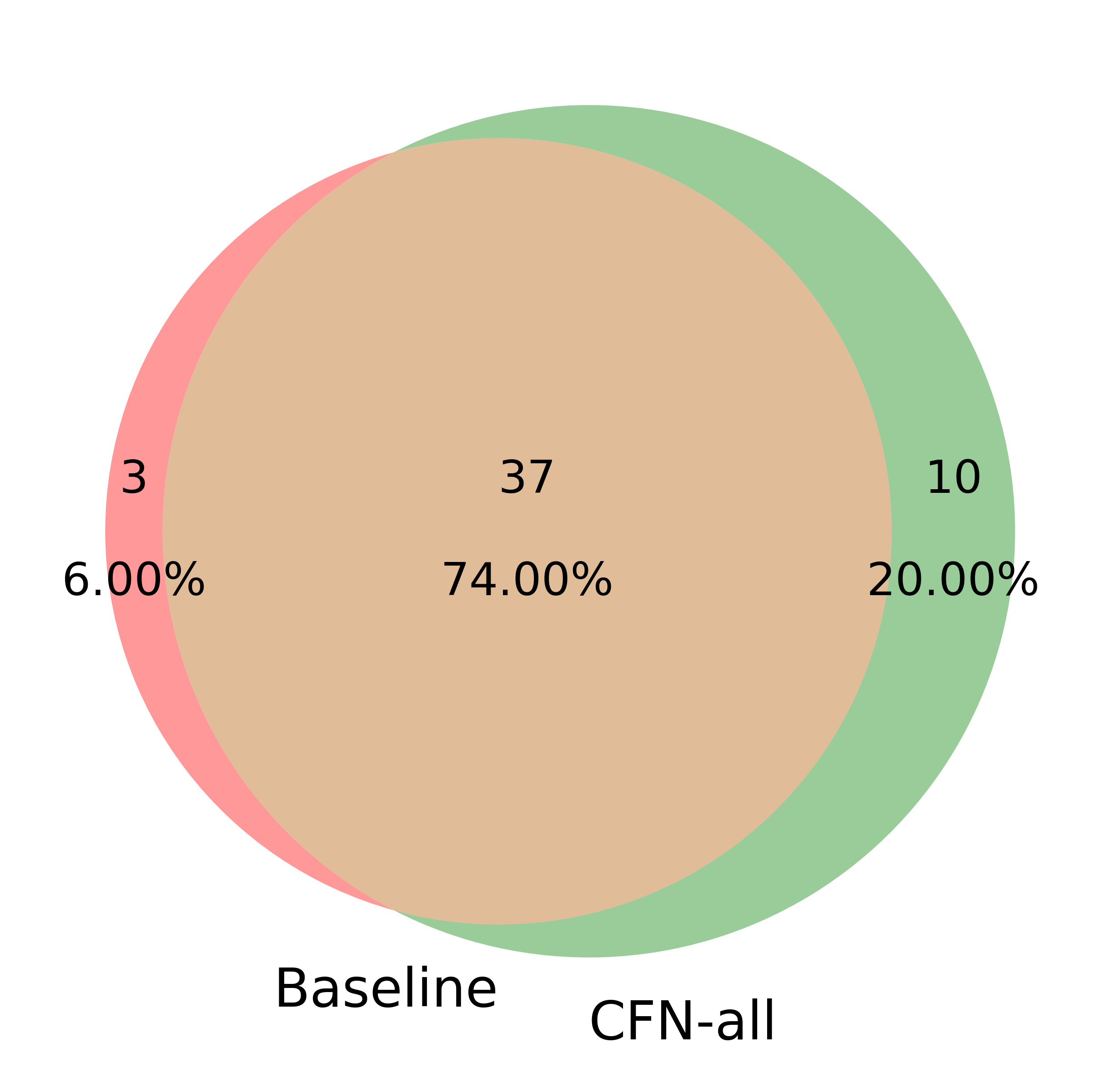} 
        \caption{CFN-all fixes 10 more bugs compared to the baseline.}
        \label{venn-diagram-DeepSeek-Coder-Instruct-6.7B-Defects4J:b}
    \end{subfigure}
    \vspace{-0.1cm}
    \\
    \begin{subfigure}[b]{0.4\textwidth} 
        \centering
        \includegraphics[width=\textwidth]{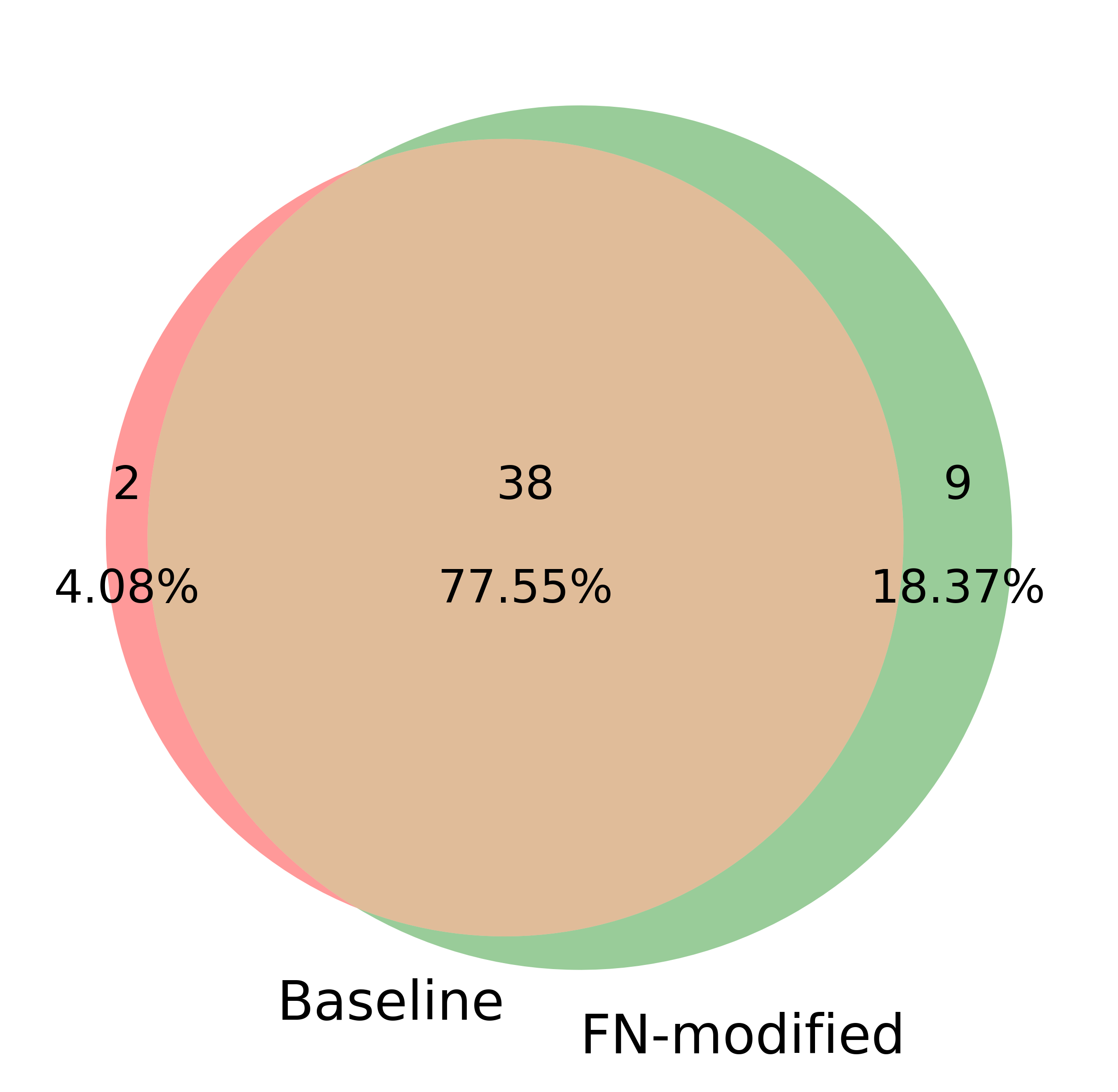} 
        \caption{FN-modified fixes 9 more bugs compared to the baseline.}
        \label{venn-diagram-DeepSeek-Coder-Instruct-6.7B-Defects4J:c}
    \end{subfigure}
    \hfill
    \begin{subfigure}[b]{0.4\textwidth} 
        \centering
        \includegraphics[width=\textwidth]{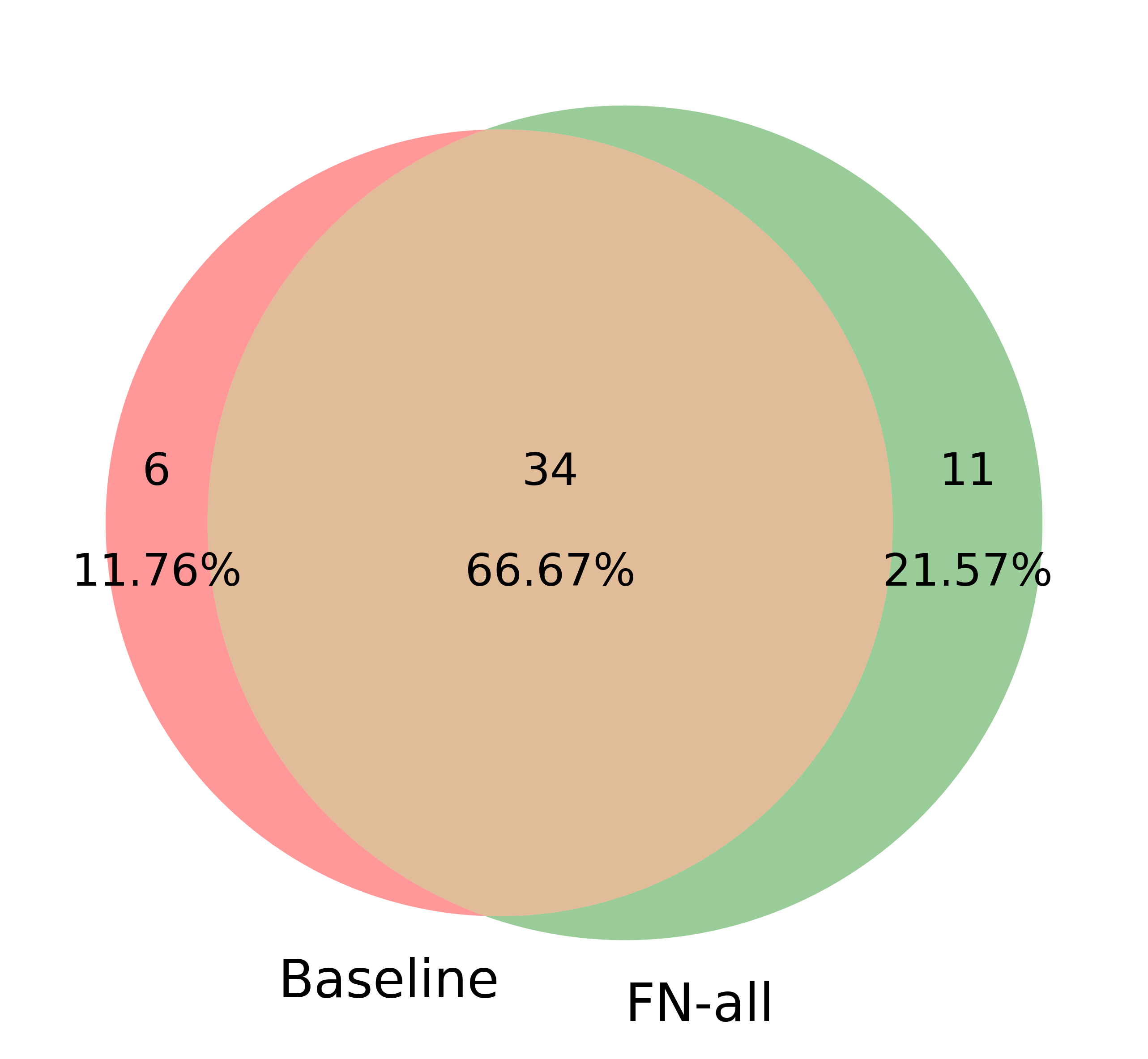} 
        \caption{FN-all fixes 11 more bugs compared to the baseline.}
        \label{venn-diagram-DeepSeek-Coder-Instruct-6.7B-Defects4J:d}
    \end{subfigure}
    \vspace{-0.1cm}
    \\
    \begin{subfigure}[b]{0.4\textwidth} 
        \centering
        \includegraphics[width=\textwidth]{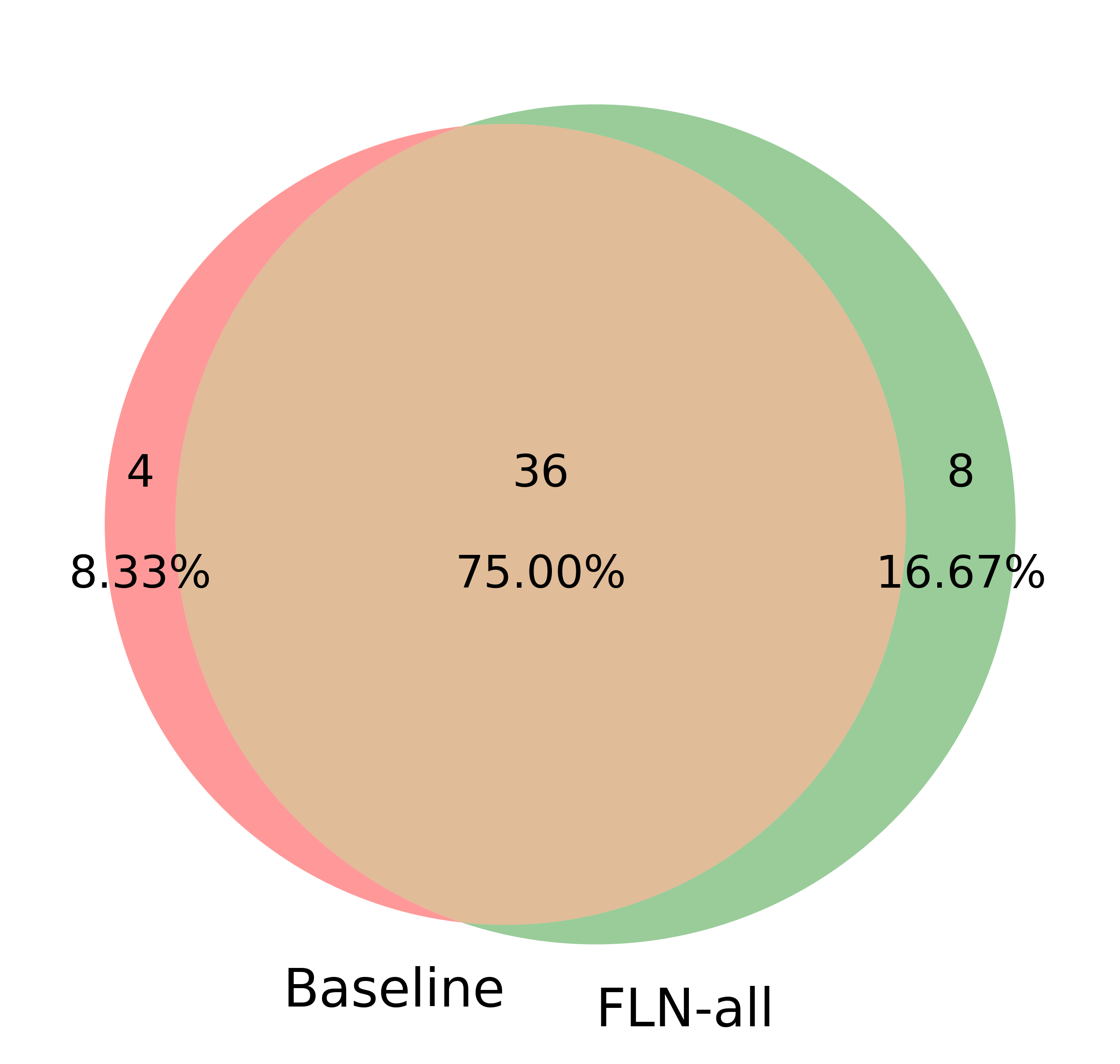} 
        \caption{FLN-all fixes 8 more bugs compared to the baseline.}
        \label{venn-diagram-DeepSeek-Coder-Instruct-6.7B-Defects4J:e}
    \end{subfigure}
    \hfill
    \begin{subfigure}[b]{0.4\textwidth} 
        \centering
        \includegraphics[width=\textwidth]{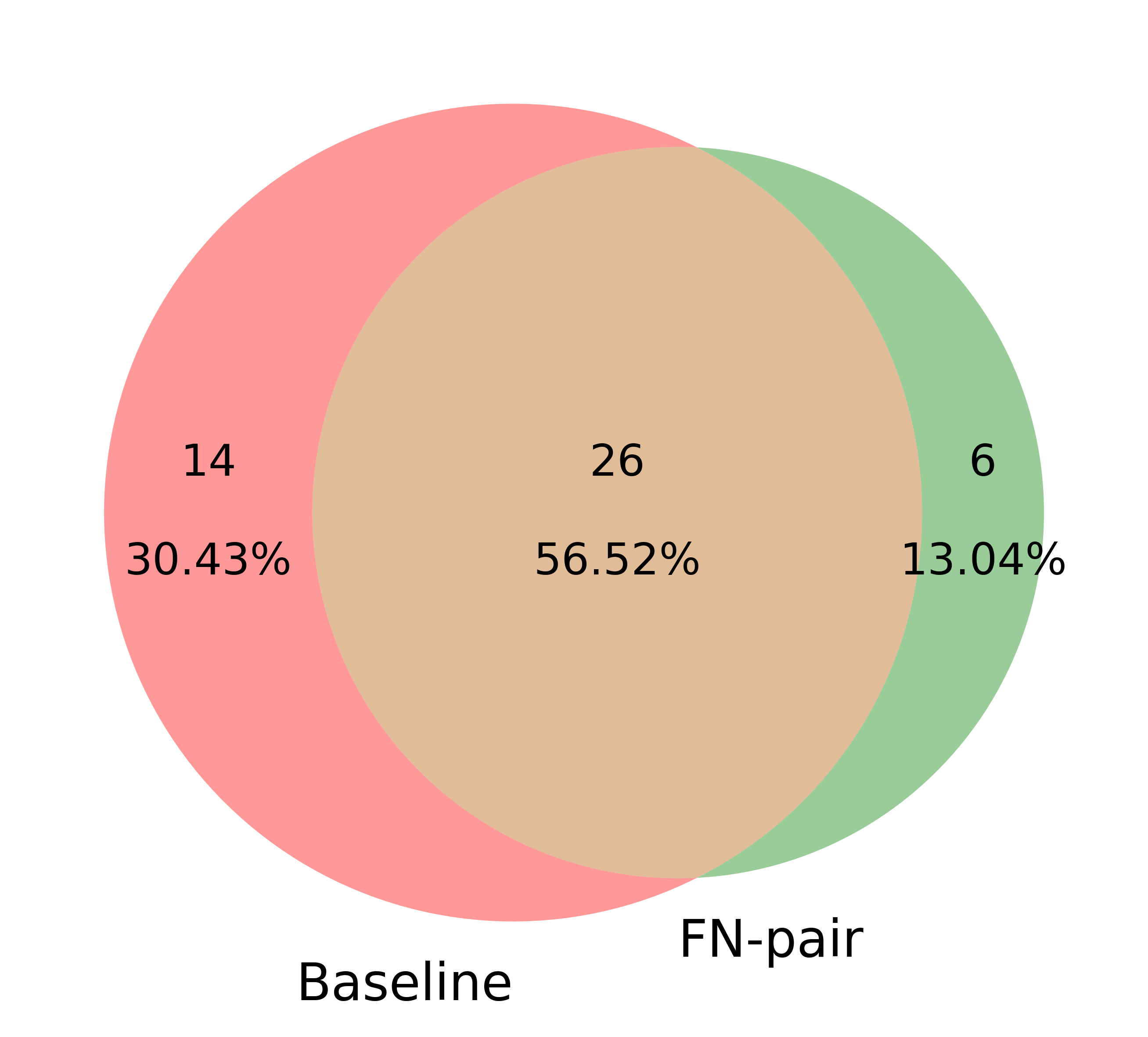} 
        \caption{FN-pair fixes 6 more bugs compared to the baseline.}
        \label{venn-diagram-DeepSeek-Coder-Instruct-6.7B-Defects4J:f}
    \end{subfigure}
    \vspace{-0.1cm}
    \\
    \begin{subfigure}[b]{0.4\textwidth} 
        \centering
        \includegraphics[width=\textwidth]{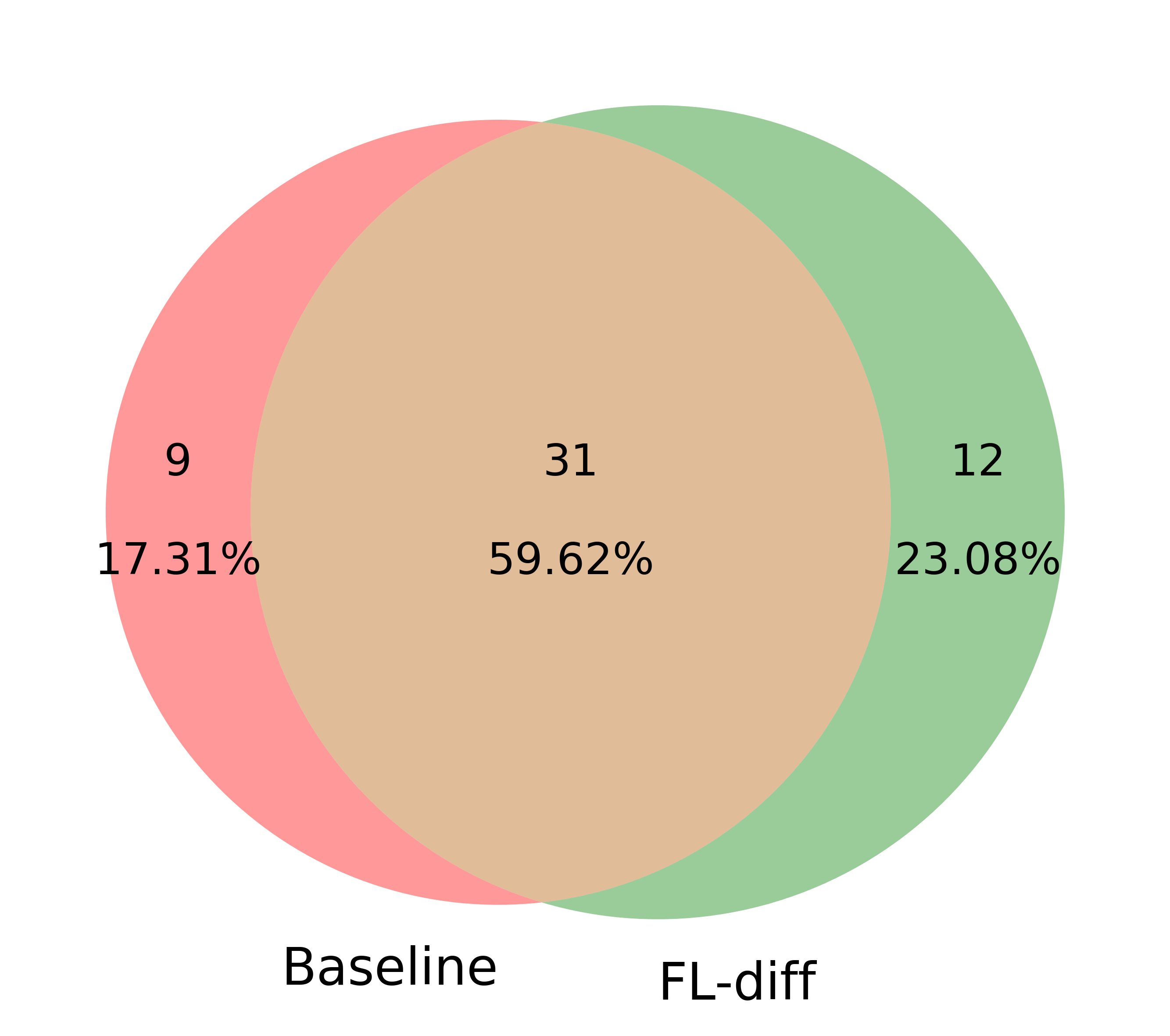} 
        \caption{FL-diff fixes 12 more bugs compared to the baseline.}
        \label{venn-diagram-DeepSeek-Coder-Instruct-6.7B-Defects4J:g}
    \end{subfigure}
    \caption{DeepSeek-Coder-Instruct-6.7B on Defects4J. Venn diagrams comparing the number of bugs fixed by the baseline (red) and the seven individual HAFix heuristics (green), with the overlapping region (brown) indicating bugs fixed by both the baseline and the heuristic. Numbers and percentages within each region denote the count and proportion of bugs fixed.}
    \label{fig:venn-diagram-DeepSeek-Coder-Instruct-6.7B-Defects4J}
\end{figure}

\textbf{Surprisingly, history-based heuristics with lower Pass@k performance can still provide strong complementary benefits by uniquely fixing bugs missed by the baseline.} Although FN-pair and FL-diff consistently underperform the baseline in Pass@k (Figure \ref{fig:rq1_passk_baseline_heuristics_2_datasets_3_models} and Table \ref{RQ1_passk}), they recover a notable number of uniquely fixed bugs. For instance, in DeepSeek-Coder-Instruct-6.7B on BugsInPy, FN-pair and FL-diff fix 3 and 4 additional bugs, respectively. In DeepSeek-Coder-V2-Lite-Instruct-16B on Defects4J, they uniquely resolve 11 and 13 bugs.
This complementary behavior is illustrated in Figure \ref{fig:venn-diagram-CodeLlama-Instruct-7B-BugsInPy} and Figure \ref{fig:venn-diagram-DeepSeek-Coder-Instruct-6.7B-Defects4J}, which show Venn diagrams comparing the bugs fixed by the baseline and each heuristic for selected model-dataset configurations. Each Venn diagram consists of three overlapping regions: the red area represents the bugs fixed exclusively by the baseline, the green area represents the bugs fixed exclusively by the heuristic under evaluation, and the brown area in the middle represents the bugs fixed by both the baseline and the heuristic. One diagram is generated per model-dataset configuration, for a total of six. Here, we present two representative sets of Venn diagrams that correspond to the lowest (\ref{fig:venn-diagram-CodeLlama-Instruct-7B-BugsInPy}) and highest (\ref{fig:venn-diagram-DeepSeek-Coder-Instruct-6.7B-Defects4J}) BugsU\# values observed. The remaining four sets of Venn diagrams are included in Appendix \ref{appendix:VennDiagrams}.

\begin{figure}[!htbp]
    \centering
    \begin{subfigure}[b]{0.48\textwidth} 
        \centering
        \includegraphics[width=\textwidth]{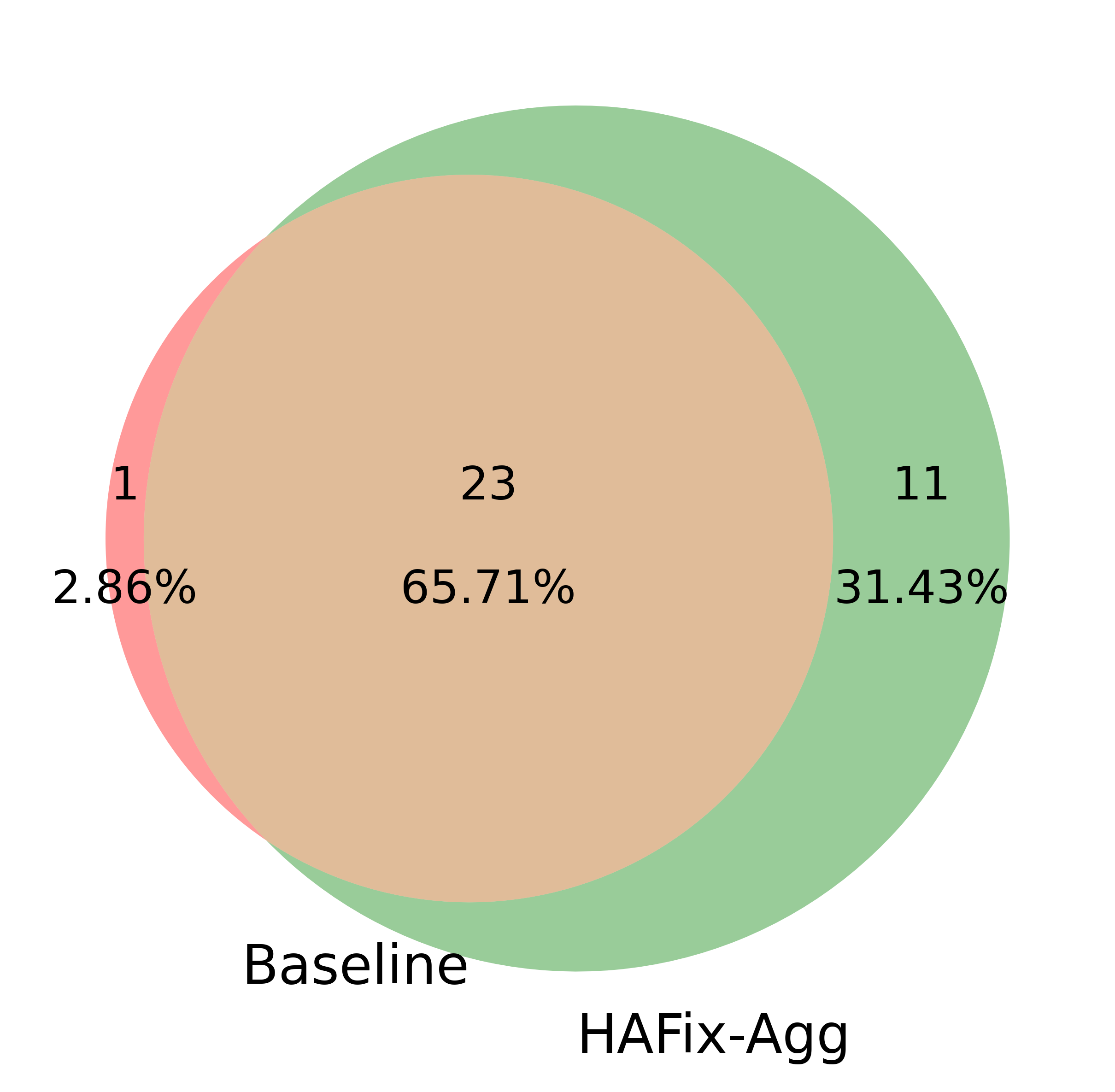} 
        \caption{CodeLlama-Instruct-7B on BugsInPy:\\HAFix-Agg fixes 11 more bugs (41.67\% improvement) compared to the baseline.}
        \label{venn-baseline-hafix-agg:sub1}
    \end{subfigure}
    \hfill
    \begin{subfigure}[b]{0.48\textwidth} 
        \centering
        \includegraphics[width=\textwidth]{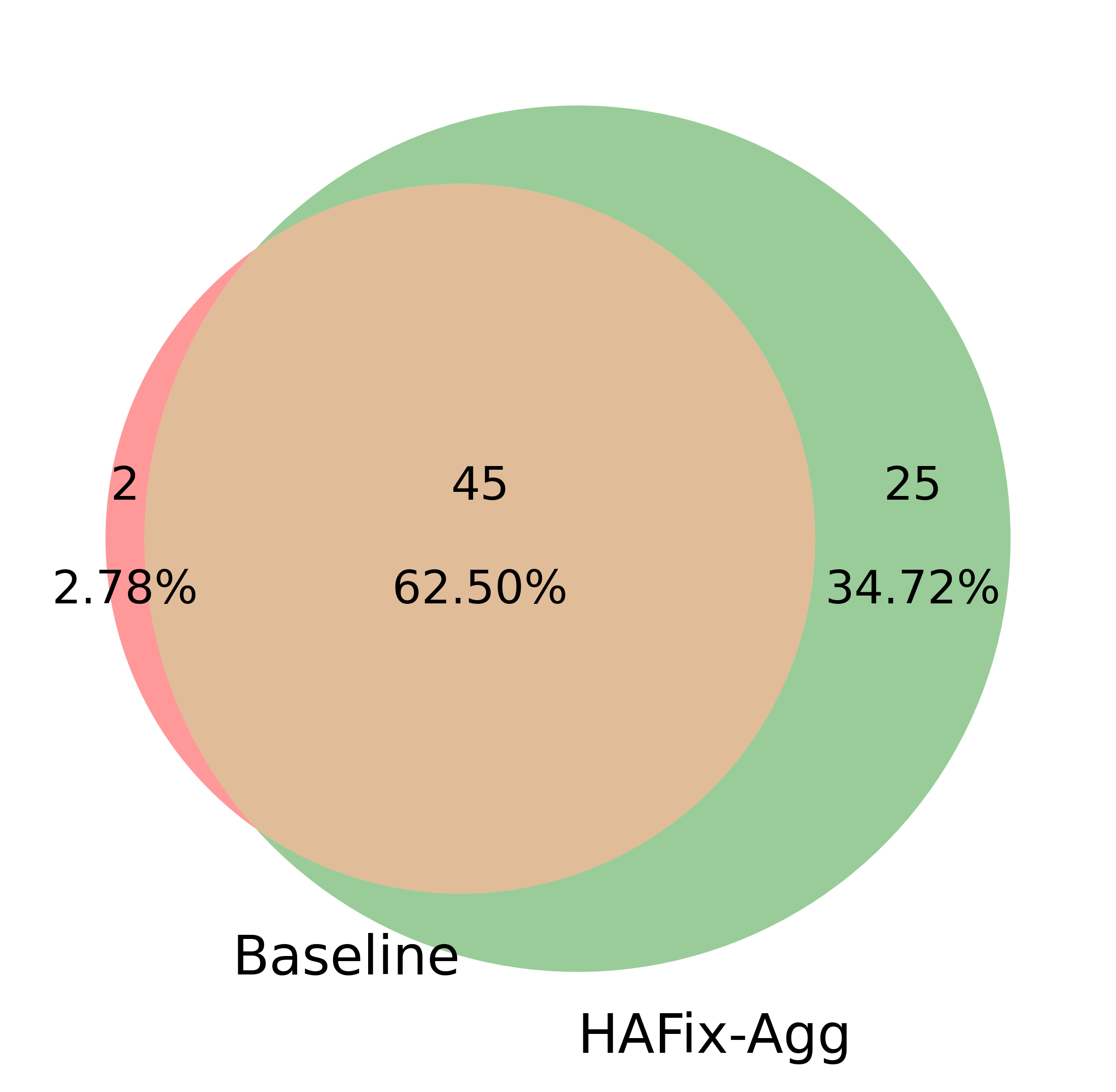} 
        \caption{CodeLlama-Instruct-7B on Defects4J:\\HAFix-Agg fixes 25 more bugs (48.94\% improvement) compared to the baseline.}
        \label{venn-baseline-hafix-agg:sub2}
    \end{subfigure}
    \vspace{-0.1cm}
    \\
    \begin{subfigure}[b]{0.48\textwidth} 
        \centering
        \includegraphics[width=\textwidth]{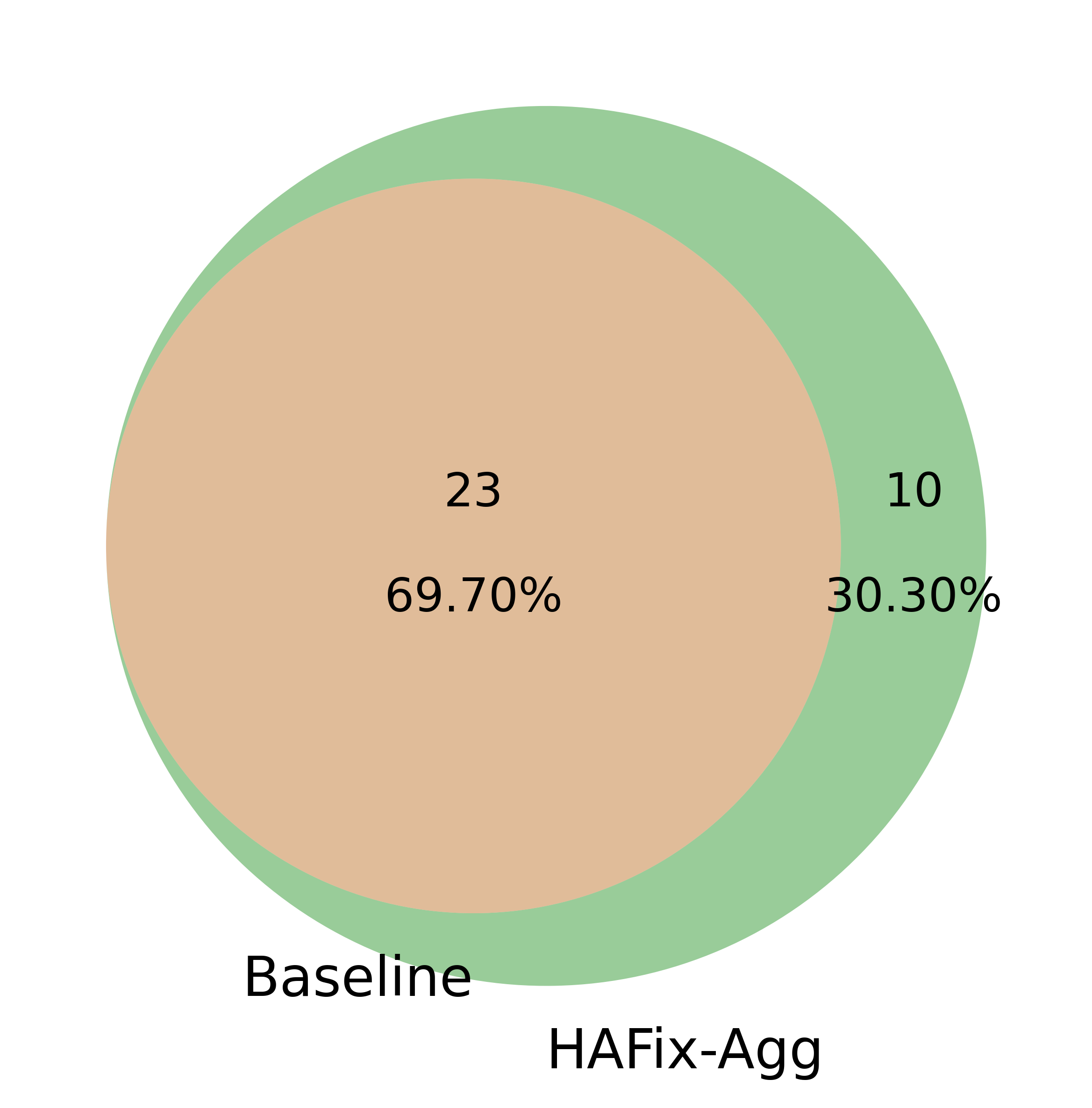} 
        \caption{DeepSeek-Coder-Instruct-6.7B on \\BugsInPy: HAFix-Agg fixes 10 more bugs (43.48\% improvement) compared to the baseline.}
        \label{venn-baseline-hafix-agg:sub3}
    \end{subfigure}
    \hfill
    \begin{subfigure}[b]{0.48\textwidth} 
        \centering
        \includegraphics[width=\textwidth]{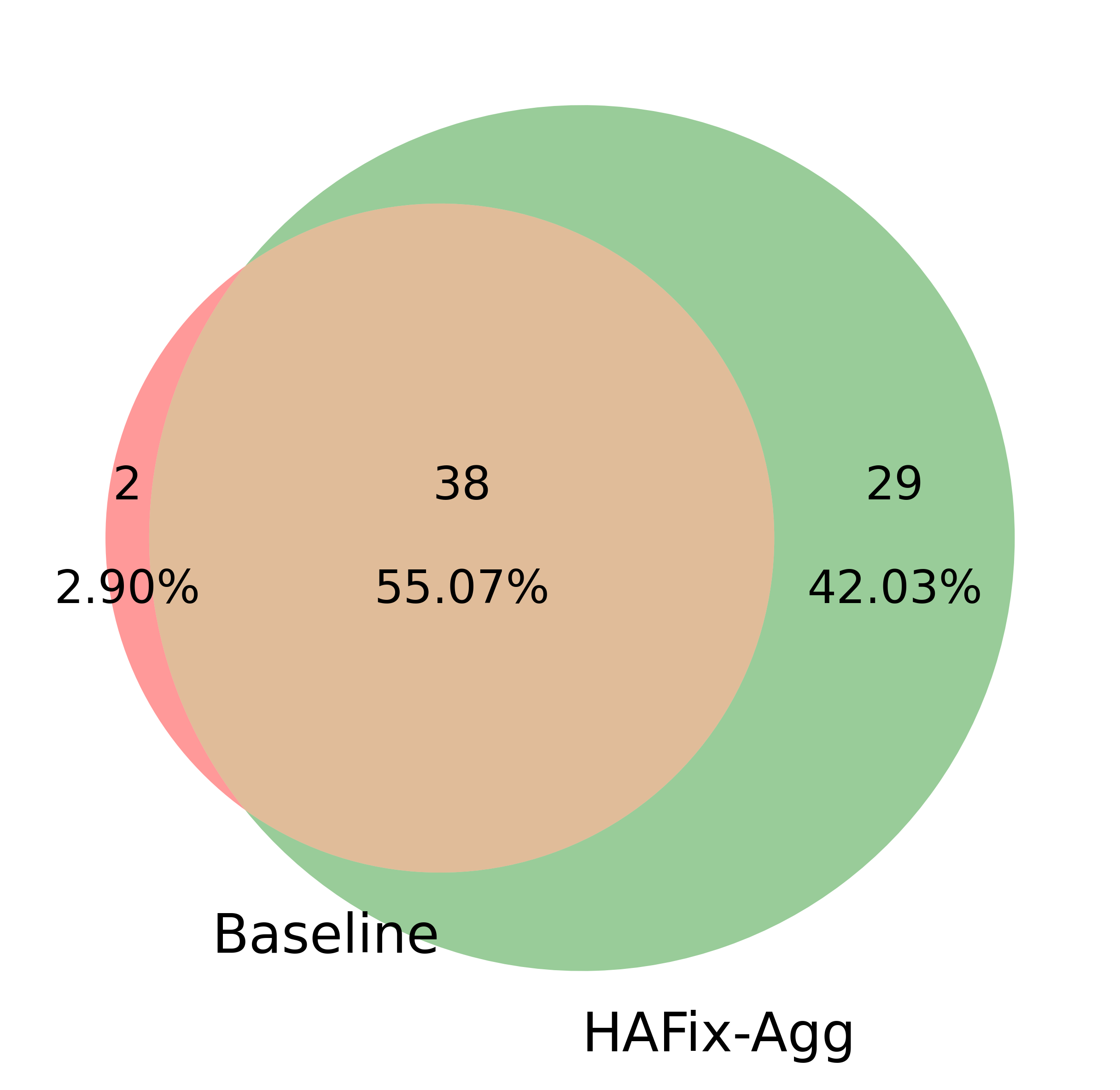} 
        \caption{DeepSeek-Coder-Instruct-6.7B on \\Defects4J: HAFix-Agg fixes 29 more bugs (67.50\% improvement) compared to the baseline.}
        \label{venn-baseline-hafix-agg:sub4}
    \end{subfigure}
    \vspace{-0.1cm}
    \\
    \begin{subfigure}[b]{0.48\textwidth} 
        \centering
        \includegraphics[width=\textwidth]{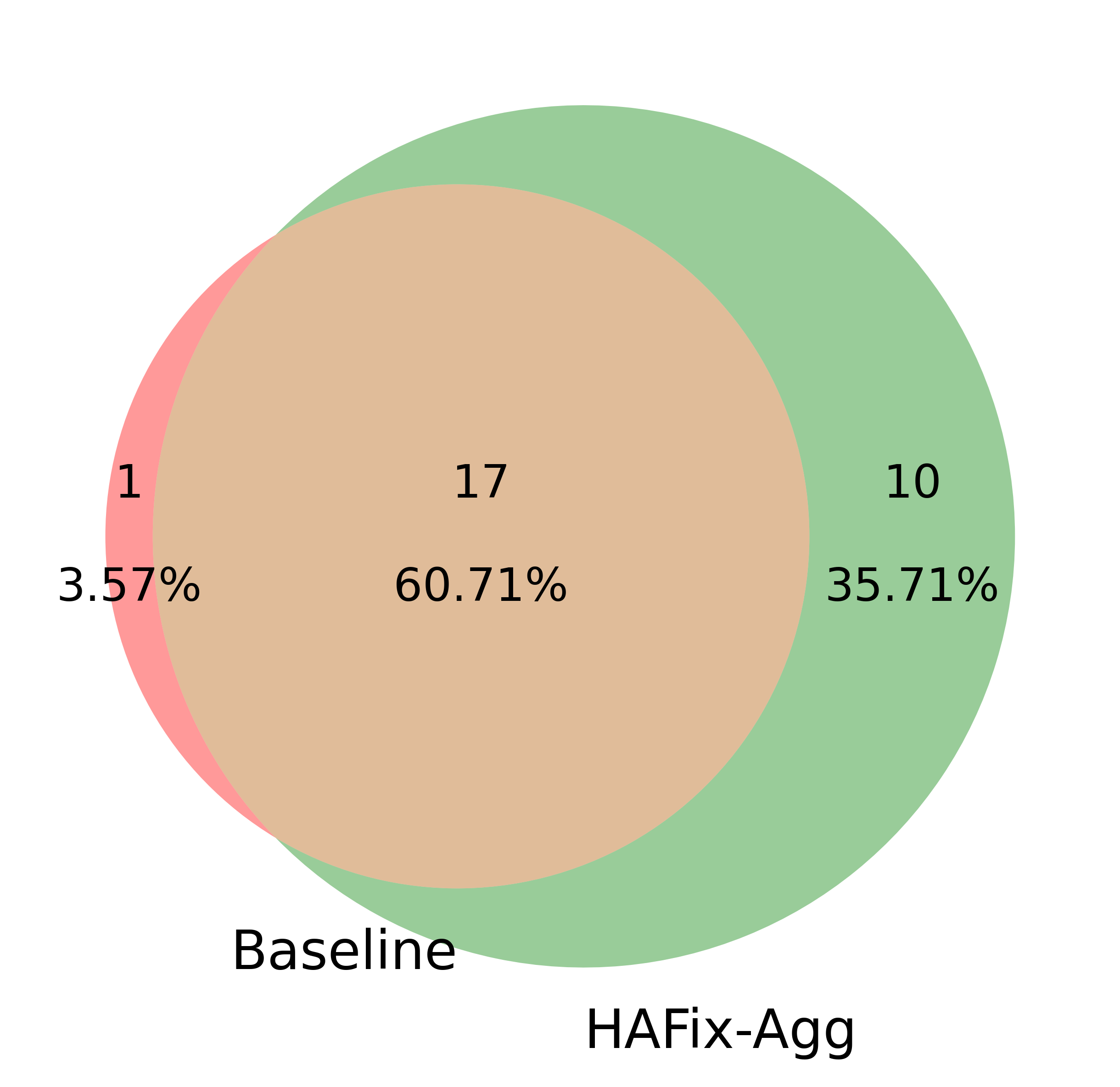} 
        \caption{DeepSeek-Coder-V2-Lite-Instruct-16B \\on BugsInPy: HAFix-Agg fixes 10 more bugs (50.00\% improvement) compared to the baseline.}
        \label{venn-baseline-hafix-agg:sub5}
    \end{subfigure}
    \hfill
    \begin{subfigure}[b]{0.48\textwidth} 
        \centering
        \includegraphics[width=\textwidth]{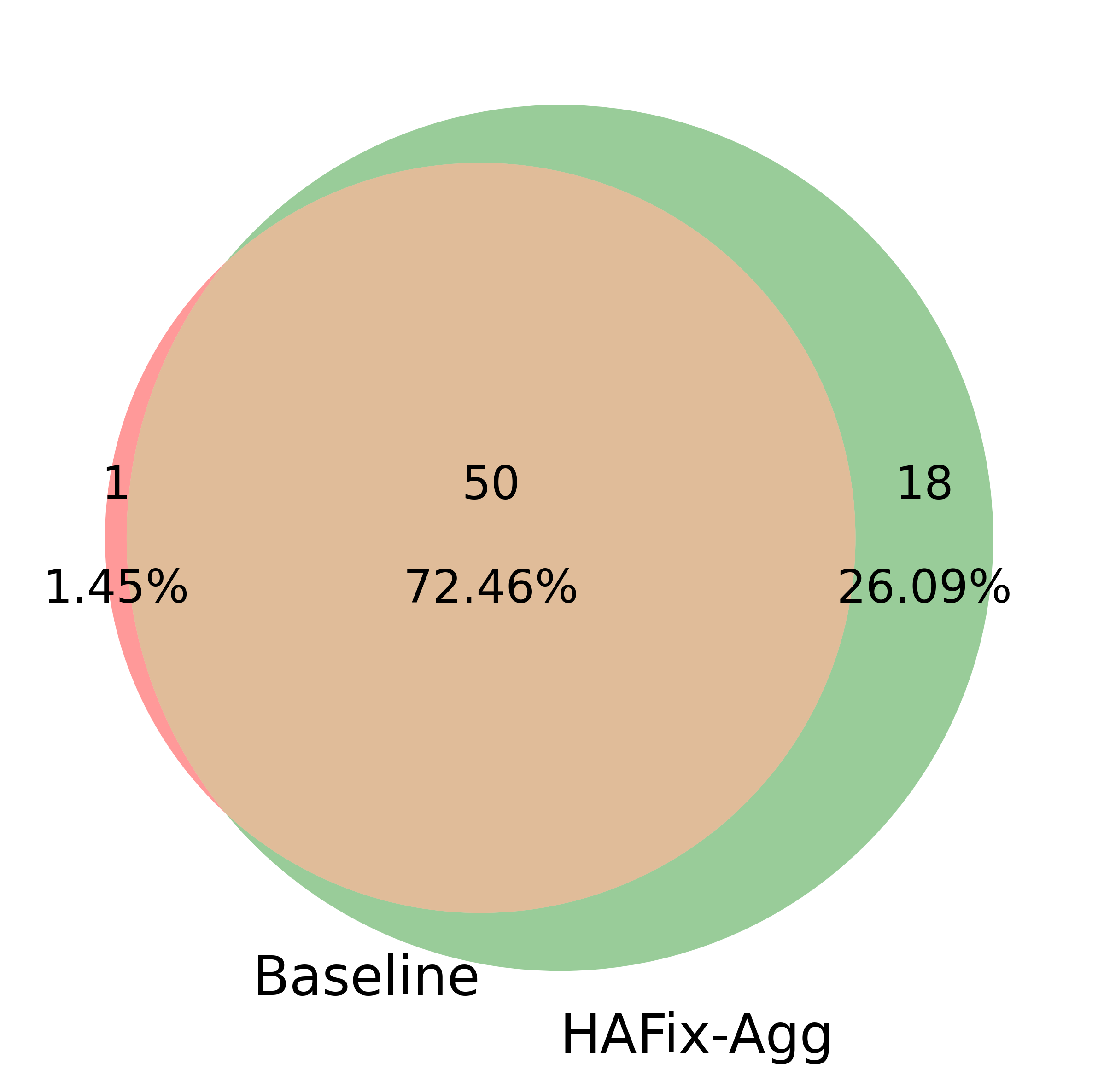} 
        \caption{DeepSeek-Coder-V2-Lite-Instruct-16B \\on Defects4J: HAFix-Agg fixes 18 more bugs (33.33\% improvement) compared to the baseline.}
        \label{venn-baseline-hafix-agg:sub6}
    \end{subfigure}
    \caption{Venn diagram comparing the number of bugs fixed by the baseline (brown) and HAFix-Agg (green) across three models and two datasets. The overlapping region represents bugs fixed by both the baseline and HAFix-Agg.}
    \label{fig:venn-baseline-hafix-agg}
\end{figure}

\textbf{Building on the complementary strengths of individual heuristics, HAFix-Agg further boosts performance by combining their unique contributions, fixing an average of 10 additional bugs (+45.05\%) on BugsInPy and 22 (+49.92\%) on Defects4J.} Motivated by the observation that individual HAFix heuristics can each fix different subsets of bugs missed by the baseline, we introduce an aggregated variant, HAFix-Agg, which integrates the results of different history heuristics to enhance bug-fixing performance. Figure \ref{fig:venn-baseline-hafix-agg} shows Venn diagrams comparing the number of bugs fixed by the baseline (brown circle) and HAFix-Agg (green circle) across three models and two datasets. In all six model-dataset configurations, HAFix-Agg fixes nearly all bugs solved by the baseline and consistently adds a substantial number of unique fixes. On BugsInPy (51 subject bugs), HAFix-Agg achieves relative improvements of 41.67\%, 43.48\%, and 50.00\% for CodeLlama-Instruct-7B, DeepSeek-Coder-Instruct-6.7B, and DeepSeek-Coder-V2-Lite-Instruct-16B, respectively, corresponding to 11, 10, and 10 additional bugs fixed. On Defects4J (116 subject bugs), the gains are similarly pronounced, with improvements of 48.94\%, 67.50\%, and 33.33\%, corresponding to 25, 29, and 18 additional bugs fixed by the same models. On average, HAFix-Agg improves over the baseline by 45.05\% on BugsInPy and 49.92\% on Defects4J. These results demonstrate that aggregating HAFix heuristics leads to significantly better performance than any individual heuristic or the baseline alone. This underscores the importance of incorporating diverse historical contexts, as they provide the model with a richer understanding and broader perspective for addressing bugs.

\begin{figure}[!htbp]
     \begin{center}
        \includegraphics[scale=0.472]{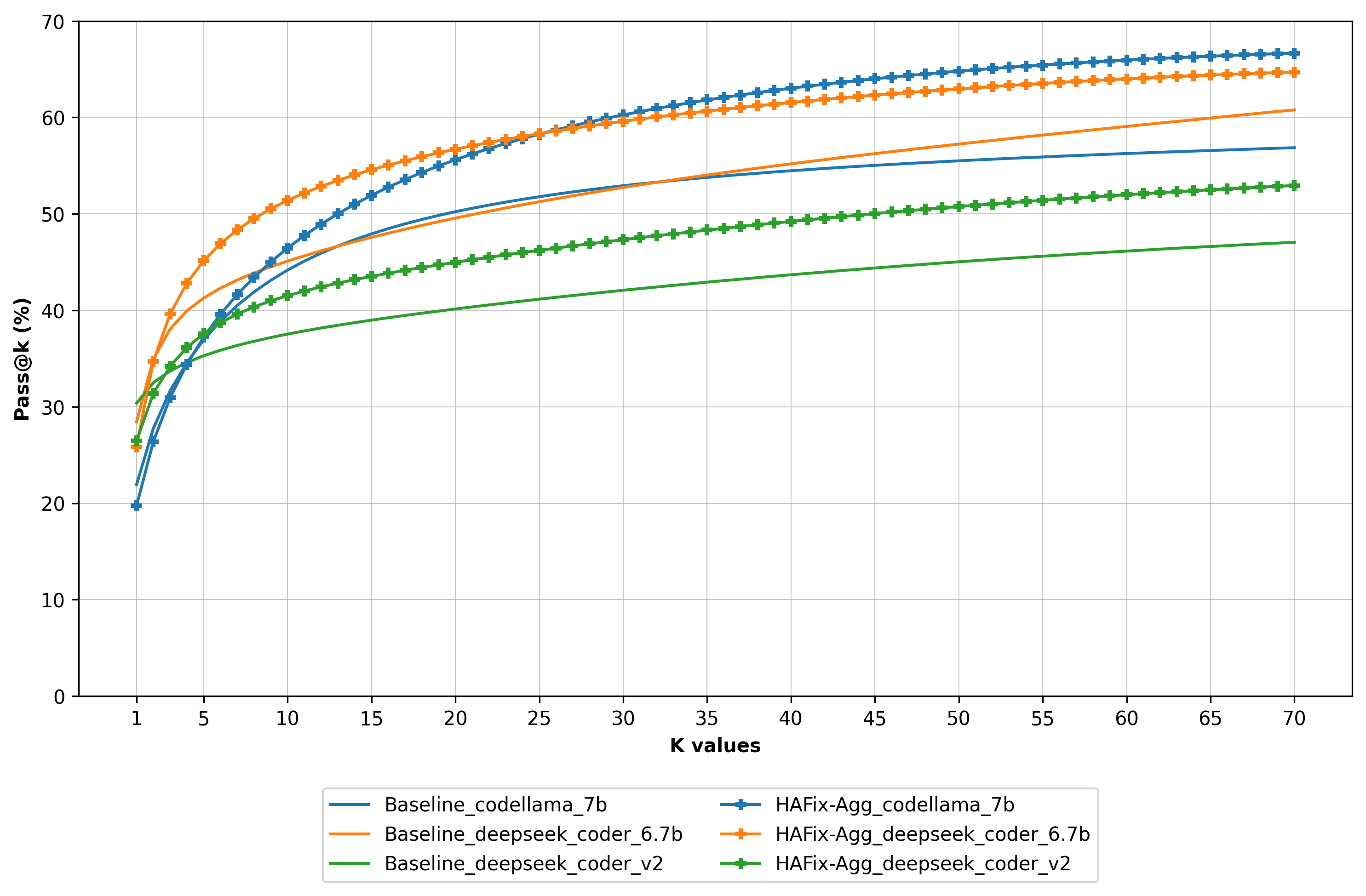}
        \end{center}
        \caption{Pass@k comparison of Baseline and HAFix-Agg (n=70) on BugsInPy across three models.}
        \label{fig:rq1_passk_baseline_hafix_2_datasets_3_models_BugsInPy}
\end{figure}

\begin{table}[!htbp]
  \centering
  \begin{threeparttable}
  \small
  \caption{Pairwise comparisons of baselines and HAFix-Agg on BugsInPy across three models using Wilcoxon signed-rank test. Values shown are $p$-values. A Bonferroni-corrected significance threshold of $\alpha = 0.0033$ ($0.05/15$) is applied.}
  \label{tab:rq1_wilcoxon_baseline_vs_hafix_agg_BugsInPy}
  \begin{tabular}{l c c c c c}
    \toprule
    & \textbf{B-CL} & \textbf{B-DS} & \textbf{B-DSV2} & \textbf{H-CL} & \textbf{H-DS} \\
    \midrule
    \textbf{B-DS}     & $6.6\e{-8}$ & - & - & - & - \\
    \textbf{B-DSV2}   & $9.8\e{-13}$ & $3.8\e{-13}$ & - & - & - \\
    \textbf{H-CL}      & $8.2\e{-13}$ & $1.6\e{-9}$ & $9.0\e{-13}$ & - & - \\
    \textbf{H-DS}     & $3.6\e{-13}$ & $4.3\e{-13}$ & $4.0\e{-13}$ & 0.53 & - \\
    \textbf{H-DSV2}   & $3.9\e{-12}$ & $3.6\e{-13}$ & $5.8\e{-13}$ & $1.4\e{-12}$ & $3.8\e{-13}$ \\
    \bottomrule
  \end{tabular}
  \begin{tablenotes}[flushleft]
    \small
    \item \textbf{Abbreviations:}
    \begin{tabularx}{\textwidth}{@{} l@{\hspace{0.6em}}l @{\hskip 1em} l@{\hspace{0.6em}}l @{}}
    B-CL    & \mbox{Baseline\_codellama\_7b}             & H-CL    & \mbox{HAFix-Agg\_codellama\_7b} \\
    B-DS    & \mbox{Baseline\_deepseek\_coder\_6.7b}     & H-DS    & \mbox{HAFix-Agg\_deepseek\_coder\_6.7b} \\
    B-DSV2  & \mbox{Baseline\_deepseek\_coder\_v2}       & H-DSV2  & \mbox{HAFix-Agg\_deepseek\_coder\_v2} \\
    \end{tabularx}
  \end{tablenotes}
  \end{threeparttable}
\end{table}

\begin{figure}[!htbp]
     \begin{center}
        \includegraphics[scale=0.472]{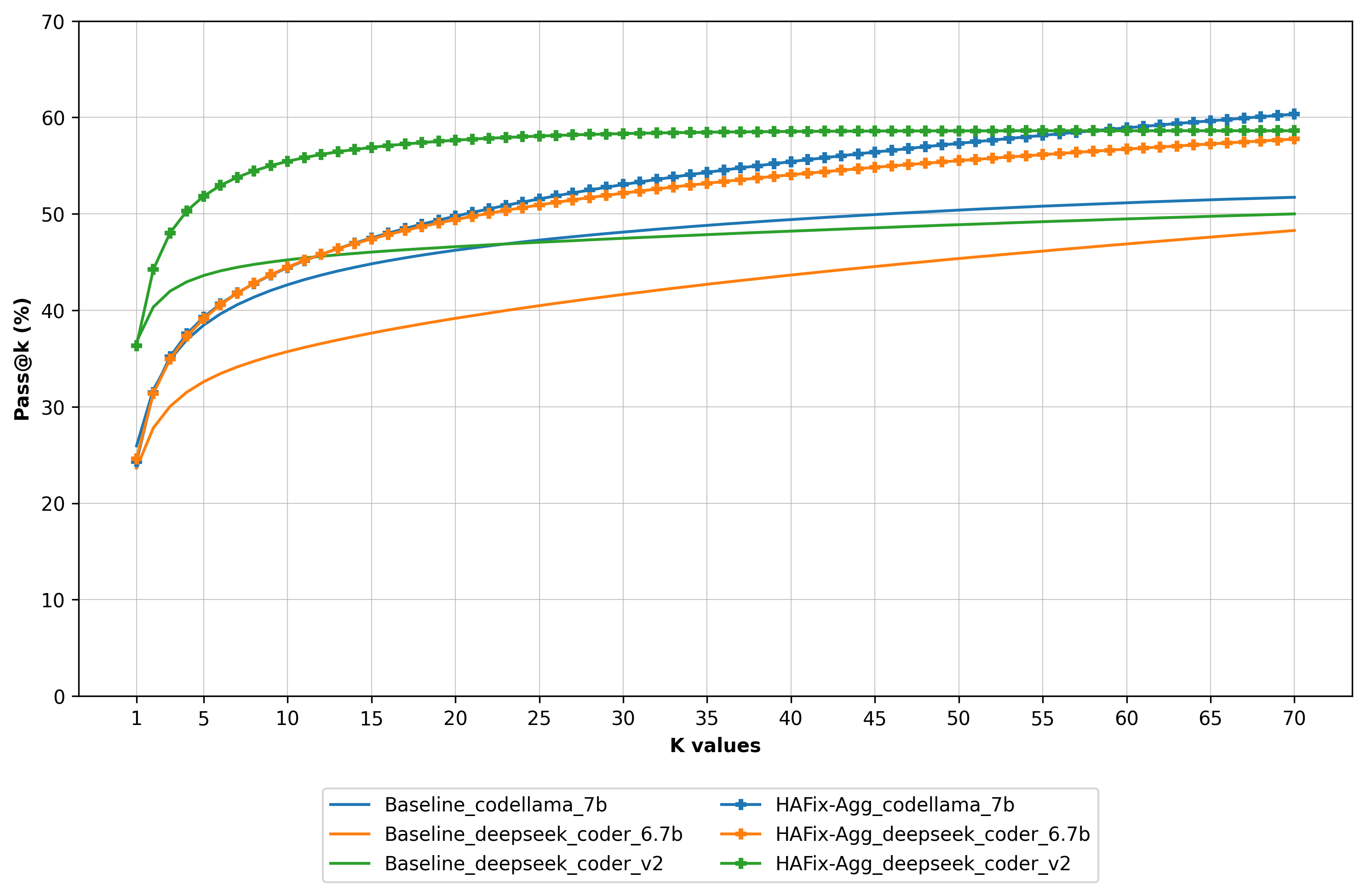}
        \end{center}
        \caption{Pass@k comparison of Baseline and HAFix-Agg (n=70) on Defects4J across three models.}
        \label{fig:rq1_passk_baseline_hafix_2_datasets_3_models_Defects4J}
\end{figure}

\begin{table}[!htbp]
  \centering
  \begin{threeparttable}
  \small
  \caption{Pairwise comparisons of baselines and HAFix-Agg on Defects4J across three models using Wilcoxon signed-rank test. Values shown are $p$-values. A Bonferroni-corrected significance threshold of $\alpha = 0.0033$ ($0.05/15$) is applied.}
  \label{tab:rq1_wilcoxon_baseline_vs_hafix_agg_Defects4J}
  \begin{tabular}{l c c c c c}
    \toprule
    & \textbf{B-CL} & \textbf{B-DS} & \textbf{B-DSV2} & \textbf{H-CL} & \textbf{H-DS} \\
    \midrule
    \textbf{B-DS}     & $3.6\e{-13}$ & - & - & - & - \\
    \textbf{B-DSV2}   & 0.094 & $3.6\e{-13}$ & - & - & - \\
    \textbf{H-CL}      & $5.6\e{-13}$ & $3.6\e{-13}$ & $6.9\e{-9}$ & - & - \\
    \textbf{H-DS}     & $5.6\e{-13}$ & $3.6\e{-13}$ & $3.9\e{-8}$ & $2.3\e{-12}$ & - \\
    \textbf{H-DSV2}   & $3.6\e{-13}$ & $3.6\e{-13}$ & $3.8\e{-13}$ & $1.8\e{-10}$ & $3.6\e{-13}$ \\
    \bottomrule
  \end{tabular}
  \begin{tablenotes}[flushleft]
    \small
    \item \textbf{Abbreviations:}
    \begin{tabularx}{\textwidth}{@{} l@{\hspace{0.6em}}l @{\hskip 1em} l@{\hspace{0.6em}}l @{}}
    B-CL    & \mbox{Baseline\_codellama\_7b}             & H-CL    & \mbox{HAFix-Agg\_codellama\_7b} \\
    B-DS    & \mbox{Baseline\_deepseek\_coder\_6.7b}     & H-DS    & \mbox{HAFix-Agg\_deepseek\_coder\_6.7b} \\
    B-DSV2  & \mbox{Baseline\_deepseek\_coder\_v2}       & H-DSV2  & \mbox{HAFix-Agg\_deepseek\_coder\_v2} \\
    \end{tabularx}
  \end{tablenotes}
  \end{threeparttable}
\end{table}

To further assess HAFix-Agg’s performance relative to the baseline in terms of Pass@k, we conducted a dedicated experiment with expanded sample sizes. HAFix-Agg's results were obtained by combining seven heuristics, each generating 10 samples per bug, resulting in 70 samples per bug. To ensure a fair comparison, the baseline was executed seven times with identical configurations, also generating 70 samples per bug. This setup provided a consistent basis for calculating Pass@k across an expanded range of k values. 
\textbf{HAFix-Agg not only significantly outperforms the corresponding baseline across all model-dataset configurations when $n = 70$ but also demonstrates nuanced strengths that vary with k and model type.} Figures \ref{fig:rq1_passk_baseline_hafix_2_datasets_3_models_BugsInPy} and \ref{fig:rq1_passk_baseline_hafix_2_datasets_3_models_Defects4J} illustrate the Pass@k trends for the baseline and HAFix-Agg on BugsInPy and Defects4J, respectively, with k increasing in steps of 1. Friedman tests revealed significant differences among configurations for both datasets ($p < 2.2 \times 10^{-16}$). Post-hoc Wilcoxon signed-rank tests confirmed that HAFix-Agg significantly outperforms the baseline across all models, as shown in Tables \ref{tab:rq1_wilcoxon_baseline_vs_hafix_agg_BugsInPy} for BugsInPy and \ref{tab:rq1_wilcoxon_baseline_vs_hafix_agg_Defects4J} for Defects4J. On BugsInPy, the p-values for CodeLlama-Instruct-7B, DeepSeek-Coder-Instruct-6.7B, and DeepSeek-Coder-V2-Lite-Instruct-16B were $8.2 \times 10^{-13}$, $4.3 \times 10^{-13}$, and $5.8 \times 10^{-13}$, with corresponding large effect sizes of 0.98, 1.00, and 0.99. On Defects4J, the p-values were $5.6 \times 10^{-13}$, $3.6 \times 10^{-13}$, and $3.8 \times 10^{-13}$, with large effect sizes of 0.99, 1.00, and 1.00. After applying Bonferroni correction ($\alpha = 0.0033$), all comparisons remained statistically significant.
Interestingly, model-specific trends emerge as k increases. On BugsInPy (Figure \ref{fig:rq1_passk_baseline_hafix_2_datasets_3_models_BugsInPy}), HAFix-Agg with CodeLlama-Instruct-7B begins to outperform DeepSeek-Coder-V2-Lite-Instruct-16B after $k > 5$ ($p = 1.4 \times 10^{-12}$, $r_{\mathrm{rb}} = 0.98$ (large)), and surpasses DeepSeek-Coder-Instruct-6.7B after $k \approx 25$ (though not significantly, $p = 0.53$), indicating promising long-tail performance. Conversely, on Defects4J (Figure \ref{fig:rq1_passk_baseline_hafix_2_datasets_3_models_Defects4J}), DeepSeek-Coder-V2-Lite-Instruct-16B achieves significantly better performance than CodeLlama-Instruct-7B across most k values ($p = 1.8 \times 10^{-10}$, $r_{\mathrm{rb}} = 0.89$ (large)), even though CodeLlama briefly overtakes it after $k \approx 60$. Both models consistently outperform DeepSeek-Coder-Instruct-6.7B throughout the k range.

\begin{tcolorbox}
\textbf{Summary for RQ1:}
\begin{enumerate}
    \item Multiple HAFix heuristics (e.g., FN-modified, FN-all, CFN-all, FLN-all) achieve statistically significant improvements with large effect sizes over the baseline, especially on Defects4J with DeepSeek-Coder models, demonstrating the effectiveness of the HAFix approach in incorporating historical context.
    
    \item Each HAFix heuristic provides strong complementary strengths, fixing an average of about 3 unique bugs on BugsInPy and 9 on Defects4J that the baseline misses, even when their overall Pass@k is lower, highlighting diverse strengths across different historical contexts.
    
    \item HAFix-Agg achieves substantial performance gains by combining individual heuristics, improving bug-fixing rates by an average of 45.05\% on BugsInPy and 49.92\% on Defects4J, while fixing nearly all bugs solved by the baseline, plus significant additional unique fixes.
    
    \item Statistical analysis with expanded sample sizes ($n = 70$) confirms HAFix-Agg significantly outperforms the baseline with a large effect size across all model-dataset configurations.
\end{enumerate}
\end{tcolorbox}

\subsection{RQ2: How Do Different Prompt Styles Impact the Bug-Fixing Performance of History-Augmented LLMs?}\label{RQ2}

\subsubsection{Motivation}\label{RQ2Motivation}

The effectiveness of LLMs can vary significantly depending on the structure and presentation of prompts \citep{xia2022less}. This research question aims to investigate how different prompt styles influence the bug-fixing performance of the individual and aggregated HAFix variant HAFix-Agg. The primary motivation is to explore the potential for optimizing LLM performance by refining prompt structures to enhance bug-fixing outcomes. We aim to investigate how varying prompt styles affect the performance of HAFix and HAFix-Agg, the most effective approach identified in RQ1. By analyzing the impact of these styles, we seek to identify the most effective approach for fixing a greater number of bugs.

\subsubsection{Approach}\label{RQ2Approach}

We utilize our baseline alongside the most promising bug-fixing approaches identified in RQ1, HAFix-Agg, combined with three prompt styles: \Instruction, \InstructionLabel, and \InstructionInfill. These prompt styles differ in specificity and context structure, providing potential to enhance the bug-fixing performance of our approach. Specifically, the \Instruction prompt explicitly mentions the buggy line after showing the entire context snippet of the line, the \InstructionLabel prompt indicates the buggy line within the snippet using labels, and the \InstructionInfill prompt replaces the buggy line by a placeholder, asking the model to fill in the missing code.

Since we already obtained the \Instruction prompt results in RQ1, here we experimented with the \InstructionLabel and \InstructionInfill prompts using nucleus sampling to identify the most effective prompt style for the baseline, and HAFix-Agg approaches, then compared to the RQ1 \Instruction prompt results. For HAFix-Agg, we aggregated the 70 samples generated from the seven individual history heuristics (10 samples per heuristic, with each sample representing a single inference result from the LLM). Finally, we determined the best-performing prompt style for each approach and compared it with the baseline and HAFix-Agg to identify the most effective prompt style.

To analyze the impact of different prompt styles on the baseline and HAFix-Agg, we refer to the combination of an approach (baseline or HAFix-Agg) with a specific prompt style (\Instruction, \InstructionLabel, or \InstructionInfill) as a configuration. Using the Pass@k metric, we evaluated and compared the performance of these configurations. Below, we present the results for each configuration, beginning with the baseline.

\subsubsection{Results}\label{RQ2Result}

\begin{figure}[!htbp]
  \centering
  \begin{subfigure}[b]{0.48\textwidth}
    \includegraphics[width=\textwidth]{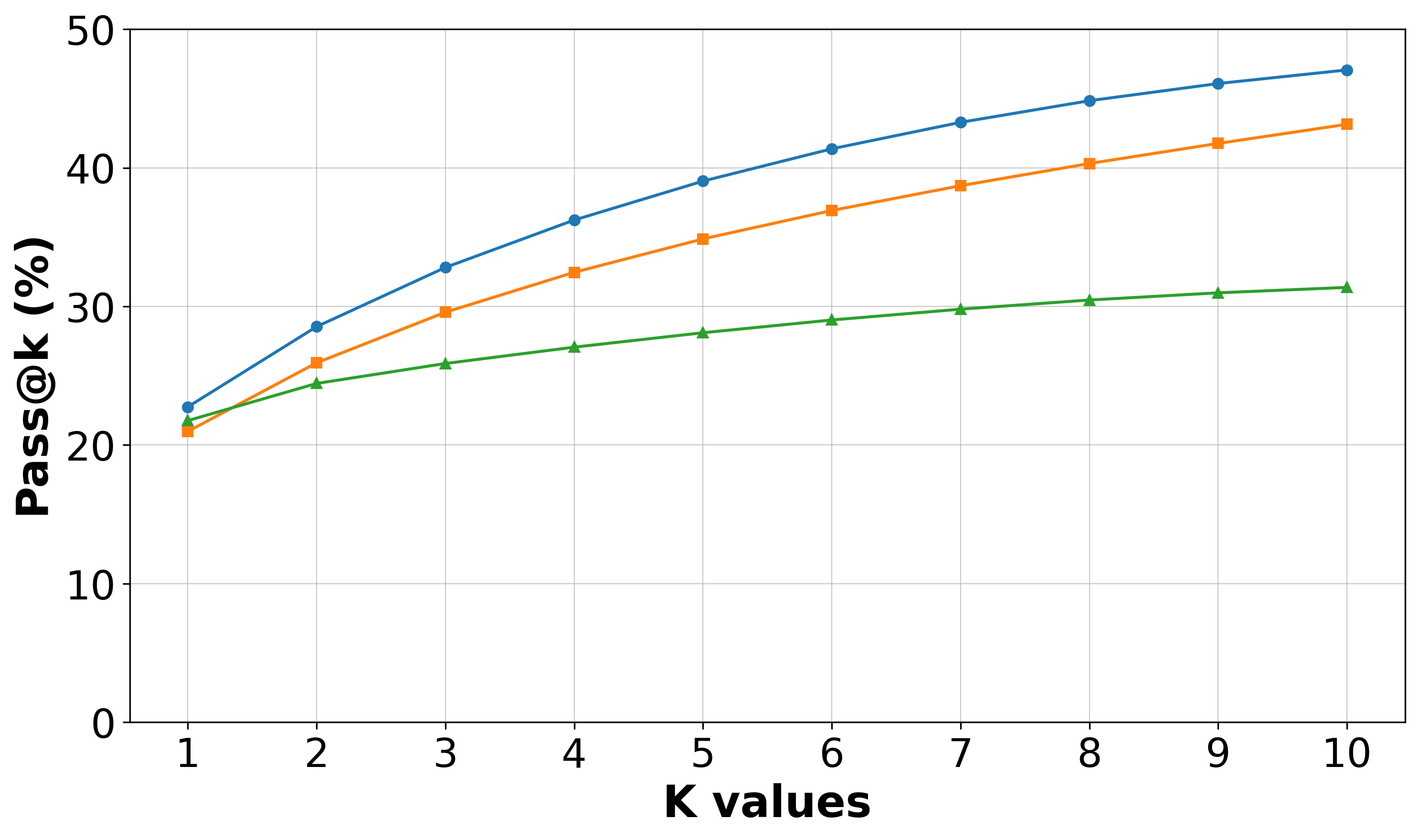}
    \caption{CodeLlama-Instruct-7B on BugsInPy}
    \label{baseline_3_prompts_comparison:a}
  \end{subfigure}
  \hfill
  \begin{subfigure}[b]{0.48\textwidth}
    \includegraphics[width=\textwidth]{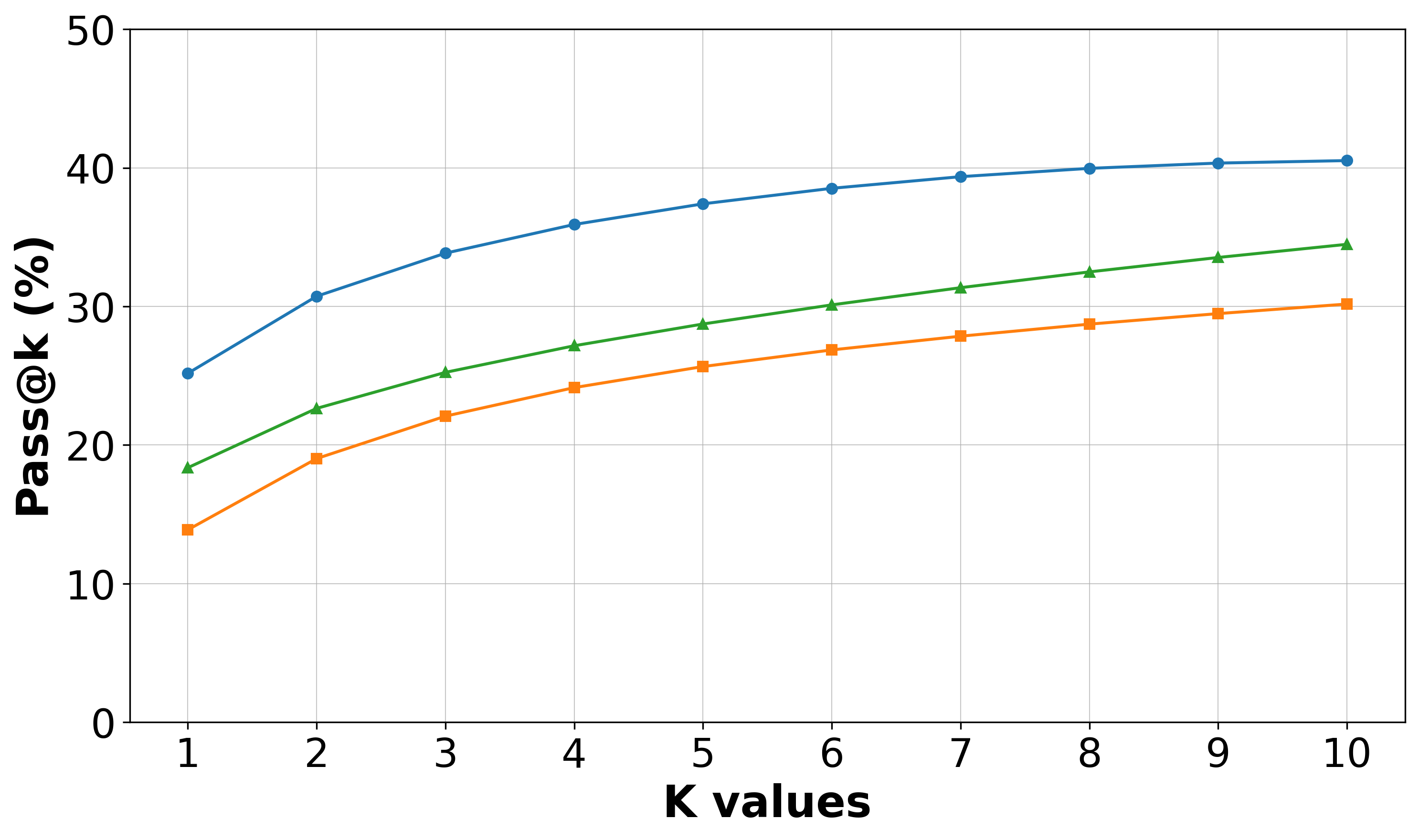}
    \caption{CodeLlama-Instruct-7B on Defects4J}
    \label{baseline_3_prompts_comparison:b}
  \end{subfigure}

  \vspace{1em} 
    
  \begin{subfigure}[b]{0.48\textwidth}
    \includegraphics[width=\textwidth]{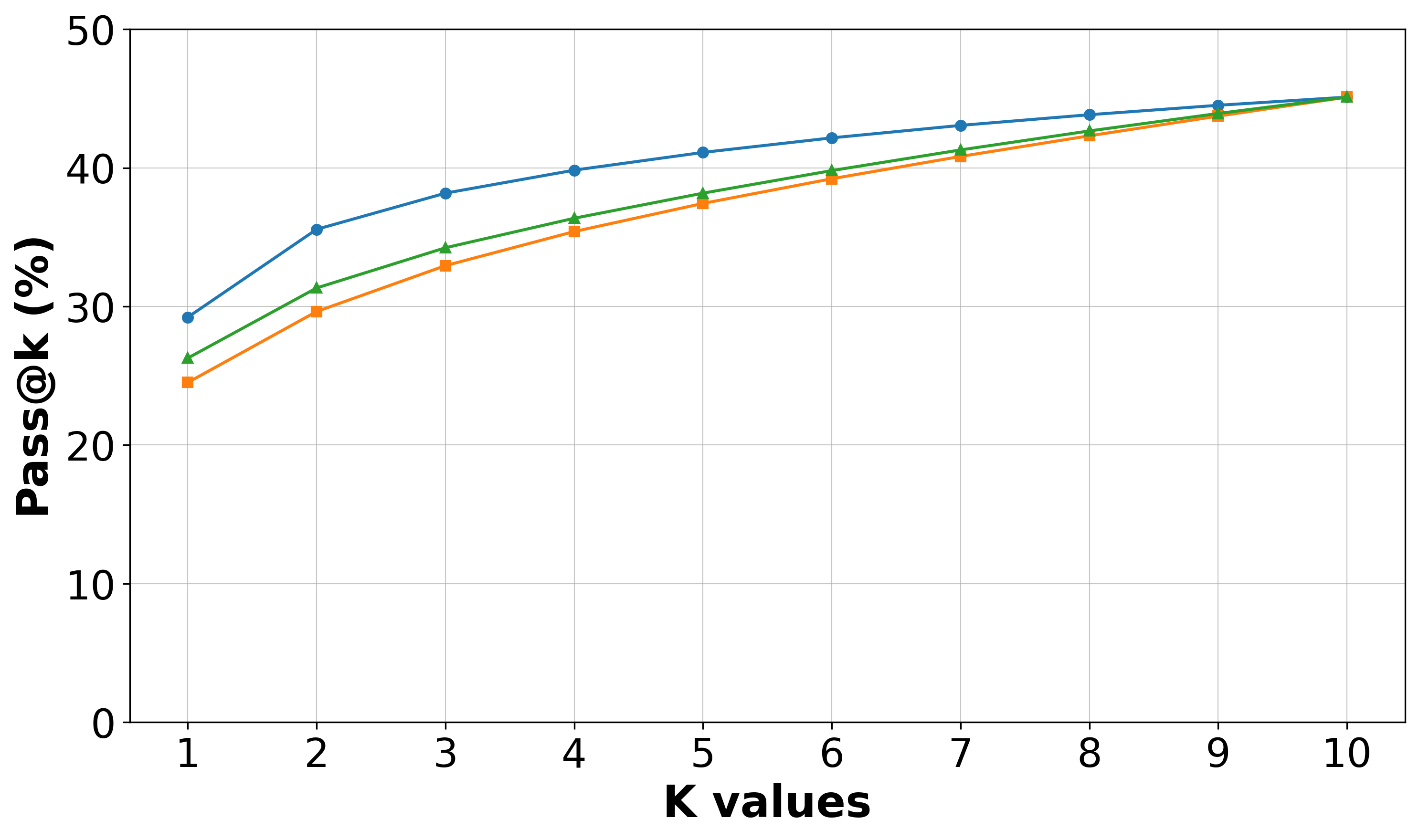}
    \caption{DeepSeek-Coder-Instruct-6.7B on \\BugsInPy}
    \label{baseline_3_prompts_comparison:c}
  \end{subfigure}
  \hfill
  \begin{subfigure}[b]{0.48\textwidth}
    \includegraphics[width=\textwidth]{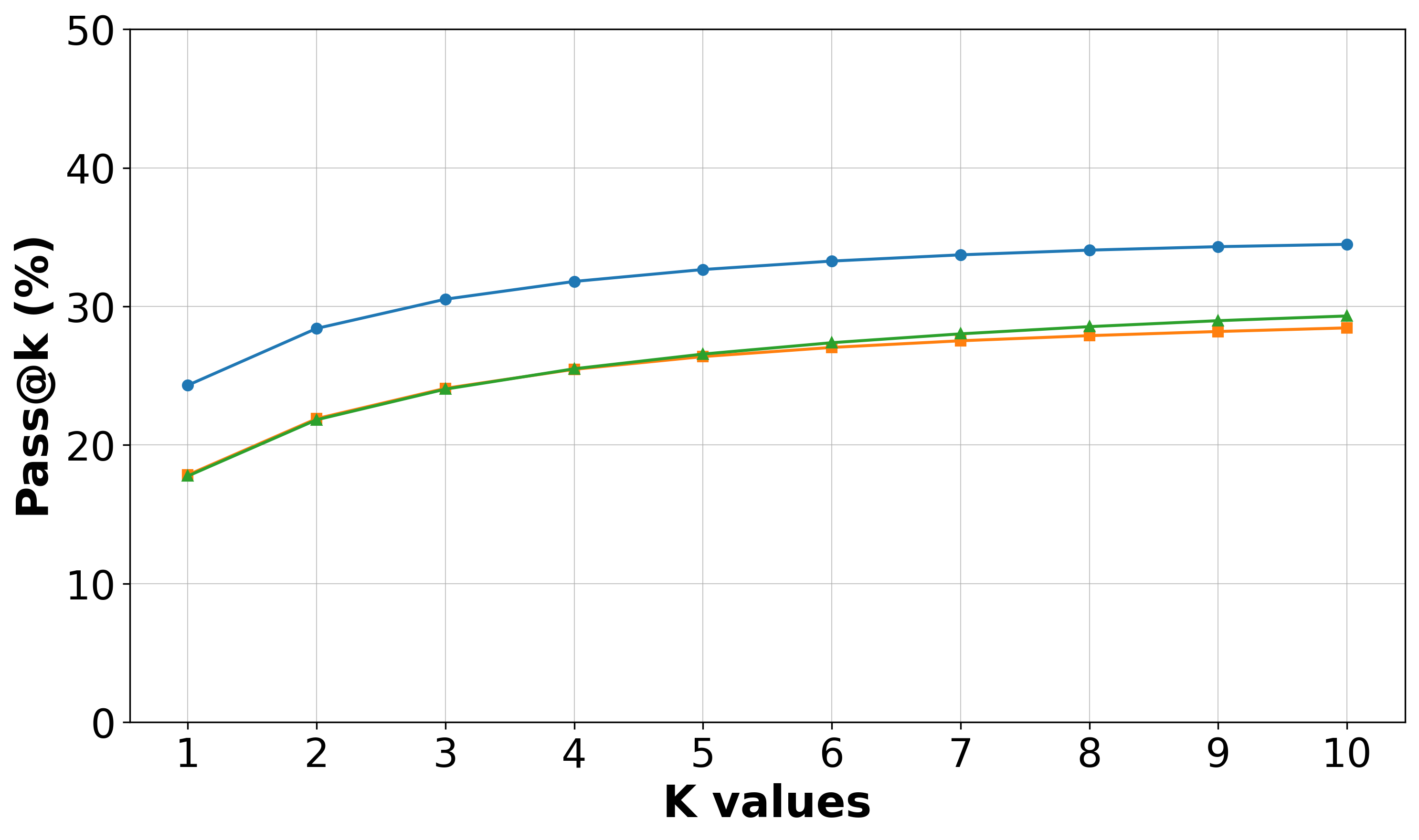}
    \caption{DeepSeek-Coder-Instruct-6.7B on \\Defects4J}
    \label{baseline_3_prompts_comparison:d}
  \end{subfigure}

  \vspace{1em} 
    
  \begin{subfigure}[b]{0.48\textwidth}
    \includegraphics[width=\textwidth]{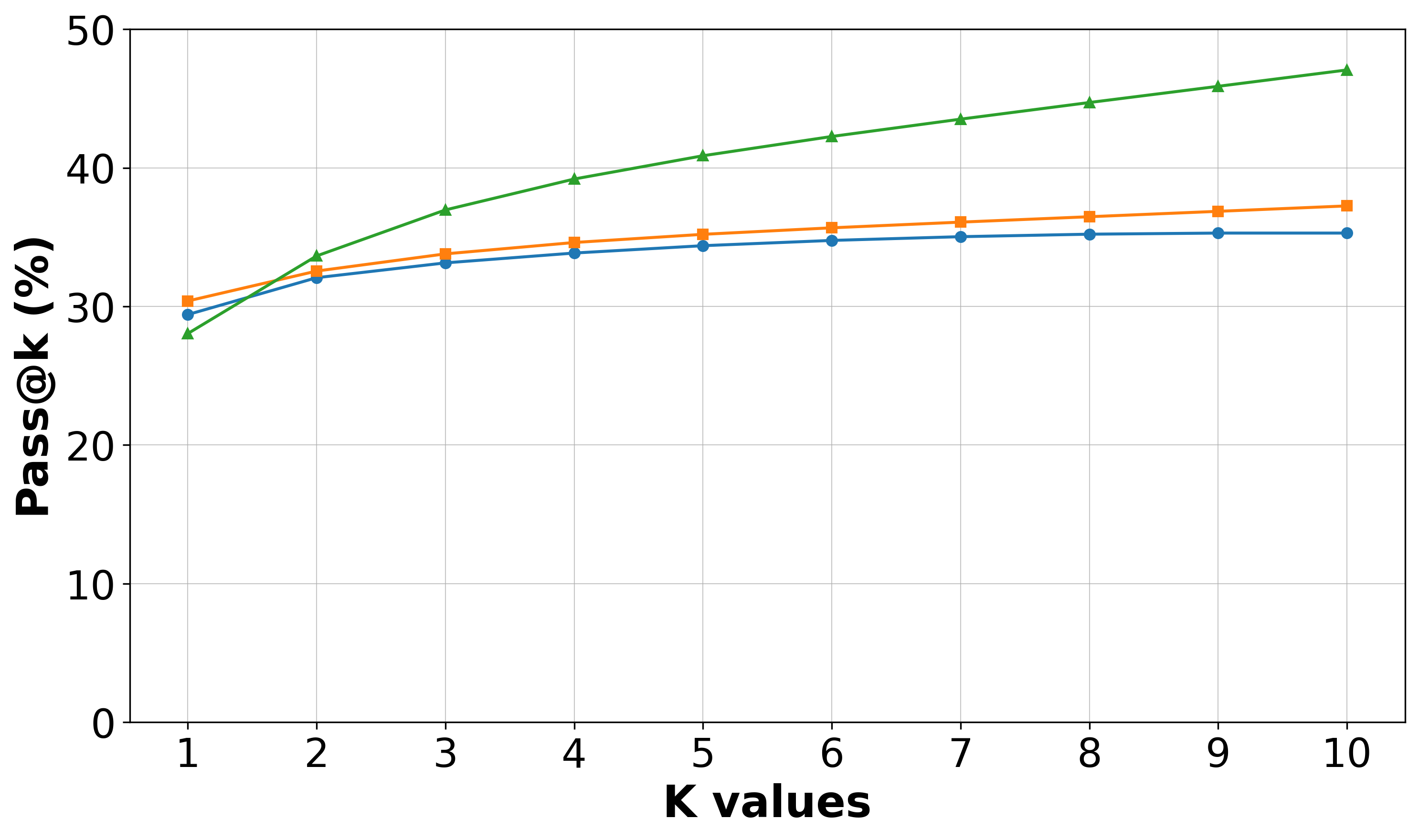}
    \caption{DeepSeek-Coder-V2-Lite-Instruct-16B \\on BugsInPy}
    \label{baseline_3_prompts_comparison:e}
  \end{subfigure}
  \hfill
  \begin{subfigure}[b]{0.48\textwidth}
    \includegraphics[width=\textwidth]{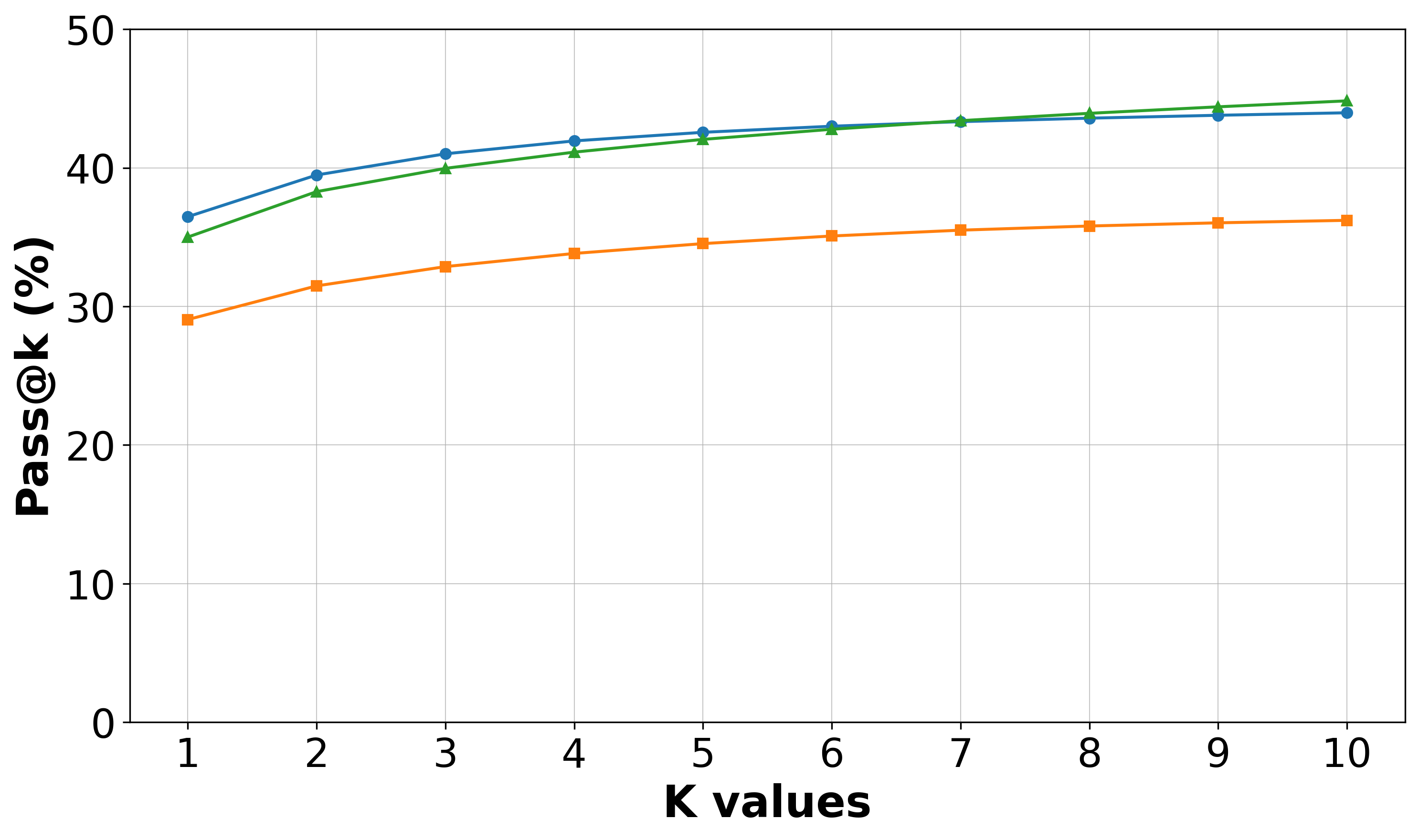}
    \caption{DeepSeek-Coder-V2-Lite-Instruct-16B \\on Defects4J}
    \label{baseline_3_prompts_comparison:f}
  \end{subfigure}
  
  \vspace{1em}
  \begin{adjustbox}{width=0.8\textwidth}
    \includegraphics{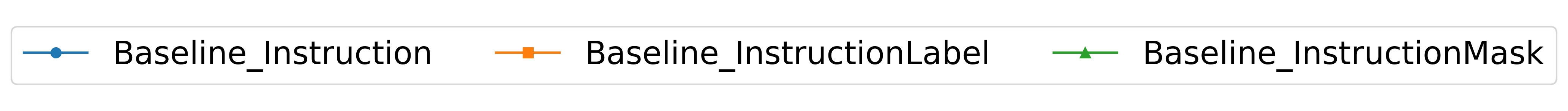}
  \end{adjustbox}

  \caption{Comparison of Pass@k for the Baseline across different prompt styles, evaluated over three models and two datasets (n = 10).}
  \label{fig:baseline_3_prompts_comparison}
\end{figure}

\begin{table}[!htbp]
  \centering
  \caption{Pairwise comparisons of baseline across different prompt styles using Wilcoxon signed-rank test over three models and two datasets. Instr, InstrLabel, and InstrMask correspond to the \Instruction, \InstructionLabel and \InstructionInfill prompt styles, respectively. Each cell reports the $p$-value and the corresponding effect size $r_{\mathrm{rb}}$ (Rank-Biserial Correlation). A Bonferroni-corrected significance threshold of $\alpha = 0.0167$ ($0.05/3$) is applied for pairwise comparisons.}
  \label{tab:wilcox_rq2_baseline_3_prompts_comparison}

  \begin{subtable}[t]{0.48\textwidth}
    \centering
    \caption{CodeLlama-Instruct-7B on BugsInPy}
    \label{wilcox_rq2_baseline_3_prompts_comparison:a}
    \begin{tabular}{lcc}
      \toprule
      & Instr & InstrLabel \\
      \midrule
      InstrLabel &
      \makecell{0.0020 \\ $r_{\mathrm{rb}} = 1.00$} & - \\
      \midrule
      InstrMask &
      \makecell{0.0020 \\ $r_{\mathrm{rb}} = 1.00$} & \makecell{0.0039 \\ $r_{\mathrm{rb}} = 0.96$} \\
      \bottomrule
    \end{tabular}
  \end{subtable}
  \hfill
  \begin{subtable}[t]{0.48\textwidth}
    \centering
    \caption{CodeLlama-Instruct-7B on Defects4J}
    \label{wilcox_rq2_baseline_3_prompts_comparison:b}
    \begin{tabular}{lcc}
      \toprule
      & Instr & InstrLabel \\
      \midrule
      InstrLabel &
      \makecell{0.0020 \\ $r_{\mathrm{rb}} = 1.00$} & - \\
      \midrule
      InstrMask &
      \makecell{0.0059 \\ $r_{\mathrm{rb}} = 1.00$} & \makecell{0.0020 \\ $r_{\mathrm{rb}} = -1.00$} \\
      \bottomrule
    \end{tabular}
  \end{subtable}

\vspace{2em}

  \begin{subtable}[t]{0.48\textwidth}
    \centering
    \caption{DeepSeek-Coder-Instruct-6.7B on \\BugsInPy}
    \label{wilcox_rq2_baseline_3_prompts_comparison:c}
    \begin{tabular}{lcc}
      \toprule
      & Instr & InstrLabel \\
      \midrule
      InstrLabel &
      \makecell{0.0092 \\ $r_{\mathrm{rb}} = 0.64$} & - \\
      \midrule
      InstrMask &
      \makecell{0.0092 \\ $r_{\mathrm{rb}} = 0.64$} & \makecell{0.0092 \\ $r_{\mathrm{rb}} = -1.00$} \\
      \bottomrule
    \end{tabular}
  \end{subtable}
  \hfill
  \begin{subtable}[t]{0.48\textwidth}
    \centering
    \caption{DeepSeek-Coder-Instruct-6.7B on \\Defects4J}
    \label{wilcox_rq2_baseline_3_prompts_comparison:d}
    \begin{tabular}{lcc}
      \toprule
      & Instr & InstrLabel \\
      \midrule
      InstrLabel &
      \makecell{0.0020 \\ $r_{\mathrm{rb}} = 1.00$} & - \\
      \midrule
      InstrMask &
      \makecell{0.0020 \\ $r_{\mathrm{rb}} = 1.00$} & \makecell{0.0645 \\ $r_{\mathrm{rb}} = -0.67$} \\
      \bottomrule
    \end{tabular}
  \end{subtable}

\vspace{2em}

  \begin{subtable}[t]{0.48\textwidth}
    \centering
    \caption{DeepSeek-Coder-V2-Lite-Instruct-16B \\on BugsInPy}
    \label{wilcox_rq2_baseline_3_prompts_comparison:e}
    \begin{tabular}{lcc}
      \toprule
      & Instr & InstrLabel \\
      \midrule
      InstrLabel &
      \makecell{0.0020 \\ $r_{\mathrm{rb}} = -1.00$} & - \\
      \midrule
      InstrMask &
      \makecell{0.0039 \\ $r_{\mathrm{rb}} = -0.96$} & \makecell{0.0059 \\ $r_{\mathrm{rb}} = -0.93$} \\
      \bottomrule
    \end{tabular}
  \end{subtable}
  \hfill
  \begin{subtable}[t]{0.48\textwidth}
    \centering
    \caption{DeepSeek-Coder-V2-Lite-Instruct-16B \\on Defects4J}
    \label{wilcox_rq2_baseline_3_prompts_comparison:f}
    \begin{tabular}{lcc}
      \toprule
      & Instr & InstrLabel \\
      \midrule
      InstrLabel &
      \makecell{0.0020 \\ $r_{\mathrm{rb}} = 1.00$} & - \\
      \midrule
      InstrMask &
      \makecell{0.2754 \\ $r_{\mathrm{rb}} = 0.42$} & \makecell{0.0020 \\ $r_{\mathrm{rb}} = -1.00$} \\
      \bottomrule
    \end{tabular}
  \end{subtable}

\end{table}


\textbf{The \Instruction prompt style significantly outperforms \InstructionLabel and \InstructionInfill in most baseline configurations}. Figure \ref{fig:baseline_3_prompts_comparison} presents the Pass@k comparison for the baseline across three prompt styles, evaluated over three models and two datasets ($n = 10$). For CodeLlama-Instruct-7B and DeepSeek-Coder-Instruct-6.7B on both BugsInPy and Defects4J, Figures \subfig{fig:baseline_3_prompts_comparison}{baseline_3_prompts_comparison:a}, \subfig{fig:baseline_3_prompts_comparison}{baseline_3_prompts_comparison:b}, \subfig{fig:baseline_3_prompts_comparison}{baseline_3_prompts_comparison:c} and \subfig{fig:baseline_3_prompts_comparison}{baseline_3_prompts_comparison:d} show that the \Instruction consistently achieves higher Pass@k scores than \InstructionLabel and \InstructionInfill. However, DeepSeek-Coder-V2-Lite-Instruct-16B exhibits different behavior: on BugsInPy (Figure \subfig{fig:baseline_3_prompts_comparison}{baseline_3_prompts_comparison:e}), \InstructionInfill achieves the highest performance among the three prompt styles, while on Defects4J (Figure \subfig{fig:baseline_3_prompts_comparison}{baseline_3_prompts_comparison:f}), both \Instruction and \InstructionInfill outperform \InstructionLabel.

To assess the statistical significance of these differences, we conducted the Friedman test for each model-dataset configuration. All configurations yielded $p$-values $< 0.001$, confirming the presence of significant differences among the three prompt styles. Subsequently, post-hoc pairwise comparisons were performed using the Wilcoxon signed-rank test, with Bonferroni correction applied to account for multiple comparisons ($0.05/3=0.0167$). The detailed pairwise results are presented in Table \ref{tab:wilcox_rq2_baseline_3_prompts_comparison}. In five out of six configurations as shown in Tables \ref{wilcox_rq2_baseline_3_prompts_comparison:a}, \ref{wilcox_rq2_baseline_3_prompts_comparison:b}, \ref{wilcox_rq2_baseline_3_prompts_comparison:c}, \ref{wilcox_rq2_baseline_3_prompts_comparison:d} and \ref{wilcox_rq2_baseline_3_prompts_comparison:f}, \Instruction significantly outperforms the other prompt styles, with $p$-values ranging from 0.0020 to 0.0092 (all below the threshold) and large effect sizes ($r_{\mathrm{rb}} = 1.00$ or $r_{\mathrm{rb}} = 0.64$). Notably, for DeepSeek-Coder-V2-Lite-Instruct-16B on Defects4J (Table \ref{wilcox_rq2_baseline_3_prompts_comparison:f}), both \Instruction and \InstructionInfill perform comparably and both significantly outperform \InstructionLabel. The only configuration where \InstructionInfill significantly outperforms all others is DeepSeek-Coder-V2-Lite-Instruct-16B on BugsInPy (Table \ref{wilcox_rq2_baseline_3_prompts_comparison:e}).
When comparing \InstructionInfill and \InstructionLabel directly, the latter is significantly better only in CodeLlama-Instruct-7B on BugsInPy (Table \ref{wilcox_rq2_baseline_3_prompts_comparison:a}). In contrast, \InstructionInfill significantly outperforms \InstructionLabel in four configurations (Tables \ref{wilcox_rq2_baseline_3_prompts_comparison:b}, \ref{wilcox_rq2_baseline_3_prompts_comparison:c}, \ref{wilcox_rq2_baseline_3_prompts_comparison:e} and \ref{wilcox_rq2_baseline_3_prompts_comparison:f}), with the remaining configuration showing comparable performance (Table \ref{wilcox_rq2_baseline_3_prompts_comparison:d}).


\begin{figure}[!htbp]
  \centering
  \begin{subfigure}[b]{0.48\textwidth}
    \includegraphics[width=\textwidth]{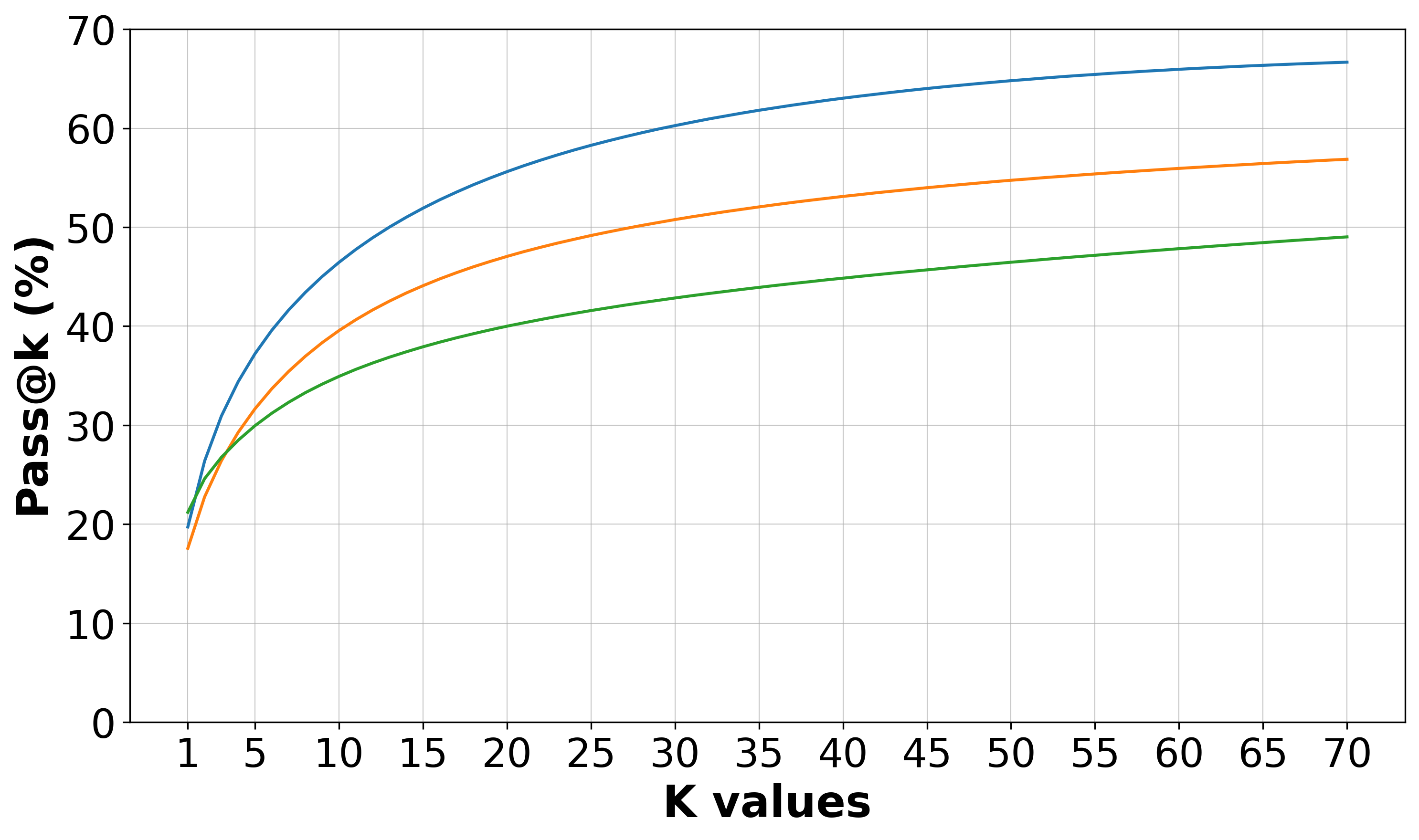}
    \caption{CodeLlama-Instruct-7B on BugsInPy}
    \label{hafix_3_prompts_comparison:a}
  \end{subfigure}
  \hfill
  \begin{subfigure}[b]{0.48\textwidth}
    \includegraphics[width=\textwidth]{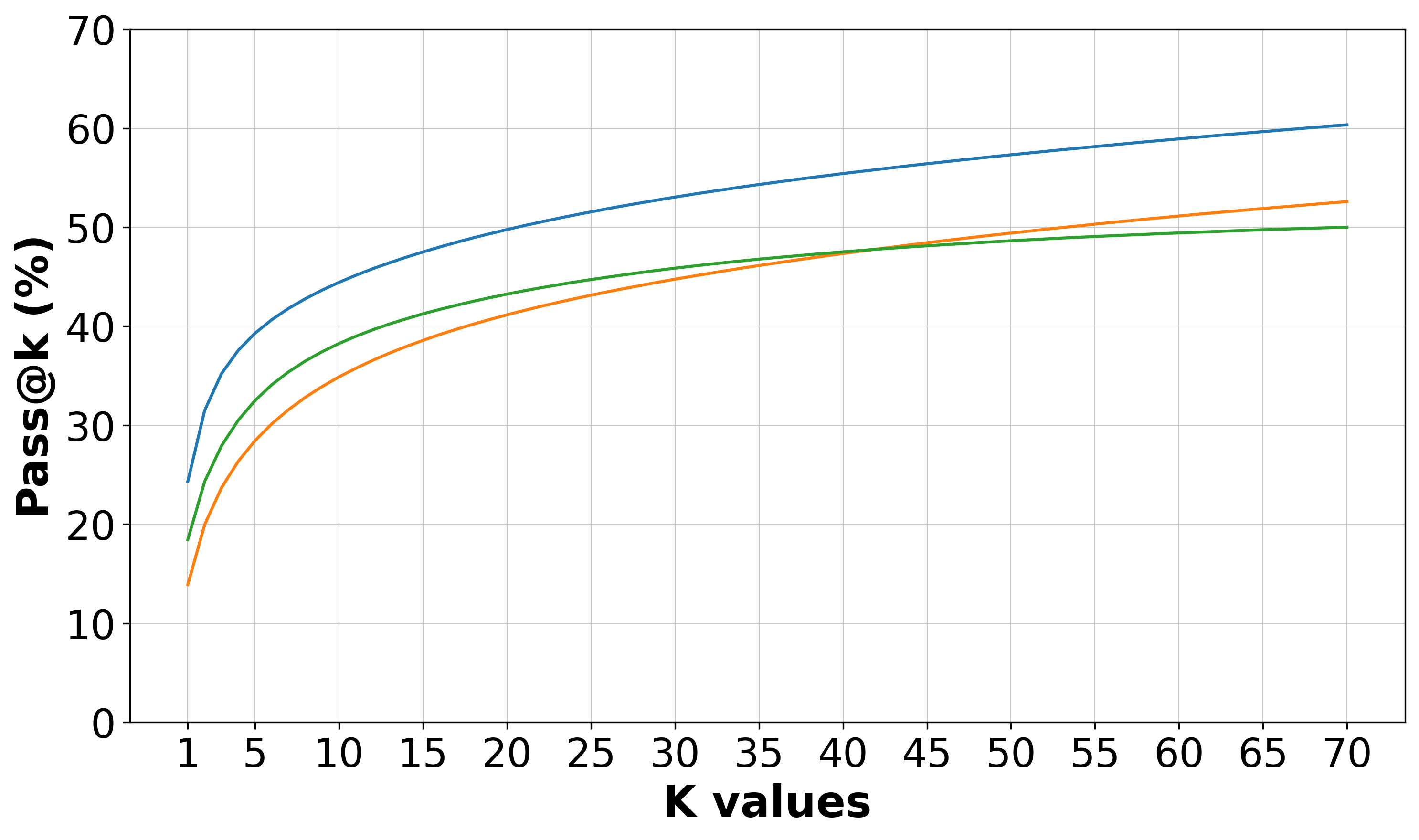}
    \caption{CodeLlama-Instruct-7B on Defects4J}
    \label{hafix_3_prompts_comparison:b}
  \end{subfigure}

  \vspace{1em} 
    
  \begin{subfigure}[b]{0.48\textwidth}
    \includegraphics[width=\textwidth]{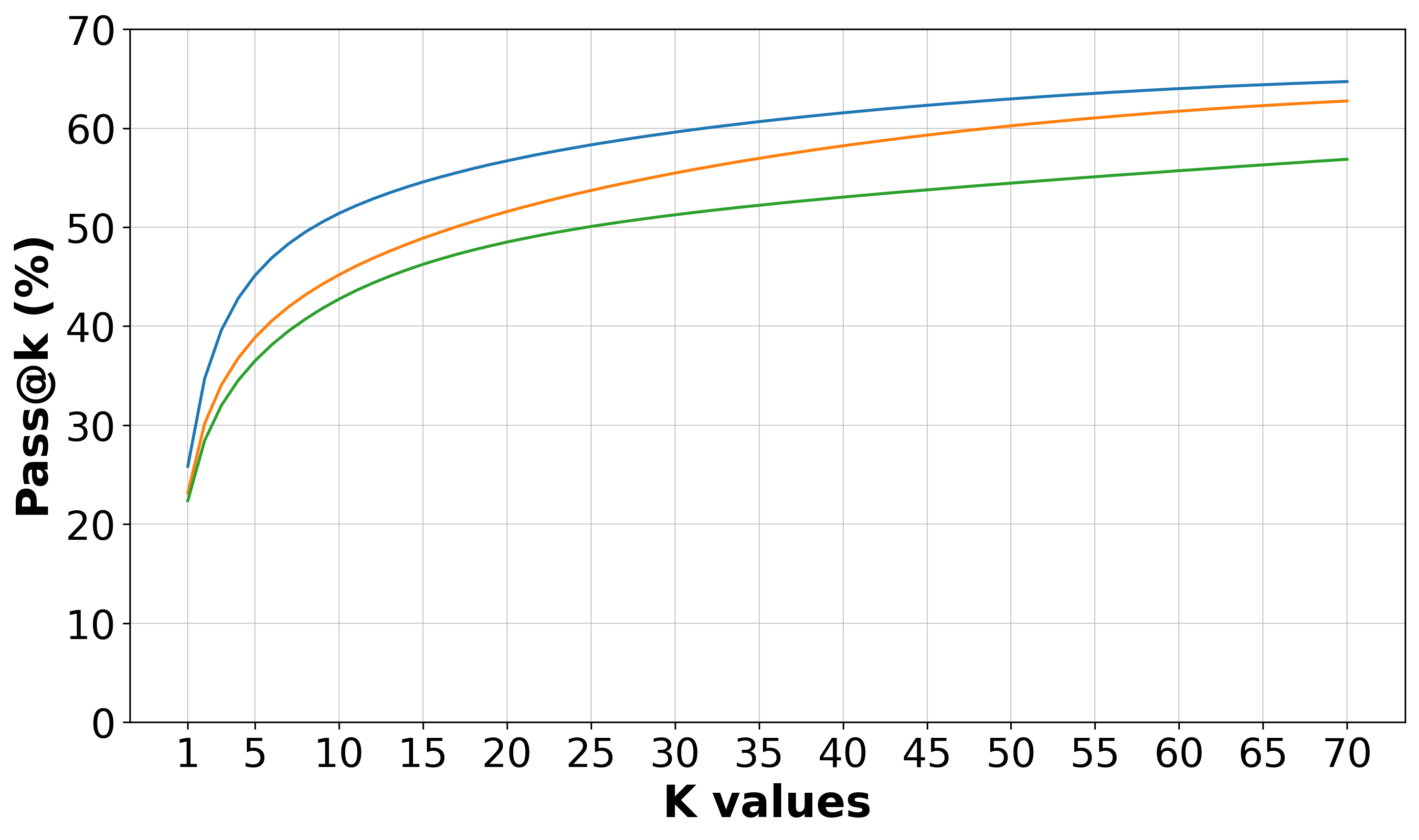}
    \caption{DeepSeek-Coder-Instruct-6.7B on \\BugsInPy}
    \label{hafix_3_prompts_comparison:c}
  \end{subfigure}
  \hfill
  \begin{subfigure}[b]{0.48\textwidth}
    \includegraphics[width=\textwidth]{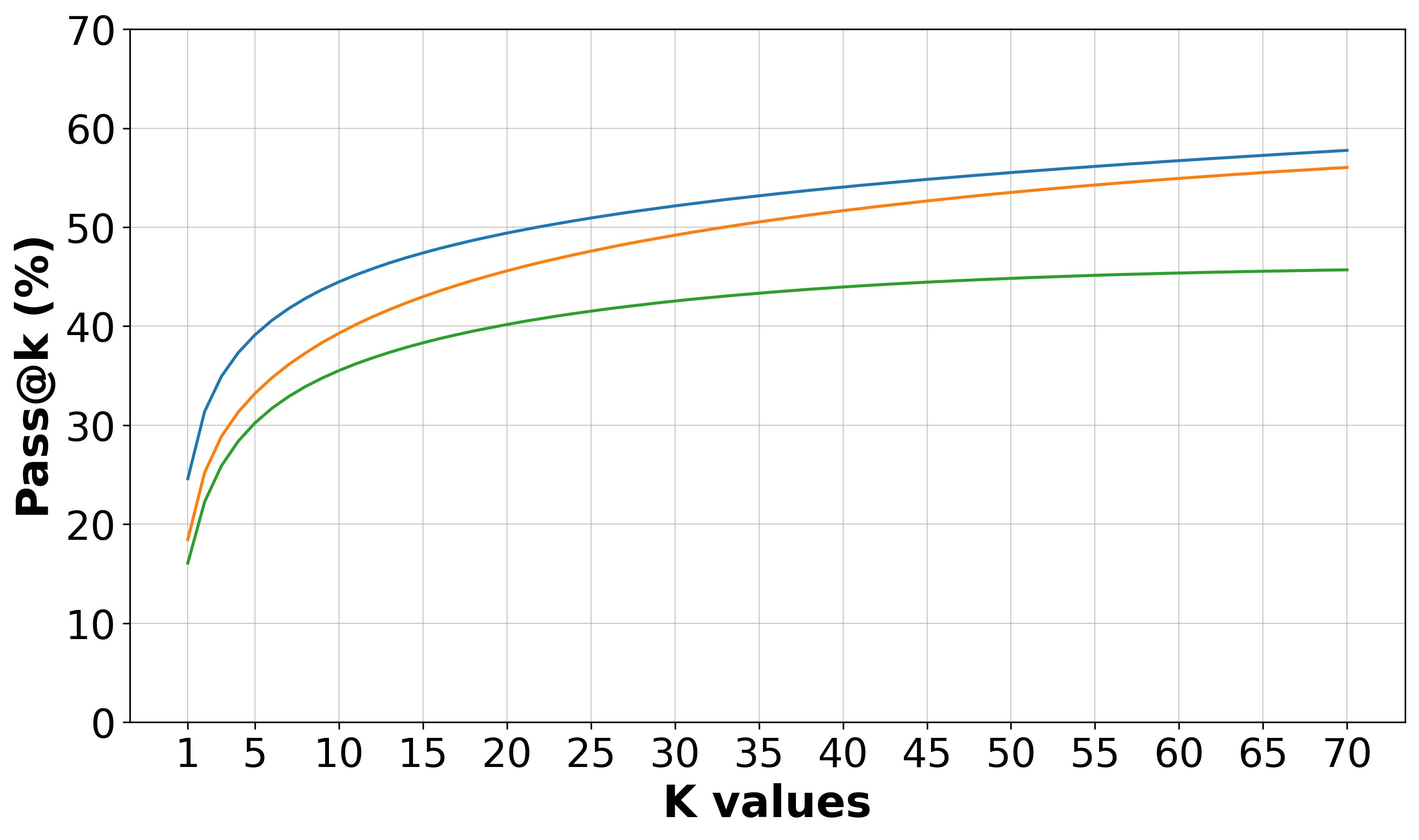}
    \caption{DeepSeek-Coder-Instruct-6.7B on \\Defects4J}
    \label{hafix_3_prompts_comparison:d}
  \end{subfigure}

  \vspace{1em} 
    
  \begin{subfigure}[b]{0.48\textwidth}
    \includegraphics[width=\textwidth]{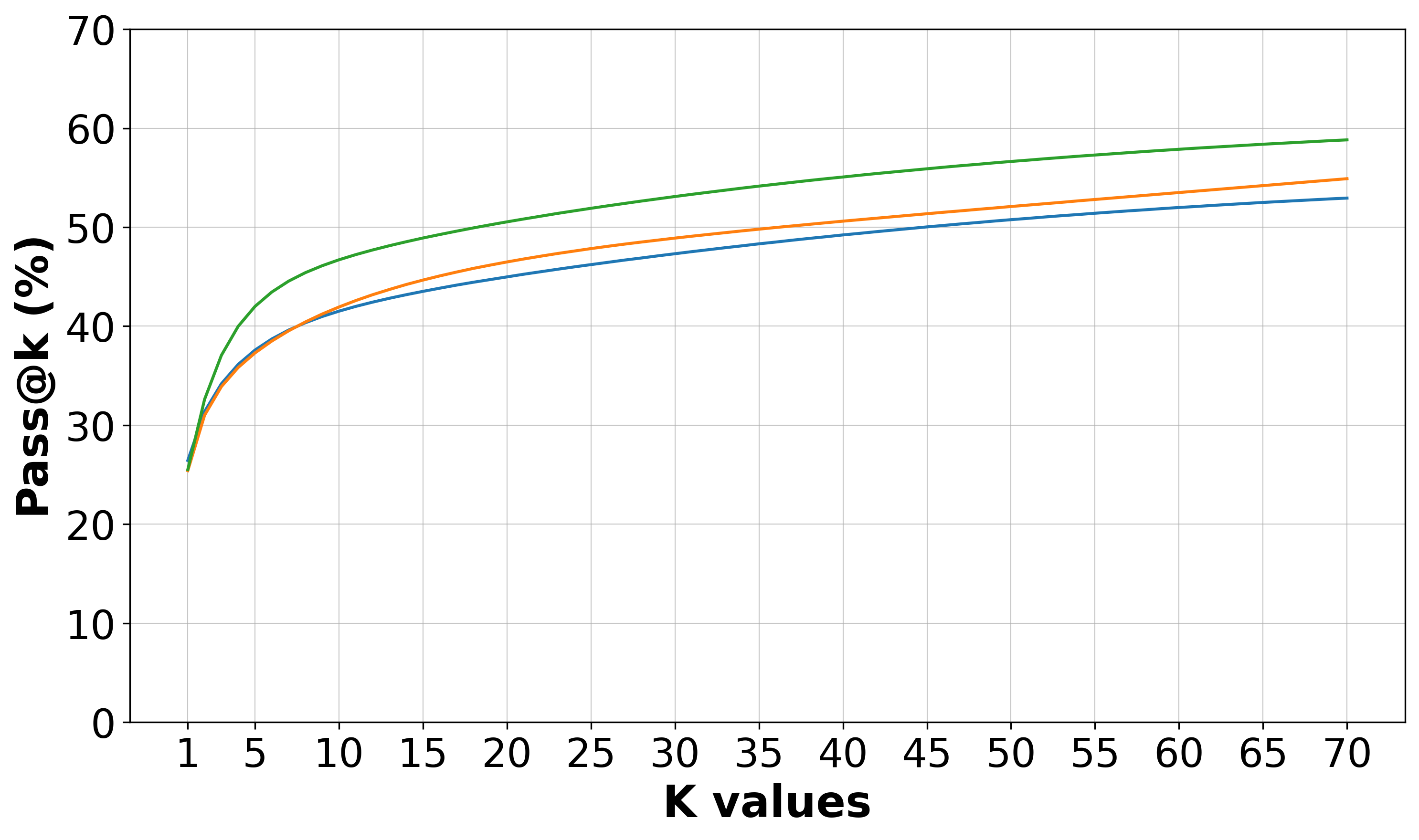}
    \caption{DeepSeek-Coder-V2-Lite-Instruct-16B \\on BugsInPy}
    \label{hafix_3_prompts_comparison:e}
  \end{subfigure}
  \hfill
  \begin{subfigure}[b]{0.48\textwidth}
    \includegraphics[width=\textwidth]{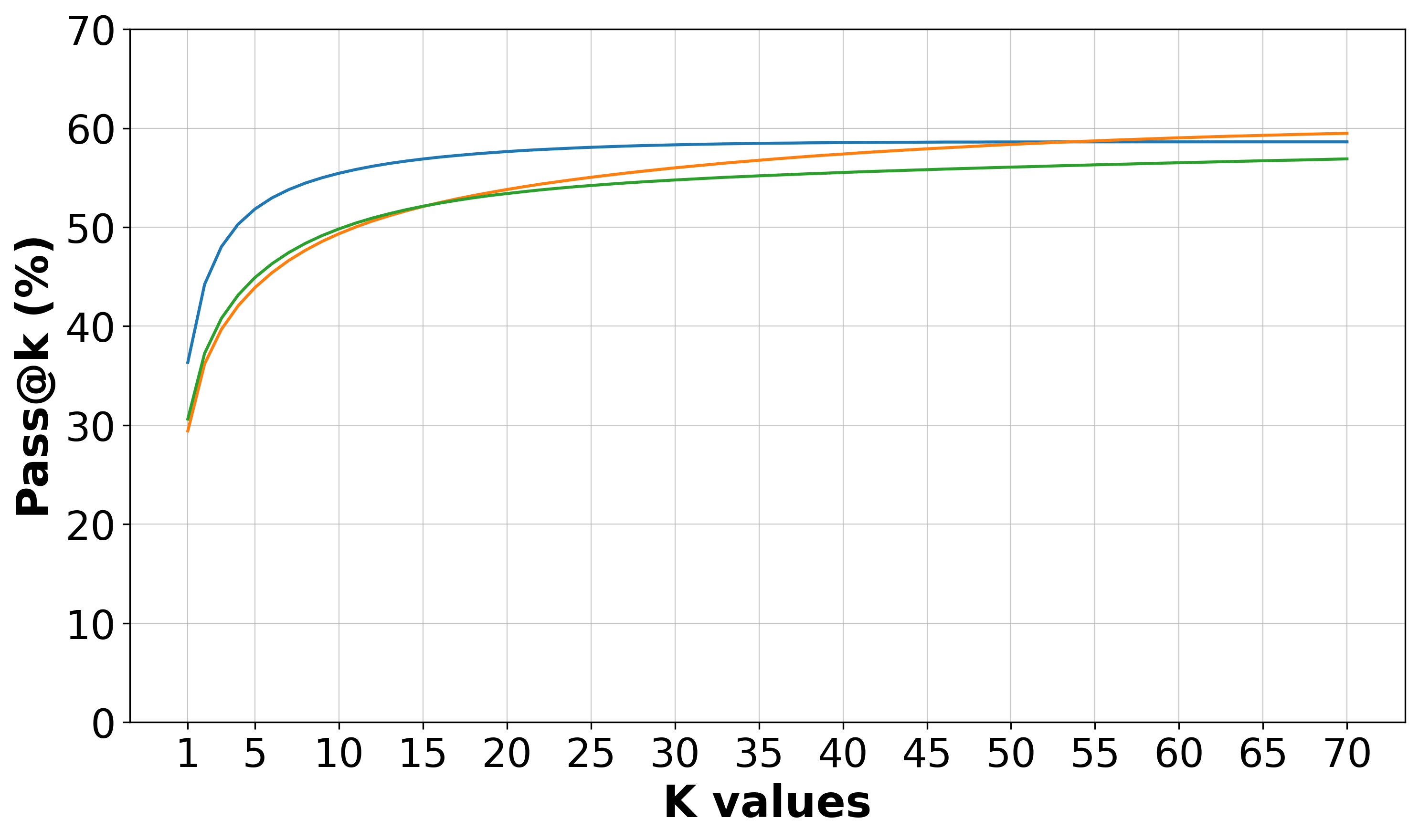}
    \caption{DeepSeek-Coder-V2-Lite-Instruct-16B \\on Defects4J}
    \label{hafix_3_prompts_comparison:f}
  \end{subfigure}

  \vspace{1em}
  \begin{adjustbox}{width=0.8\textwidth}
    \includegraphics{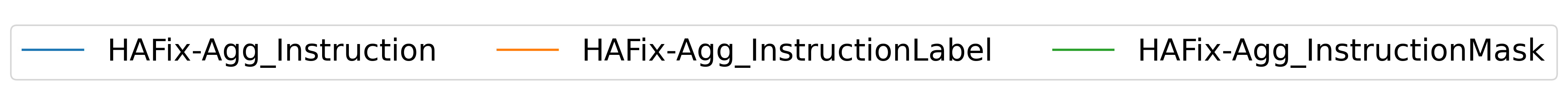}
  \end{adjustbox}
  
  \caption{Comparison of Pass@k for the HAFix-Agg across different prompt styles, evaluated over three models and two datasets (n = 70).}
  \label{fig:hafix_3_prompts_comparison}
\end{figure}

\begin{table}[!htbp]
  \centering
  \caption{Pairwise comparisons of HAFix-Agg across different prompt styles using Wilcoxon signed-rank test over three models and two datasets. Instr, InstrLabel, and InstrMask correspond to the \Instruction, \InstructionLabel and \InstructionInfill prompt styles, respectively. Each cell reports the $p$-value and the corresponding effect size $r_{\mathrm{rb}}$ (Rank-Biserial Correlation). A Bonferroni-corrected significance threshold of $\alpha = 0.0167$ ($0.05/3$) is applied for pairwise comparisons.}
  \label{tab:wilcox_rq2_hafix_agg_3_prompts_comparison}
  \begin{subtable}[t]{0.48\textwidth}
    \centering
    \caption{CodeLlama-Instruct-7B on BugsInPy}
    \label{wilcox_rq2_hafix_agg_3_prompts_comparison:a}
    \begin{tabular}{lcc}
      \toprule
      & Instr & InstrLabel \\
      \midrule
      InstrLabel &
      \makecell{$3.6\e{-13}$ \\ $r_{\mathrm{rb}} = 1.00$} & - \\
      \midrule
      InstrMask &
      \makecell{$3.8\e{-13}$ \\ $r_{\mathrm{rb}} = 1.00$} & \makecell{$6.1\e{-13}$ \\ $r_{\mathrm{rb}} = 0.99$} \\
      \bottomrule
    \end{tabular}
  \end{subtable}
  \hfill
  \begin{subtable}[t]{0.48\textwidth}
    \centering
    \caption{CodeLlama-Instruct-7B on Defects4J}
    \label{wilcox_rq2_hafix_agg_3_prompts_comparison:b}
    \begin{tabular}{lcc}
      \toprule
      & Instr & InstrLabel \\
      \midrule
      InstrLabel &
      \makecell{$3.6\e{-13}$ \\ $r_{\mathrm{rb}} = 1.00$} & - \\
      \midrule
      InstrMask &
      \makecell{$3.6\e{-13}$ \\ $r_{\mathrm{rb}} = 1.00$} & \makecell{$0.0159$ \\ $r_{\mathrm{rb}} = -0.33$} \\
      \bottomrule
    \end{tabular}
  \end{subtable}

\vspace{2em}

  \begin{subtable}[t]{0.48\textwidth}
    \centering
    \caption{DeepSeek-Coder-Instruct-6.7B on \\BugsInPy}
    \label{wilcox_rq2_hafix_agg_3_prompts_comparison:c}
    \begin{tabular}{lcc}
      \toprule
      & Instr & InstrLabel \\
      \midrule
      InstrLabel &
      \makecell{$3.6\e{-13}$ \\ $r_{\mathrm{rb}} = 1.00$} & - \\
      \midrule
      InstrMask &
      \makecell{$3.6\e{-13}$ \\ $r_{\mathrm{rb}} = 1.00$} & \makecell{$3.6\e{-13}$ \\ $r_{\mathrm{rb}} = 1.00$} \\
      \bottomrule
    \end{tabular}
  \end{subtable}
  \hfill
  \begin{subtable}[t]{0.48\textwidth}
    \centering
    \caption{DeepSeek-Coder-Instruct-6.7B on \\Defects4J}
    \label{wilcox_rq2_hafix_agg_3_prompts_comparison:d}
    \begin{tabular}{lcc}
      \toprule
      & Instr & InstrLabel \\
      \midrule
      InstrLabel &
      \makecell{$3.6\e{-13}$ \\ $r_{\mathrm{rb}} = 1.00$} & - \\
      \midrule
      InstrMask &
      \makecell{$3.6\e{-13}$ \\ $r_{\mathrm{rb}} = 1.00$} & \makecell{$3.6\e{-13}$ \\ $r_{\mathrm{rb}} = 1.00$} \\
      \bottomrule
    \end{tabular}
  \end{subtable}

\vspace{2em}

  \begin{subtable}[t]{0.48\textwidth}
    \centering
    \caption{DeepSeek-Coder-V2-Lite-Instruct-16B \\on BugsInPy}
    \label{wilcox_rq2_hafix_agg_3_prompts_comparison:e}
    \begin{tabular}{lcc}
      \toprule
      & Instr & InstrLabel \\
      \midrule
      InstrLabel &
      \makecell{$2.5\e{-12}$ \\ $r_{\mathrm{rb}} = -0.96$} & - \\
      \midrule
      InstrMask &
      \makecell{$3.5\e{-13}$ \\ $r_{\mathrm{rb}} = -1.00$} & \makecell{$3.6\e{-13}$ \\ $r_{\mathrm{rb}} = -1.00$} \\
      \bottomrule
    \end{tabular}
  \end{subtable}
  \hfill
  \begin{subtable}[t]{0.48\textwidth}
    \centering
    \caption{DeepSeek-Coder-V2-Lite-Instruct-16B \\on Defects4J}
    \label{wilcox_rq2_hafix_agg_3_prompts_comparison:f}
    \begin{tabular}{lcc}
      \toprule
      & Instr & InstrLabel \\
      \midrule
      InstrLabel &
      \makecell{$9.4\e{-9}$ \\ $r_{\mathrm{rb}} = 0.79$} & - \\
      \midrule
      InstrMask &
      \makecell{$3.6\e{-13}$ \\ $r_{\mathrm{rb}} = 1.00$} & \makecell{$3.5\e{-9}$ \\ $r_{\mathrm{rb}} = 0.81$} \\
      \bottomrule
    \end{tabular}
  \end{subtable}

\end{table}


\textbf{HAFix-Agg with \Instruction prompt style also outperforms \InstructionLabel and \InstructionInfill in most configurations}. Figure \ref{fig:hafix_3_prompts_comparison} presents the Pass@k comparison for the HAFix-Agg across three prompt styles, evaluated over three models and two datasets ($n = 70$). For CodeLlama-Instruct-7B and DeepSeek-Coder-Instruct-6.7B on both BugsInPy and Defects4J, Figures \subfig{fig:hafix_3_prompts_comparison}{hafix_3_prompts_comparison:a}, \subfig{fig:hafix_3_prompts_comparison}{hafix_3_prompts_comparison:b}, \subfig{fig:hafix_3_prompts_comparison}{hafix_3_prompts_comparison:c} and \subfig{fig:hafix_3_prompts_comparison}{hafix_3_prompts_comparison:d} show that the \Instruction consistently achieves higher Pass@k scores than \InstructionLabel and \InstructionInfill. The same trend is observed for DeepSeek-Coder-V2-Lite-Instruct-16B on Defects4J (Figure \subfig{fig:hafix_3_prompts_comparison}{hafix_3_prompts_comparison:f}). In contrast, DeepSeek-Coder-V2-Lite-Instruct-16B on BugsInPy (Figure \subfig{fig:hafix_3_prompts_comparison}{hafix_3_prompts_comparison:e}) exhibits different behavior: \InstructionInfill achieves the highest performance among the three.

To assess the statistical significance of these differences, we conducted the Friedman test for each model-dataset configuration. All configurations yielded $p$-values $< 2.2 \times 10^{-16}$, confirming the presence of significant differences among the three prompt styles. Subsequently, post-hoc pairwise comparisons were performed using the Wilcoxon signed-rank test, with Bonferroni correction applied to account for multiple comparisons ($0.05/3=0.0167$). The detailed pairwise results are presented in Table \ref{tab:wilcox_rq2_hafix_agg_3_prompts_comparison}. In five out of six configurations as shown in Tables \ref{wilcox_rq2_hafix_agg_3_prompts_comparison:a}, \ref{wilcox_rq2_hafix_agg_3_prompts_comparison:b}, \ref{wilcox_rq2_hafix_agg_3_prompts_comparison:c}, \ref{wilcox_rq2_hafix_agg_3_prompts_comparison:d} and \ref{wilcox_rq2_hafix_agg_3_prompts_comparison:f}, \Instruction significantly outperforms the other prompt styles, with $p$-values ranging from $3.6 \times 10^{-13}$ to $9.4 \times 10^{-9}$ (all below the threshold) and large effect sizes ($r_{\mathrm{rb}} = 1.00$ or $r_{\mathrm{rb}} = 0.79$). The only configuration where \InstructionInfill significantly outperforms all others is DeepSeek-Coder-V2-Lite-Instruct-16B on BugsInPy (Table \ref{wilcox_rq2_hafix_agg_3_prompts_comparison:e}). In the direct comparison between \InstructionInfill and \InstructionLabel directly, the latter is significantly better in four configurations (Tables \ref{wilcox_rq2_hafix_agg_3_prompts_comparison:a}, \ref{wilcox_rq2_hafix_agg_3_prompts_comparison:c}, \ref{wilcox_rq2_hafix_agg_3_prompts_comparison:d} and \ref{wilcox_rq2_hafix_agg_3_prompts_comparison:f}). In contrast, \InstructionInfill significantly outperforms \InstructionLabel in two configurations (Tables \ref{wilcox_rq2_hafix_agg_3_prompts_comparison:b} and \ref{wilcox_rq2_hafix_agg_3_prompts_comparison:e}). These results highlight that the clarity and explicitness of the \Instruction prompt enable the model to leverage historical heuristics more effectively.

\begin{tcolorbox}
\textbf{Summary for RQ2:}
\begin{enumerate}
    \item Across both the baseline and HAFix-Agg, the \Instruction prompt style consistently outperforms \InstructionLabel and \InstructionInfill in the majority of model-dataset configurations, with statistically significant improvements and large effect sizes.

    \item While \Instruction prompt style is generally the most effective, \InstructionInfill shows competitive or superior performance in specific settings, particularly for DeepSeek-Coder-V2-Lite-Instruct-16B on BugsInPy, indicating that different prompt styles may better suit different model-dataset contexts.

    \item These results highlight the importance of prompt design in maximizing the performance of HAFix in enabling models to leverage historical context effectively for bug fixing.
\end{enumerate}
\end{tcolorbox}

\subsection{RQ3: What Is the Cost of History-Augmented LLMs on Bug Fixing?}\label{RQ3}

\subsubsection{Motivation}\label{RQ3Motivation}
While HAFix-Agg with the \Instruction prompt style significantly improves bug-fixing performance, it is essential to examine the trade-offs between performance, inference cost of time and price. Unlike single-heuristic approaches, HAFix-Agg requires the inference results from seven heuristics, each producing 10 samples, which proportionally increases inference time and price. This raises important questions about the practicality of deploying HAFix-Agg at scale. Specifically, RQ3 investigates two dimensions of cost: inference time and inference price. Inference time measures the model's time efficiency, indicating how quickly it can generate results. For inference price, we use the total number of input and output tokens as a proxy for financial cost. Although we use open-source, locally hosted models (incurring no direct monetary cost), this metric allows for comparison across scenarios, as higher token usage typically implies higher cost \citep{hidvegi2024cigar}.

\subsubsection{Approach}\label{RQ3Approach}

We measure cost from two perspectives. First, we report inference time, defined as the elapsed time from submitting a prompt to receiving ten samples. All runs use the same GPU Docker environment and network and are executed in one continuous computation period to reduce external variance.

Second, we use the total number of inference tokens (input + output) as a proxy for monetary cost. Since output tokens are often priced higher than input tokens \citep{OpenAIPricing,AnthropicPricing,DeepSeekPricing}, we also explored a weighted variant, but observed similar trends; therefore, we report the unweighted total token count for simplicity and interpretability.

Finally, HAFix-Agg executes multiple heuristics, so its total cost depends on the heuristic order and whether execution stops once a valid patch is found. We define four scenarios to estimate inference time and tokens; in summary, Exhaustive estimates an upper bound cost, while the three ES variants estimate early-stopping costs under different heuristic orders (ES, ES-AccSorted, and ES-UniSorted). The four scenarios are defined as follows:

\begin{itemize}
    \item \textbf{Exhaustive}: Executes baseline and all heuristic methods for every bug, regardless of order. This means that all seven individual heuristics of HAFix yield a separate inference request, each resulting in ten sample bug fixes. This represents the upper bound in cost.

    \item \textbf{ES} (EarlyStop): Sequentially executes the baseline and individual heuristics in a fixed order (see Table \ref{rq1_acc_percentage}, column \textit{Heuristic}), stopping as soon as one good fix (i.e., passing the test case) has been generated. The execution sequence for all model-dataset configurations is the same: Baseline, CFN-modified, CFN-all, FN-modified, FN-all, FLN-all, FN-pair, FL-diff.

    \item \textbf{ES-AccSorted}: Based on the idea of ES, the order is determined based on the number of bugs fixed by each heuristic for a specific model-dataset configuration (see Table \ref{rq1_acc_percentage}, column \textit{Bugs\#}). Heuristics that fix more bugs earlier are prioritized. For example, on CodeLlama-Instruct-7B with BugsInPy, the execution sequence is: CFN-modified, Baseline, CFN-all, FLN-all, FN-modified, FN-all, FL-diff, FN-pair. Other model-dataset configurations are similar to this.
    
    \item \textbf{ES-UniSorted}: Also based on the idea of ES, but prioritizes heuristics by the number of uniquely fixed bugs compared to the baseline (see Table \ref{rq1_acc_percentage}, column \textit{BugsU\#}), with the baseline always executed first. For example, on CodeLlama-Instruct-7B with BugsInPy, the execution sequence is: Baseline, CFN-modified, FN-all, FN-modified, CFN-all, FLN-all, FN-pair, FL-diff. Other model-dataset configurations are similar to this.
\end{itemize}

We apply these four scenarios to calculate the inference time and tokens, aiming to identify any trade-offs between performance, inference time, and inference tokens across the different scenarios. This RQ focuses exclusively on HAFix in the \Instruction prompt style across three models and two datasets, as they demonstrate the best performance overall.

\subsubsection{Results}\label{RQ3Result}


\begin{figure}[!htbp]
  \centering
  \begin{subfigure}[b]{0.48\textwidth}
    \includegraphics[width=\textwidth]{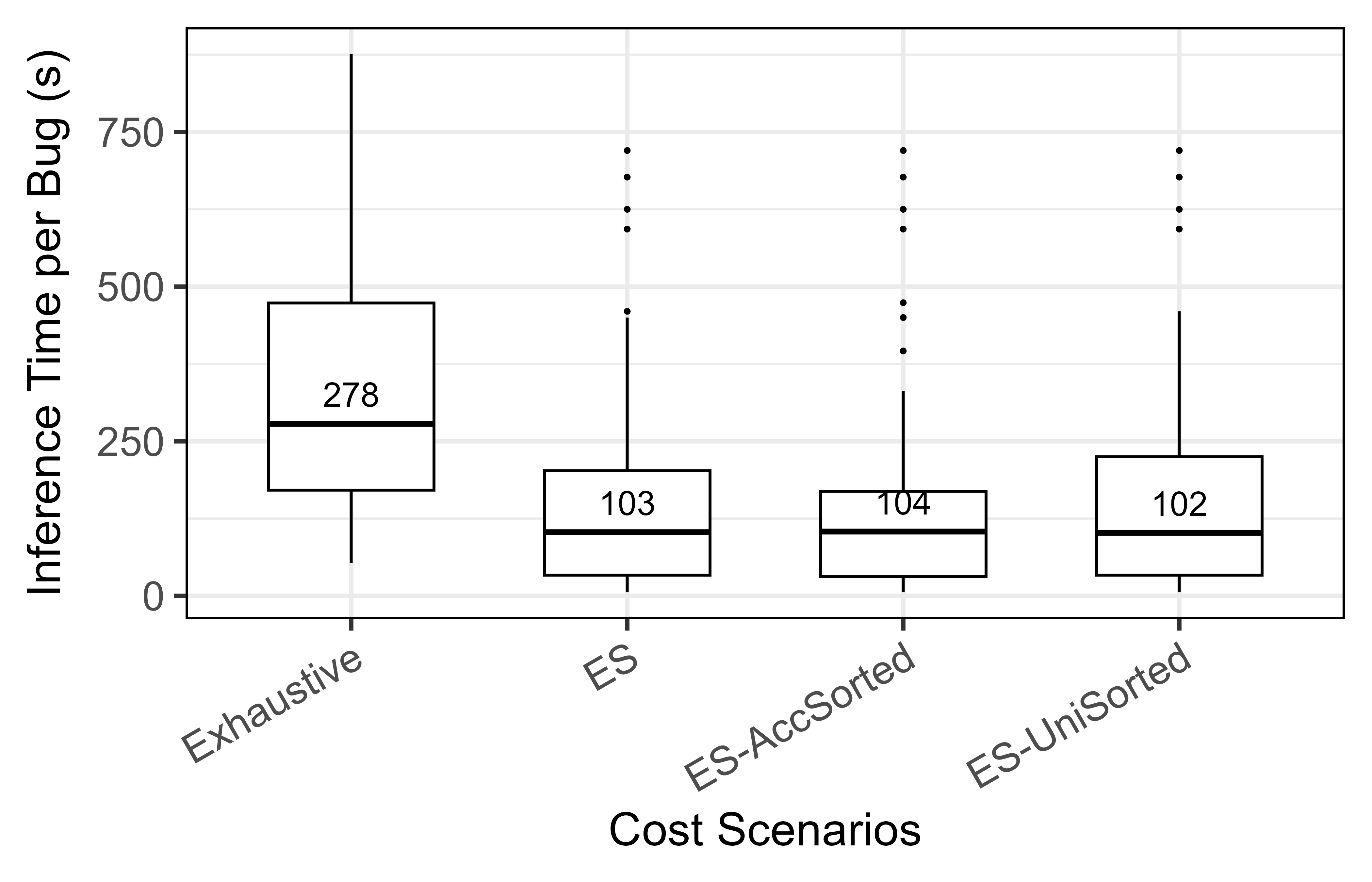}
    \caption{CodeLlama-Instruct-7B on BugsInPy}
    \label{rq3_box_plot_time_2_datasets_3_models:a}
  \end{subfigure}
  \hfill
  \begin{subfigure}[b]{0.48\textwidth}
    \includegraphics[width=\textwidth]{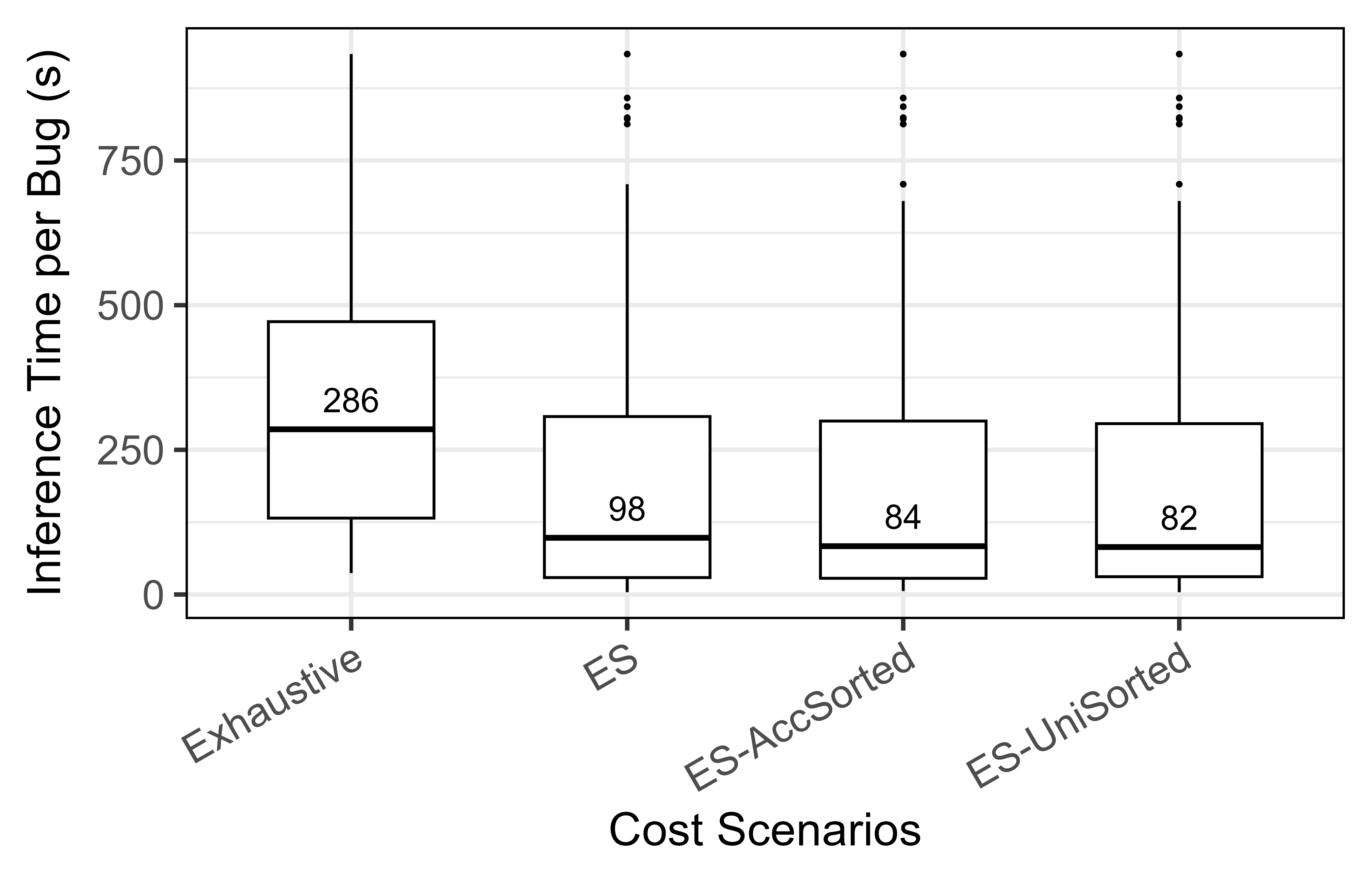}
    \caption{CodeLlama-Instruct-7B on Defects4J}
    \label{rq3_box_plot_time_2_datasets_3_models:b}
  \end{subfigure}

  \vspace{1em} 
    
  \begin{subfigure}[b]{0.48\textwidth}
    \includegraphics[width=\textwidth]{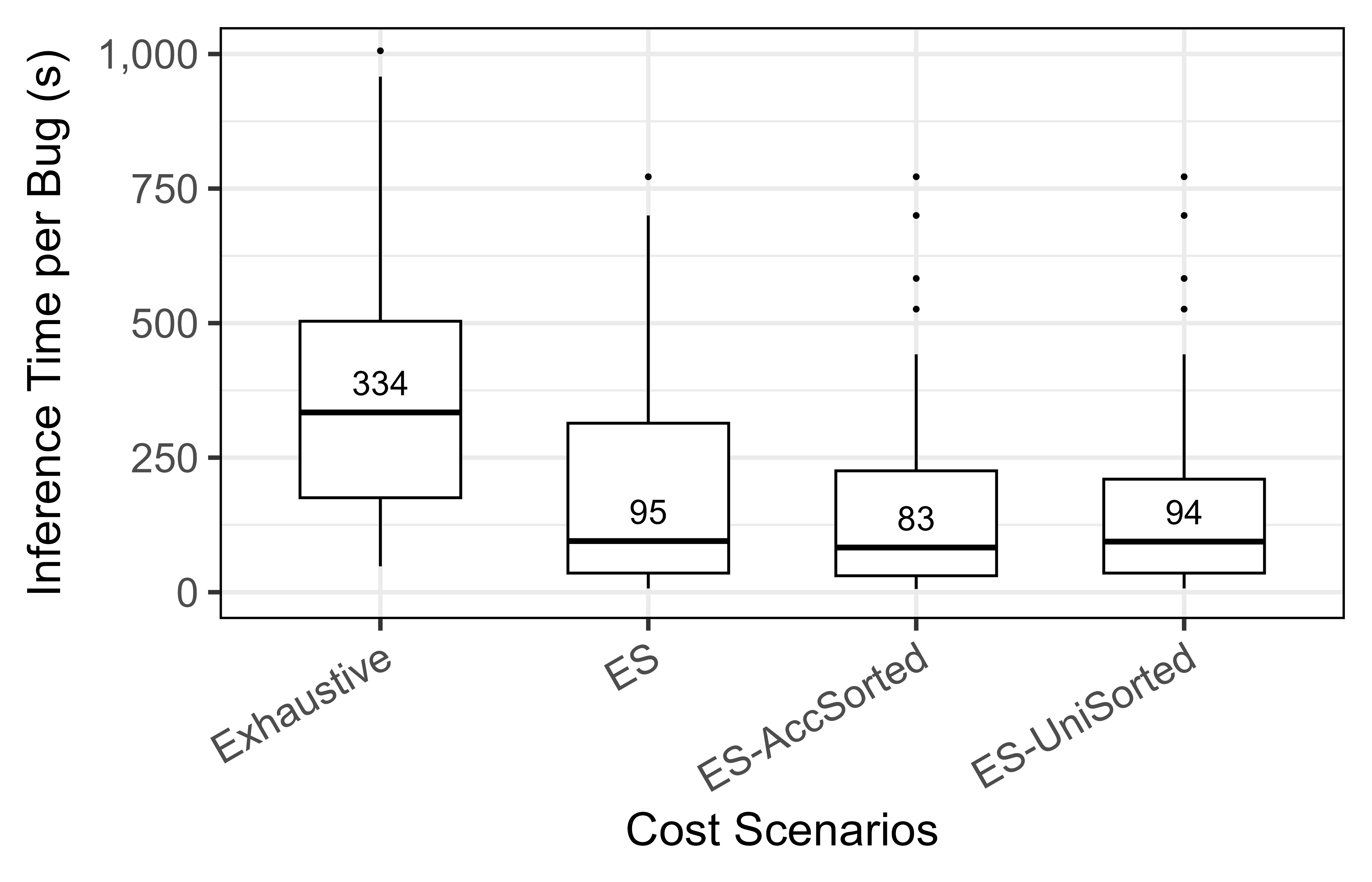}
    \caption{DeepSeek-Coder-Instruct-6.7B on \\BugsInPy}
    \label{rq3_box_plot_time_2_datasets_3_models:c}
  \end{subfigure}
  \hfill
  \begin{subfigure}[b]{0.48\textwidth}
    \includegraphics[width=\textwidth]{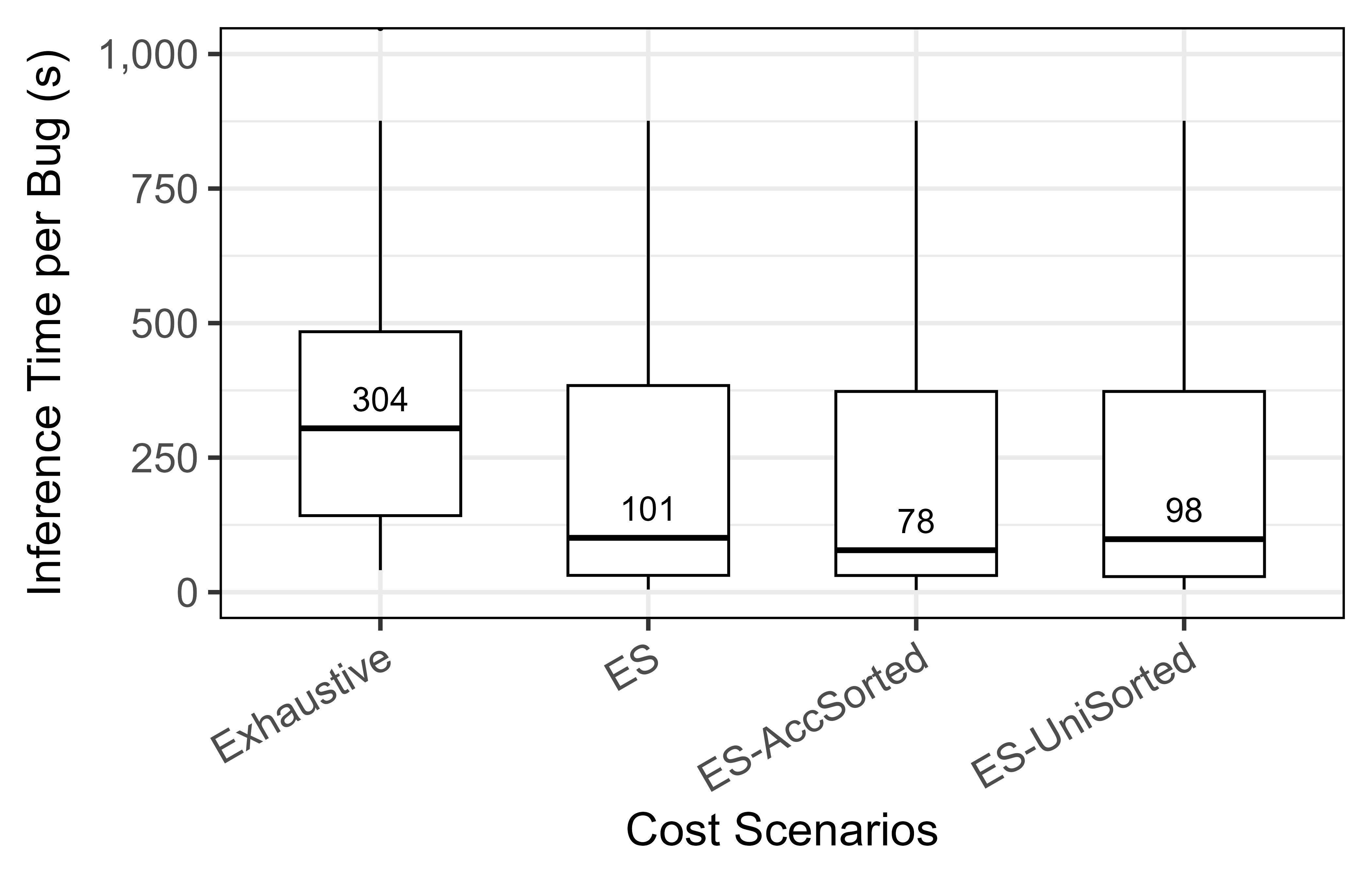}
    \caption{DeepSeek-Coder-Instruct-6.7B on \\Defects4J}
    \label{rq3_box_plot_time_2_datasets_3_models:d}
  \end{subfigure}

  \vspace{1em} 
    
  \begin{subfigure}[b]{0.48\textwidth}
    \includegraphics[width=\textwidth]{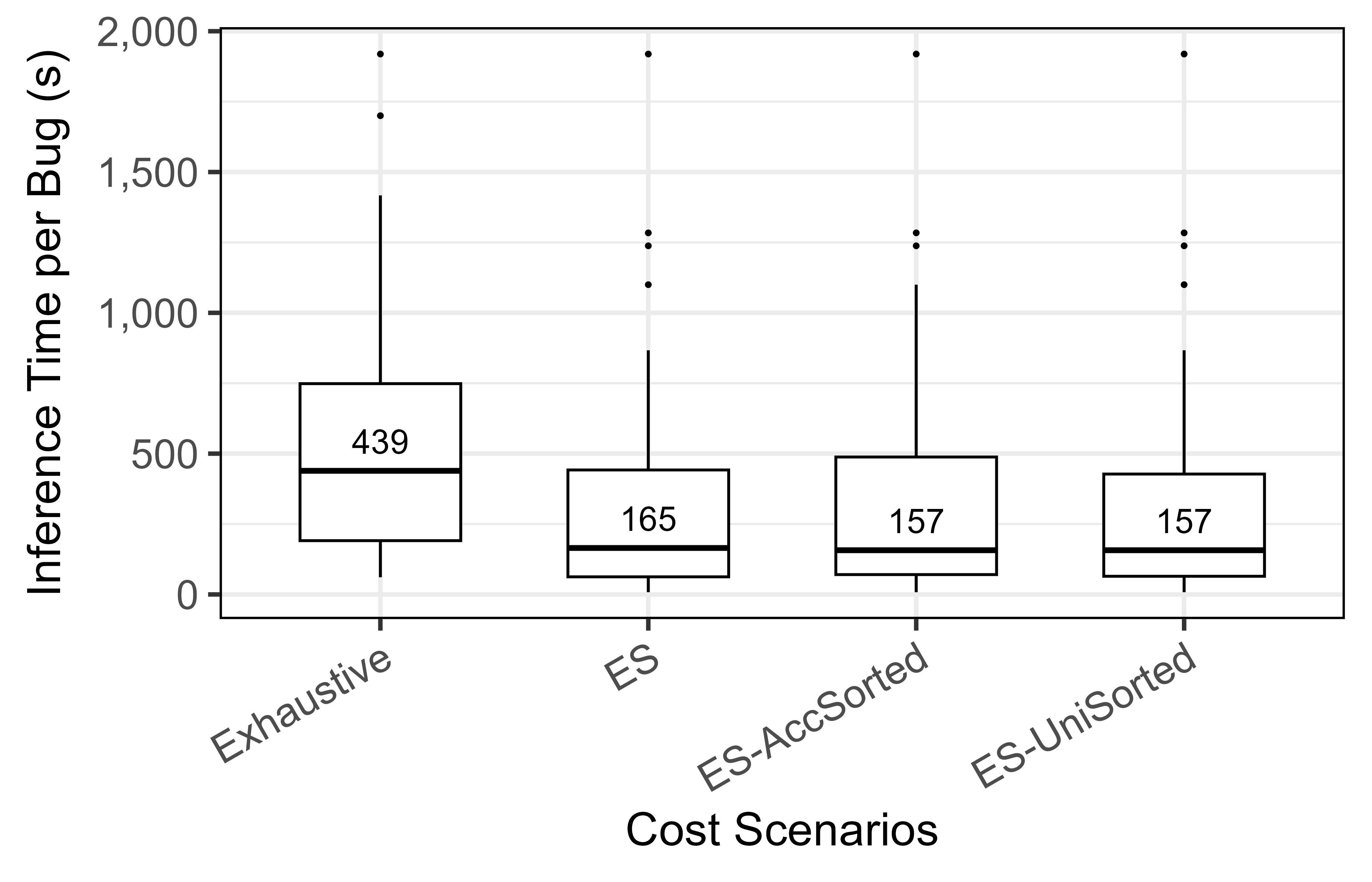}
    \caption{DeepSeek-Coder-V2-Lite-Instruct-16B \\on BugsInPy}
    \label{rq3_box_plot_time_2_datasets_3_models:e}
  \end{subfigure}
  \hfill
  \begin{subfigure}[b]{0.48\textwidth}
    \includegraphics[width=\textwidth]{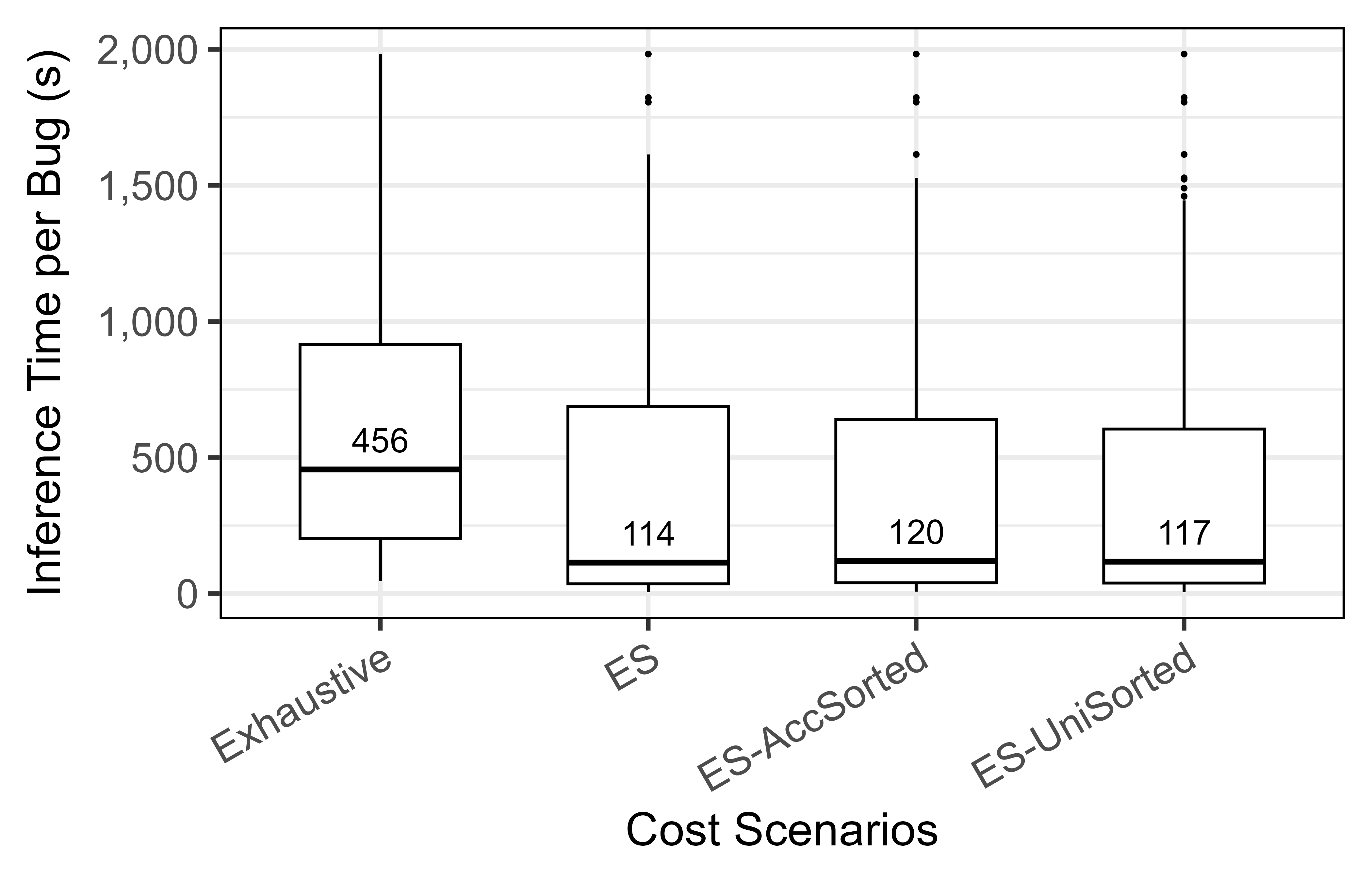}
    \caption{DeepSeek-Coder-V2-Lite-Instruct-16B \\on Defects4J}
    \label{rq3_box_plot_time_2_datasets_3_models:f}
  \end{subfigure}
  \caption{Distribution of inference time of bugs across different cost scenarios, evaluated on two datasets and three models.}
  \label{fig:rq3_box_plot_time_2_datasets_3_models}
\end{figure}

\begin{table}[!htbp]
  \centering
  \caption{Pairwise comparisons of inference time of bugs using Wilcoxon signed-rank test over three models and two datasets. E, ES, ES-A, and ES-U correspond to the Exhaustive, EarlyStop, ES-AccSorted, and ES-UniSorted cost scenarios, respectively. Each cell reports the $p$-value and the corresponding effect size $r_{\mathrm{rb}}$ (Rank-Biserial Correlation). A Bonferroni-corrected significance threshold of $\alpha = 0.0083$ ($0.05/6$) is applied for pairwise comparisons.}
  \label{tab:wilcox_rq3_box_plot_time_2_datasets_3_models}
  \begin{subtable}[t]{0.46\textwidth}
    \centering
    \caption{CodeLlama-Instruct-7B on BugsInPy}
    \label{wilcox_rq3_box_plot_time_2_datasets_3_models:a}
    \begin{tabular}{lccc}
      \toprule
      & ES & E & ES-A \\
      \midrule
      E & \makecell{$2.6 \times 10^{-7}$ \\ $r_{\mathrm{rb}} = 0.05$} & - & - \\
      \midrule
      ES-A & \makecell{0.4840 \\ $r_{\mathrm{rb}} = 0.76$} & \makecell{$2.6 \times 10^{-7}$ \\ $r_{\mathrm{rb}} = 0.05$} & - \\
      \midrule
      ES-U & \makecell{0.8551 \\ $r_{\mathrm{rb}} = 0.99$} & \makecell{$2.6 \times 10^{-7}$ \\ $r_{\mathrm{rb}} = 0.05$} & \makecell{0.5874 \\ $r_{\mathrm{rb}} = 0.63$} \\
      \bottomrule
    \end{tabular}
  \end{subtable}
  \hfill
  \begin{subtable}[t]{0.46\textwidth}
    \centering
    \caption{CodeLlama-Instruct-7B on Defects4J}
    \label{wilcox_rq3_box_plot_time_2_datasets_3_models:b}
    \begin{tabular}{lccc}
      \toprule
      & ES & E & ES-A \\
      \midrule
      E & \makecell{$3.6 \times 10^{-13}$ \\ $r_{\mathrm{rb}} = 0.27$} & - & - \\
      \midrule
      ES-A & \makecell{0.2813 \\ $r_{\mathrm{rb}} = 0.72$} & \makecell{$3.6 \times 10^{-13}$ \\ $r_{\mathrm{rb}} = 0.27$} & - \\
      \midrule
      ES-U & \makecell{0.0979 \\ $r_{\mathrm{rb}} = 0.93$} & \makecell{$2.5 \times 10^{-13}$ \\ $r_{\mathrm{rb}} = 0.25$} & \makecell{0.4174 \\ $r_{\mathrm{rb}} = 0.59$} \\
      \bottomrule
    \end{tabular}
  \end{subtable}

\vspace{2em}

  \begin{subtable}[t]{0.46\textwidth}
    \centering
    \caption{DeepSeek-Coder-Instruct-6.7B on \\BugsInPy}
    \label{wilcox_rq3_box_plot_time_2_datasets_3_models:c}
    \begin{tabular}{lccc}
      \toprule
      & ES & E & ES-A \\
      \midrule
      E & \makecell{$8.3 \times 10^{-7}$ \\ $r_{\mathrm{rb}} = 0.20$} & - & - \\
      \midrule
      ES-A & \makecell{0.0254 \\ $r_{\mathrm{rb}} = 0.49$} & \makecell{$5.6 \times 10^{-7}$ \\ $r_{\mathrm{rb}} = 0.15$} & - \\
      \midrule
      ES-U & \makecell{0.0283 \\ $r_{\mathrm{rb}} = 0.93$} & \makecell{$5.6 \times 10^{-7}$ \\ $r_{\mathrm{rb}} = 0.15$} & \makecell{0.6874 \\ $r_{\mathrm{rb}} = 0.67$} \\
      \bottomrule
    \end{tabular}
  \end{subtable}
  \hfill
  \begin{subtable}[t]{0.46\textwidth}
    \centering
    \caption{DeepSeek-Coder-Instruct-6.7B on \\Defects4J}
    \label{wilcox_rq3_box_plot_time_2_datasets_3_models:d}
    \begin{tabular}{lccc}
      \toprule
      & ES & E & ES-A \\
      \midrule
      E & \makecell{$3.6 \times 10^{-12}$ \\ $r_{\mathrm{rb}} = 0.39$} & - & - \\
      \midrule
      ES-A & \makecell{0.4135 \\ $r_{\mathrm{rb}} = 0.70$} & \makecell{$2.5 \times 10^{-12}$ \\ $r_{\mathrm{rb}} = 0.37$} & - \\
      \midrule
      ES-U & \makecell{0.4945 \\ $r_{\mathrm{rb}} = 0.93$} & \makecell{$2.5 \times 10^{-12}$ \\ $r_{\mathrm{rb}} = 0.37$} & \makecell{0.5189 \\ $r_{\mathrm{rb}} = 0.74$} \\
      \bottomrule
    \end{tabular}
  \end{subtable}

\vspace{2em}

  \begin{subtable}[t]{0.46\textwidth}
    \centering
    \caption{DeepSeek-Coder-V2-Lite-Instruct-16B \\on BugsInPy}
    \label{wilcox_rq3_box_plot_time_2_datasets_3_models:e}
    \begin{tabular}{lccc}
      \toprule
      & ES & E & ES-A \\
      \midrule
      E & \makecell{$4 \times 10^{-6}$ \\ $r_{\mathrm{rb}} = 0.39$} & - & - \\
      \midrule
      ES-A & \makecell{0.2751 \\ $r_{\mathrm{rb}} = 0.82$} & \makecell{$4 \times 10^{-6}$ \\ $r_{\mathrm{rb}} = 0.39$} & - \\
      \midrule
      ES-U & \makecell{0.6781 \\ $r_{\mathrm{rb}} = 0.96$} & \makecell{$4 \times 10^{-6}$ \\ $r_{\mathrm{rb}} = 0.39$} & \makecell{0.5199 \\ $r_{\mathrm{rb}} = 0.74$} \\
      \bottomrule
    \end{tabular}
  \end{subtable}
  \hfill
  \begin{subtable}[t]{0.46\textwidth}
    \centering
    \caption{DeepSeek-Coder-V2-Lite-Instruct-16B \\on Defects4J}
    \label{wilcox_rq3_box_plot_time_2_datasets_3_models:f}
    \begin{tabular}{lccc}
      \toprule
      & ES & E & ES-A \\
      \midrule
      E & \makecell{$7.8 \times 10^{-13}$ \\ $r_{\mathrm{rb}} = 0.31$} & - & - \\
      \midrule
      ES-A & \makecell{0.1226 \\ $r_{\mathrm{rb}} = 0.79$} & \makecell{$5.3 \times 10^{-13}$ \\ $r_{\mathrm{rb}} = 0.29$} & - \\
      \midrule
      ES-U & \makecell{0.3942 \\ $r_{\mathrm{rb}} = 0.97$} & \makecell{$5.3 \times 10^{-13}$ \\ $r_{\mathrm{rb}} = 0.29$} & \makecell{0.0063 \\ $r_{\mathrm{rb}} = 0.62$} \\
      \bottomrule
    \end{tabular}
  \end{subtable}
\end{table}


\textbf{Early stopping reduces inference time by an average of 69\% and is significantly more efficient than the Exhaustive scenario.}
Figure \ref{fig:rq3_box_plot_time_2_datasets_3_models} presents the distributions of inference time across all bugs over three models and two datasets under the four cost scenarios: Exhaustive, ES, ES-AccSorted, and ES-UniSorted. Among them, the Exhaustive scenario has the highest median inference time (e.g., 278 seconds for CodeLlama-Instruct-7B on BugsInPy), compared to 103, 104, and 102 seconds for the other three early stopping scenarios, respectively. Early stopping reduces inference time by an average of 69\%, demonstrating substantial efficiency gains over the Exhaustive strategy. Additionally, the difference between the sorted (ES-AccSorted and ES-UniSorted) and unsorted (ES) approaches is minimal.

To assess the statistical significance of these differences across scenarios, we conducted the Friedman test for each model-dataset configuration. All configurations yielded $p$-values $< 0.001$, confirming the presence of significant differences among the four cost scenarios. Subsequently, post-hoc pairwise comparisons were performed using the Wilcoxon signed-rank test, with Bonferroni correction applied to account for multiple comparisons ($0.05/6=0.0083$). The detailed pairwise results are presented in Table \ref{tab:wilcox_rq3_box_plot_time_2_datasets_3_models}. Across all six configurations, the Exhaustive scenario is significantly more time-consuming than all other cost scenarios, with $p$-values ranging from $2.5 \times 10^{-13}$ to $4 \times 10^{-6}$ (all below the $\alpha$ threshold). The effect size is negligible in exactly one configuration (CodeLlama-Instruct-7B on BugsInPy), whereas the other five configurations show small to medium effect sizes ranging from 0.15 to 0.39. When comparing ES, ES-AccSorted, and ES-UniSorted directly, their performance is largely comparable in five out of six configurations. The only exception is for DeepSeek-Coder-V2-Lite-Instruct-16B on Defects4J (Table \ref{wilcox_rq3_box_plot_time_2_datasets_3_models:f}), where ES-UniSorted (ES-U) is significantly better than ES-AccSorted (ES-A) with $p = 0.0063$ and $r_{\mathrm{rb}} = 0.62$ (large). These results highlight that early stopping significantly reduces inference time compared to the Exhaustive scenario, with minimal differences between sorted and unsorted scenarios.



\textbf{FL-diff and FN-all are consistently the most time-consuming heuristics.} To gain deeper insights into the impact of individual heuristics of HAFix-Agg on inference time, we specifically examine the Exhaustive scenario, as it involves executing all individual heuristics. In this scenario, each heuristic produces distinct inference times for each bug, enabling clear identification of time-intensive and efficient heuristics. Figure \ref{fig:rq3_box_plot_time_exhaustive_2_datasets_3_models} (Appendix \ref{appendix:AdditionalRQ3Results}) shows the distribution of inference time per heuristic in the Exhaustive scenario across two datasets and three models. For example, with CodeLlama-Instruct-7B on BugsInPy (Figure \subfig{fig:rq3_box_plot_time_exhaustive_2_datasets_3_models}{rq3_box_plot_time_exhaustive_2_datasets_3_models:a}), median inference time ranges from 31 seconds (FN-pair) to 38 seconds (FN-all). This narrow range is similarly observed in the next three configurations (Figures \subfig{fig:rq3_box_plot_time_exhaustive_2_datasets_3_models}{rq3_box_plot_time_exhaustive_2_datasets_3_models:a}-\subfig{fig:rq3_box_plot_time_exhaustive_2_datasets_3_models}{rq3_box_plot_time_exhaustive_2_datasets_3_models:d}). In contrast, the inference time variations are broader for DeepSeek-Coder-V2-Lite-Instruct-16B: 43 seconds (CFN-all) to 76 seconds (FN-all) on BugsInPy, and 41 seconds (Baseline) to 104 seconds (FL-diff) on Defects4J.

To evaluate statistical significance, we conducted Friedman tests followed by pairwise Wilcoxon signed-rank tests, with detailed pairwise results reported in Appendix \ref{appendix:AdditionalRQ3Results} (Table \ref{tab:wilcox_rq3_box_plot_time_exhaustive_2_datasets_3_models}). For CodeLlama-Instruct-7B on BugsInPy (Table \ref{wilcox_rq3_box_plot_time_exhaustive_2_datasets_3_models:a}), no heuristic pairs exhibited significant differences, indicating comparable inference time across heuristics. However, for CodeLlama-Instruct-7B on Defects4J (Table \ref{wilcox_rq3_box_plot_time_exhaustive_2_datasets_3_models:b}), FN-pair significantly takes less time than the other six heuristics (except the baseline), with $p$-values ranging from $3.4 \times 10^{-9}$ (medium effect size $r_{\mathrm{rb}} = 0.41$) to 0.0004 (effect size $r_{\mathrm{rb}} = 0.05$). Conversely, FN-all and CFN-all significantly take more time than Baseline and CFN-modified.
Similar trends emerged with DeepSeek-Coder-Instruct-6.7B. On BugsInPy (Table \ref{wilcox_rq3_box_plot_time_exhaustive_2_datasets_3_models:c}), FN-all significantly exceeds Baseline, CFN-all, and FLN-all inference time ($p \leq 0.0018$). On Defects4J (Table \ref{wilcox_rq3_box_plot_time_exhaustive_2_datasets_3_models:d}), FL-diff significantly takes more time than all seven other heuristics, and FN-all also takes significantly more time than Baseline and CFN-modified.
For DeepSeek-Coder-V2-Lite-Instruct-16B, similar findings are observed. On BugsInPy (Table \ref{wilcox_rq3_box_plot_time_exhaustive_2_datasets_3_models:e}), FN-all consistently requires more inference time compared to five other heuristics, and FL-diff is significantly slower than two heuristics. On Defects4J (Table \ref{wilcox_rq3_box_plot_time_exhaustive_2_datasets_3_models:f}), FN-all significantly exceeds inference time for six other heuristics, and FL-diff is slower than all seven remaining heuristics.
Collectively, these results confirm that FL-diff and FN-all are the most time-consuming heuristics, while FN-pair is notably efficient, particularly on CodeLlama-Instruct-7B with Defects4J. Other heuristics exhibit comparable inference time across most model-dataset combinations.


\begin{figure}[!htbp]
  \centering
  \begin{subfigure}[b]{0.48\textwidth}
    \includegraphics[width=\textwidth]{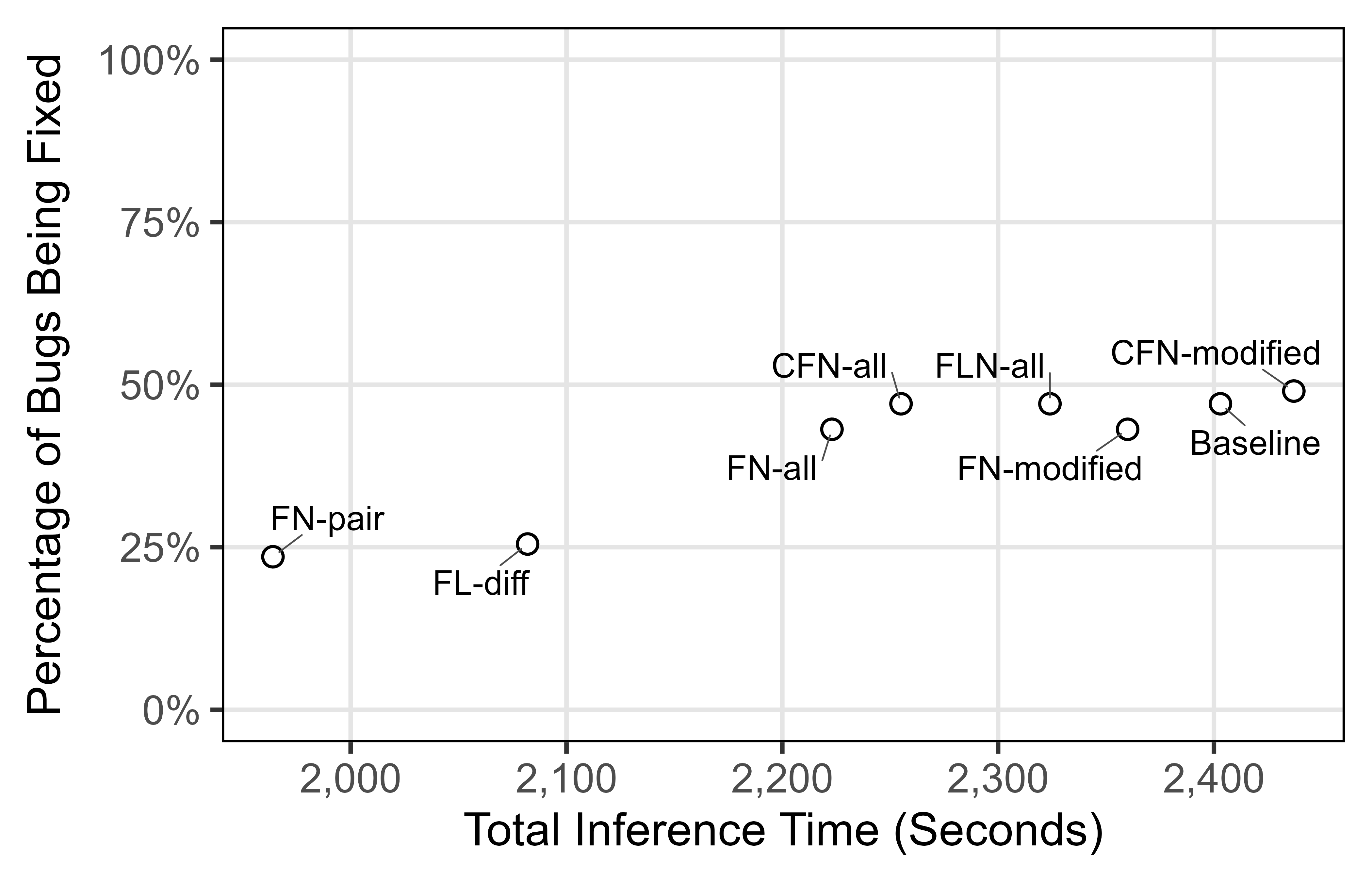}
    \caption{CodeLlama-Instruct-7B on BugsInPy}
    \label{rq3_scatter_time_2_datasets_3_models:a}
  \end{subfigure}
  \hfill
  \begin{subfigure}[b]{0.48\textwidth}
    \includegraphics[width=\textwidth]{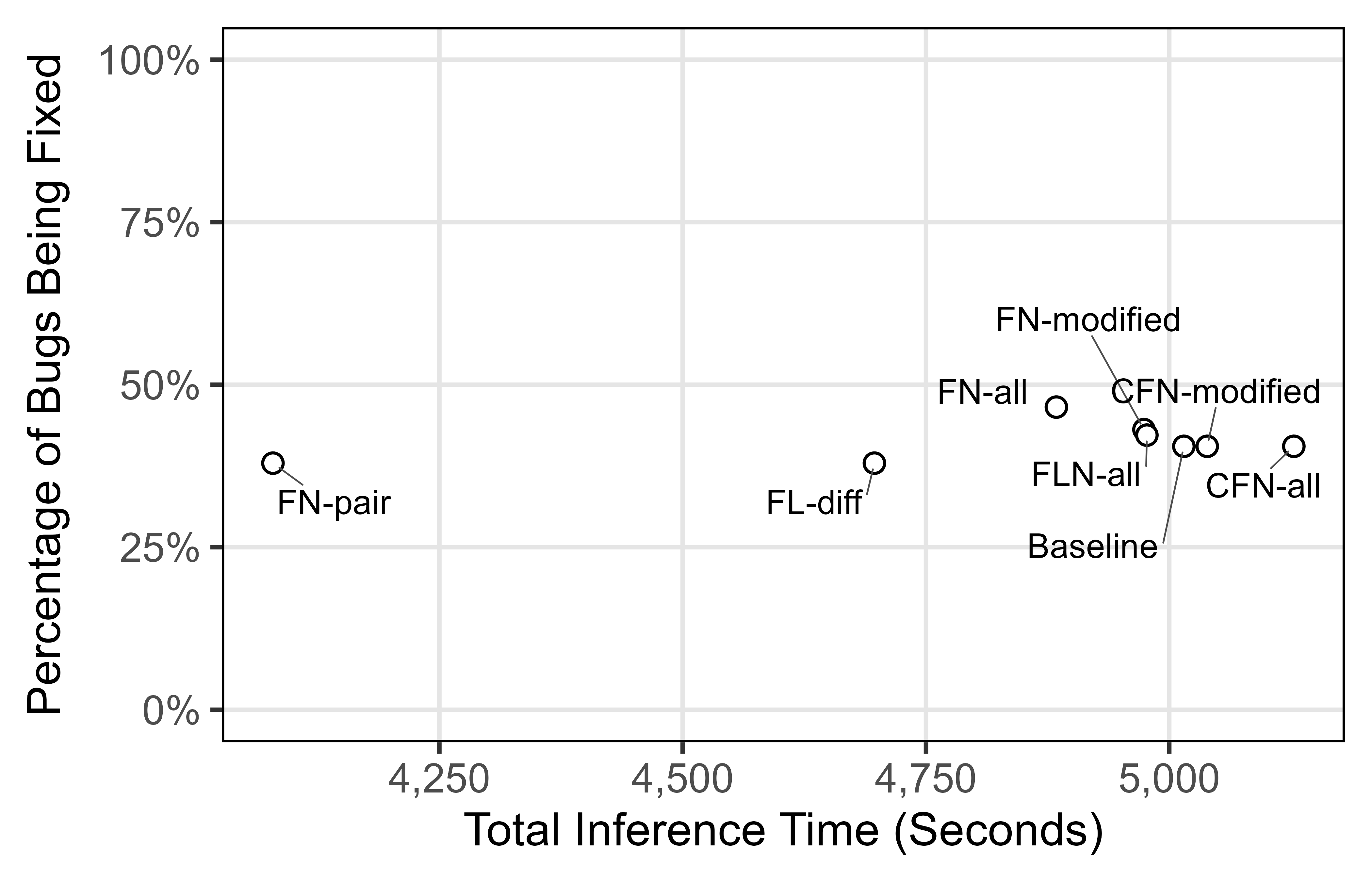}
    \caption{CodeLlama-Instruct-7B on Defects4J}
    \label{rq3_scatter_time_2_datasets_3_models:b}
  \end{subfigure}

  \vspace{1em} 
    
  \begin{subfigure}[b]{0.48\textwidth}
    \includegraphics[width=\textwidth]{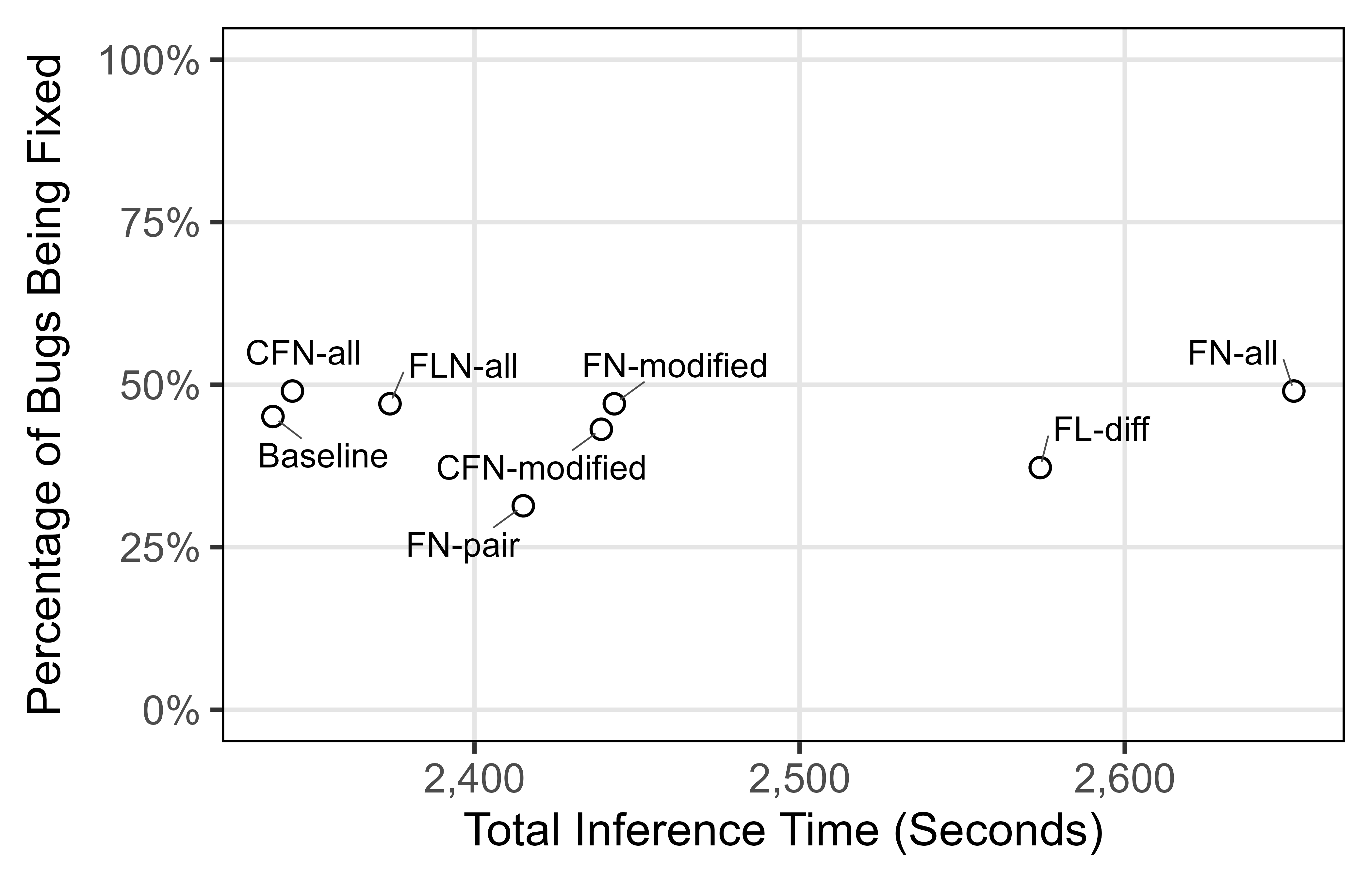}
    \caption{DeepSeek-Coder-Instruct-6.7B on \\BugsInPy}
    \label{rq3_scatter_time_2_datasets_3_models:c}
  \end{subfigure}
  \hfill
  \begin{subfigure}[b]{0.48\textwidth}
    \includegraphics[width=\textwidth]{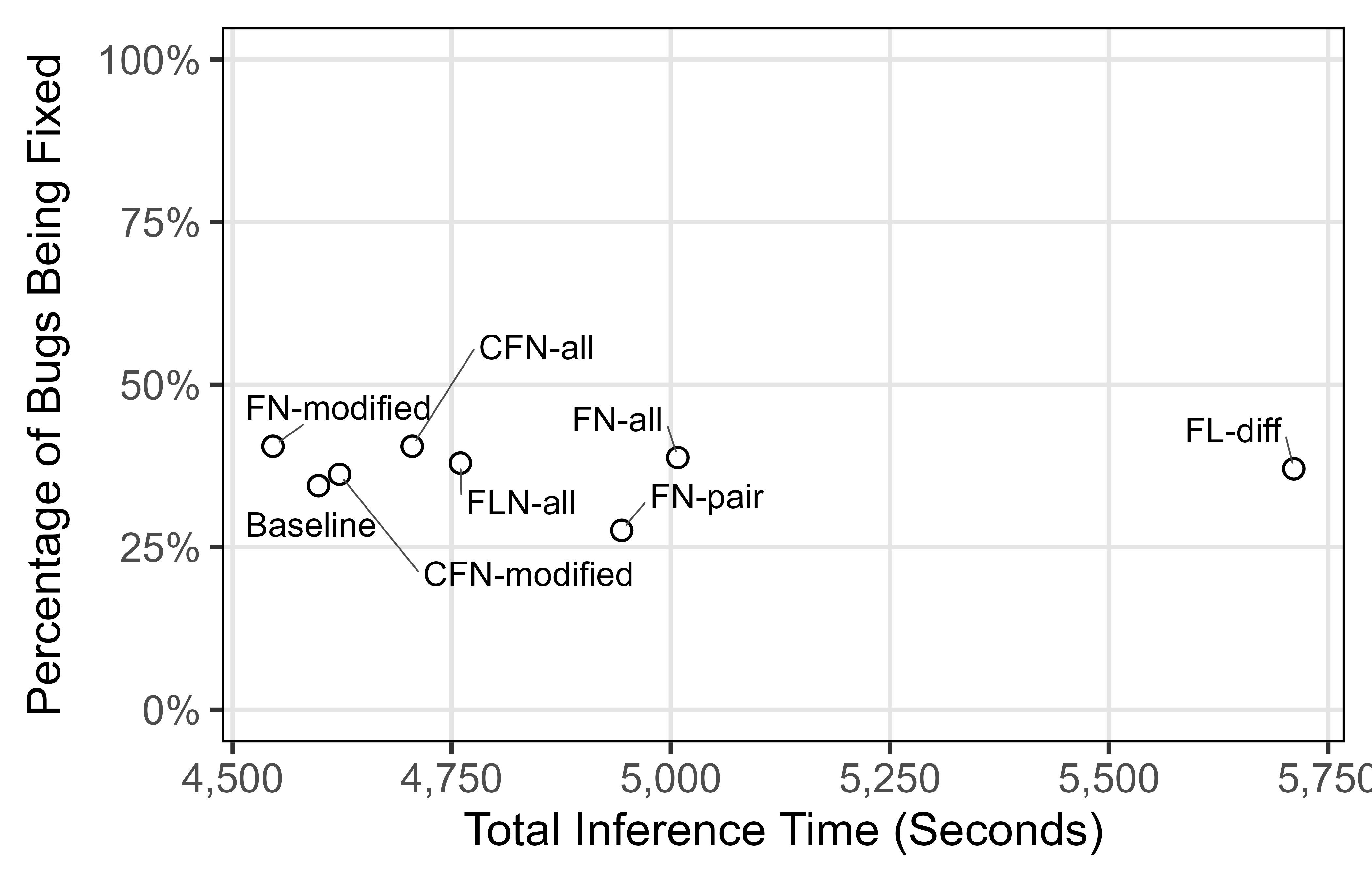}
    \caption{DeepSeek-Coder-Instruct-6.7B on \\Defects4J}
    \label{rq3_scatter_time_2_datasets_3_models:d}
  \end{subfigure}

  \vspace{1em} 
    
  \begin{subfigure}[b]{0.48\textwidth}
    \includegraphics[width=\textwidth]{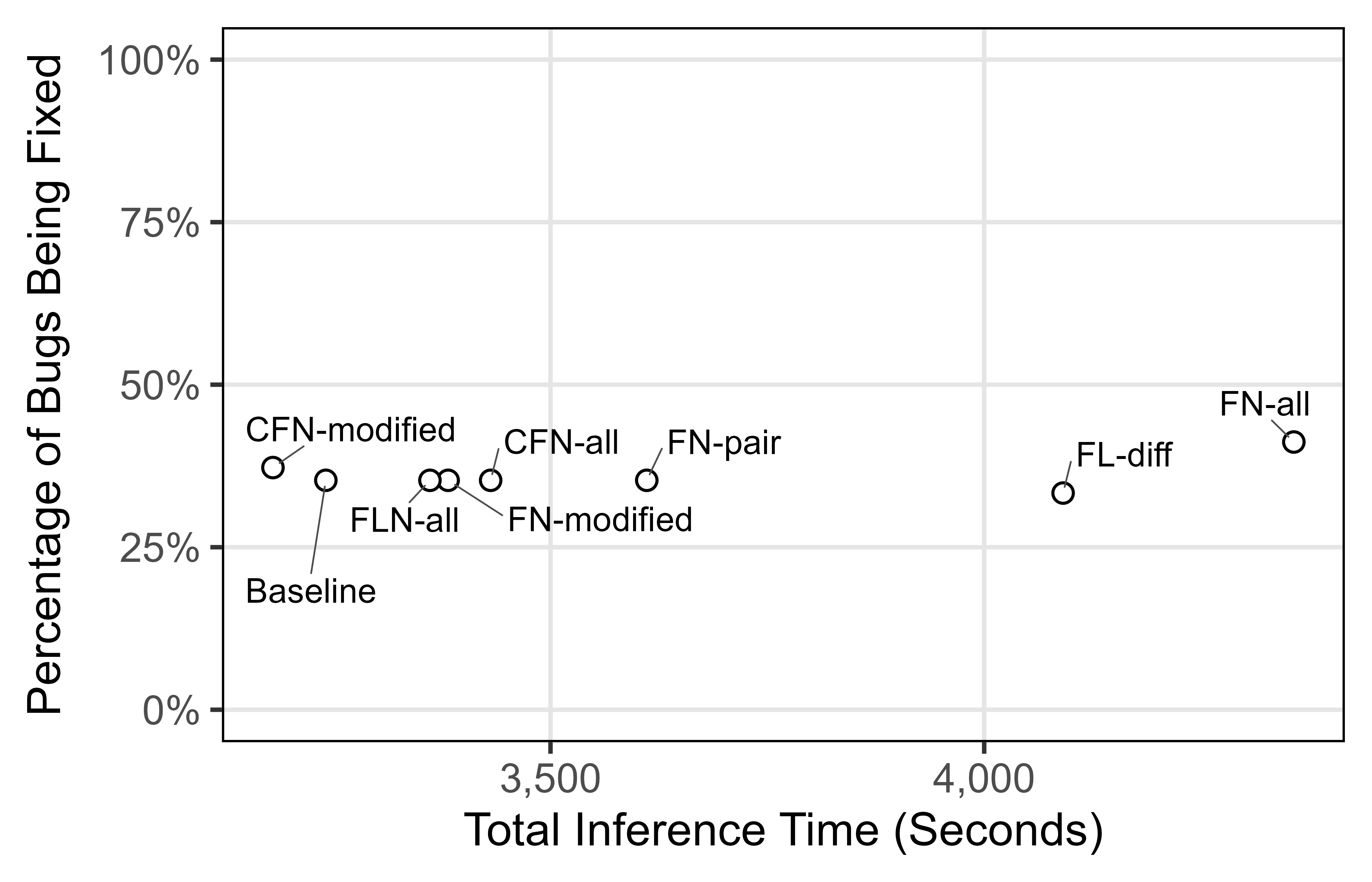}
    \caption{DeepSeek-Coder-V2-Lite-Instruct-16B \\on BugsInPy}
    \label{rq3_scatter_time_2_datasets_3_models:e}
  \end{subfigure}
  \hfill
  \begin{subfigure}[b]{0.48\textwidth}
    \includegraphics[width=\textwidth]{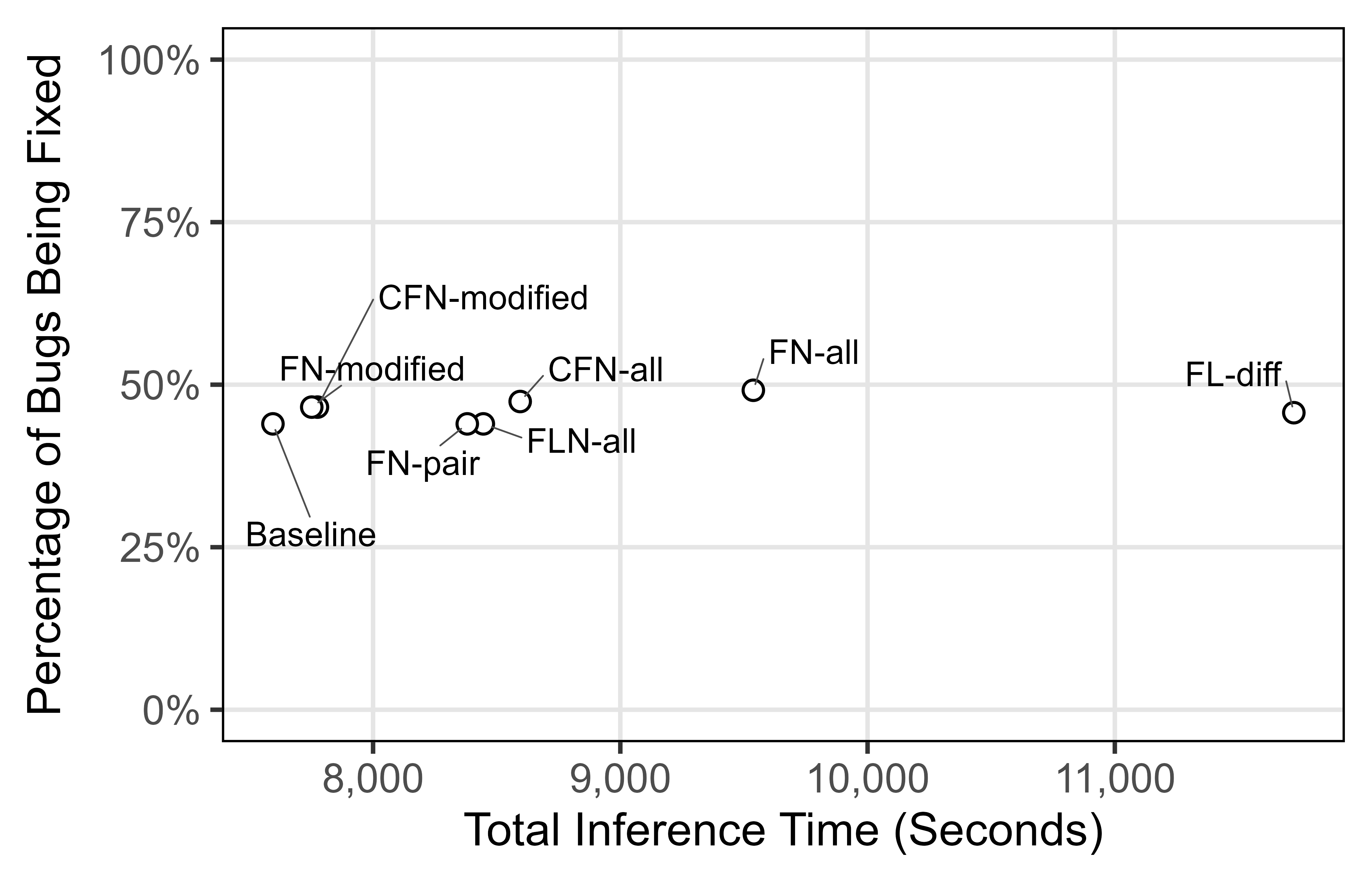}
    \caption{DeepSeek-Coder-V2-Lite-Instruct-16B \\on Defects4J}
    \label{rq3_scatter_time_2_datasets_3_models:f}
  \end{subfigure}
  \caption{Percentage of bugs being fixed (Table \ref{rq1_acc_percentage}) vs. total inference time across different heuristics in the Exhaustive scenario. The inference time reflects the total time required to process all bugs for each heuristic.}
  \label{fig:rq3_scatter_time_2_datasets_3_models}
\end{figure}

\textbf{FL-diff consistently exhibits the poorest trade-off between inference time and bug-fixing performance, while FN-modified, CFN-modified, and CFN-all generally achieve more favorable time performance balances across models and datasets.}
To further explore the relationship between heuristic performance and inference time efficiency, Figure \ref{fig:rq3_scatter_time_2_datasets_3_models} illustrates the percentage of bugs fixed in relation to inference time for different heuristics in the Exhaustive scenario. The y-axis represents the percentage of bugs being fixed, while the x-axis denotes inference time in seconds. 
For CodeLlama-Instruct-7B, the time-effectiveness trade-offs vary significantly between datasets. On BugsInPy (Figure \subfig{fig:rq3_scatter_time_2_datasets_3_models}{rq3_scatter_time_2_datasets_3_models:a}), CFN-modified achieves the highest performance but requires the longest inference time, while FN-pair demonstrates the fastest execution but delivers the lowest time effectiveness. CFN-all strikes a balanced trade-off with moderate time and performance. On Defects4J (Figure \subfig{fig:rq3_scatter_time_2_datasets_3_models}{rq3_scatter_time_2_datasets_3_models:b}), FN-pair maintains its time efficiency advantage while achieving performance comparable to other heuristics, suggesting dataset-dependent optimization potential.
DeepSeek-Coder-Instruct-6.7B reveals more pronounced efficiency disparities. On BugsInPy (Figure \subfig{fig:rq3_scatter_time_2_datasets_3_models}{rq3_scatter_time_2_datasets_3_models:c}), CFN-all emerges as the optimal choice, delivering the highest performance with relatively fast execution, while FN-all achieves comparable effectiveness but requires significantly more time. FL-diff exhibits poor time effectiveness with extended execution time and suboptimal performance. On Defects4J (Figure \subfig{fig:rq3_scatter_time_2_datasets_3_models}{rq3_scatter_time_2_datasets_3_models:d}), FN-modified demonstrates superior efficiency, combining the fastest execution with the highest performance, while FL-diff continues to underperform. 
For DeepSeek-Coder-V2-Lite-Instruct-16B, similar patterns persist across both datasets. On BugsInPy (Figure \subfig{fig:rq3_scatter_time_2_datasets_3_models}{rq3_scatter_time_2_datasets_3_models:e}), FN-modified achieves strong time efficiency with competitive performance, while FN-all achieves the highest performance at the cost of the longest time. On Defects4J (Figure \subfig{fig:rq3_scatter_time_2_datasets_3_models}{rq3_scatter_time_2_datasets_3_models:f}), FN-modified and CFN-modified achieve favorable time-performance balances, whereas FN-all maintains peak effectiveness but demands the higher inference time. Notably, FL-diff consistently exhibits the worst time-effectiveness profile across both datasets. These results highlight that FL-diff consistently demonstrates poor time-performance balance across all model-dataset configurations. Conversely, heuristics like FN-modified, CFN-modified and CFN-all generally provide more favorable trade-offs between inference time and bug-fixing effectiveness.


\begin{figure}[!htbp]
  \centering
  \begin{subfigure}[b]{0.48\textwidth}
    \includegraphics[width=\textwidth]{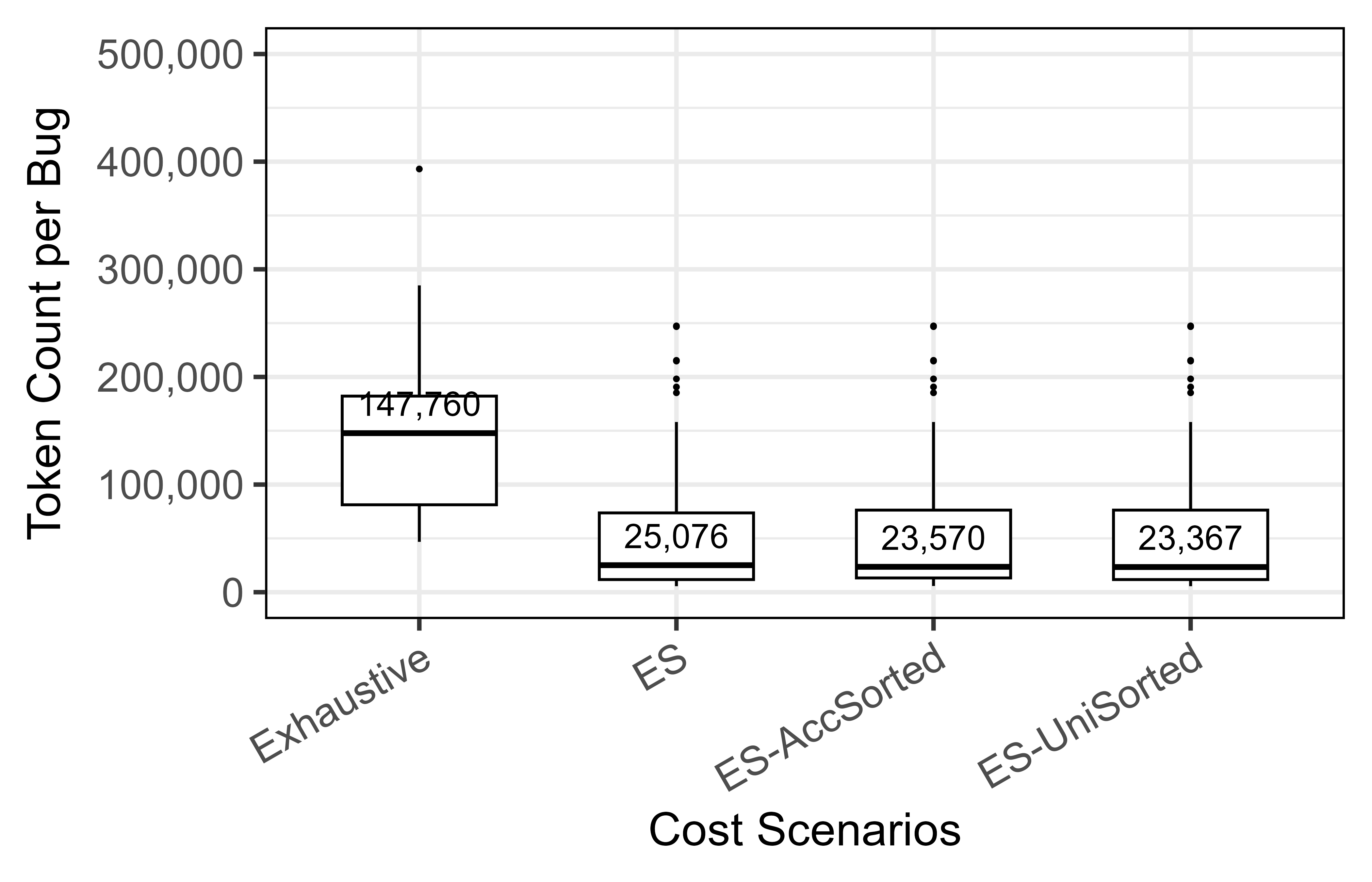}
    \caption{CodeLlama-Instruct-7B on BugsInPy}
    \label{rq3_box_plot_token_2_datasets_3_models:a}
  \end{subfigure}
  \hfill
  \begin{subfigure}[b]{0.48\textwidth}
    \includegraphics[width=\textwidth]{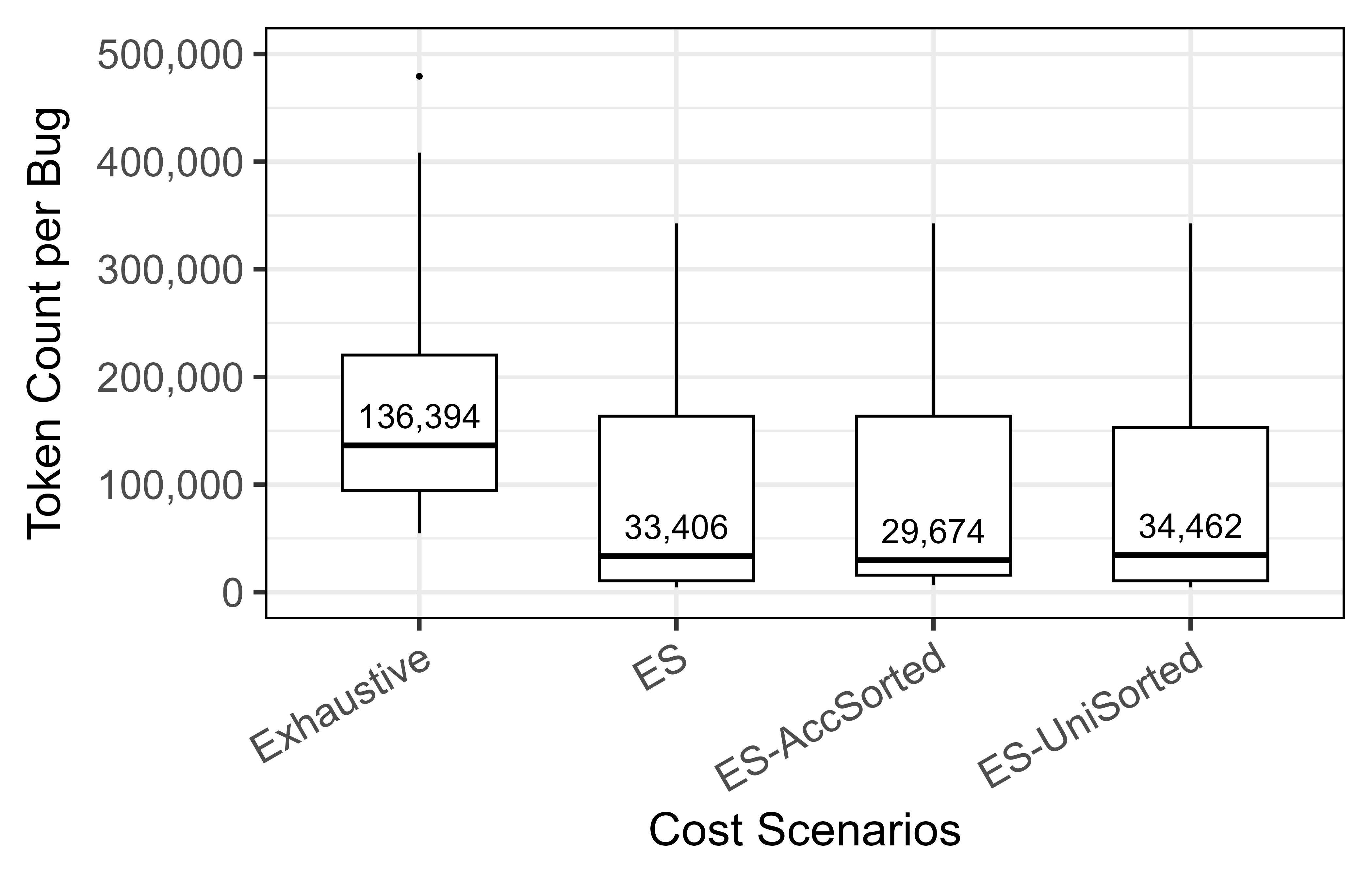}
    \caption{CodeLlama-Instruct-7B on Defects4J}
    \label{rq3_box_plot_token_2_datasets_3_models:b}
  \end{subfigure}

  \vspace{1em} 
    
  \begin{subfigure}[b]{0.48\textwidth}
    \includegraphics[width=\textwidth]{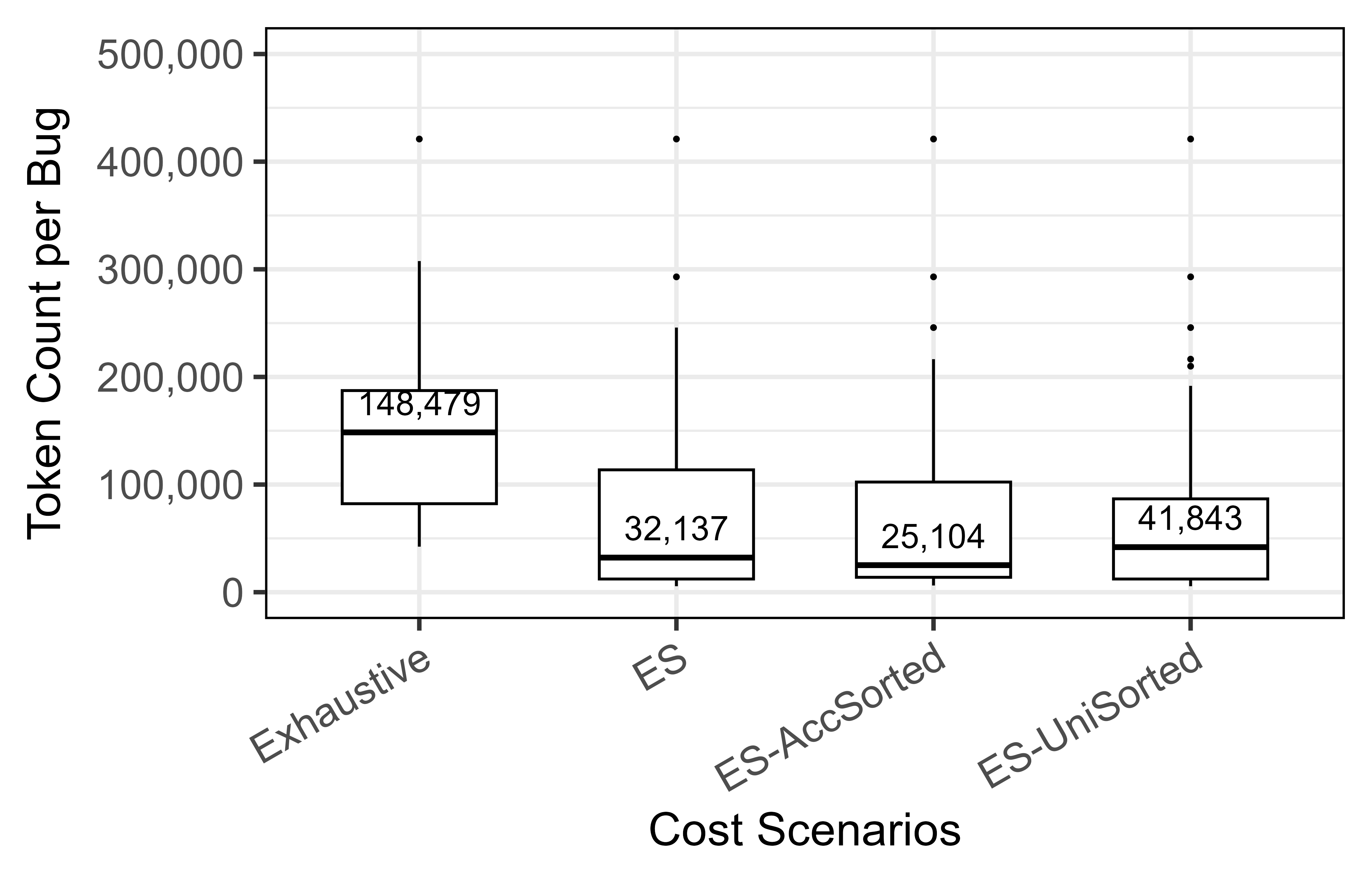}
    \caption{DeepSeek-Coder-Instruct-6.7B on \\BugsInPy}
    \label{rq3_box_plot_token_2_datasets_3_models:c}
  \end{subfigure}
  \hfill
  \begin{subfigure}[b]{0.48\textwidth}
    \includegraphics[width=\textwidth]{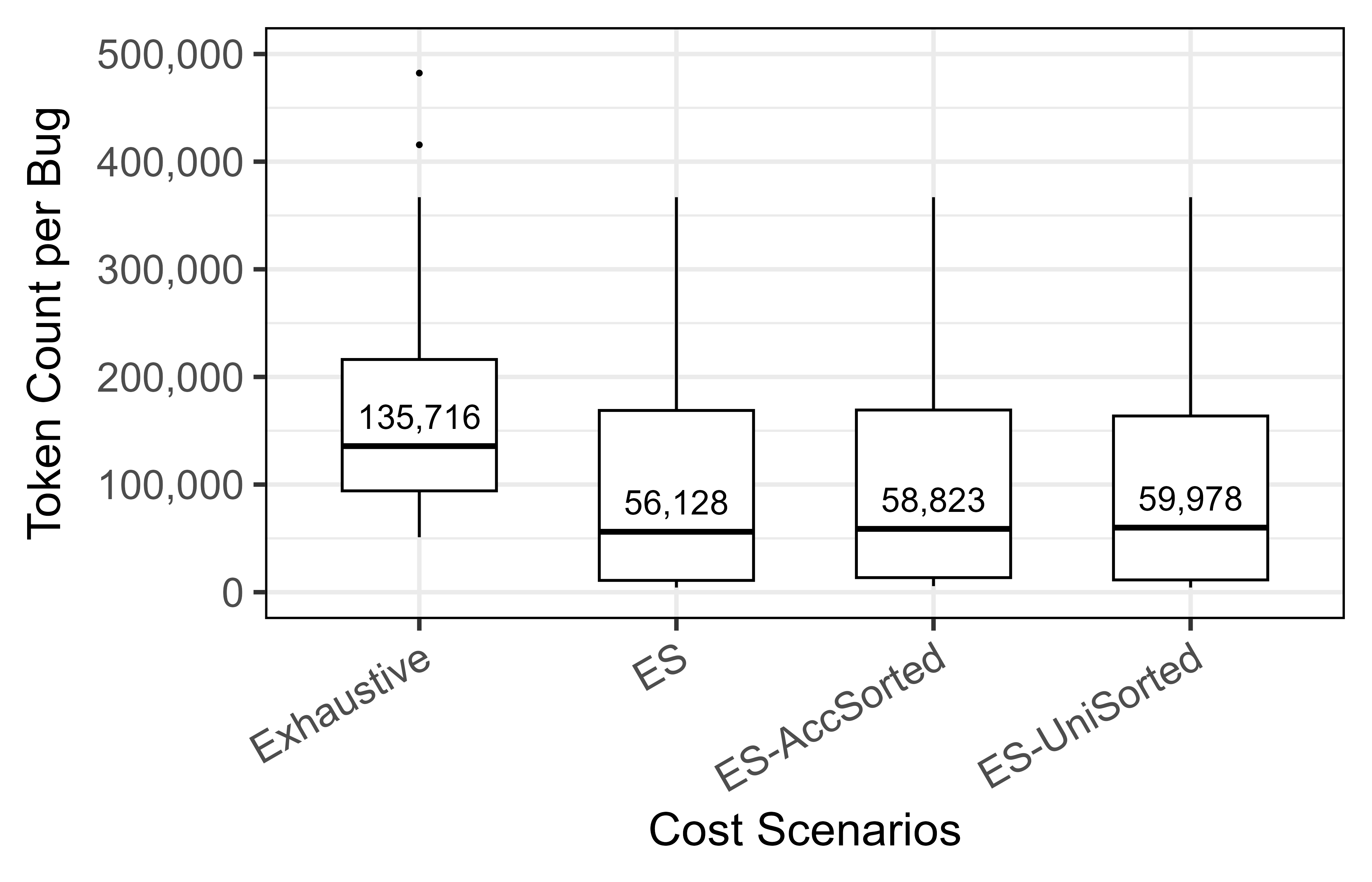}
    \caption{DeepSeek-Coder-Instruct-6.7B on \\Defects4J}
    \label{rq3_box_plot_token_2_datasets_3_models:d}
  \end{subfigure}

  \vspace{1em} 
    
  \begin{subfigure}[b]{0.48\textwidth}
    \includegraphics[width=\textwidth]{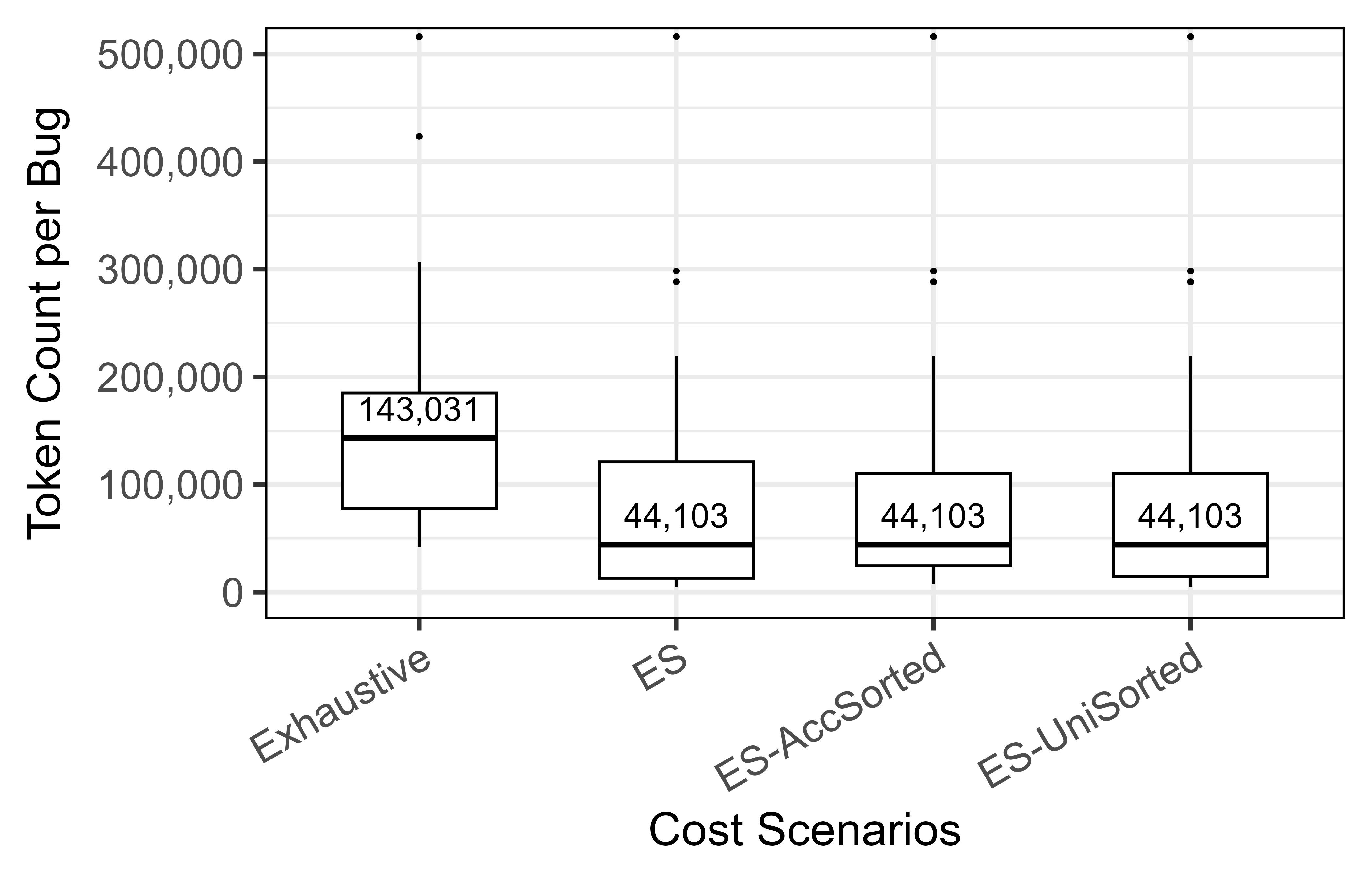}
    \caption{DeepSeek-Coder-V2-Lite-Instruct-16B \\on BugsInPy}
    \label{rq3_box_plot_token_2_datasets_3_models:e}
  \end{subfigure}
  \hfill
  \begin{subfigure}[b]{0.48\textwidth}
    \includegraphics[width=\textwidth]{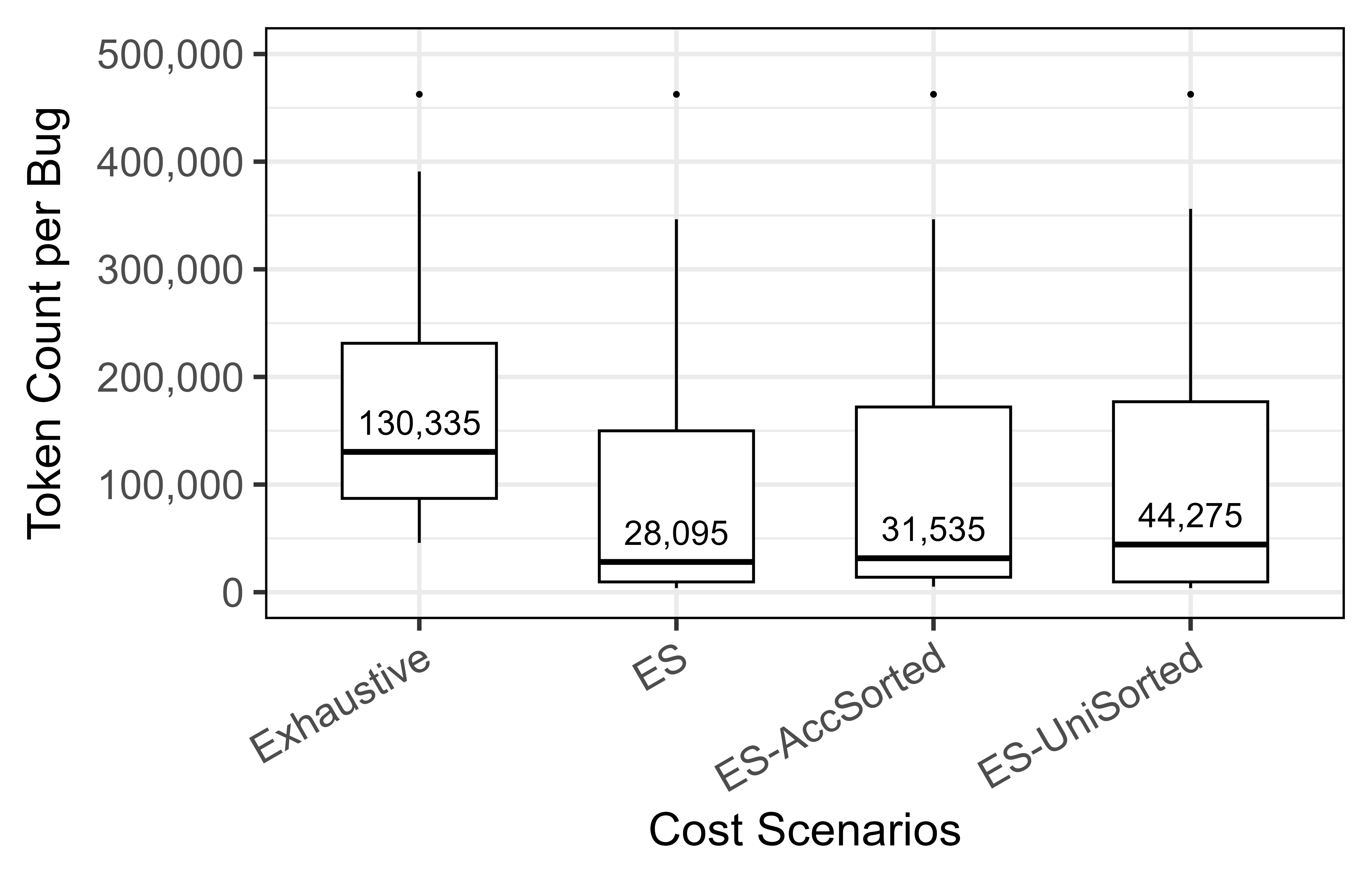}
    \caption{DeepSeek-Coder-V2-Lite-Instruct-16B \\on Defects4J}
    \label{rq3_box_plot_token_2_datasets_3_models:f}
  \end{subfigure}
  \caption{Distribution of the number of inference tokens across different cost scenarios, evaluated on two datasets and three models.}
  \label{fig:rq3_box_plot_token_2_datasets_3_models}
\end{figure}


\textbf{Early stopping reduces inference tokens by an average of 73\% and is significantly more efficient than the Exhaustive scenario.}
Figure \ref{fig:rq3_box_plot_token_2_datasets_3_models} presents the distributions of the number of inference tokens across all bugs over three models and two datasets under the four cost scenarios: Exhaustive, ES, ES-AccSorted, and ES-UniSorted. The Exhaustive scenario consistently incurs the highest median token usage. For instance, CodeLlama-Instruct-7B on BugsInPy consumes a median of 148479 tokens, compared to 32137, 25104, and 41843 for the other three early stopping scenarios, respectively. Early stopping reduces inference tokens by an average of 73\%, demonstrating substantial efficiency gains over the Exhaustive strategy. Differences between sorted (ES-AccSorted and ES-UniSorted) and unsorted (ES) variants remain generally minimal.

The statistical analysis reinforces this observation. For each model-dataset configuration, the Friedman test revealed significant differences among the four scenarios ($p < 0.001$). Subsequent pairwise Wilcoxon signed-rank tests, adjusted using Bonferroni correction ($\alpha = 0.0083$), confirmed that the Exhaustive scenario consumes significantly more tokens than all early stopping variants across all six configurations. The detailed pairwise results are reported in Appendix \ref{appendix:AdditionalRQ3Results} (Table \ref{tab:wilcox_rq3_box_plot_token_2_datasets_3_models}). The $p$-values range from $2.5 \times 10^{-13}$ to $4.6 \times 10^{-6}$. The effect size is negligible (0.05) in one configuration (CodeLlama-Instruct-7B on BugsInPy), while the remaining five configurations show non-negligible effect sizes with $r_{\mathrm{rb}}$ ranging from 0.15 to 0.39 (small to medium). This trend is consistent with the inference time results, as both aspects reflect the computational cost of model inference. These findings underscore the effectiveness of early stopping strategies in not only reducing inference latency but also minimizing inference tokens.


\textbf{Early stopping avoids inefficient heuristics, and further prioritization by the ES-UniSorted approach offers additional benefits in some configurations.}
With the overall differences in inference tokens consumption established, we now explore how the token consumption evolves during the bug-fixing process within each scenario. Figure \ref{fig:rq3_symmetry_bar_2_datasets_3_models} in Appendix \ref{appendix:AdditionalRQ3Results} further visualizes inference tokens consumption trends across different cost scenarios for HAFix-Agg, evaluated on the two datasets and three models, showing the cost of successfully fixed bugs (above the x-axis) and unfixed bugs (below the x-axis). Across all six model-dataset configurations, the Exhaustive scenario consistently incurs the highest token costs with substantial negative contributions from failed attempts. For CodeLlama-Instruct-7B on BugsInPy (Figure \subfig{fig:rq3_symmetry_bar_2_datasets_3_models}{rq3_symmetry_bar_2_datasets_3_models:a}), prioritizing heuristics such as Baseline and CFN-modified helps achieve more fixes early on. However, token usage remains comparable across ES, ES-AccSorted, and ES-UniSorted, indicating limited benefit from heuristic reordering in this configuration. In contrast, the Exhaustive scenario applies all heuristics regardless of efficiency, leading to significantly higher token consumption without proportional gains in bug fixes. This confirms the cost-saving advantage of early stopping.
Notably, in several other configurations, including DeepSeek-Coder-Instruct-6.7B on BugsInPy and all three models on Defects4J (Figures \subfig{fig:rq3_symmetry_bar_2_datasets_3_models}{rq3_symmetry_bar_2_datasets_3_models:b}, \subfig{fig:rq3_symmetry_bar_2_datasets_3_models}{rq3_symmetry_bar_2_datasets_3_models:c}, \subfig{fig:rq3_symmetry_bar_2_datasets_3_models}{rq3_symmetry_bar_2_datasets_3_models:d} and \subfig{fig:rq3_symmetry_bar_2_datasets_3_models}{rq3_symmetry_bar_2_datasets_3_models:f}), the ES-UniSorted strategy shows a reduced token footprint for unfixed cases. This suggests that reordering heuristics to prioritize more effective ones can further mitigate the cost of failed attempts through earlier termination.
These findings highlight the effectiveness of early stopping as a core strategy and suggest that further heuristic prioritization via ES-UniSorted may yield additional benefits in configurations where failure costs are high.


\begin{figure}[!htbp]
  \centering
  \begin{subfigure}[b]{0.48\textwidth}
    \includegraphics[width=\textwidth]{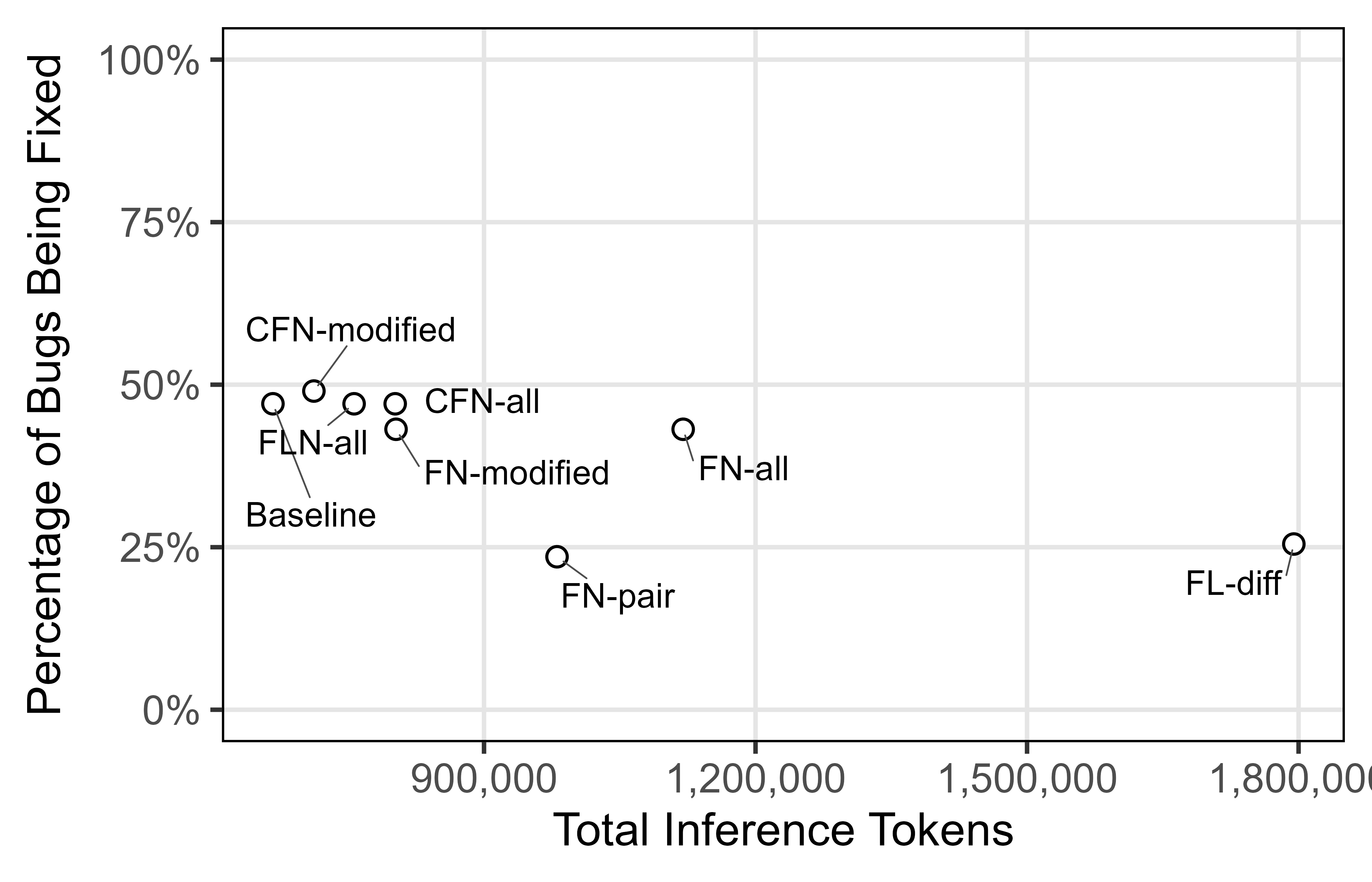}
    \caption{CodeLlama-Instruct-7B on BugsInPy}
    \label{rq3_scatter_token_2_datasets_3_models:a}
  \end{subfigure}
  \hfill
  \begin{subfigure}[b]{0.48\textwidth}
    \includegraphics[width=\textwidth]{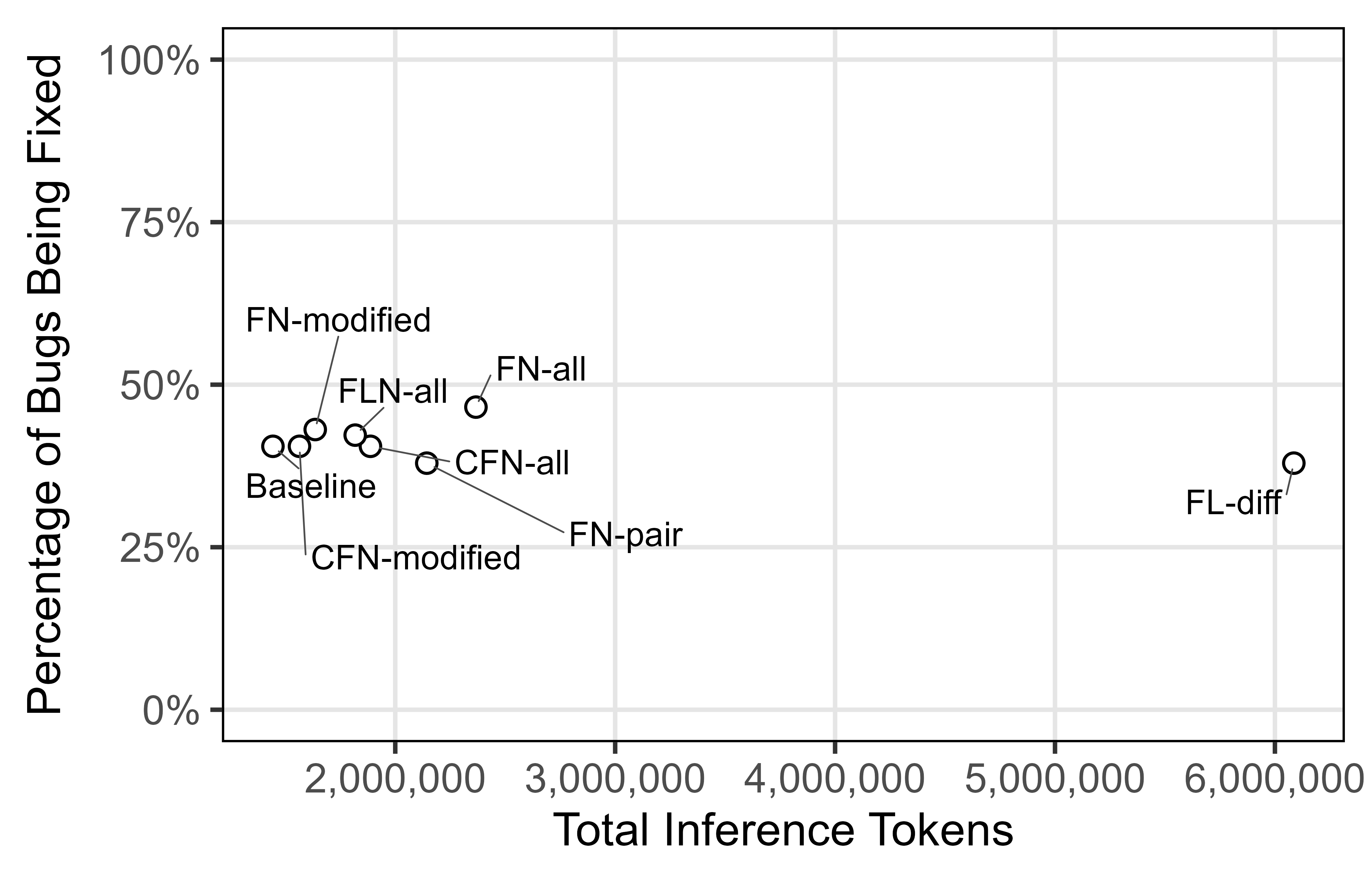}
    \caption{CodeLlama-Instruct-7B on Defects4J}
    \label{rq3_scatter_token_2_datasets_3_models:b}
  \end{subfigure}

  \vspace{1em} 
    
  \begin{subfigure}[b]{0.48\textwidth}
    \includegraphics[width=\textwidth]{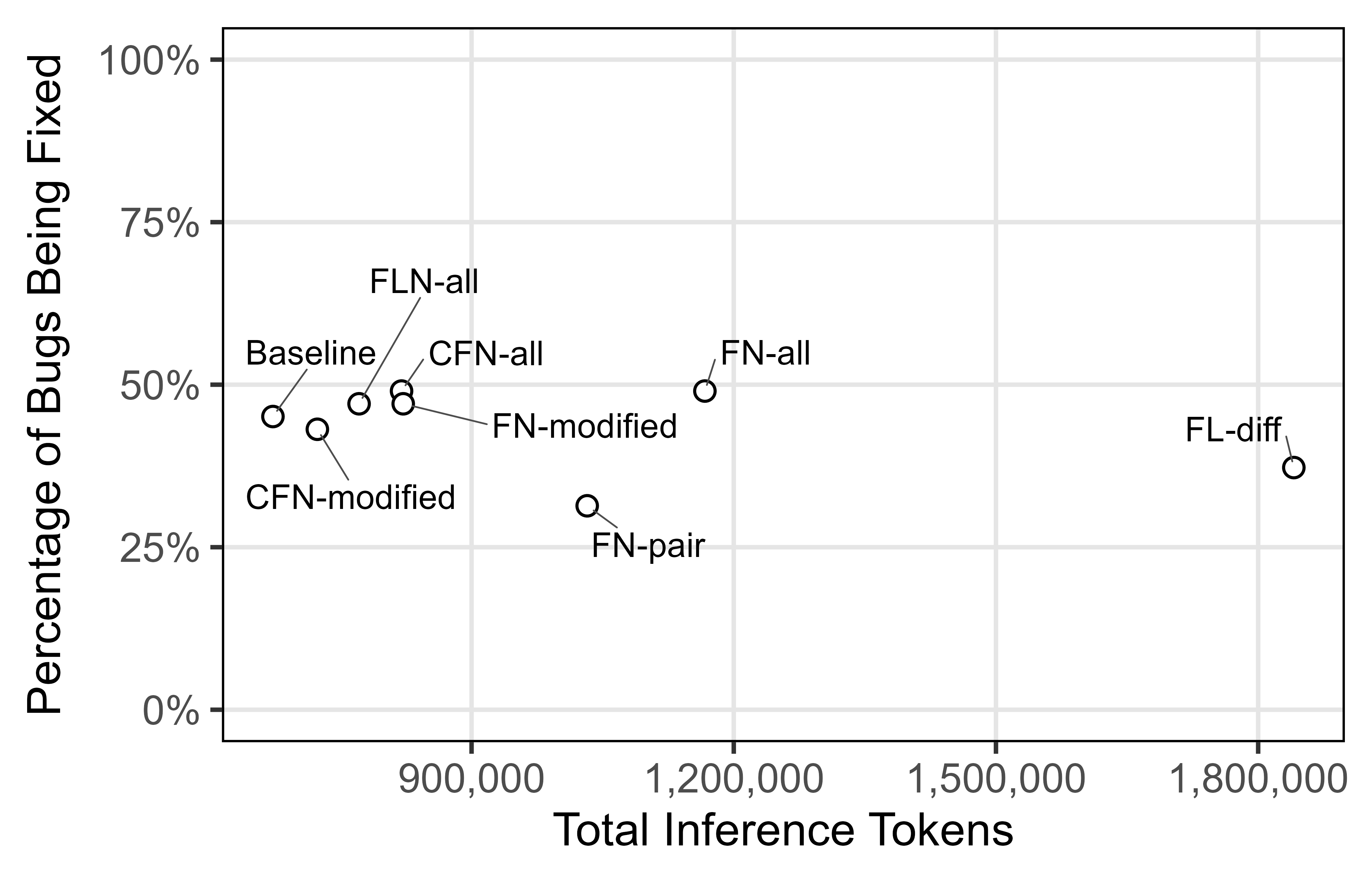}
    \caption{DeepSeek-Coder-Instruct-6.7B on \\BugsInPy}
    \label{rq3_scatter_token_2_datasets_3_models:c}
  \end{subfigure}
  \hfill
  \begin{subfigure}[b]{0.48\textwidth}
    \includegraphics[width=\textwidth]{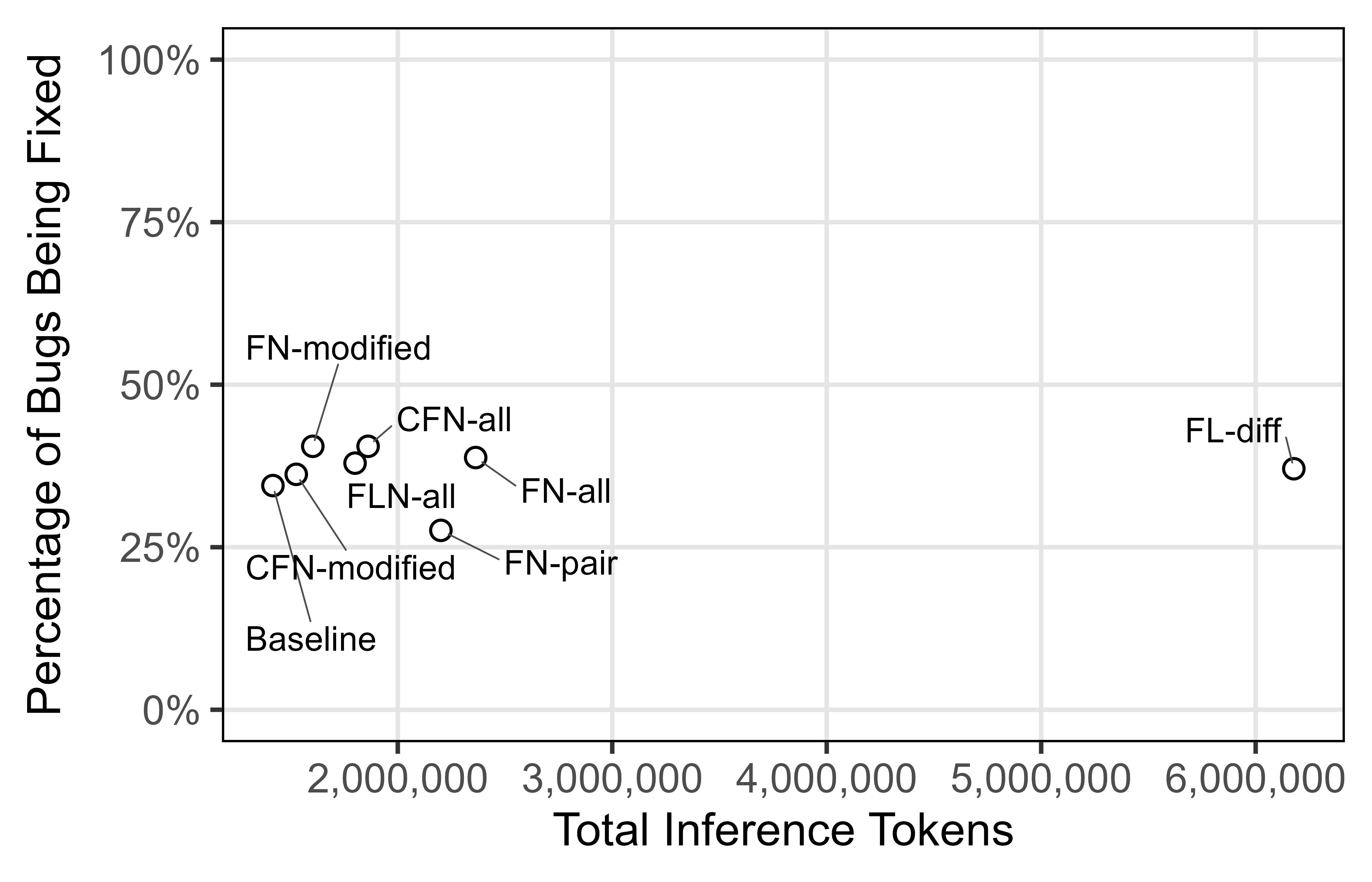}
    \caption{DeepSeek-Coder-Instruct-6.7B on \\Defects4J}
    \label{rq3_scatter_token_2_datasets_3_models:d}
  \end{subfigure}

  \vspace{1em} 
    
  \begin{subfigure}[b]{0.48\textwidth}
    \includegraphics[width=\textwidth]{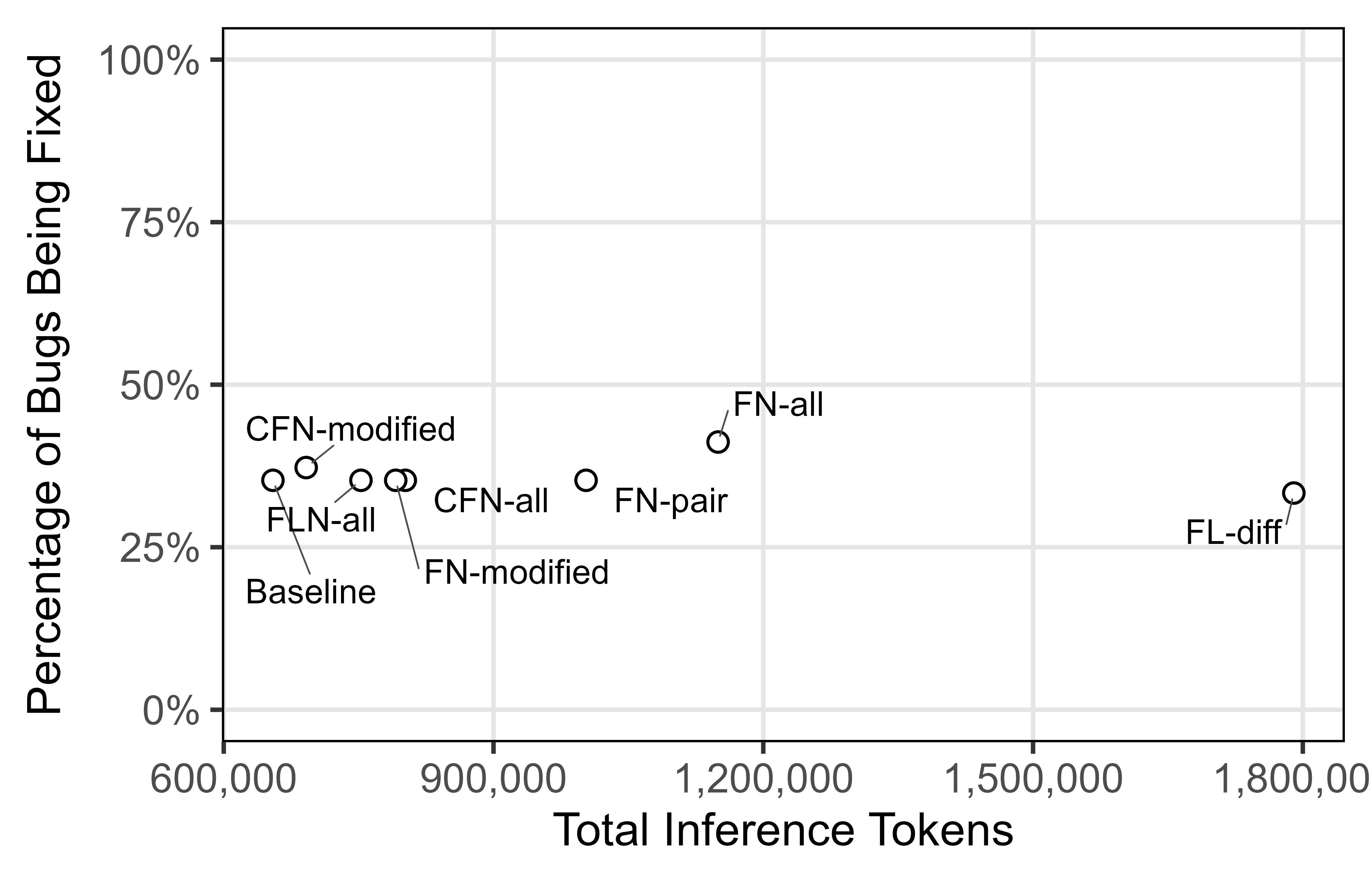}
    \caption{DeepSeek-Coder-V2-Lite-Instruct-16B \\on BugsInPy}
    \label{rq3_scatter_token_2_datasets_3_models:e}
  \end{subfigure}
  \hfill
  \begin{subfigure}[b]{0.48\textwidth}
    \includegraphics[width=\textwidth]{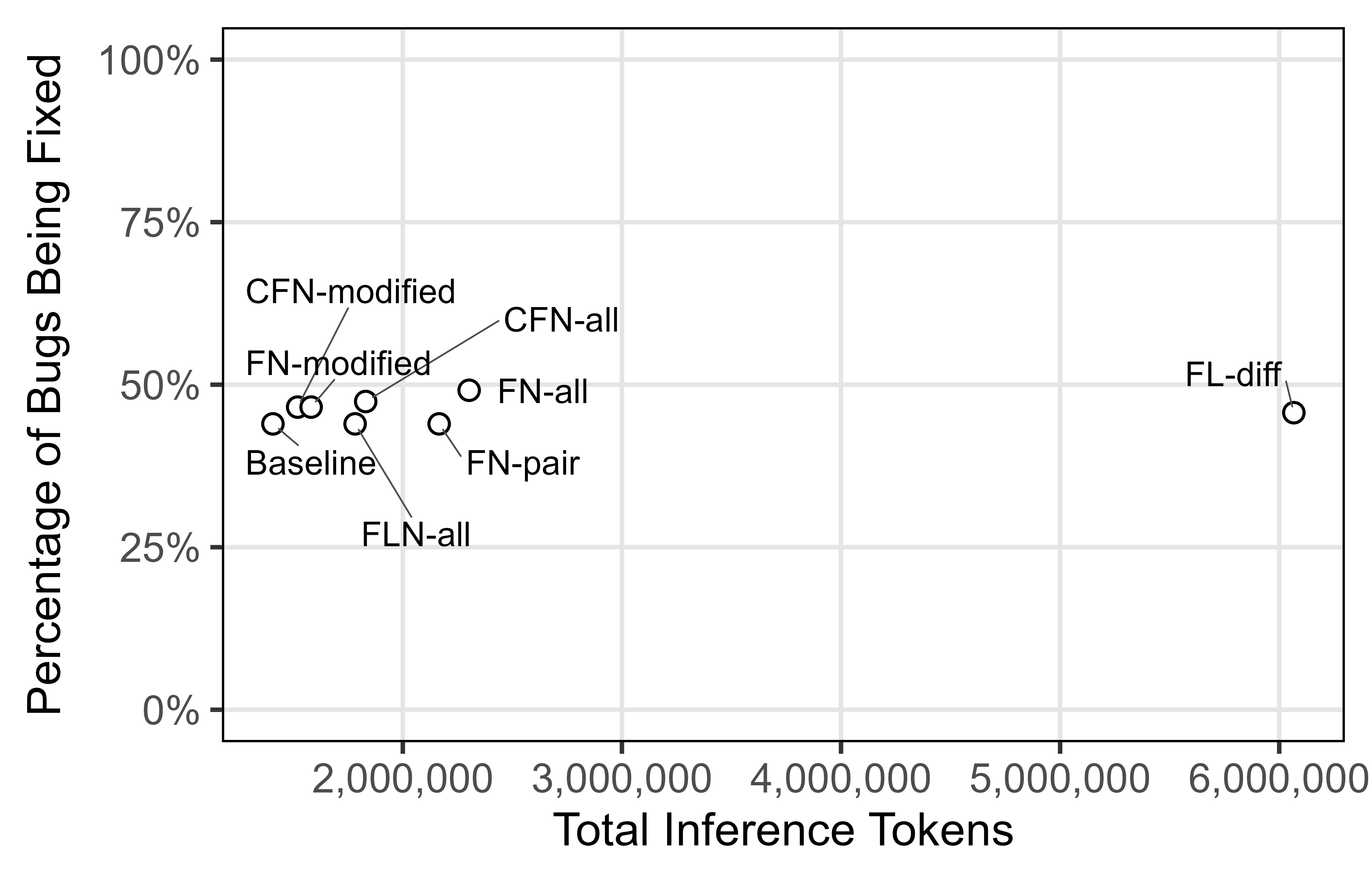}
    \caption{DeepSeek-Coder-V2-Lite-Instruct-16B \\on Defects4J}
    \label{rq3_scatter_token_2_datasets_3_models:f}
  \end{subfigure}
  \caption{Percentage of bugs being fixed (Table \ref{rq1_acc_percentage}) vs. total inference tokens across different heuristics in the Exhaustive scenario. The inference token reflects the total number of tokens required to process all bugs for each heuristic.}
  \label{fig:rq3_scatter_token_2_datasets_3_models}
\end{figure}

\textbf{FL-diff and FN-all incur high token costs with minimal performance gains, while other heuristics achieve more favorable trade-offs between tokens and performance.}
After analyzing overall cost trends, it is also important to examine how inference token usage correlates with heuristic effectiveness. Figure \ref{fig:rq3_scatter_token_2_datasets_3_models} illustrates the relationship between the percentage of bugs fixed and inference tokens consumed by each heuristic under the Exhaustive scenario.
Across all six model-dataset configurations, the results mirror those observed for inference time: FL-diff consistently requires the highest number of tokens yet achieves only moderate bug-fixing performance, with FN-all exhibiting a similar pattern. In contrast, Baseline, CFN-modified, FLN-all, FN-modified, and CFN-all demonstrate better token-performance balances, making them more efficient choices across both datasets and models.

\begin{tcolorbox}
\textbf{Summary for RQ3:}
\begin{enumerate}
    \item The Exhaustive scenario incurs significantly higher inference time and token usage than all early stopping strategies (ES, ES-AccSorted, ES-UniSorted) across all configurations. The corresponding effect sizes are mostly non-negligible (small to medium), with one negligible configuration. Overall, early stopping reduces inference time and tokens by an average of 69\% and 73\%, respectively.
    
    \item FL-diff and FN-all consistently emerge as the most time-consuming and token-intensive heuristics across model-dataset configurations, while other heuristics (Baseline, CFN-modified, FLN-all, FN-modified, CFN-all) exhibit more comparable and efficient resource requirements.
    
    \item Time-performance and token-performance trade-off analysis reveals that FL-diff and FN-all exhibit poor cost-effectiveness, while FN-modified, CFN-modified, and CFN-all achieve more favorable balances between inference cost and bug-fixing effectiveness.
    
    \item ES-UniSorted offers additional benefits in certain configurations by reducing inference tokens for failed attempts, suggesting that heuristic prioritization can provide incremental efficiency gains beyond early stopping alone.
\end{enumerate}
\end{tcolorbox}

\section{Threats to Validity}\label{Threats to Validity}

\subsection{Internal Validity}\label{sec:InternalValidity}

One threat involves potential data leakage when training the Large Language Models. The commit messages and bug descriptions mined from GitHub might contain explicit hints or detailed information on how the bug was fixed, which could influence the LLM's ability to generate correct repairs. To address this, we manually reviewed all mined texts and two people discussed and removed those that contained repair-specific information. While this manual filtering reduces the risk of leakage in the model inference stage, the data leakage during training may still exist, i.e., the LLM might have seen the issue report from collected open-source training datasets. However, we believe our manual work mitigates the issue in the context of our experiments during model inference.

Regarding model selection, we employed CodeLlama-Instruct-7B, DeepSeek-Coder-Instruct-6.7B and DeepSeek-Coder-V2-Lite-Instruct-16B due to their balanced trade-off between performance and resource efficiency \citep{roziere2023code,zhu2024deepseek}, with model sizes ranging from approximately 13GB to 31GB. While larger models such as CodeLlama-70B or DeepSeek-Coder-V2-236B may yield improved performance, they demand substantially more computational resources and memory. This limitation may constrain the baseline’s performance, as larger models could produce higher-quality fixes. As a result, this poses a potential threat to internal validity by not reflecting the optimal capability of available models.

Furthermore, in RQ3, we calculate the inference time and token usage of HAFix solely during the model inference stage, excluding the cost of running test cases. We observed that most of the time required for running test cases is consumed by installing the dependent libraries, such as averaging around 6 minutes per bug on BugsInPy, while the actual execution of the test cases typically takes no more than a second, given that they usually consist of just a few lines of code. In practice, setting up the test environment is generally a one-time action, as developers typically maintain a stable development environment and do not reinstall libraries for each test. Regarding inference token usage, resources for model inference on GPUs are usually much more expensive than running test cases on a local CPU environment. Therefore, we believe that the cost and efficiency analyzed in RQ3 were underestimated but cover the main aspects of the real cost. 

For RQ3, we report token usage rather than inference price because API pricing varies across models, time periods, and providers. Even for the same model, pricing may differ depending on the time of day or commercial factors \citep{OpenAIPricing,AnthropicPricing,DeepSeekPricing}. Additionally, input and output tokens are priced differently, with output tokens typically costing around four times more. To account for this, we experimented with a weighted token cost that gives more weight to output tokens. The results were consistent with our main findings. To maintain focus and avoid overloading the evaluation, we do not include the weighted cost plots in the paper.

\subsection{External Validity}\label{sec:ExternalValidity}

One threat to external validity in our study lies in the focus on single-line bugs. This research does not aim to address complex bugs beyond state-of-the-art approaches, but rather to demonstrate the potential of incorporating historical commit data into the bug-fixing process of LLMs. Most of the recent existing works focus on single-line bugs \citep{xia2022less,ye2022neural,jiang2023impact,prenner2024out}. 

Additionally, we evaluate 51 Python bugs and 116 Java bugs, which could be further extended. We investigated the recent works and found that \citet{prenner2022can} evaluate 40 bugs in both Java and Python, \citet{kolak2022patch} evaluate 72 Python bugs and \citet{chen2024large} evaluate 124 bugs. While our dataset covers 27 open-source projects and provides diversity, it does not fully capture the complexity of real-world bug-fixing scenarios, particularly those involving multi-line or multi-hunk bugs. Extending our study to other programming languages and larger, more varied datasets could yield additional insights and broaden the applicability of our approach. Future work should explore the scalability of history-augmented LLMs across different programming languages and more complex bug scenarios, such as multi-line or multi-hunk bugs, to validate their effectiveness in diverse contexts.

\subsection{Construct Validity}\label{sec:ConstructValidity}

Regarding construct validity, a notable limitation stems from the inherent non-determinism in the outputs of LLMs \cite{hassan2024rethinking}. As LLMs generate results probabilistically, more repeated runs might yield slightly different outputs, potentially leading to variations in pass@k results. To address this, we ran the baseline experiment 7 times, with each run obtaining 10 fixes (obtained using 10 separate queries) per bug across two datasets and three models. Following this paradigm, we also run the baseline experiment under different model-dataset configurations 7 times, with each time there are 10 samples generated. We conduct a stability analysis of these repeated experiments in Subsection \ref{Stability Analysis of Inference Results} of the Discussion section. These approaches help reduce the impact of randomness in the generation process and provide a more comprehensive evaluation of the model’s capabilities, although some variation in outcomes may still occur.

Another potential limitation relates to the use of test suites to measure correctness. While we rely on the official test suites from BugsInPy and Defects4J, which are widely adopted in the literature, we acknowledge that passing all test cases may not fully guarantee semantic correctness, as some incorrect fixes might coincidentally pass the tests. A manual review of all generated fixes would be impractical at our scale of over 120k solutions (167 bugs × 3 models × 8 configurations × 10 samples × 3 prompt styles). However, to validate that this limitation does not compromise our conclusions, we strategically selected a representative subset for manual inspection. Specifically, based on Table \ref{rq1_acc_percentage}, we identified the model-dataset configuration with the highest number of unique fixes by a HAFix heuristic over the baseline (CodeLlama on Defects4J with FN-all, 14 unique bugs). Our execution yielded 36 different solutions for the 14 unique bugs. We manually inspected these 36 solutions, out of which 12 matched the developer fix exactly, 11 were semantically equivalent, and 13 were neither identical nor semantically equivalent. Representative examples of these cases are provided in Appendix \ref{appendix:BugFixExample}. This limitation underscores the need for correctness metrics beyond test-case-dependent measures like pass@k or manual checks, especially given the scale of outputs in LLM-based approaches. To mitigate this limitation, we use the pass@k metric, which measures the likelihood of generating at least one fix passing test suites among multiple samples, a popular practice in recent LLM-based bug-fixing studies \citep{campos2025empirical}.

\section{Discussion and Future Work}\label{Future Work and Discussion}

\subsection{Comparison with SOTA Bug-Fixing Tools}\label{Comparison with SOTA Bug-Fixing Tools}

\begin{figure}[!t]
  \centering
  \begin{subfigure}[b]{0.48\textwidth}
    \includegraphics[width=\textwidth]{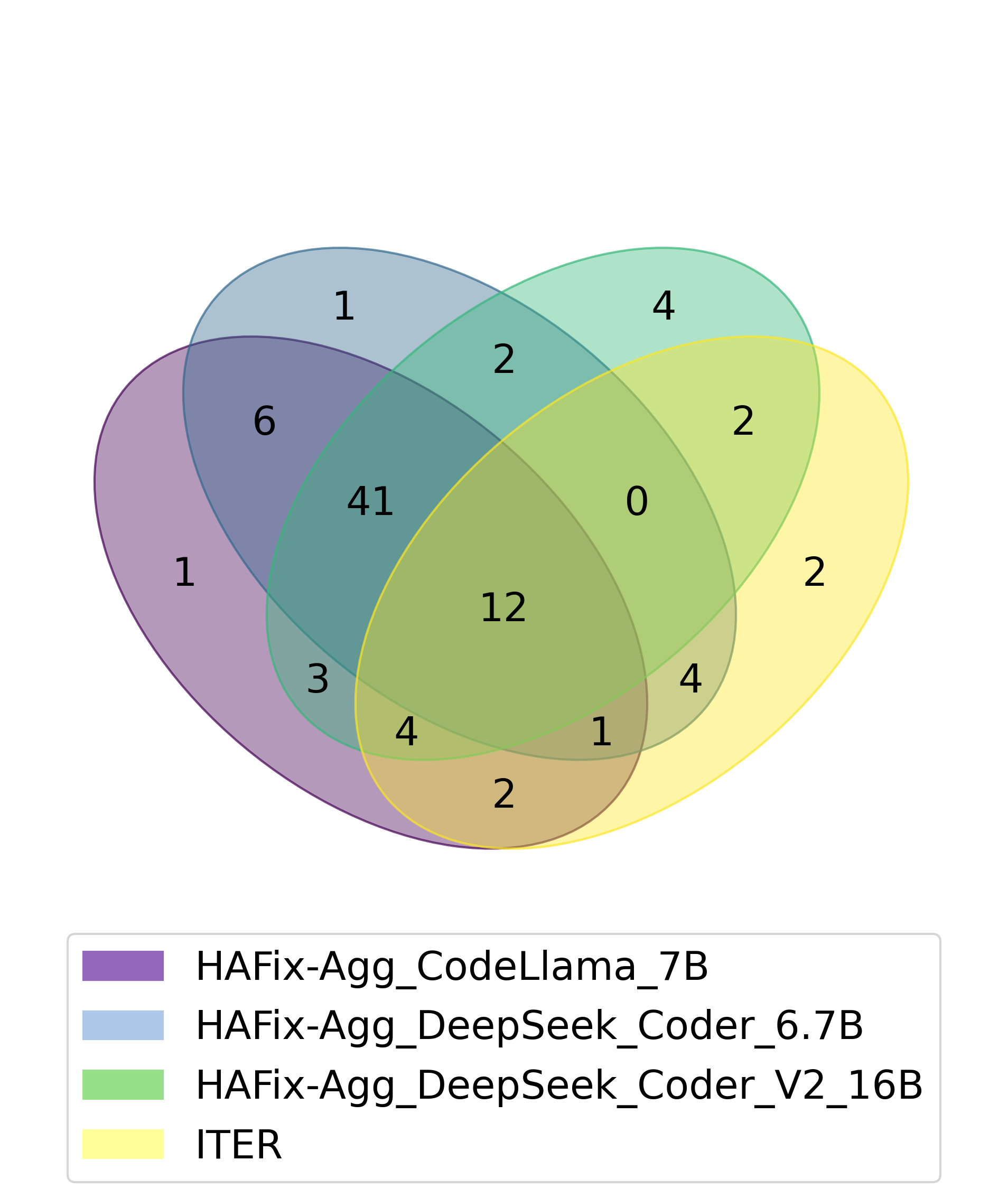}
    \caption{HAFix-Agg vs ITER}
    \label{compare_sota_tools:a}
  \end{subfigure}
  \hfill
  \begin{subfigure}[b]{0.48\textwidth}
    \includegraphics[width=\textwidth]{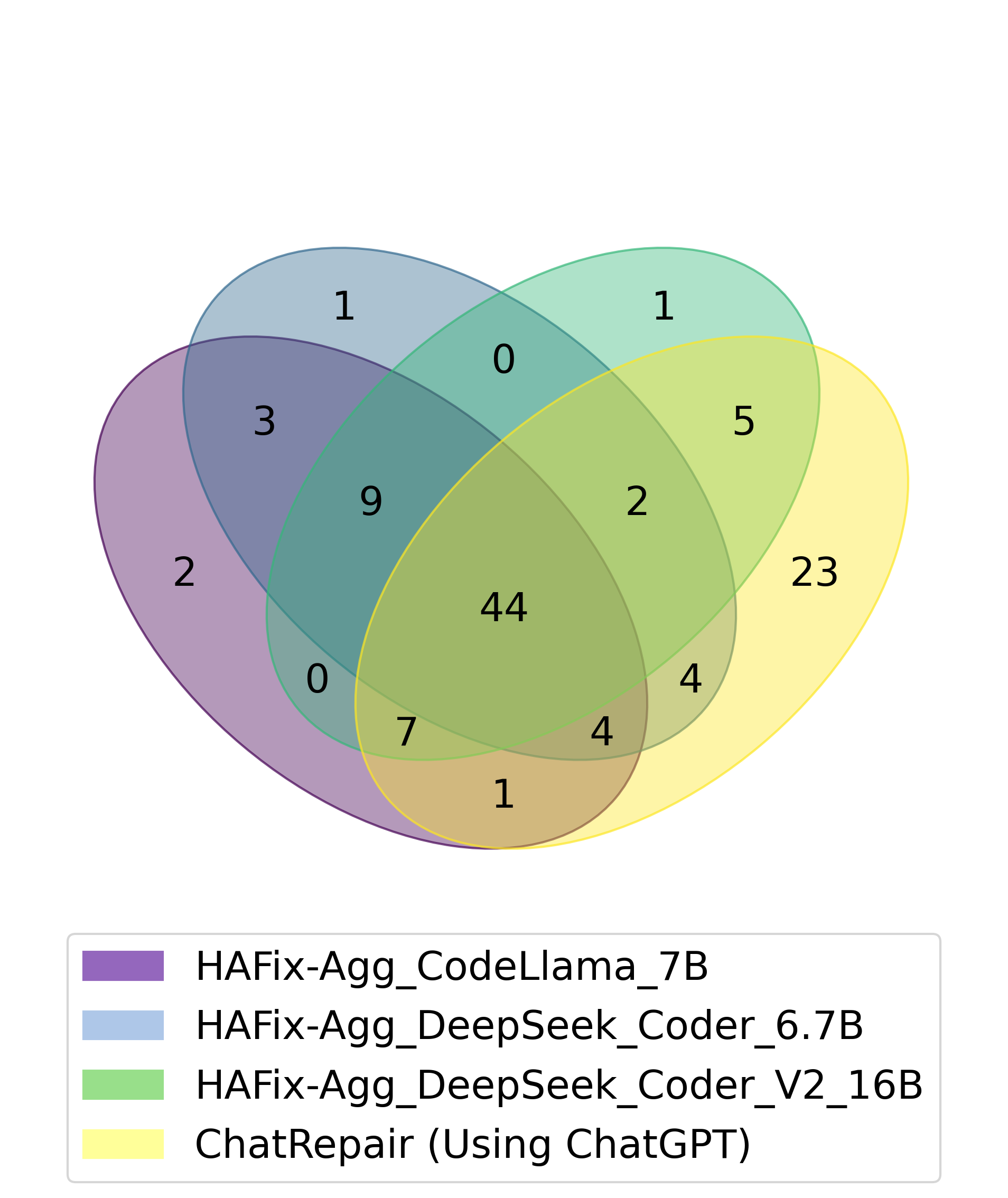}
    \caption{HAFix-Agg vs ChatRepair}
    \label{compare_sota_tools:b}
  \end{subfigure}
  \caption{Comparison of HAFix-Agg with ITER and ChatRepair on Defects4J. Each Venn diagram shows the overlap of bugs fixed.}
  \label{fig:compare_sota_tools}
\end{figure}

Beyond our main results in RQ1, we also compare HAFix with state-of-the-art (SOTA) bug-fixing tools to further contextualize its effectiveness. Specifically, we examine ChatRepair \citep{xia2024automated} and ITER \citep{ye2024iter}, two recent approaches that represent the current SOTA in LLM-based bug fixing. Both tools employ multi-turn interactive inference, with ChatRepair leveraging closed-source ChatGPT and ITER using a specialized iterative training process. Using evaluation results from \citep{bouzenia2024repairagent}, we compare HAFix-Agg with these tools on their common set of single-line bugs in Defects4J. Figure \ref{fig:compare_sota_tools} illustrates the overlap of bugs fixed by the three approaches. 

HAFix-Agg with CodeLlama, DeepSeek-Coder, and DeepSeek-Coder-V2-Lite fixes 51, 50, and 50 more bugs than ITER, and 14, 13, and 10 more bugs than ChatRepair, respectively. Conversely, ChatRepair fixes 23 additional bugs, likely due to its larger closed-source model (i.e., ChatGPT) and multi-turn interaction, whereas HAFix relies on smaller open-source models (from 6.7B to 16B). These findings reinforce that historical context offers substantial benefits, even when using smaller models, and highlight opportunities for integrating HAFix with stronger models and agentic frameworks in future work.

\subsection{Stability Analysis of Inference Results}\label{Stability Analysis of Inference Results}

To assess the stability of the model's inference results, we used the evaluation data from RQ1 (Subsection \ref{RQ1}), which involved repeatedly running the baseline 7 times over two datasets and three models. We measure stability using the coefficient of variation (CV), the ratio of the standard deviation to the mean, which quantifies relative variability in repeated runs.

As shown in Table \ref{tab:baseline_stability_analysis}, the coefficient of variation (CV) on BugsInPy ranged from 2.00\% to 7.69\%, indicating low to moderate variability, while on Defects4J it ranged from 1.17\% to 3.60\%, indicating high stability. This level of consistency is expected in stochastic processes like LLM inference, confirming the robustness of our reported findings throughout three RQs. Historical signals can improve bug fixing by reducing uncertainty in the prompt, narrowing the space of plausible fixes toward edits that align with recent code evolution, while irrelevant history can add noise and hurt performance. More broadly, this aligns with long-standing evidence from the MSR community that well-chosen historical information improves core SE tasks, such as identifying bug-inducing commits (SZZ) \citep{sliwerski2005changes} or improving regression test selection \citep{elbaum2014techniques}.

\begin{table}[t!]
\centering
\caption{Stability analysis of pass@k performance across 7 repeated baseline runs in the \Instruction prompt style over two datasets and three models. Pass@k\_A represents the averaged pass@k values, and CV represents the coefficient of variation.}
\label{tab:baseline_stability_analysis}
\renewcommand{\arraystretch}{1.1}
\begin{subtable}{\textwidth}
\centering
\caption{BugsInPy}
\begin{tabular}{ccccccc}
\toprule
\multirow{2}{*}{k} & \multicolumn{2}{c}{\shortstack{CodeLlama\\[2pt]-Instruct-7B}} & \multicolumn{2}{c}{\shortstack{DeepSeek-Coder\\[2pt]-Instruct-6.7B}} & \multicolumn{2}{c}{\shortstack{DeepSeek-Coder-V2\\[2pt]-Lite-Instruct-16B}} \\
\cline{2-7}
& Pass@k\_A & CV & Pass@k\_A & CV & Pass@k\_A & CV \\
\cline{1-7}
1  &   21.93\% & 4.29\% &   28.43\% & 3.76\% &   30.36\% & 2.08\%   \\
2  &   27.68\% & 3.94\% &   34.86\% & 2.75\% &   32.47\% & 2.00\%   \\
3  &   31.62\% & 4.05\% &   37.99\% & 2.29\% &   33.74\% & 2.64\%   \\
4  &   34.66\% & 4.41\% &   39.92\% & 2.18\% &   34.65\% & 3.32\%   \\
5  &   37.13\% & 4.93\% &   41.28\% & 2.25\% &   35.38\% & 3.96\%   \\
6  &   39.23\% & 5.48\% &   42.33\% & 2.48\% &   35.99\% & 4.53\%   \\
7  &   41.02\% & 6.02\% &   43.19\% & 2.82\% &   36.52\% & 5.09\%   \\
8  &   42.57\% & 6.58\% &   43.91\% & 3.21\% &   36.99\% & 5.60\%   \\
9  &   43.92\% & 7.10\% &   44.54\% & 3.61\% &   37.42\% & 6.12\%   \\
10 &   45.10\% & 7.69\% &   45.10\% & 4.01\% &   37.81\% & 6.64\%   \\
\bottomrule
\end{tabular}
\end{subtable}

\vspace{0.5cm} 

\begin{subtable}{\textwidth}
\centering
\caption{Defects4J}
\renewcommand{\arraystretch}{1.2}
\begin{tabular}{ccccccc}
\toprule
\multirow{2}{*}{k} & \multicolumn{2}{c}{\shortstack{CodeLlama\\[2pt]-Instruct-7B}} & \multicolumn{2}{c}{\shortstack{DeepSeek-Coder\\[2pt]-Instruct-6.7B}} & \multicolumn{2}{c}{\shortstack{DeepSeek-Coder-V2\\[2pt]-Lite-Instruct-16B}} \\
\cline{2-7}
& Pass@k\_A & CV & Pass@k\_A & CV & Pass@k\_A & CV \\
\cline{1-7}
1  &   25.93\% & 2.04\% &   23.63\% & 1.74\% &   36.75\% & 1.17\%   \\
2  &   31.80\% & 1.73\% &   27.82\% & 2.26\% &   40.33\% & 1.19\%   \\
3  &   34.95\% & 1.66\% &   30.06\% & 2.59\% &   42.01\% & 1.21\%   \\
4  &   37.02\% & 1.76\% &   31.54\% & 2.79\% &   42.97\% & 1.26\%   \\
5  &   38.54\% & 1.89\% &   32.62\% & 2.94\% &   43.61\% & 1.31\%   \\
6  &   39.75\% & 2.04\% &   33.44\% & 3.05\% &   44.08\% & 1.36\%   \\
7  &   40.73\% & 2.23\% &   34.11\% & 3.17\% &   44.46\% & 1.46\%   \\
8  &   41.56\% & 2.45\% &   34.67\% & 3.26\% &   44.78\% & 1.61\%   \\
9  &   42.27\% & 2.77\% &   35.16\% & 3.41\% &   45.06\% & 1.78\%   \\
10 &   42.86\% & 3.17\% &   35.59\% & 3.60\% &   45.32\% & 1.99\%   \\
\bottomrule
\end{tabular}
\end{subtable}
\end{table}

\subsection{Performance Regressions due to Irrelevant Context}\label{Performance Regressions from Irrelevant Context}

While most HAFix heuristics improve bug-fixing accuracy by incorporating historical context, a few (e.g., FLN-all, FN-pair) occasionally underperform compared to the baseline. For instance, in Lang-21 (Defects4J, CodeLlama-Instruct-7B), the baseline fix was semantically equivalent to the developer’s patch and passed all tests, whereas FN-all added historical context that was not directly relevant to the bug, leading to incorrect edits and test failures (Appendix \ref{appendix:BugFixExample}, Listing \ref{lst:Lang-21}).

This behavior aligns with findings in LLM research that more context does not always improve performance. \citet{shi2023large} demonstrate that LLMs can become easily distracted by irrelevant context, leading to significant accuracy drops, even in simple arithmetic reasoning tasks, if prompts include unnecessary sentences. \citet{levy2024same} show that performance often degrades when relevant information appears in the middle or deeper parts of long input contexts, meaning key signals may be overlooked in long prompts. In HAFix, some heuristics introduce extra historical content beyond the baseline, which may occasionally dilute the prominence of essential signals and reduce Pass@k. Nevertheless, these heuristics often fix unique bugs that the baseline misses (Table \ref{rq1_acc_percentage}, Figures \ref{fig:venn-diagram-CodeLlama-Instruct-7B-BugsInPy} and \ref{fig:venn-diagram-DeepSeek-Coder-Instruct-6.7B-Defects4J} and Appendix \ref{appendix:VennDiagrams}).

Potential mitigation strategies include filtering historical snippets for direct relevance and dynamically selecting heuristics based on bug characteristics. Although such regressions are infrequent, refining these heuristics could further reduce the risk of losing baseline fixes.

\subsection{Impact of Model Size}\label{Impact of Model Size}

We found that larger model capacity does not consistently improve history-augmented bug fixing. On BugsInPy (Figure \ref{fig:rq1_passk_baseline_hafix_2_datasets_3_models_BugsInPy}), DeepSeek-Coder-V2-Lite-Instruct (16B) performs worse than DeepSeek-Coder-Instruct (6.7B) for both the baseline and HAFix-Agg, whereas on Defects4J (Figure \ref{fig:rq1_passk_baseline_hafix_2_datasets_3_models_Defects4J}), the 16B model consistently outperforms the 6.7B model. This suggests that the benefits of historical context depend on the dataset and programming language, not only on parameter count.

A limitation of our study is that we only evaluate models from 6.7B to 16B parameters. Investigating HAFix across a wider range of model sizes, such as 70B, hundreds of billions (e.g., DeepSeek-Coder-V2-236B), or even trillions of parameters (such as GPT-4, size estimated from \citep{gpt4params}), is an important direction for future work.

\subsection{Prompt Style and Cost Trade-offs}\label{Prompt Style and Cost Trade-offs}

Our results show that across most model-dataset configurations, \Instruction achieves the best performance, suggesting that explicit instruction-style prompts are a strong default choice for deploying HAFix in practice. At the same time, higher performance generally requires more inference budget, creating a trade-off between bug-fixing performance, inference price, and time efficiency. Our RQ3 analysis shows that early-stopping scenarios such as ES can reduce cost while preserving competitive performance compared to exhaustive execution. Overall, these results suggest a pragmatic strategy: use \Instruction for performance, and apply early stopping to control cost under limited computational resources.

\subsection{Future Work}\label{Future Work}

For future work, our approach is currently specialized for single-line bugs, which allows us to clearly validate the core idea that historical context improves LLM-based bug fixing. While current state-of-the-art (SOTA) LLM-based APR tools primarily focus on single-line bugs \citep{xia2022less,ye2022neural,jiang2023impact,prenner2024out}, future work should aim to extend the applicability of our approach to more complex cases, such as multi-line and multi-hunk bugs. In particular, we plan to integrate HAFix into recent agentic software engineering workflows, which are better suited for addressing more complex bug scenarios.

Moreover, exploring deeper historical information beyond blame commit information is possible. Although this work employs seven historical heuristics, such as co-evolved file names identified from the blame commit, other recent commit information may provide additional insights that can further improve bug-fixing performance. For example, exploring commit histories preceding or following the bug-introducing commit may reveal patterns or relevant code changes that can inform the fix. Future research could explore methods to extract and integrate this broader historical information into the bug-fixing process, potentially improving the model’s understanding of how bugs emerge and evolve.


Additionally, although this study focused on bug-fixing, the history-based heuristics that we developed could be evaluated on other software engineering tasks, such as code generation or completion. For these tasks, the co-evolution of files and commit histories might also provide useful context, helping LLMs to generate more contextually aware and consistent code. Future work should investigate the effectiveness of these heuristics in different tasks and assess how well they generalize across various code-related challenges.

Furthermore, incorporating explainable AI techniques could help increase the transparency and trustworthiness of HAFix. For example, attention visualization \citep{zhao2024explainability} or gradient-based attribution method \citep{wang2024gradient} could provide insights into which historical signals most strongly influence the model’s bug-fixing decisions, thereby making the approach more interpretable for developers and researchers.

In addition, optimization and adaptive strategies represent promising directions for improving HAFix. Future work could explore dynamic heuristic weighting and metaheuristic-based aggregation, drawing on inspiration from recent work \citep{biswas2025integrating} to balance global exploration with local refinement, and \citep{agrawal2025quick} to manage uncertainty in historical data, adaptive deep-learning frameworks \citep{hussein2025smart} to adjust heuristic weights in real time as new bug–fix data becomes available, and optimization approach \citep{abualigah2025adaptive} to combine global-best guidance with localized search. Integrating these techniques into HAFix could make heuristic selection more responsive, robust, and effective across diverse software engineering contexts.

Last but not least, as LLMs are deployed at scale, managing the cost and efficiency of inference becomes increasingly important. Developing better cost models that account for both token usage and hardware limitations will be essential to ensure the practical scalability of LLM-based bug fixing.

\section{Conclusion}\label{Conclusion}

Inspired by the foundations of mining software repositories (MSR), this study explores the integration of seven different ways (and one aggregated variant HAFix-Agg) of adding historical context into LLM-based bug fixing, evaluates the impact of different LLM prompt styles, and investigates trade-offs between bug-fixing performance, inference time, and inference token usage. The results reveal critical insights into how LLMs can be optimized by incorporating historical information in bug fixing.

The incorporation of historical heuristics, such as FN-modified and FN-all on Defects4J, demonstrates significant improvements with large effect sizes in bug-fixing performance compared to a baseline inspired by GitHub Copilot. Our aggregated approach, HAFix-Agg, extends these improvements by leveraging the strong complementary strengths of multiple heuristics, achieving an average of 45.05\% on BugsInPy and 49.92\% on Defects4J compared to the non-history-based baseline in 6 model-dataset configurations. This finding highlights the importance of historical context as a valuable addition to LLM prompts for understanding and addressing bugs, reflecting the way in which historical software engineering data made great strides towards better software analytics early on in the mining software repositories domain.

The analysis of prompt styles underscores the critical role of prompt design in enabling models to leverage
historical context effectively for bug fixing. Among the three styles evaluated, \Instruction overall outperformed \InstructionLabel and \InstructionInfill, demonstrating its ability to provide clarity and explicitness that enables the model to make optimal use of historical heuristics. This finding establishes \Instruction as the preferred style for crafting prompts in history-augmented bug fixing.

In evaluating the trade-offs between performance, time efficiency and token usage for bug fixing, early stopping strategies emerged as a practical solution for reducing inference time and tokens by an average of 69\% and 73\%, respectively, without compromising effectiveness. Sorting approaches such as ES-UniSorted offer additional benefits in certain configurations by reducing inference tokens for failed attempts, suggesting that heuristic prioritization can provide incremental efficiency gains beyond early stopping alone.

Our findings collectively provide actionable insights into optimizing LLM-based bug fixing. By integrating historical heuristics, employing effective prompt designs, and leveraging cost-efficient execution strategies, developers can enhance both the practicality and scalability of automated bug-fixing systems over using individual, history-agnostic approaches. Future research could explore extending these methods to more complex types of bugs, additional programming languages, and other software engineering tasks such as code generation, to further validate and refine the role of historical context in LLMs for software engineering tasks.


\clearpage
\begin{appendices}
\section{Prompt Example}\label{appendix:PromptExample}

\captionsetup{type=listing}
\tcolorbox[
    colframe=black,      
    colback=white,       
    boxrule=0.5pt,       
    sharp corners,       
    left=0pt,            
    right=0pt,           
    top=0pt,             
    bottom=0pt,          
    width=\textwidth     
]

\begin{lstlisting}[style=promptstyle, escapeinside={||}]
|\textcolor{orange}{<s>[INST] <<SYS>>}|
You are a helpful and honest code assistant expert in fixing the buggy code in Python. I mined a buggy code snippet and its related information from GitHub. I will provide you with the project name, buggy file name, buggy function name, the date time, the current version of this buggy code snippet, the corresponding bug description that might indicate how this buggy code should be fixed, and the buggy line content that might suggest where this buggy code should be fixed. Please only generate the fixed code snippet of this buggy code, don't explain any other things.  Please wrap your fixed code snippet between ```python and ```
|\textcolor{orange}{<</SYS>>}|
# The project name: luigi
# The buggy file name: scheduler.py
# The buggy function name: get_pending_tasks
# The buggy code snippet:
\end{lstlisting}
\begin{lstlisting}[style=pythonstyle]
def get_pending_tasks(self, state):
    """
|    Get PENDING (and RUNNING) tasks for this worker.|
|    You have to pass in the state for optimization reasons.|
    """
    if len(self.tasks) < state.num_pending_tasks():
        return six.moves.filter(lambda task: task.status in [PENDING, RUNNING], self.tasks)
    else:
        return state.get_pending_tasks()
\end{lstlisting}
\begin{lstlisting}[style=promptstyle]
# The bug description: Filters tasks in second branch of Worker.get_pending_tasks (#1849)
When a worker has many DONE tasks, get_pending_tasks may switch to using state.get_pending_tasks in order to speed up the process. This can include pending tasks not owned by the worker, invalidating the result and causing functions like is_trivial_worker to return erroneous results. To fix this, we simply filter the results of state.get_pending_tasks to remove any tasks that don't include this worker.
# The buggy line content: return state.get_pending_tasks()
# The fixed code snippet:
|\textcolor{orange}{[/INST]}|
\end{lstlisting}
\endtcolorbox

\captionsetup{skip=0pt}
\captionof{listing}{A real example of a baseline prompt for a bug from the Luigi project.\protect\footnotemark[1] The prompt is designed based on the template from the official Code Llama documentation.\protect\footnotemark[2] The system prompt is enclosed within \texttt{<<SYS>>}, and the different components are structured by underscores.}
\label{BaselinePromptExample}

\footnotetext[1]{\url{https://github.com/spotify/luigi/commit/3c55acd2cd5cf9c6c760bec5bb3159e0bc48a614}}
\footnotetext[2]{\url{https://www.llama.com/docs/model-cards-and-prompt-formats/meta-code-llama/}}

\section{Representative Example of Dataset}\label{appendix:RepresentativeExamples}

\captionsetup{type=listing}
\begin{lstlisting}[
language=json, 
showstringspaces=false, 
breaklines=true,
frame=single,         % Add a single-line frame around the content
framerule=0.5pt,      % Set the thickness of the frame
rulesepcolor=\color{black}, % Set the frame color
]
{
  "bug_id": {
    "project_name": "luigi",
    "project_url": "https://github.com/spotify/luigi.git",
    "bugsinpy_id": "10",
    "is_single_line": true,
    "buggy_line_location": 305,
    "buggy_line_content": "            return state.get_pending_tasks()",
    "in_function": true,
    "commit": {
      "commit_id": "3c55acd2cd5cf9c6c760bec5bb3159e0bc48a614",
      "commit_message": "Filters tasks in second branch of Worker.get_pending_tasks (#1849)\n\nWhen a worker has many DONE tasks, get_pending_tasks may switch to using\r\nstate.get_pending_tasks in order to speed up the process. This can include\r\npending tasks not owned by the worker, invalidating the result and causing\r\nfunctions like is_trivial_worker to return erroneous results.\r\n\r\nTo fix this, we simply filter the results of state.get_pending_tasks to\r\nremove any tasks that don't include this worker.",
      "commit_parent": "f538d1b3d473d542a19d508e5f7e0809b1dfe5ef",
      "commit_date": "2016-09-12 09:51:39",
      "commit_file_diff": "@@ -302,7 +302,7 @@ class Worker(object):\n             return six.moves.filter(lambda task: task.status in [PENDING, RUNNING],\n                                     self.tasks)\n         else:\n-            return state.get_pending_tasks()\n+            return six.moves.filter(lambda task: self.id in task.workers, state.get_pending_tasks())\n \n     def is_trivial_worker(self, state):\n         \"\"\"\n"
    },
    "function": {
      "function_name": "get_pending_tasks",
      "function_parent": "Worker",
      "function_before_start_line": 295,
      "function_before_end_line": 305,
      "function_after_start_line": 295,
      "function_after_end_line": 305,
      "function_before_token_count": 54,
      "function_after_token_count": 72,
      "function_before": "def get_pending_tasks(self, state):\n    \"\"\"\n        Get PENDING (and RUNNING) tasks for this worker.\n\n        You have to pass in the state for optimization reasons.\n        \"\"\"\n    if len(self.tasks) < state.num_pending_tasks():\n        return six.moves.filter(lambda task: task.status in [PENDING, RUNNING], self.tasks)\n    else:\n        return state.get_pending_tasks()",
      "function_after": "def get_pending_tasks(self, state):\n    \"\"\"\n        Get PENDING (and RUNNING) tasks for this worker.\n\n        You have to pass in the state for optimization reasons.\n        \"\"\"\n    if len(self.tasks) < state.num_pending_tasks():\n        return six.moves.filter(lambda task: task.status in [PENDING, RUNNING], self.tasks)\n    else:\n        return six.moves.filter(lambda task: self.id in task.workers, state.get_pending_tasks())"
    },
    "file": {
      "file_name": "scheduler.py",
      "file_path": "luigi/scheduler.py",
      "file_nloc": 952,
      "file_complexity": 375,
      "file_token_count": 7424
    }
  }
}
\end{lstlisting}
\captionof{listing}{A representative example of a JSON-formatted dataset entry. This example shows metadata about a bug.}
\label{RepresentativeExampleDataset}

\section{Examples of Generated-Fixed Code by HAFix}\label{appendix:BugFixExample}

\begin{lstlisting}[style=javabox, caption={Example of Chart-4 in Defects4J, fixed code by HAFix is semantically different from the developer-written fix, yet still passes the test cases.}, label={lst:Chart-4}]
public String generateToolTipFragment(String toolTipText) {

    // (*@\textbf{\textcolor{red}{Buggy code}}@*)
    return " title=\"" + toolTipText + "\" alt=\"\"";
    
    // (*@\textbf{\textcolor{yellow}{Fixed code by developer}}@*)
    return " title=\"" + ImageMapUtilities.htmlEscape(toolTipText) + "\" alt=\"\"";

    // (*@\textbf{\textcolor{black}{Fixed code by baseline (CodeLlama-Instruct-7B), test failed}}@*)
    return " title=\"" + Jsoup.clean(toolTipText, Whitelist.none()) + "\" alt=\"\"";
    
    // (*@\textbf{\textcolor{green}{Fixed code by HAFix (FN-all, CodeLlama-Instruct-7B), test passed}}@*)
    return " title=\"" + escapeText(toolTipText) + "\" alt=\"\"";
}
\end{lstlisting}

\begin{lstlisting}[style=javabox, caption={Example of Jsoup-51 in Defects4J, fixed code by HAFix is semantically equivalent to the developer-written fix and passes the test cases.}, label={lst:Jsoup-51}]
boolean matchesLetter() {
    if (isEmpty())
        return false;
    char c = input[pos];
    // (*@\textbf{\textcolor{red}{Buggy code}}@*)
    return (c >= 'A' && c <= 'Z') || (c >= 'a' && c <= 'z');
    
    // (*@\textbf{\textcolor{yellow}{Fixed code by developer}}@*)
    return (c >= 'A' && c <= 'Z') || (c >= 'a' && c <= 'z') || Character.isLetter(c);
    
    // (*@\textbf{\textcolor{black}{Fixed code by baseline (CodeLlama-Instruct-7B), test failed}}@*)
    return (c >= 'A' && c <= 'Z') || (c >= 'a' && c <= 'z');
    
    // (*@\textbf{\textcolor{green}{Fixed code by HAFix (FN-all, CodeLlama-Instruct-7B), test passed}}@*)
    return Character.isLetter(c);
}
\end{lstlisting}

\begin{lstlisting}[style=javabox, caption={Example of Lang-21 in Defects4J, fixed code by baseline is semantically equivalent to the developer-written fix and passes the test cases.}, label={lst:Lang-21}]
public static boolean isSameLocalTime(Calendar cal1, Calendar cal2) {
    ... // Hidden unrelated code for saving space.
    return (cal1.get(Calendar.MILLISECOND) == cal2.get(Calendar.MILLISECOND) &&
            cal1.get(Calendar.SECOND) == cal2.get(Calendar.SECOND) &&
            cal1.get(Calendar.MINUTE) == cal2.get(Calendar.MINUTE) &&
			
			// (*@\textbf{\textcolor{red}{Buggy code}}@*)
            cal1.get(Calendar.HOUR) == cal2.get(Calendar.HOUR) &&
			
			// (*@\textbf{\textcolor{yellow}{Fixed code by developer}}@*)
            cal1.get(Calendar.HOUR_OF_DAY) == cal2.get(Calendar.HOUR_OF_DAY) &&
			
			// (*@\textbf{\textcolor{black}{Fixed code by baseline (CodeLlama-Instruct-7B), test passed}}@*)
            (cal1.get(Calendar.HOUR_OF_DAY) == cal2.get(Calendar.HOUR_OF_DAY)) &&
			
			// (*@\textbf{\textcolor{green}{Fixed code by HAFix (FN-all, CodeLlama-Instruct-7B), test failed}}@*)
            (cal1.get(Calendar.HOUR) == cal2.get(Calendar.HOUR) || (cal1.get(Calendar.HOUR) + 12) == cal2.get(Calendar.HOUR)) &&

            cal1.get(Calendar.DAY_OF_YEAR) == cal2.get(Calendar.DAY_OF_YEAR) &&
            cal1.get(Calendar.YEAR) == cal2.get(Calendar.YEAR) &&
            cal1.get(Calendar.ERA) == cal2.get(Calendar.ERA) &&
            cal1.getClass() == cal2.getClass());
}
\end{lstlisting}

\section{Venn Diagrams: HAFix heuristics vs Baseline}\label{appendix:VennDiagrams}

\begin{figure}[!htbp]
    \centering
    \begin{subfigure}[b]{0.4\textwidth} 
        \centering
        \includegraphics[width=\textwidth]{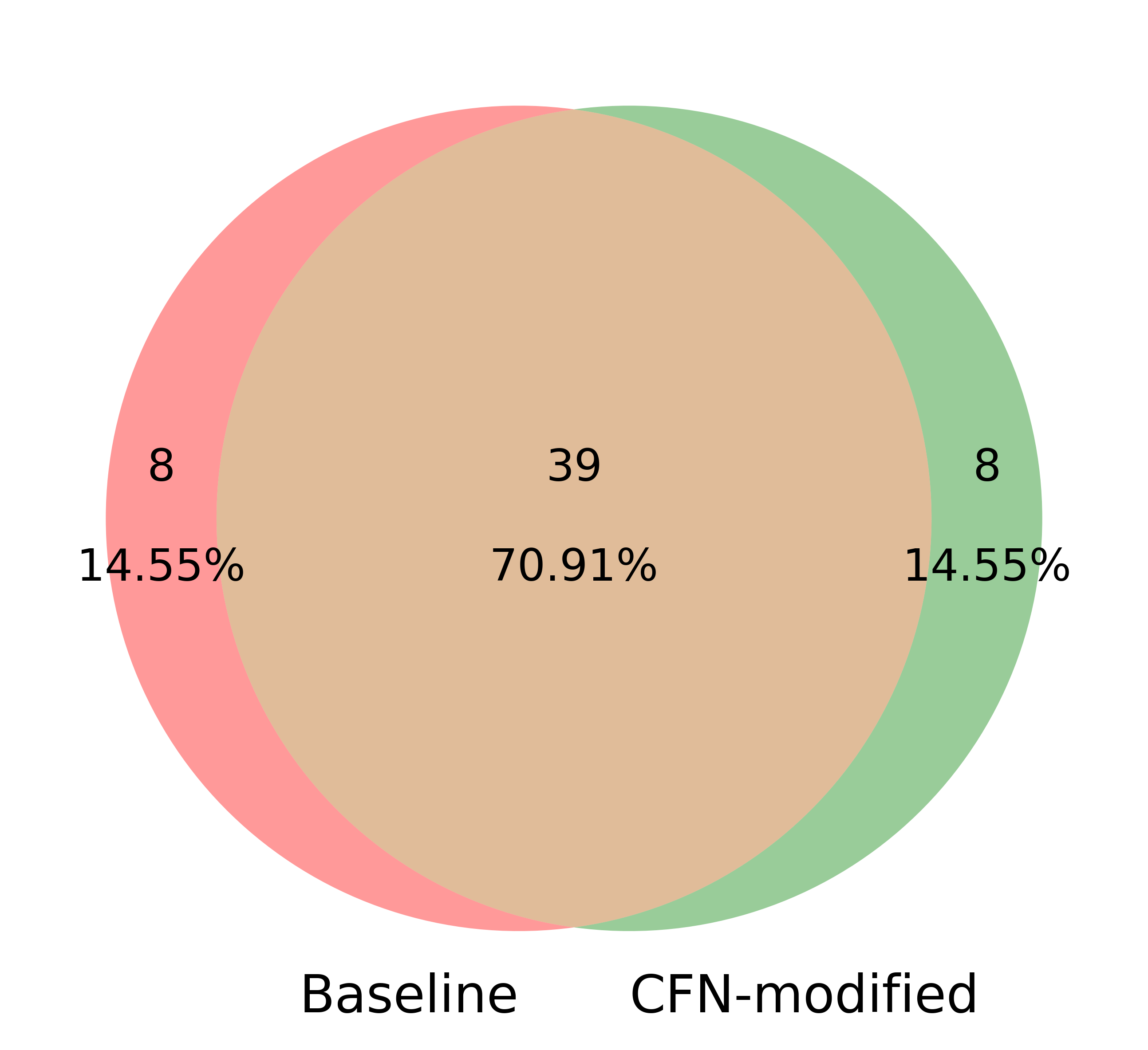} 
        \caption{CFN-modified fixes 8 more bugs compared to the baseline.}
        \label{venn-diagram-CodeLlama-Instruct-7B-Defects4J:a}
    \end{subfigure}
    \hfill 
    \begin{subfigure}[b]{0.4\textwidth} 
        \centering
        \includegraphics[width=\textwidth]{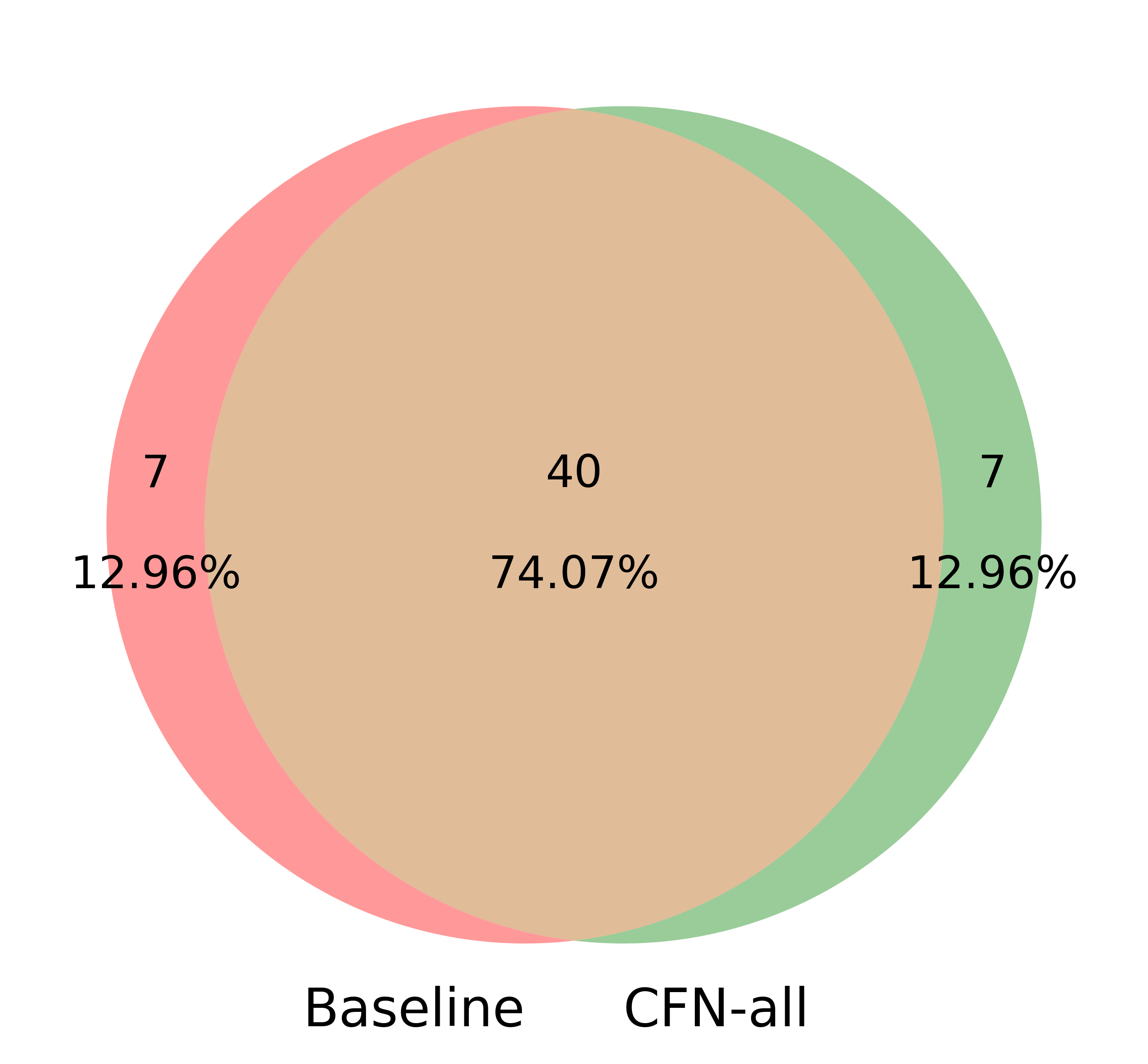} 
        \caption{CFN-all fixes 7 more bugs compared to the baseline.}
        \label{venn-diagram-CodeLlama-Instruct-7B-Defects4J:b}
    \end{subfigure}
    \vspace{-0.1cm}
    \\
    \begin{subfigure}[b]{0.4\textwidth} 
        \centering
        \includegraphics[width=\textwidth]{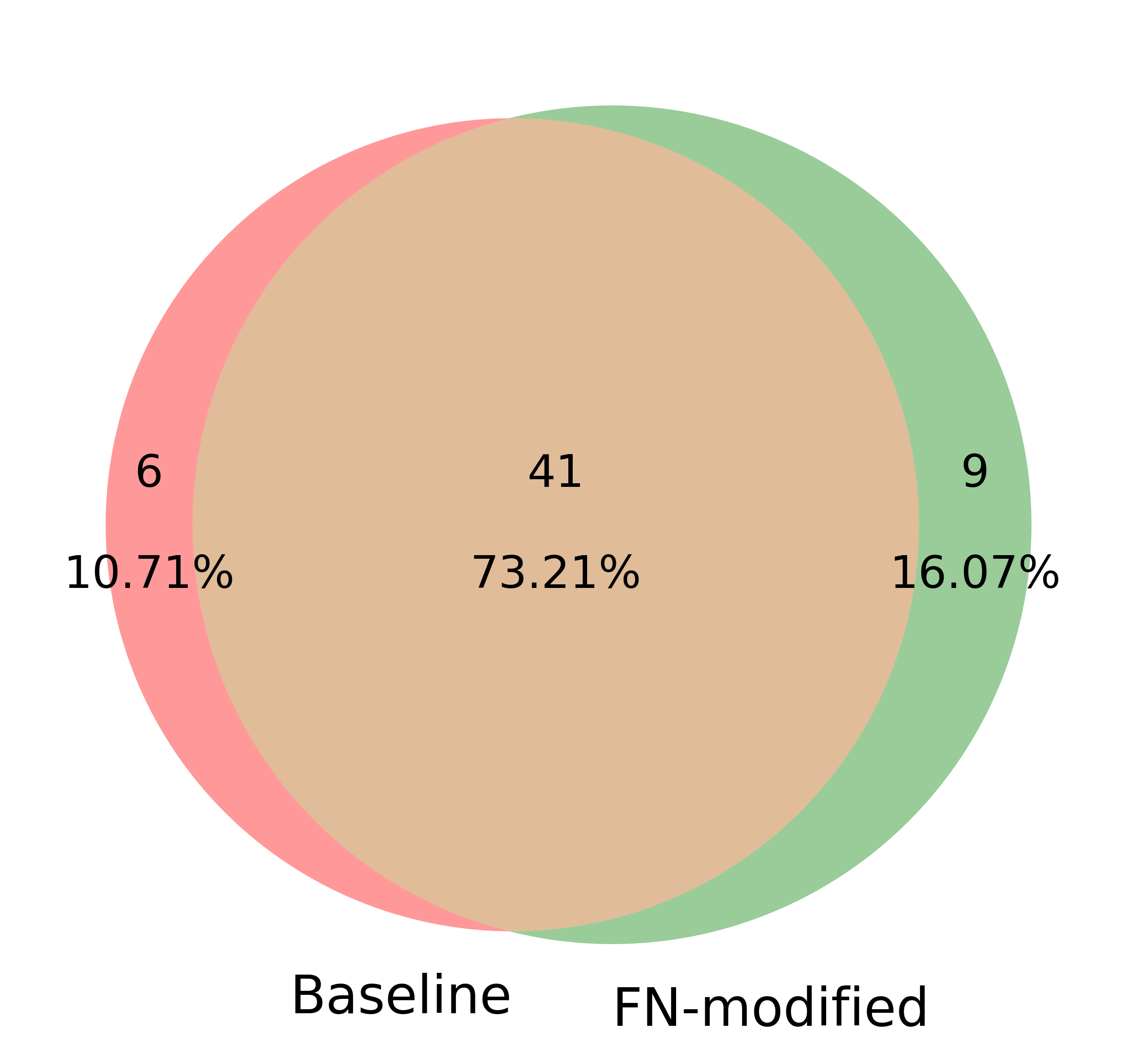} 
        \caption{FN-modified fixes 9 more bugs compared to the baseline.}
        \label{venn-diagram-CodeLlama-Instruct-7B-Defects4J:c}
    \end{subfigure}
    \hfill
    \begin{subfigure}[b]{0.4\textwidth} 
        \centering
        \includegraphics[width=\textwidth]{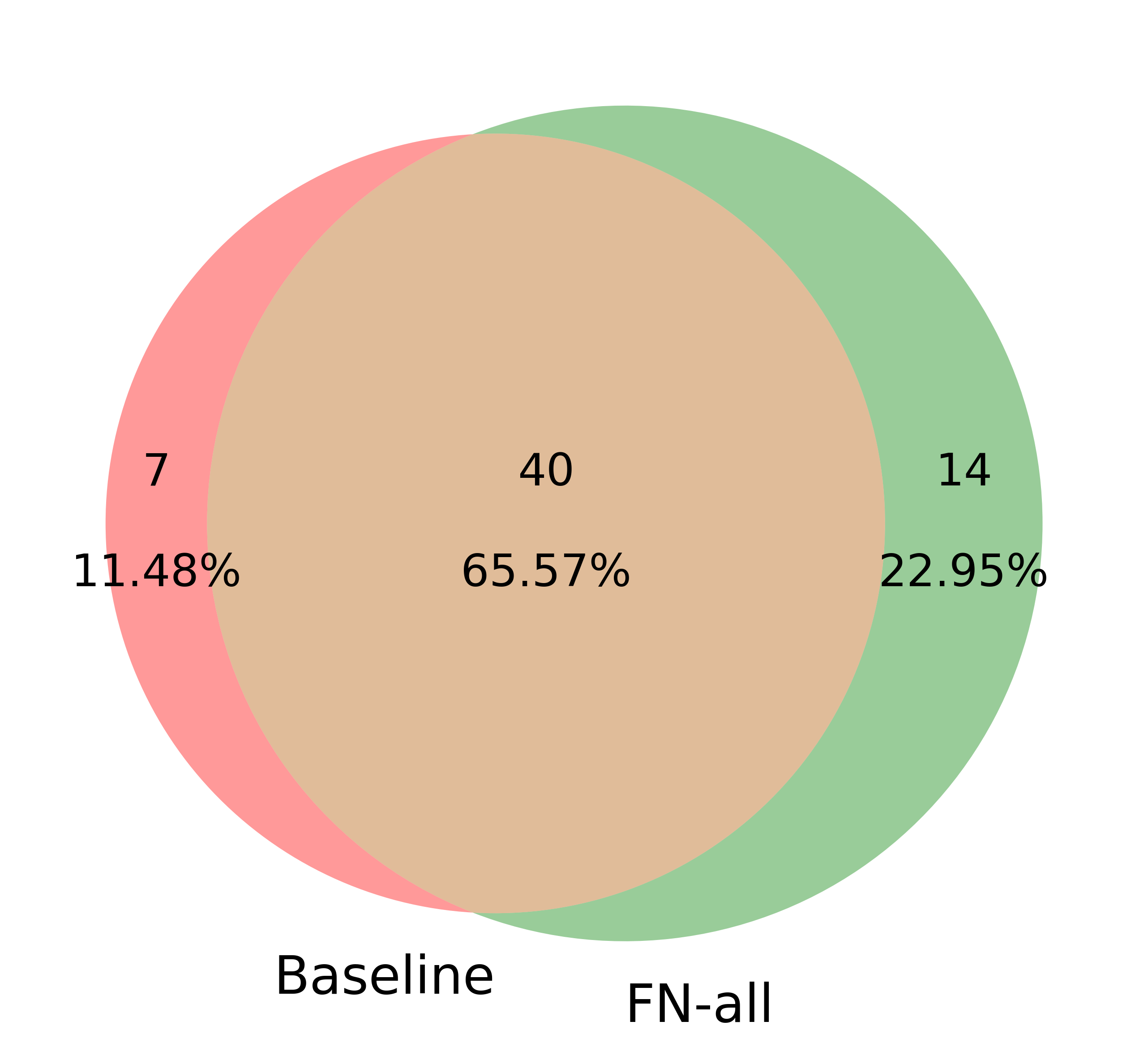} 
        \caption{FN-all fixes 14 more bugs compared to the baseline.}
        \label{venn-diagram-CodeLlama-Instruct-7B-Defects4J:d}
    \end{subfigure}
    \vspace{-0.1cm}
    \\
    \begin{subfigure}[b]{0.4\textwidth} 
        \centering
        \includegraphics[width=\textwidth]{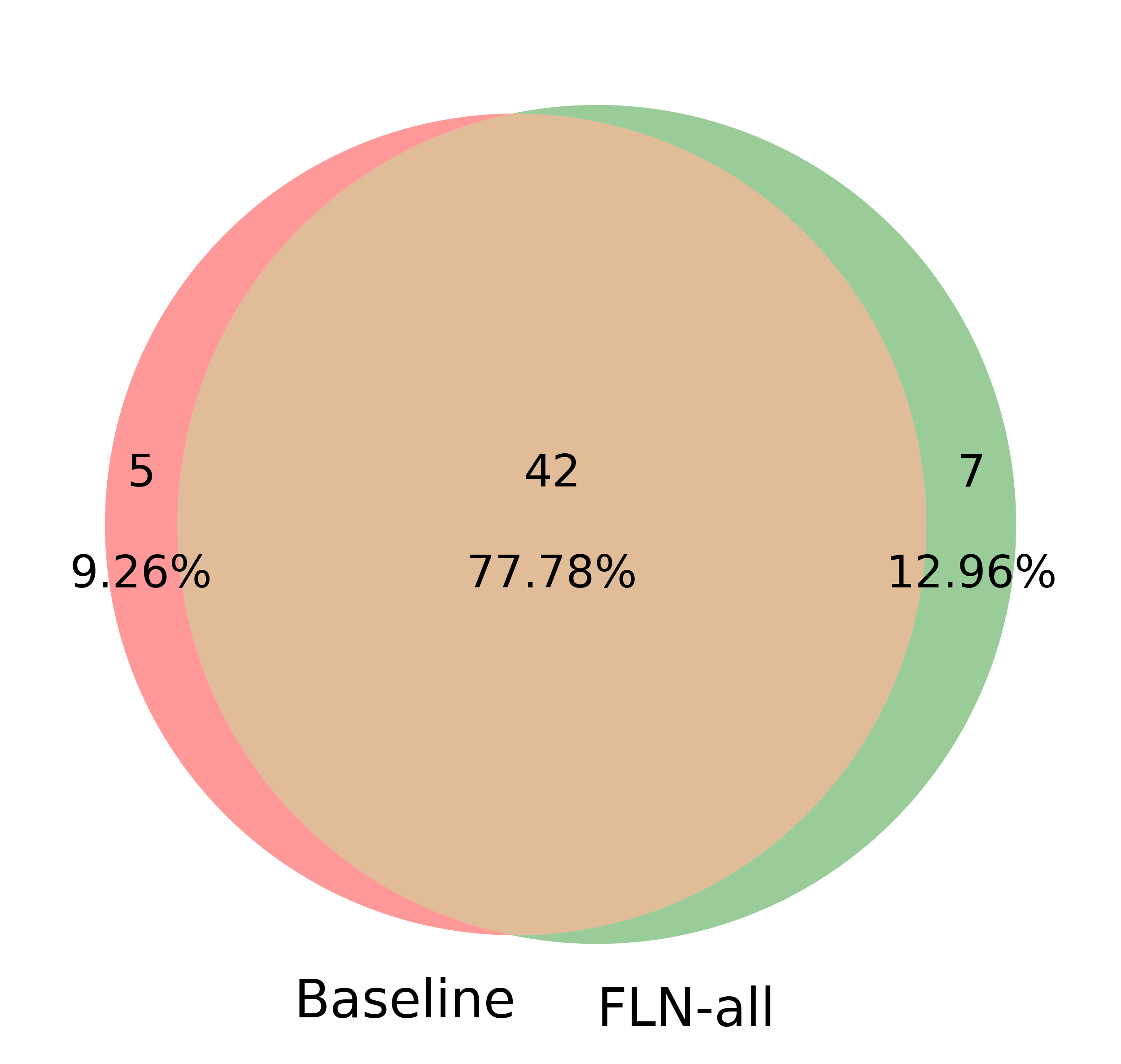} 
        \caption{FLN-all fixes 7 more bugs compared to the baseline.}
        \label{venn-diagram-CodeLlama-Instruct-7B-Defects4J:e}
    \end{subfigure}
    \hfill
    \begin{subfigure}[b]{0.4\textwidth} 
        \centering
        \includegraphics[width=\textwidth]{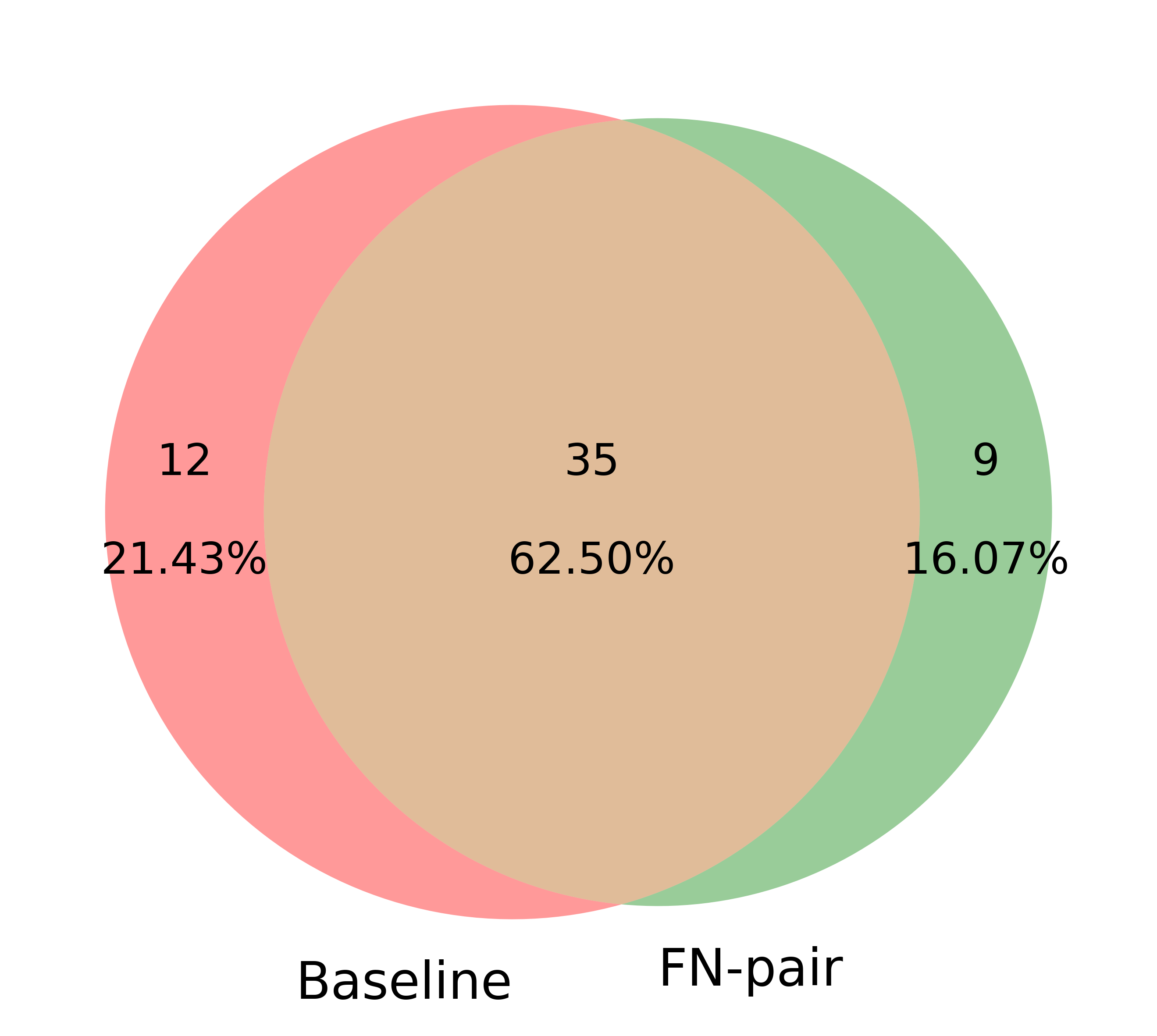} 
        \caption{FN-pair fixes 9 more bugs compared to the baseline.}
        \label{venn-diagram-CodeLlama-Instruct-7B-Defects4J:f}
    \end{subfigure}
    \vspace{-0.1cm}
    \\
    \begin{subfigure}[b]{0.4\textwidth} 
        \centering
        \includegraphics[width=\textwidth]{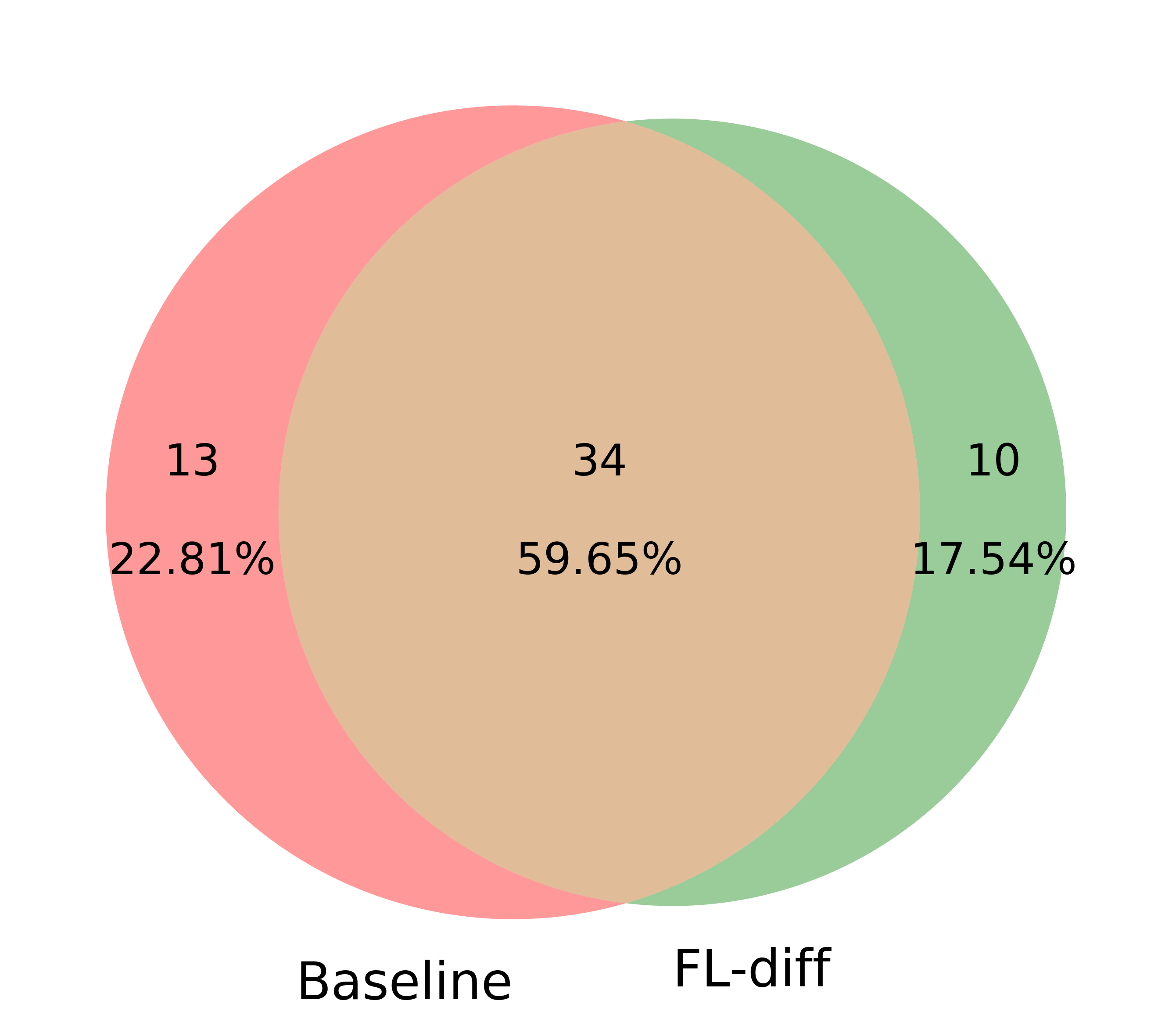} 
        \caption{FL-diff fixes 10 more bugs compared to the baseline.}
        \label{venn-diagram-CodeLlama-Instruct-7B-Defects4J:g}
    \end{subfigure}
    \caption{CodeLlama-Instruct-7B on Defects4J. Venn diagrams comparing the number of bugs fixed by the baseline (red) and the seven individual HAFix heuristics (green), with the overlapping region (brown) indicating bugs fixed by both the baseline and the heuristic. Numbers and percentages within each region denote the count and proportion of bugs fixed.}
    \label{fig:venn-diagram-CodeLlama-Instruct-7B-Defects4J}
\end{figure}

\begin{figure}[!htbp]
    \centering
    \begin{subfigure}[b]{0.4\textwidth} 
        \centering
        \includegraphics[width=\textwidth]{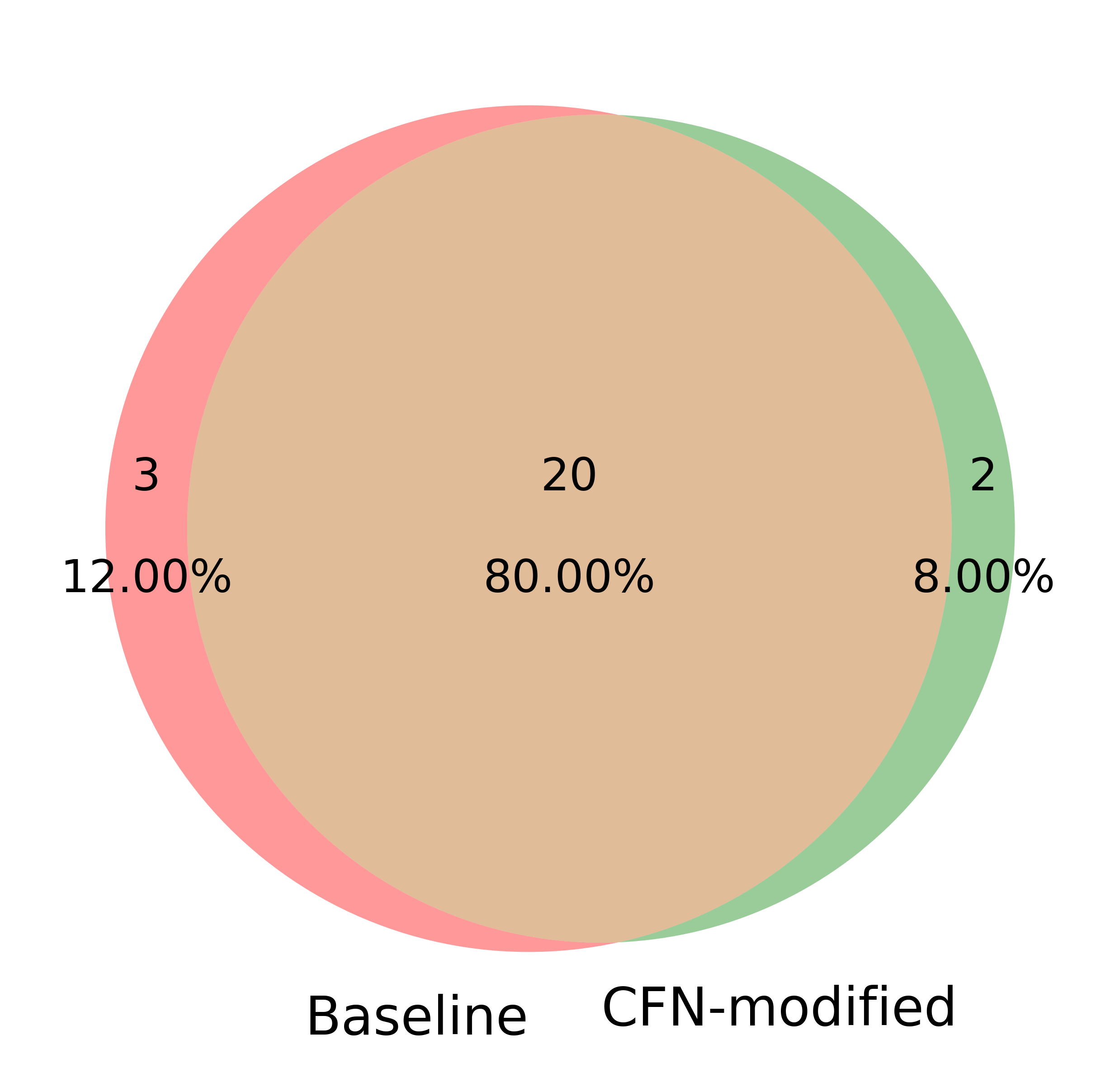} 
        \caption{CFN-modified fixes 2 more bugs compared to the baseline.}
        \label{venn-diagram-DeepSeek-Coder-Instruct-6.7B-BugsInPy:a}
    \end{subfigure}
    \hfill 
    \begin{subfigure}[b]{0.4\textwidth} 
        \centering
        \includegraphics[width=\textwidth]{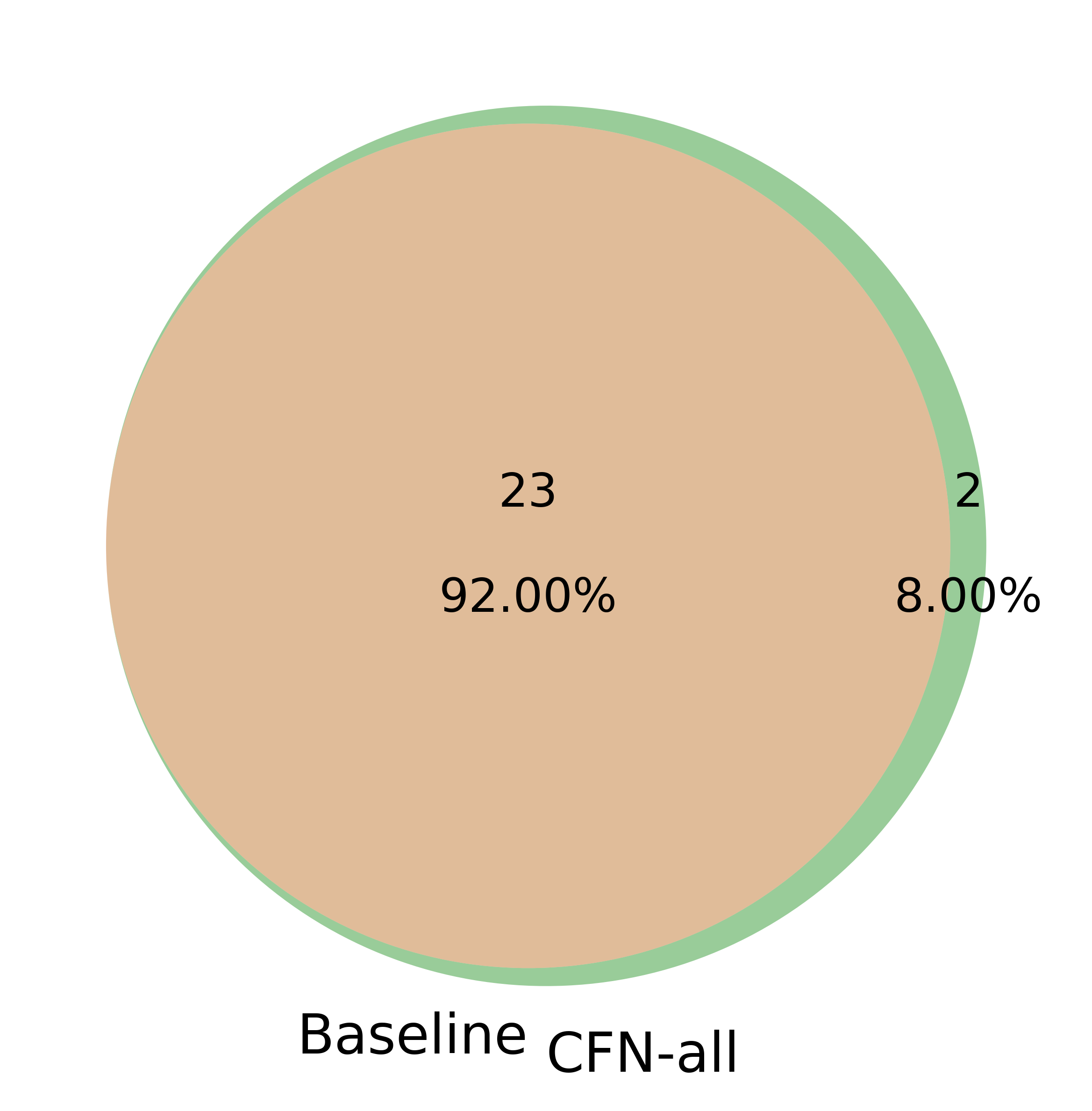} 
        \caption{CFN-all fixes 2 more bugs compared to the baseline.}
        \label{venn-diagram-DeepSeek-Coder-Instruct-6.7B-BugsInPy:b}
    \end{subfigure}
    \vspace{-0.1cm}
    \\
    \begin{subfigure}[b]{0.4\textwidth} 
        \centering
        \includegraphics[width=\textwidth]{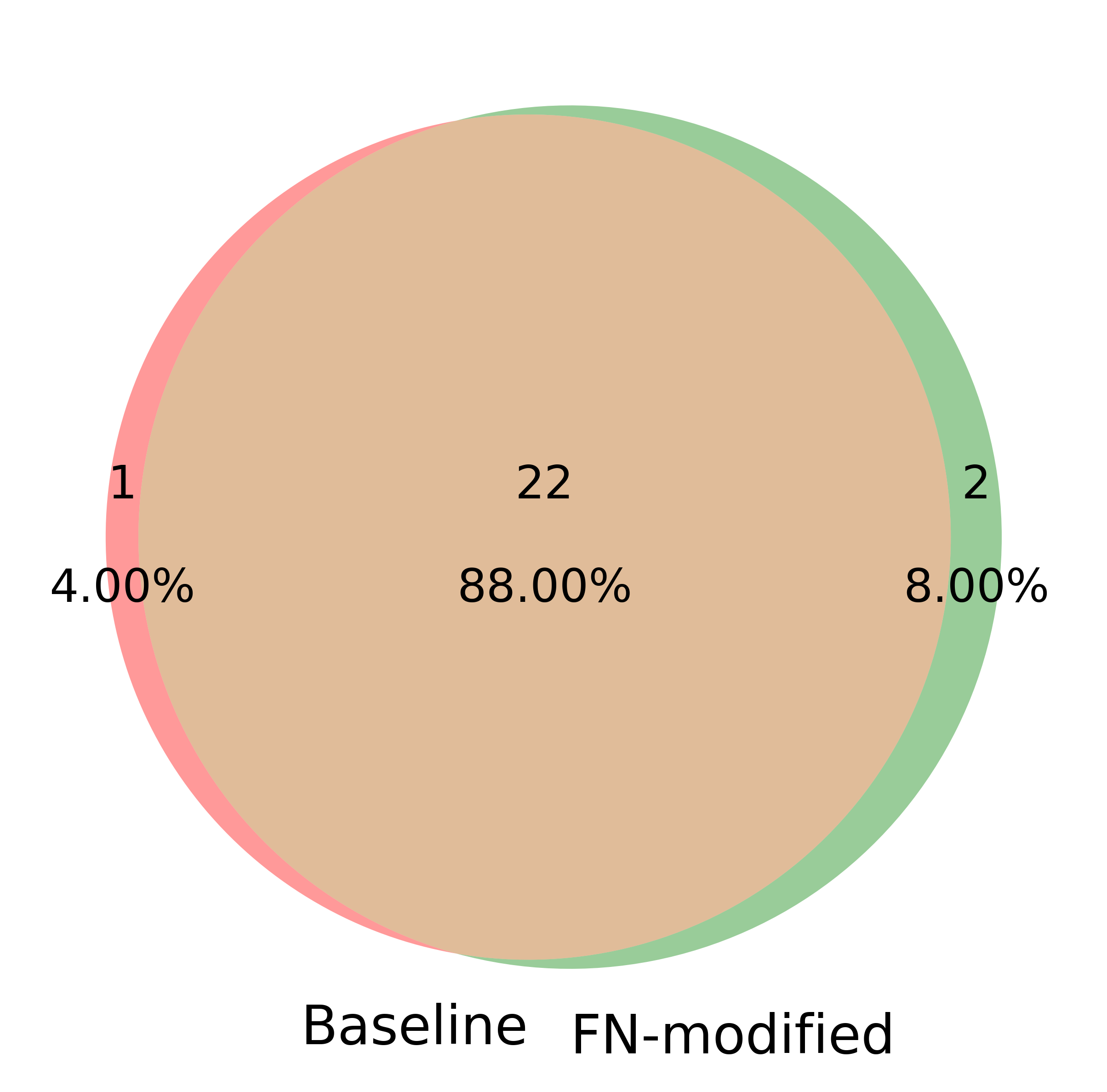} 
        \caption{FN-modified fixes 2 more bugs compared to the baseline.}
        \label{venn-diagram-DeepSeek-Coder-Instruct-6.7B-BugsInPy:c}
    \end{subfigure}
    \hfill
    \begin{subfigure}[b]{0.4\textwidth} 
        \centering
        \includegraphics[width=\textwidth]{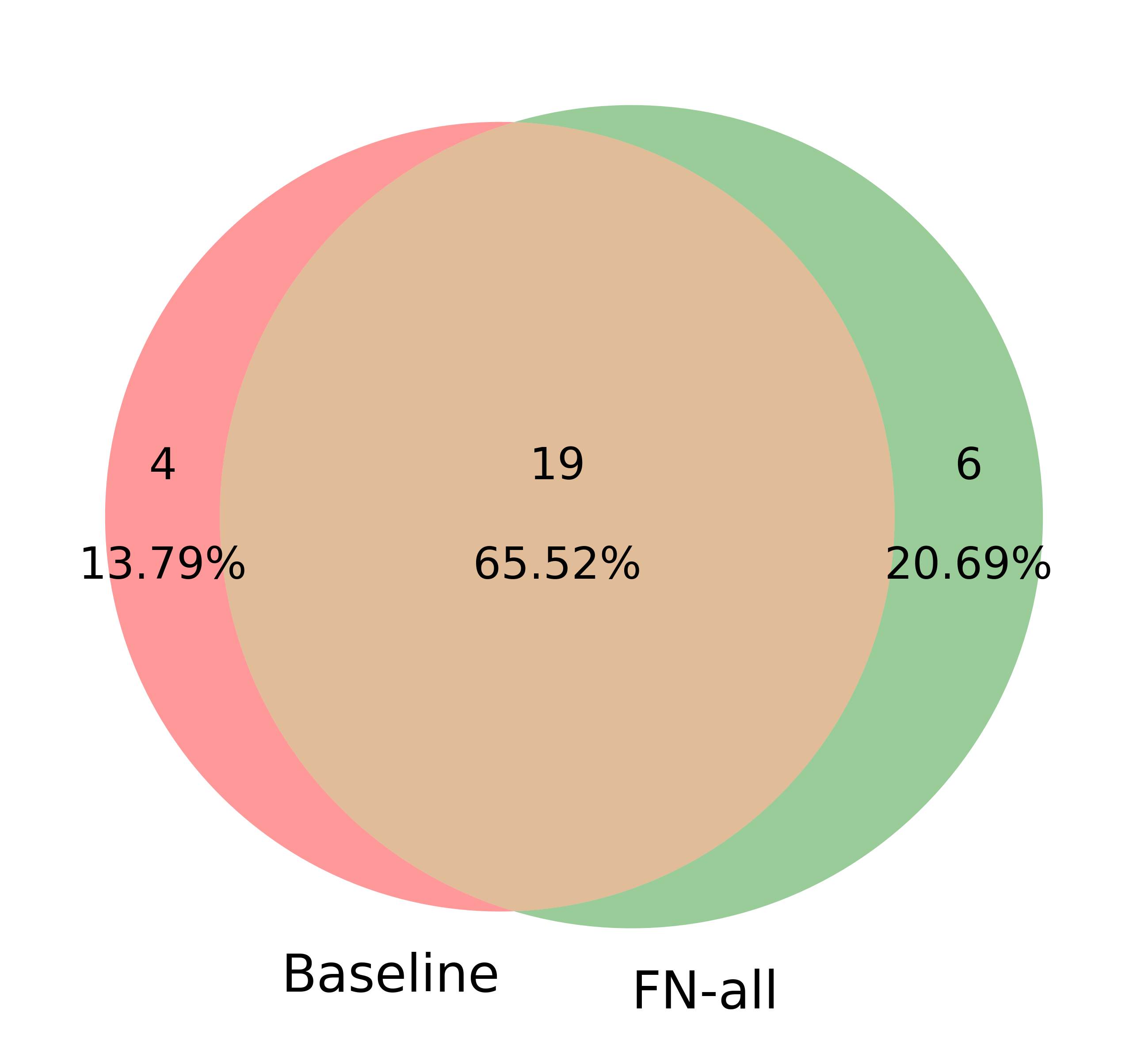} 
        \caption{FN-all fixes 6 more bugs compared to the baseline.}
        \label{venn-diagram-DeepSeek-Coder-Instruct-6.7B-BugsInPy:d}
    \end{subfigure}
    \vspace{-0.1cm}
    \\
    \begin{subfigure}[b]{0.4\textwidth} 
        \centering
        \includegraphics[width=\textwidth]{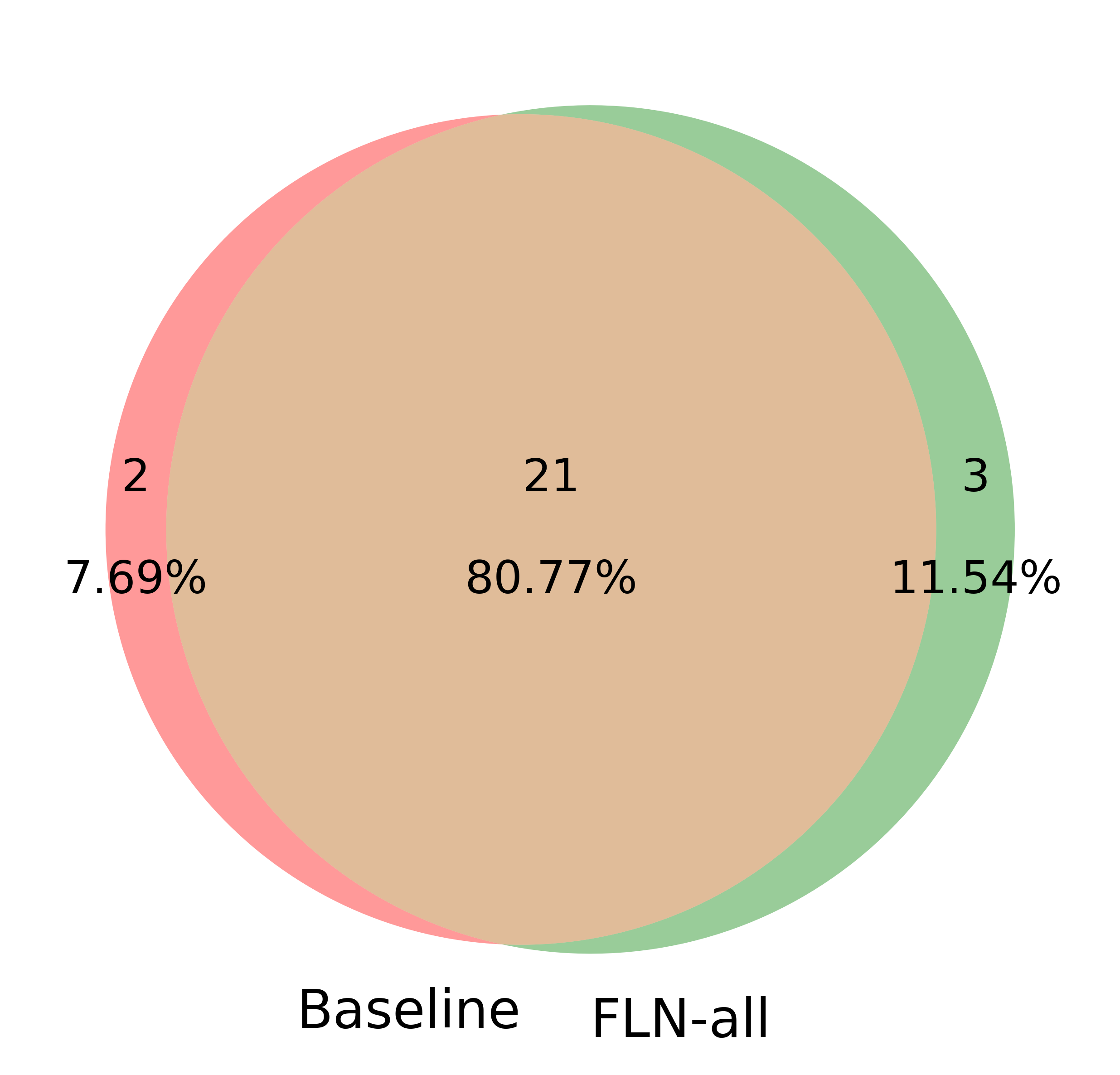} 
        \caption{FLN-all fixes 3 more bugs compared to the baseline.}
        \label{venn-diagram-DeepSeek-Coder-Instruct-6.7B-BugsInPy:e}
    \end{subfigure}
    \hfill
    \begin{subfigure}[b]{0.4\textwidth} 
        \centering
        \includegraphics[width=\textwidth]{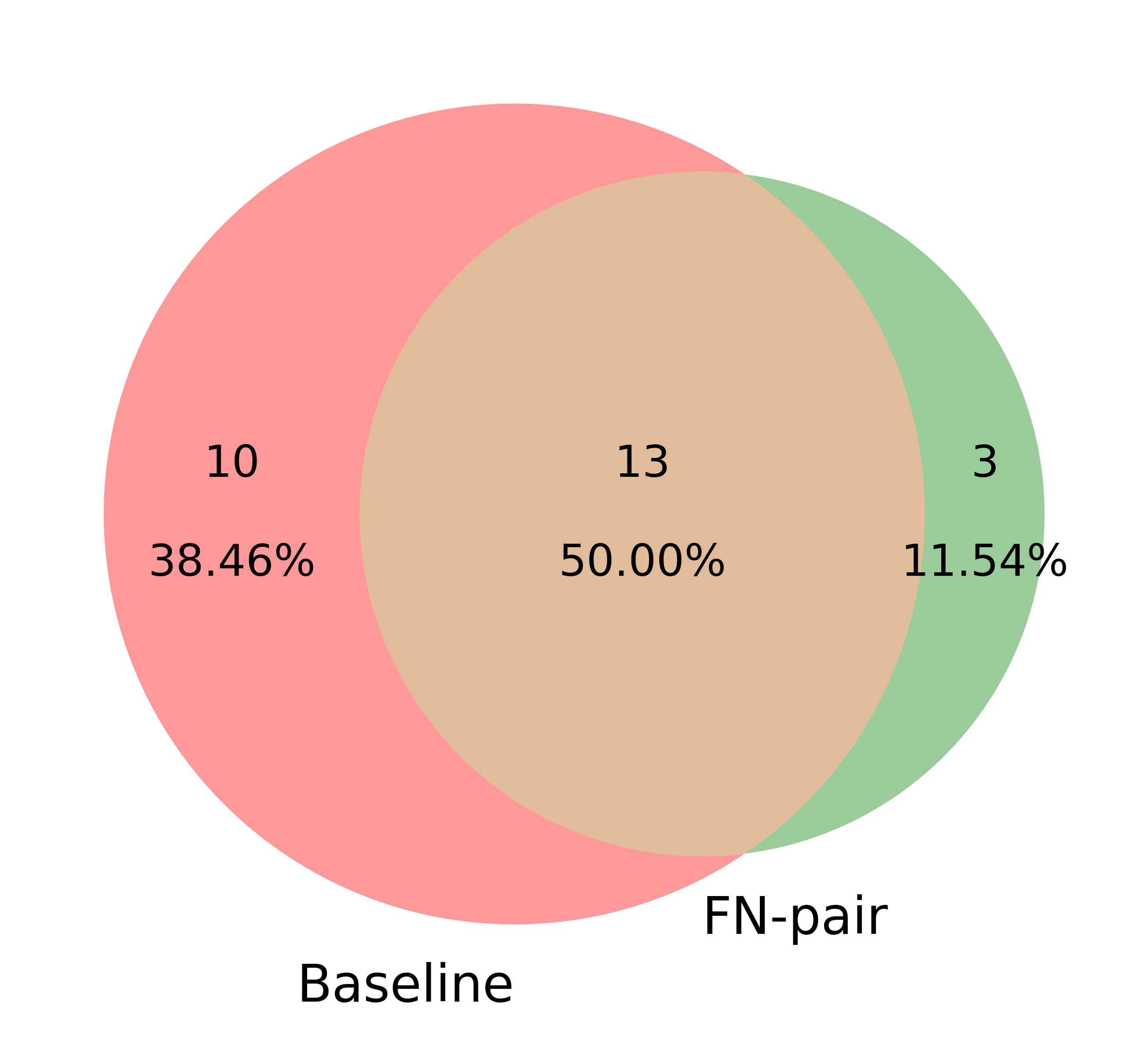} 
        \caption{FN-pair fixes 3 more bugs compared to the baseline.}
        \label{venn-diagram-DeepSeek-Coder-Instruct-6.7B-BugsInPy:f}
    \end{subfigure}
    \vspace{-0.1cm}
    \\
    \begin{subfigure}[b]{0.4\textwidth} 
        \centering
        \includegraphics[width=\textwidth]{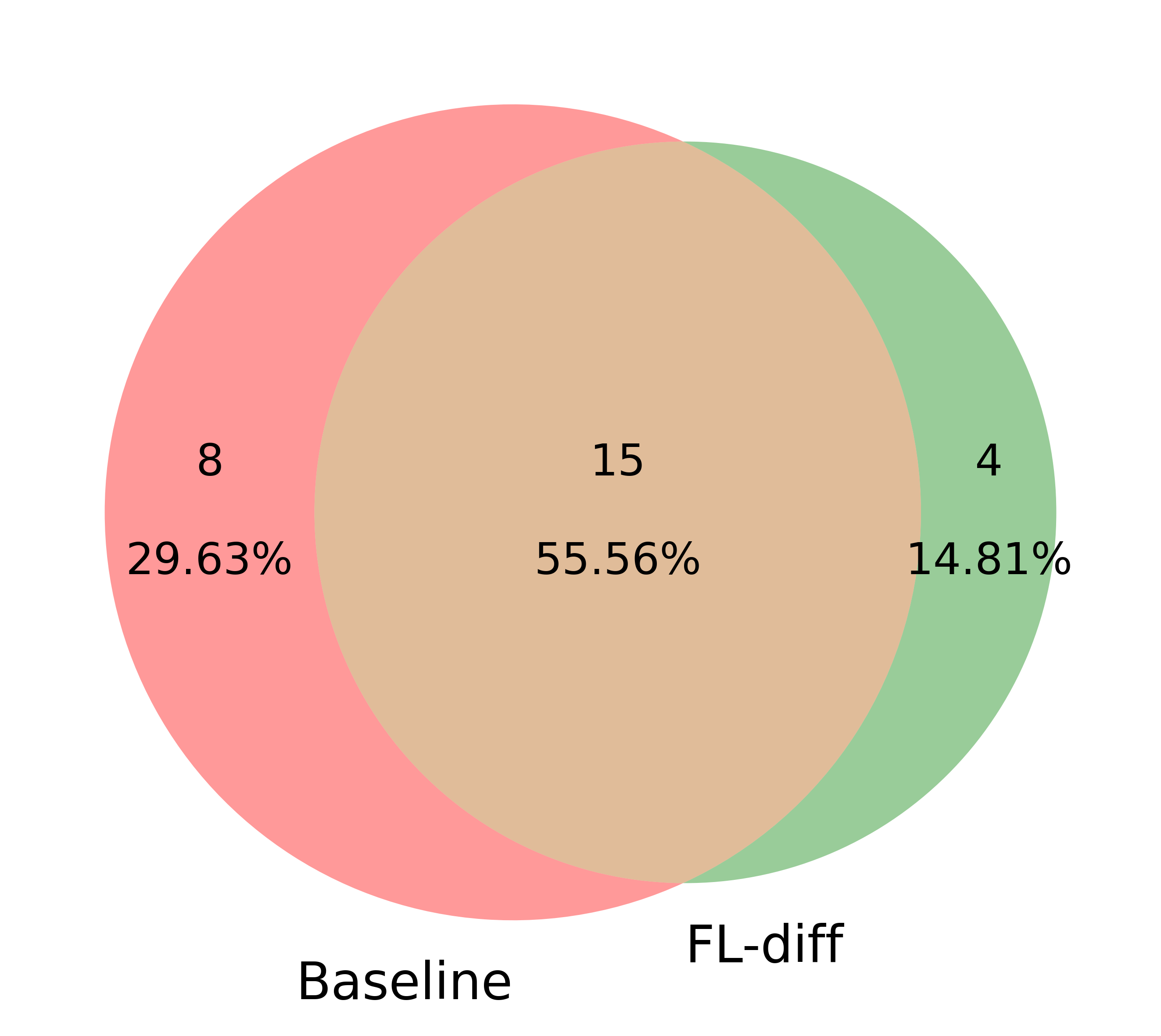} 
        \caption{FL-diff fixes 4 more bugs compared to the baseline.}
        \label{venn-diagram-DeepSeek-Coder-Instruct-6.7B-BugsInPy:g}
    \end{subfigure}
    \caption{DeepSeek-Coder-Instruct-6.7B on BugsInPy. Venn diagrams comparing the number of bugs fixed by the baseline (red) and the seven individual HAFix heuristics (green), with the overlapping region (brown) indicating bugs fixed by both the baseline and the heuristic. Numbers and percentages within each region denote the count and proportion of bugs fixed.}
    \label{fig:venn-diagram-DeepSeek-Coder-Instruct-6.7B-BugsInPy}
\end{figure}

\begin{figure}[!htbp]
    \centering
    \begin{subfigure}[b]{0.4\textwidth} 
        \centering
        \includegraphics[width=\textwidth]{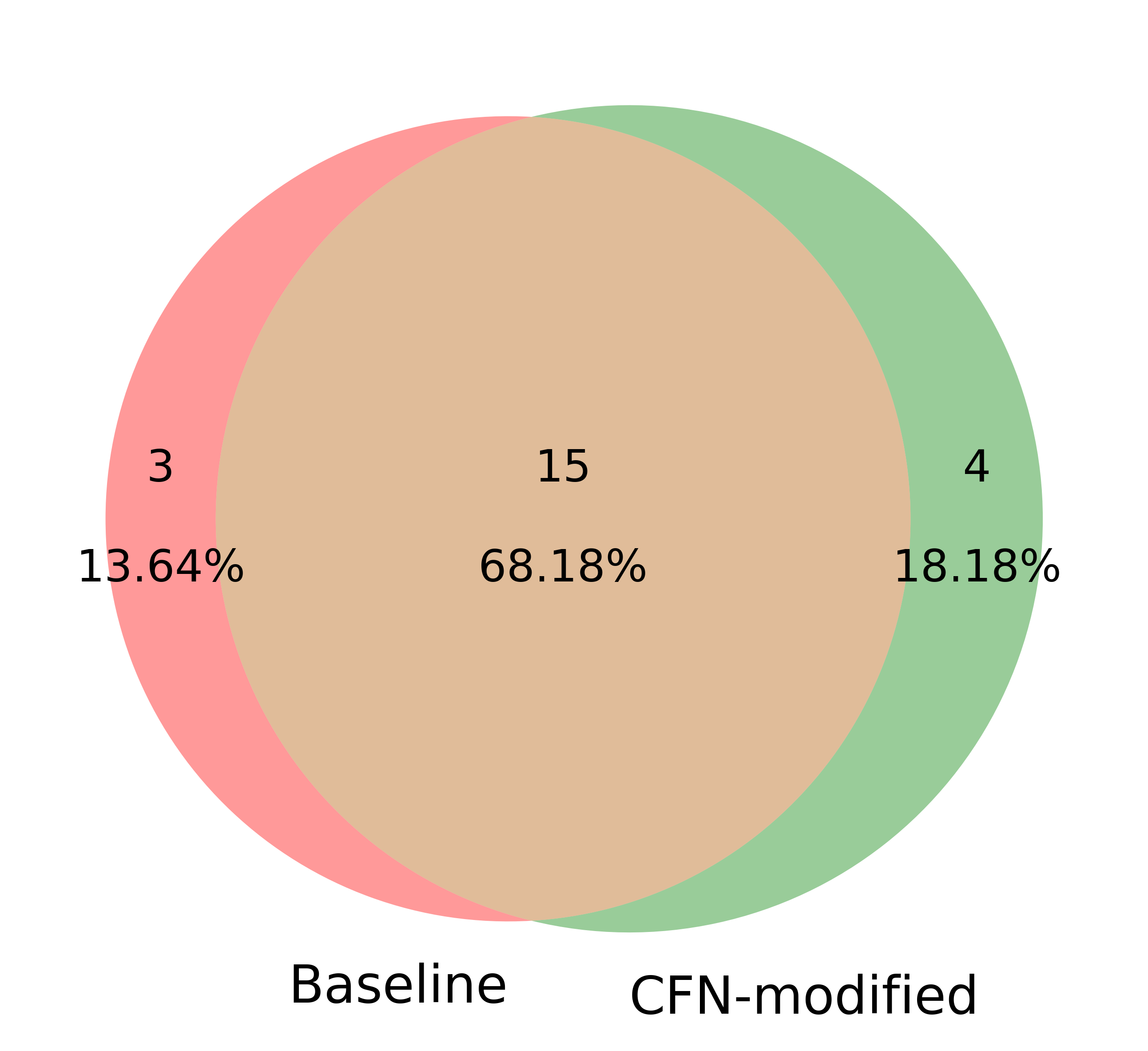} 
        \caption{CFN-modified fixes 4 more bugs compared to the baseline.}
        \label{venn-diagram-DeepSeek-Coder-V2-Lite-Instruct-16B-BugsInPy:a}
    \end{subfigure}
    \hfill 
    \begin{subfigure}[b]{0.4\textwidth} 
        \centering
        \includegraphics[width=\textwidth]{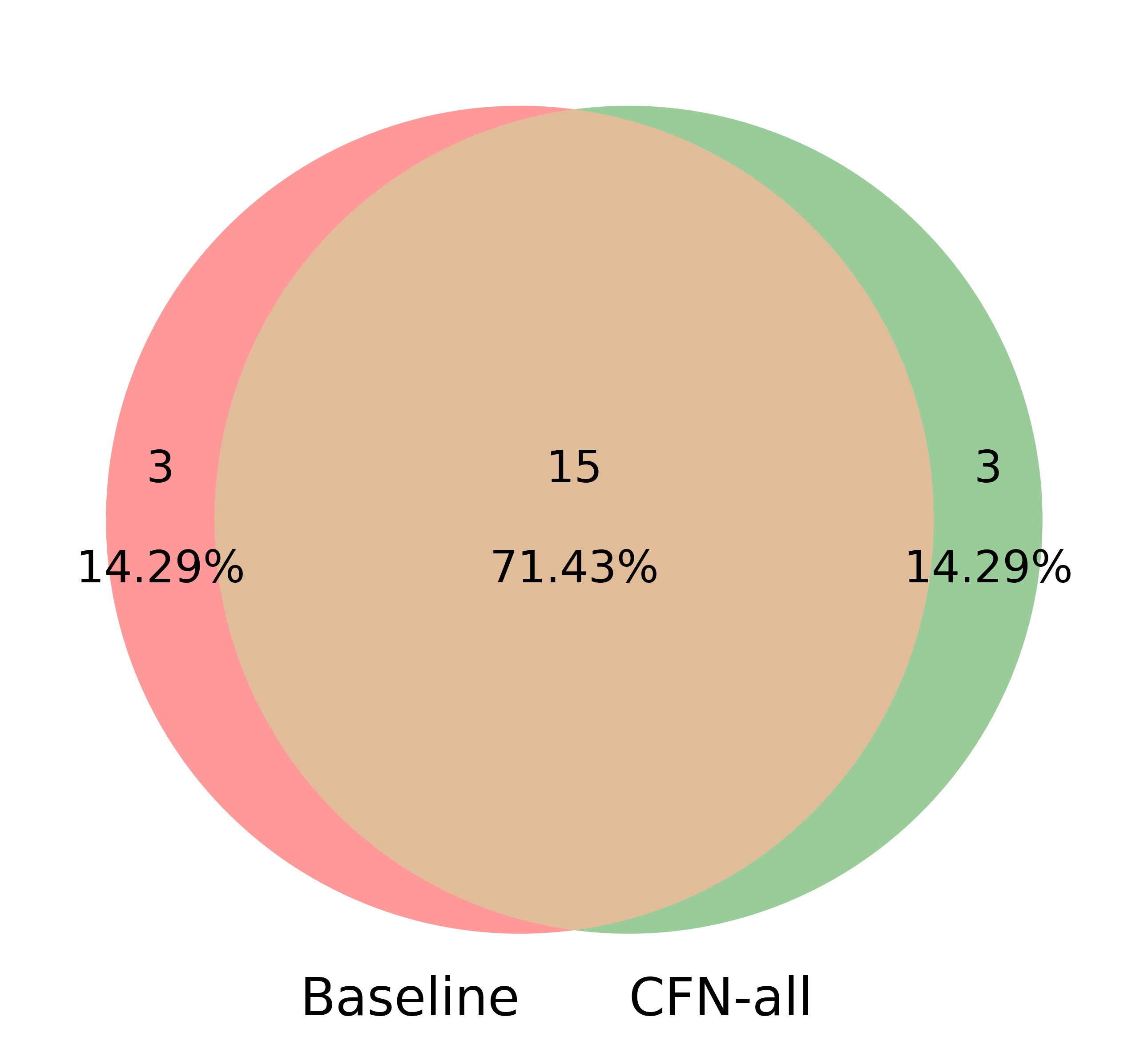} 
        \caption{CFN-all fixes 3 more bugs compared to the baseline.}
        \label{venn-diagram-DeepSeek-Coder-V2-Lite-Instruct-16B-BugsInPy:b}
    \end{subfigure}
    \vspace{-0.1cm}
    \\
    \begin{subfigure}[b]{0.4\textwidth} 
        \centering
        \includegraphics[width=\textwidth]{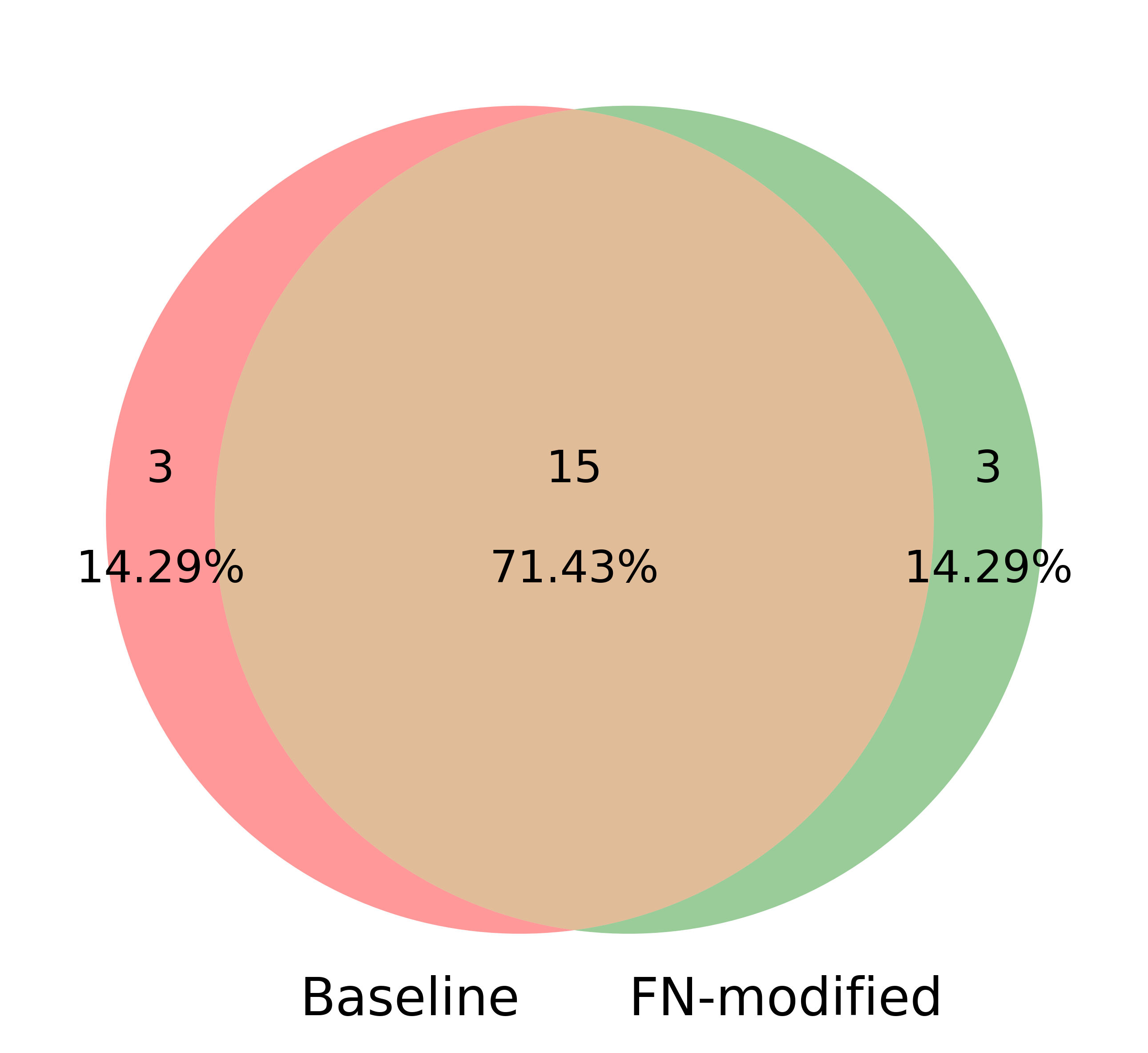} 
        \caption{FN-modified fixes 3 more bugs compared to the baseline.}
        \label{venn-diagram-DeepSeek-Coder-V2-Lite-Instruct-16B-BugsInPy:c}
    \end{subfigure}
    \hfill
    \begin{subfigure}[b]{0.4\textwidth} 
        \centering
        \includegraphics[width=\textwidth]{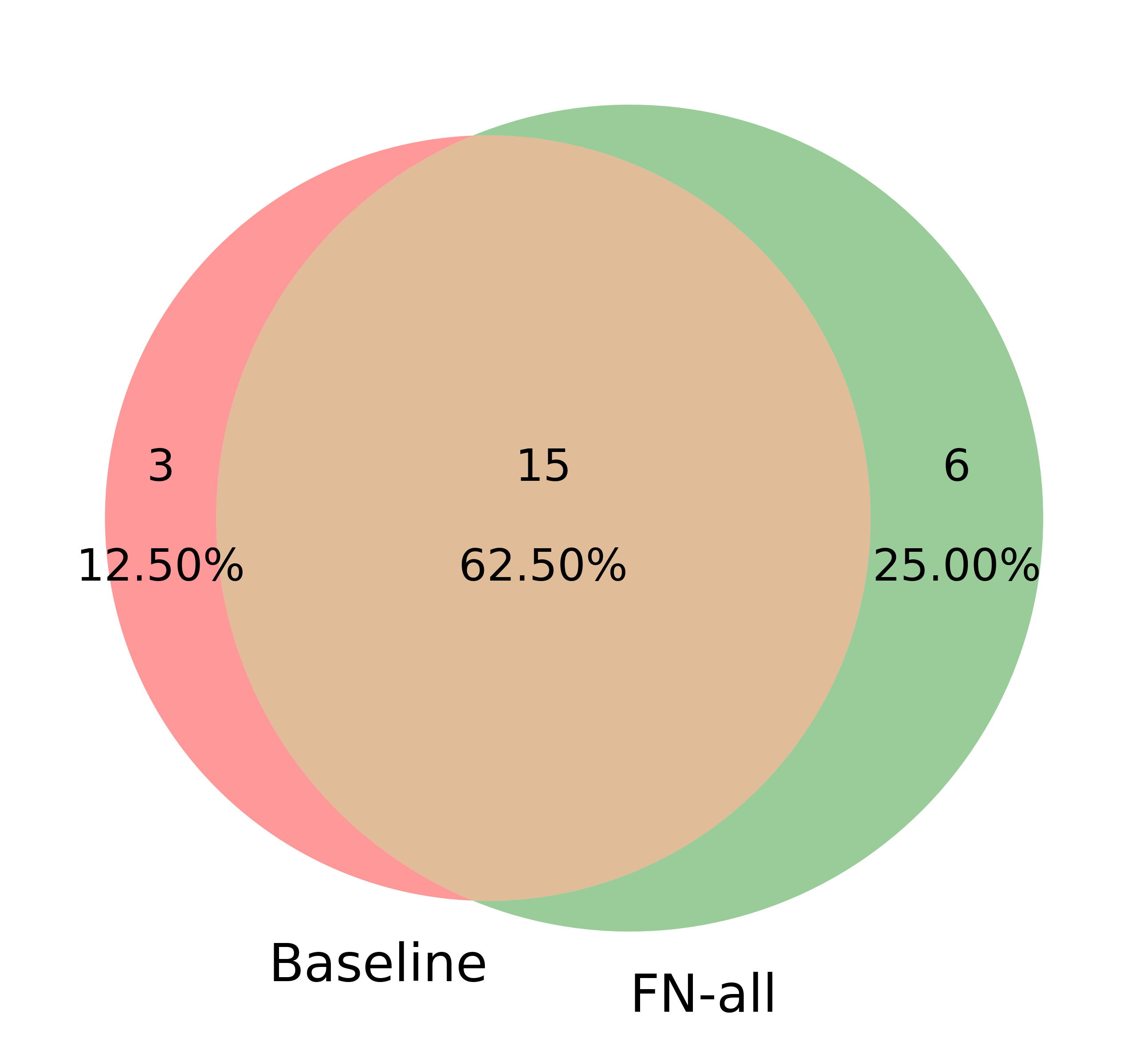} 
        \caption{FN-all fixes 6 more bugs compared to the baseline.}
        \label{venn-diagram-DeepSeek-Coder-V2-Lite-Instruct-16B-BugsInPy:d}
    \end{subfigure}
    \vspace{-0.1cm}
    \\
    \begin{subfigure}[b]{0.4\textwidth} 
        \centering
        \includegraphics[width=\textwidth]{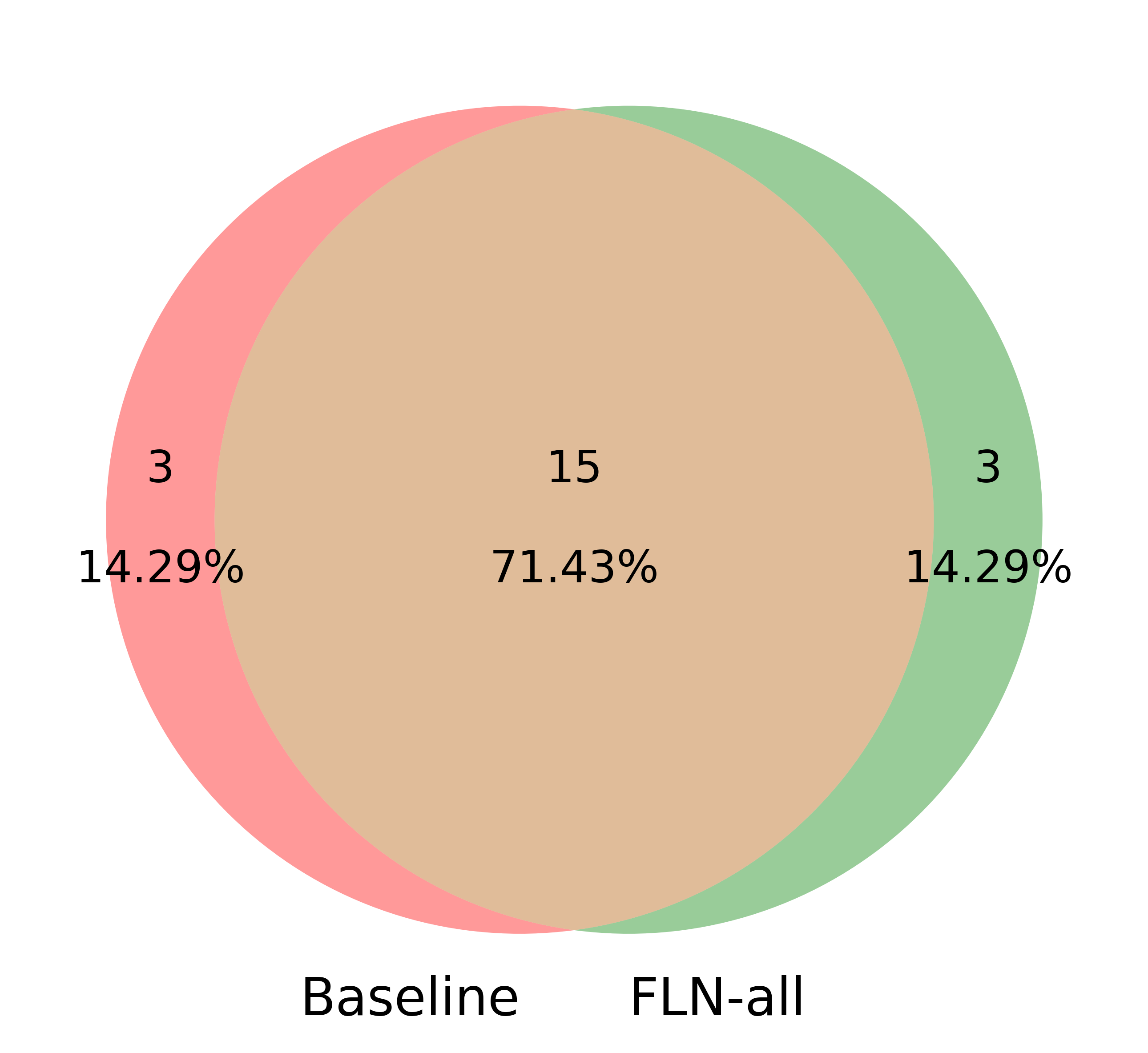} 
        \caption{FLN-all fixes 3 more bugs compared to the baseline.}
        \label{venn-diagram-DeepSeek-Coder-V2-Lite-Instruct-16B-BugsInPy:e}
    \end{subfigure}
    \hfill
    \begin{subfigure}[b]{0.4\textwidth} 
        \centering
        \includegraphics[width=\textwidth]{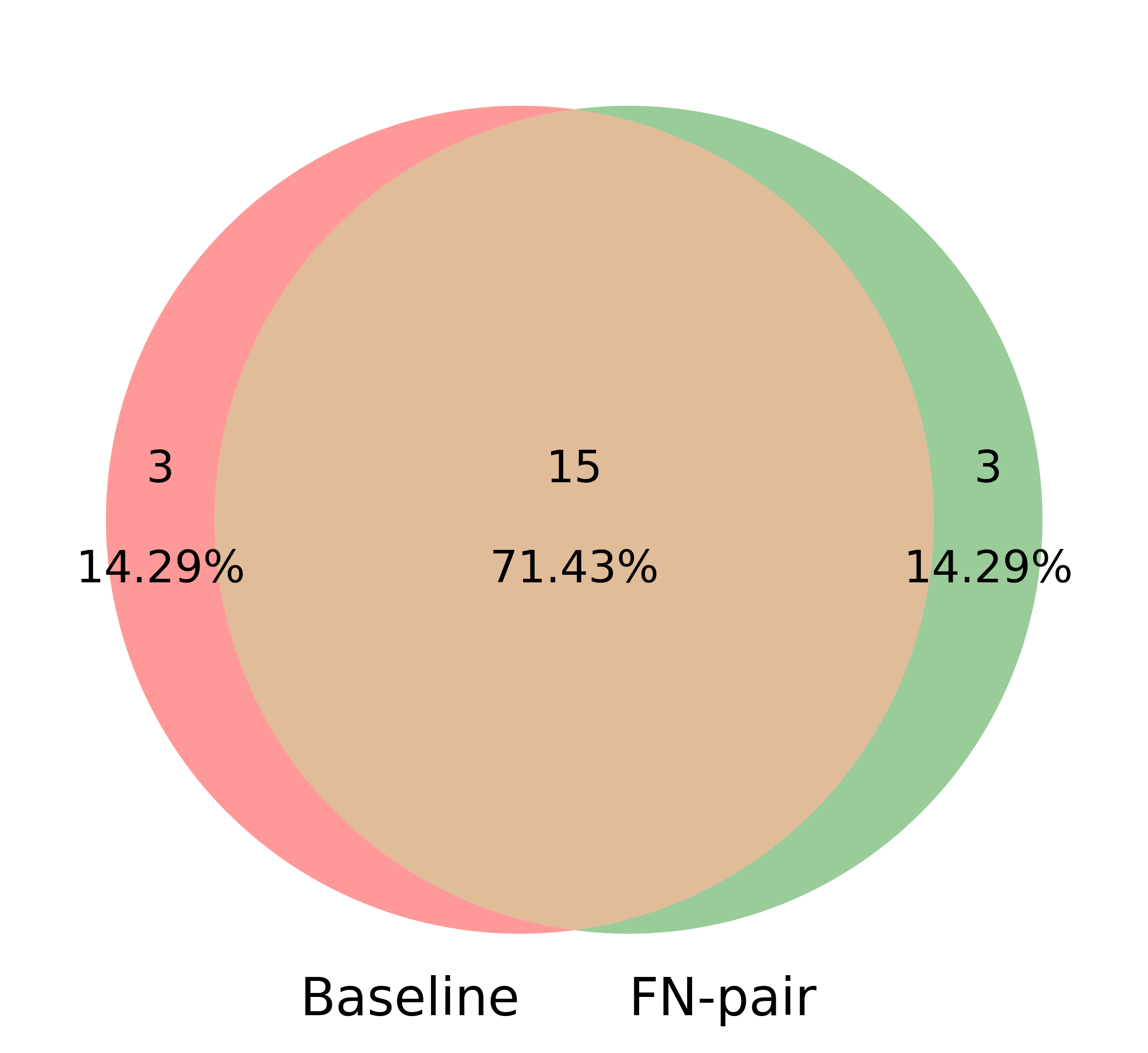} 
        \caption{FN-pair fixes 3 more bugs compared to the baseline.}
        \label{venn-diagram-DeepSeek-Coder-V2-Lite-Instruct-16B-BugsInPy:f}
    \end{subfigure}
    \vspace{-0.1cm}
    \\
    \begin{subfigure}[b]{0.4\textwidth} 
        \centering
        \includegraphics[width=\textwidth]{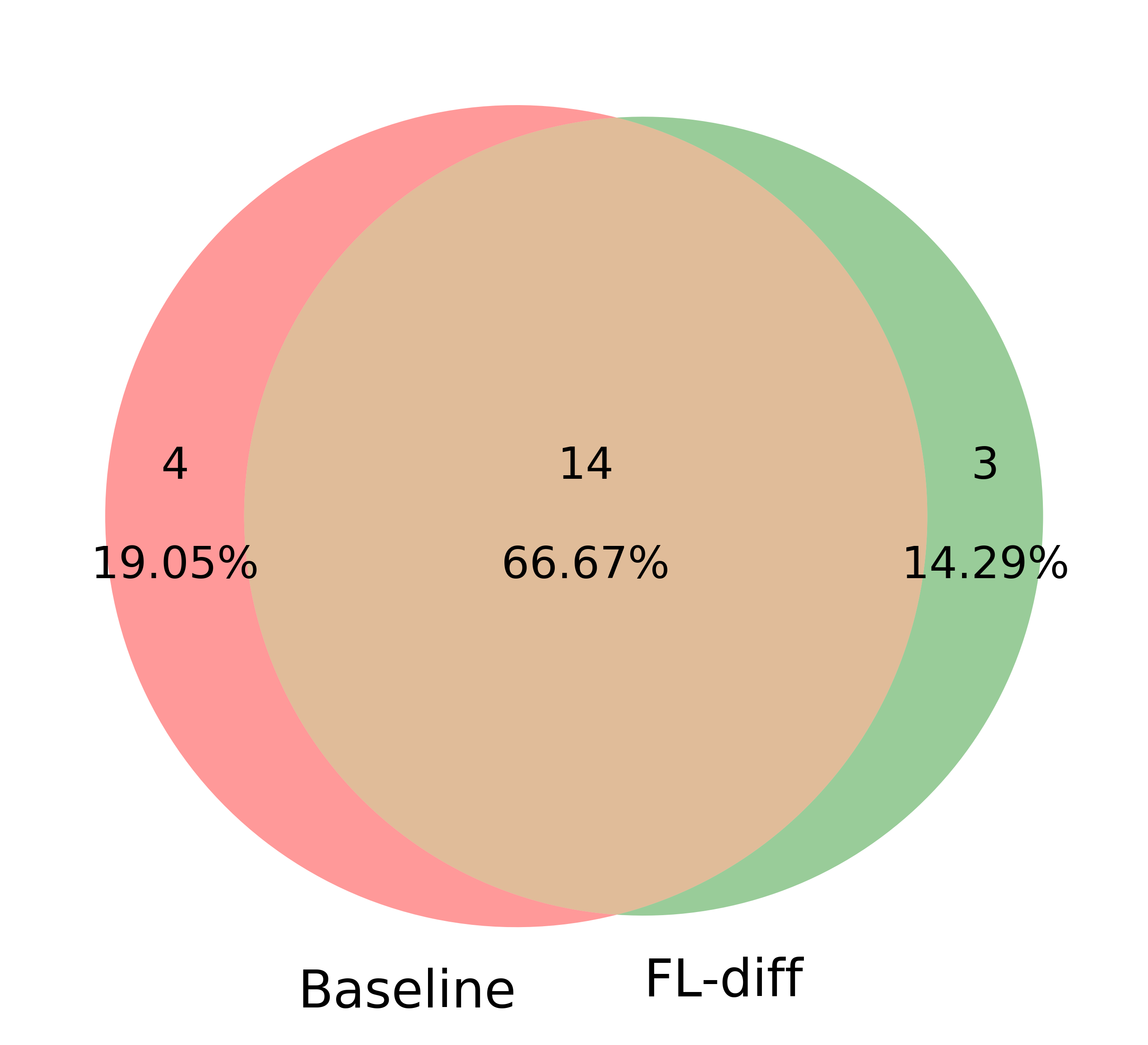} 
        \caption{FL-diff fixes 3 more bugs compared to the baseline.}
        \label{venn-diagram-DeepSeek-Coder-V2-Lite-Instruct-16B-BugsInPy:g}
    \end{subfigure}
    \caption{DeepSeek-Coder-V2-Lite-Instruct-16B on BugsInPy. Venn diagrams comparing the number of bugs fixed by the baseline (red) and the seven individual HAFix heuristics (green), with the overlapping region (brown) indicating bugs fixed by both the baseline and the heuristic. Numbers and percentages within each region denote the count and proportion of bugs fixed.}
    \label{venn-diagram-DeepSeek-Coder-V2-Lite-Instruct-16B-BugsInPy}
\end{figure}

\begin{figure}[!htbp]
    \centering
    \begin{subfigure}[b]{0.4\textwidth} 
        \centering
        \includegraphics[width=\textwidth]{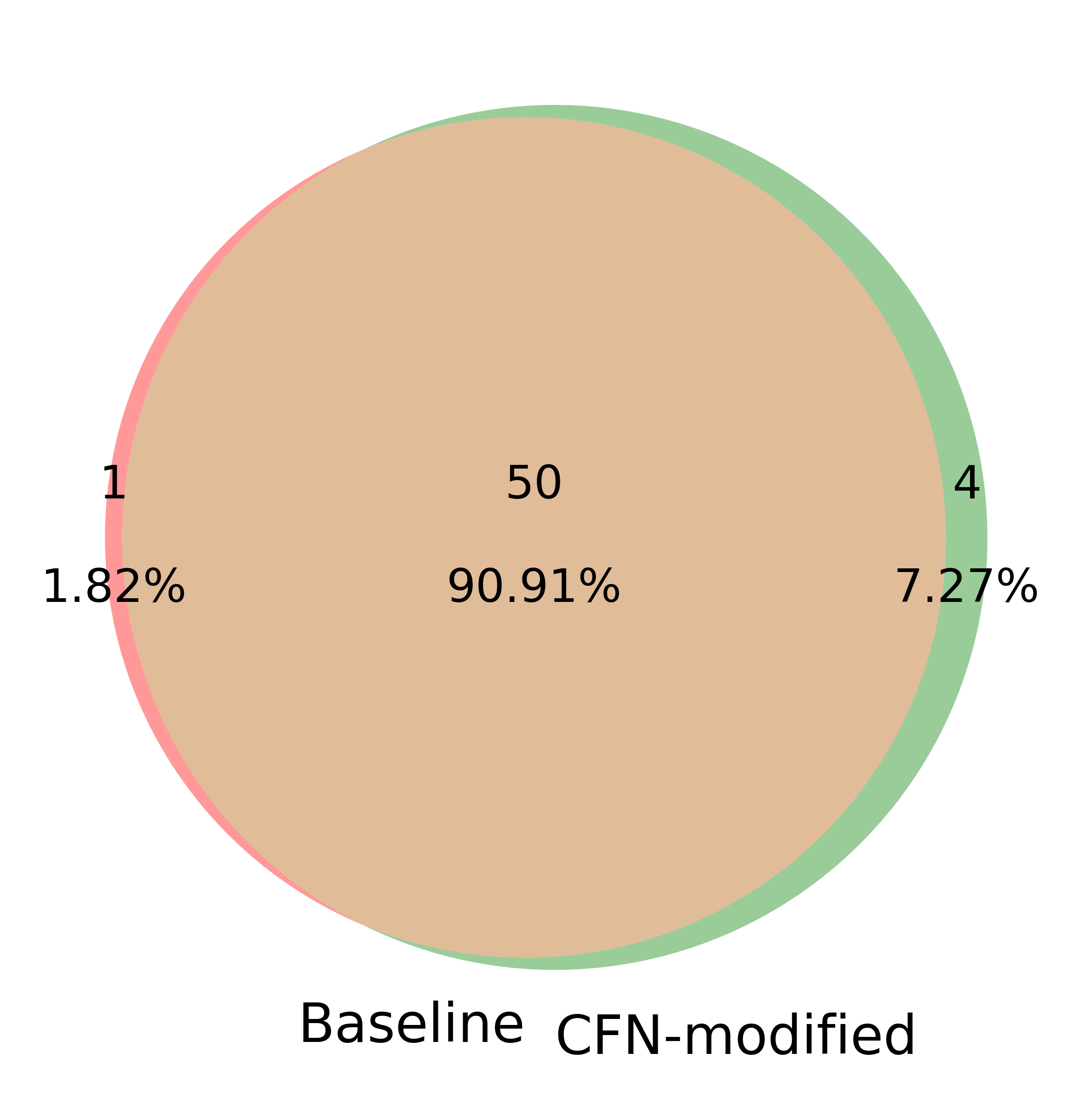} 
        \caption{CFN-modified fixes 4 more bugs compared to the baseline.}
        \label{venn-diagram-DeepSeek-Coder-V2-Lite-Instruct-16B-Defects4J:a}
    \end{subfigure}
    \hfill 
    \begin{subfigure}[b]{0.4\textwidth} 
        \centering
        \includegraphics[width=\textwidth]{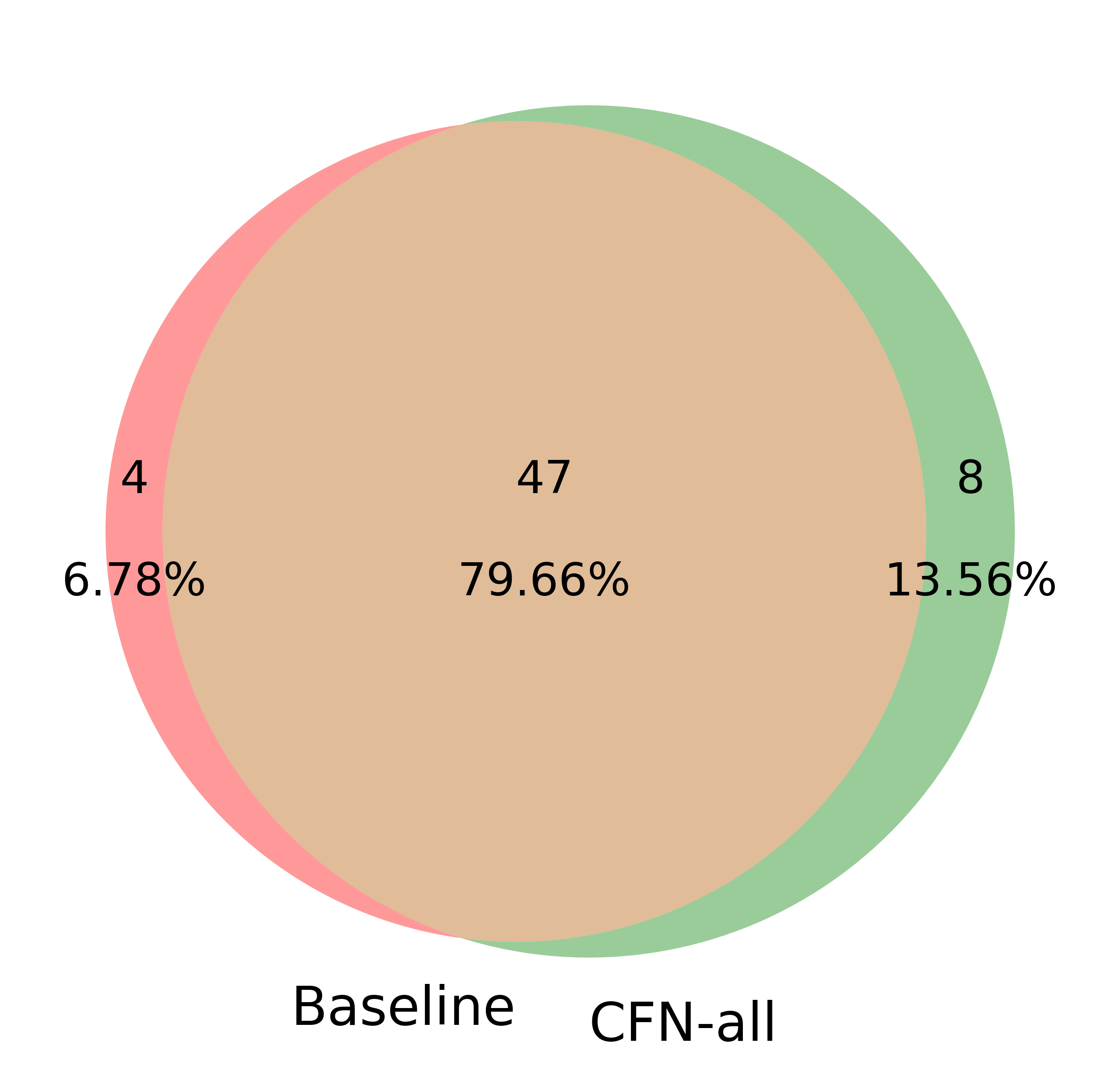} 
        \caption{CFN-all fixes 8 more bugs compared to the baseline.}
        \label{venn-diagram-DeepSeek-Coder-V2-Lite-Instruct-16B-Defects4J:b}
    \end{subfigure}
    \vspace{-0.1cm}
    \\
    \begin{subfigure}[b]{0.4\textwidth} 
        \centering
        \includegraphics[width=\textwidth]{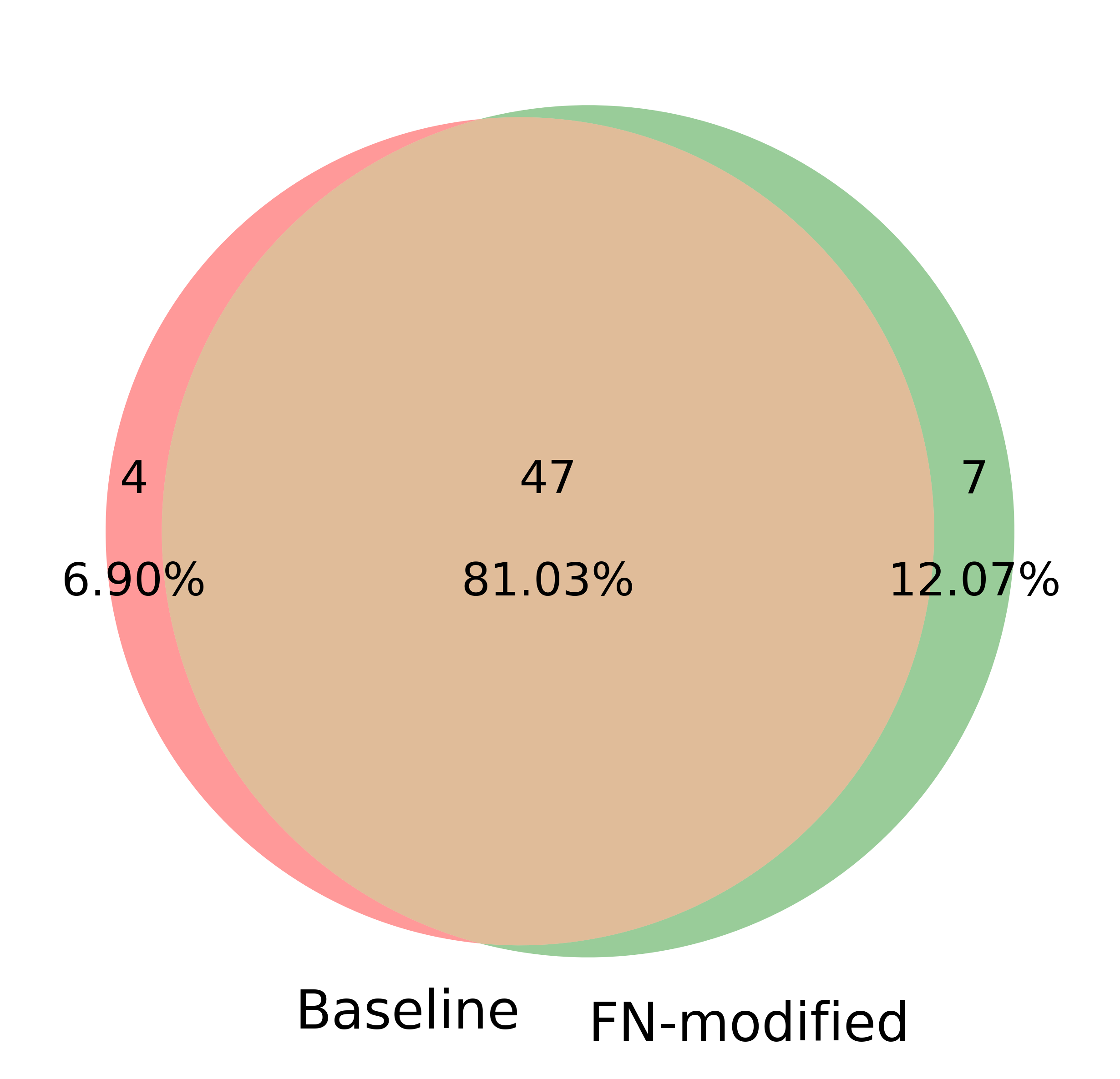} 
        \caption{FN-modified fixes 7 more bugs compared to the baseline.}
        \label{venn-diagram-DeepSeek-Coder-V2-Lite-Instruct-16B-Defects4J:c}
    \end{subfigure}
    \hfill
    \begin{subfigure}[b]{0.4\textwidth} 
        \centering
        \includegraphics[width=\textwidth]{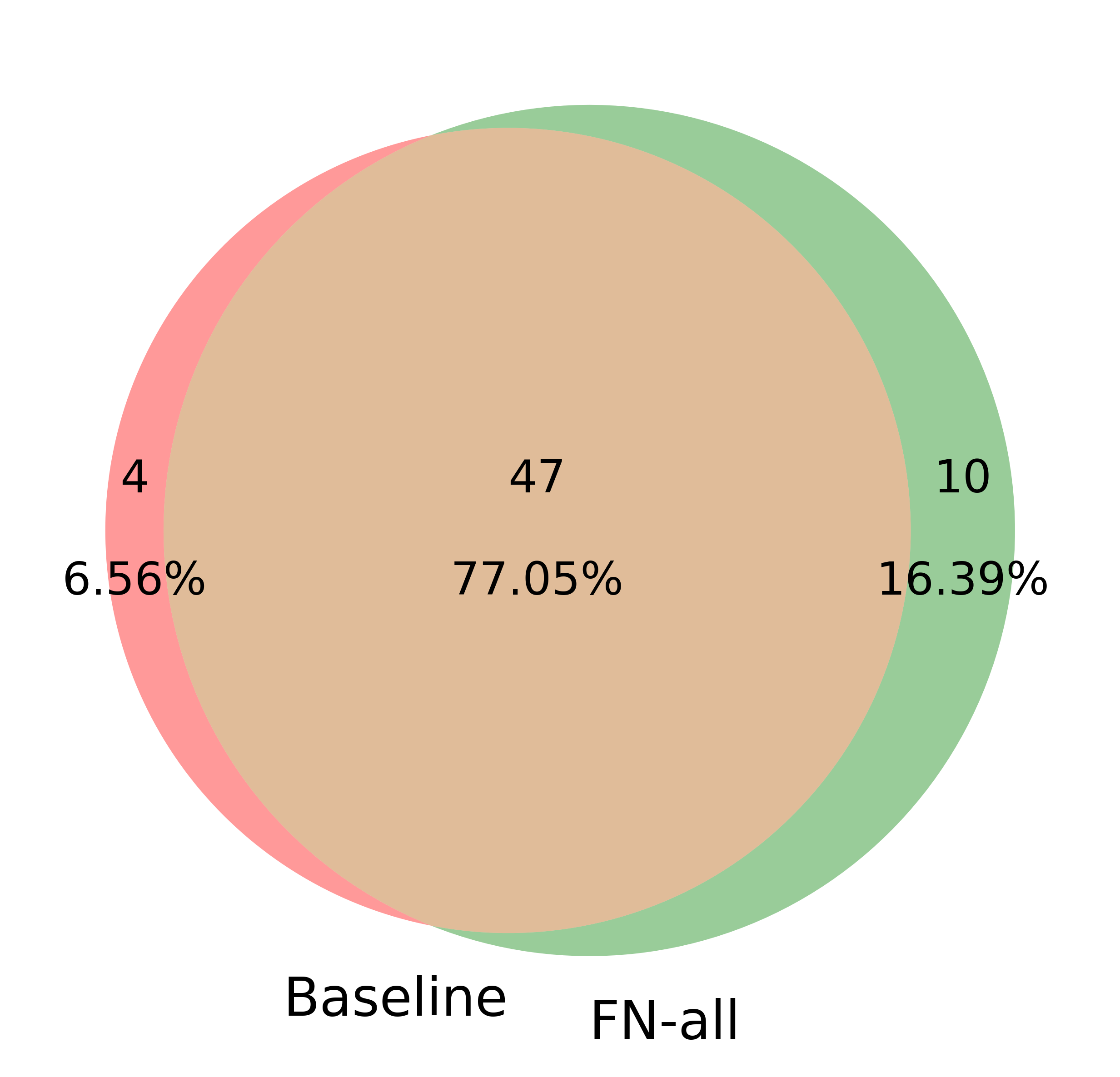} 
        \caption{FN-all fixes 10 more bugs compared to the baseline.}
        \label{venn-diagram-DeepSeek-Coder-V2-Lite-Instruct-16B-Defects4J:d}
    \end{subfigure}
    \vspace{-0.1cm}
    \\
    \begin{subfigure}[b]{0.4\textwidth} 
        \centering
        \includegraphics[width=\textwidth]{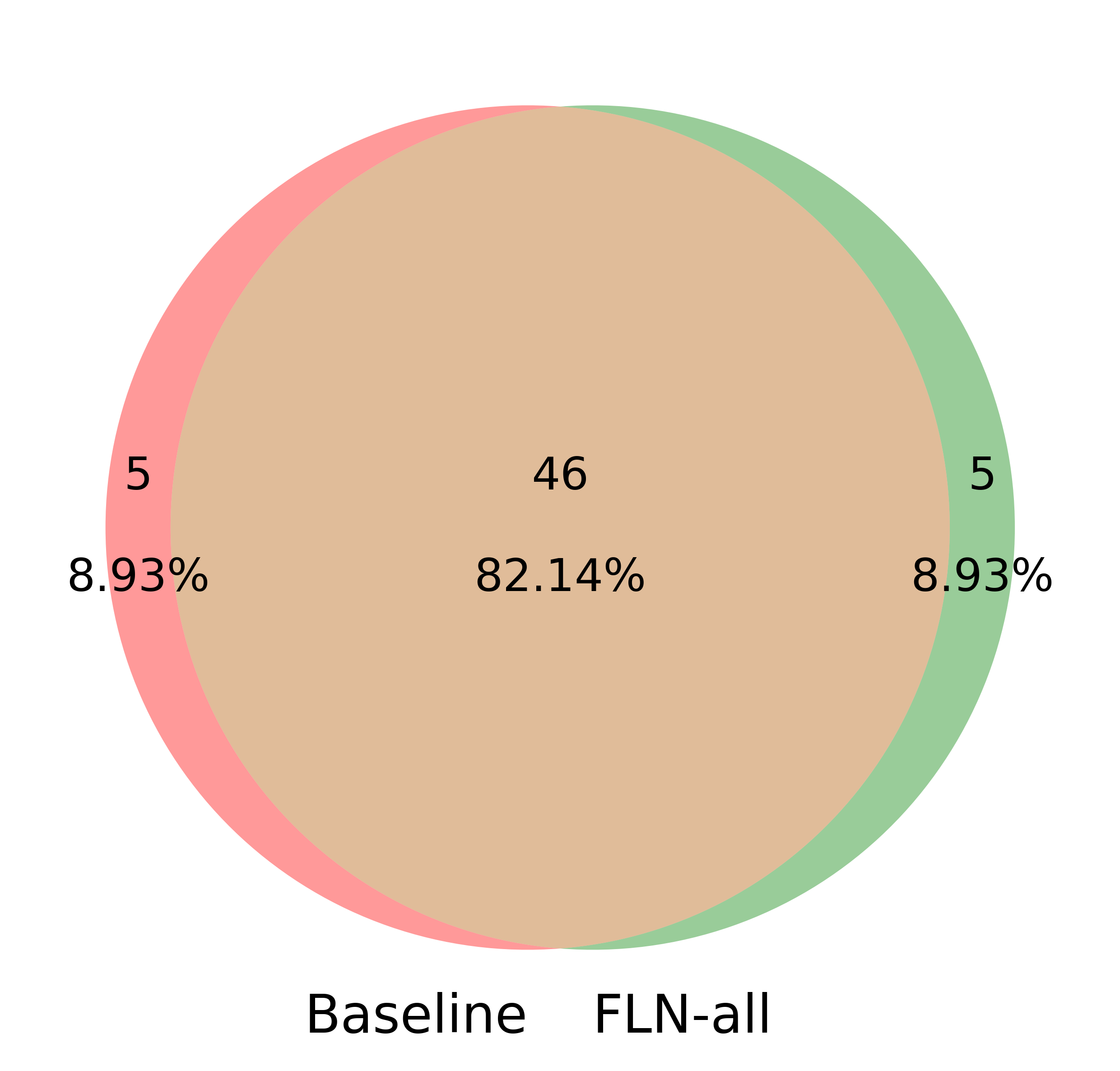} 
        \caption{FLN-all fixes 5 more bugs compared to the baseline.}
        \label{venn-diagram-DeepSeek-Coder-V2-Lite-Instruct-16B-Defects4J:e}
    \end{subfigure}
    \hfill
    \begin{subfigure}[b]{0.4\textwidth} 
        \centering
        \includegraphics[width=\textwidth]{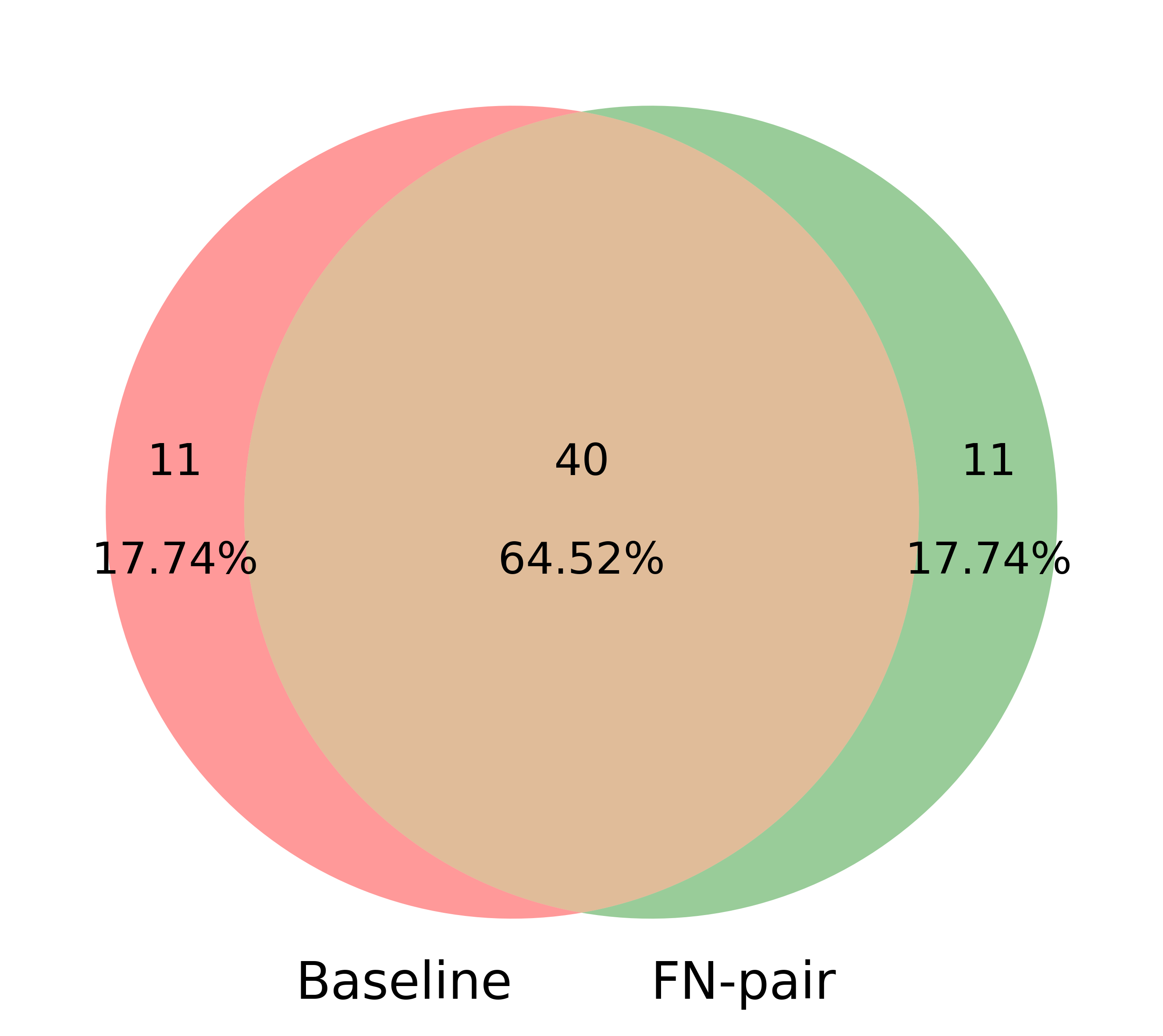} 
        \caption{FN-pair fixes 11 more bugs compared to the baseline.}
        \label{venn-diagram-DeepSeek-Coder-V2-Lite-Instruct-16B-Defects4J:f}
    \end{subfigure}
    \vspace{-0.1cm}
    \\
    \begin{subfigure}[b]{0.4\textwidth} 
        \centering
        \includegraphics[width=\textwidth]{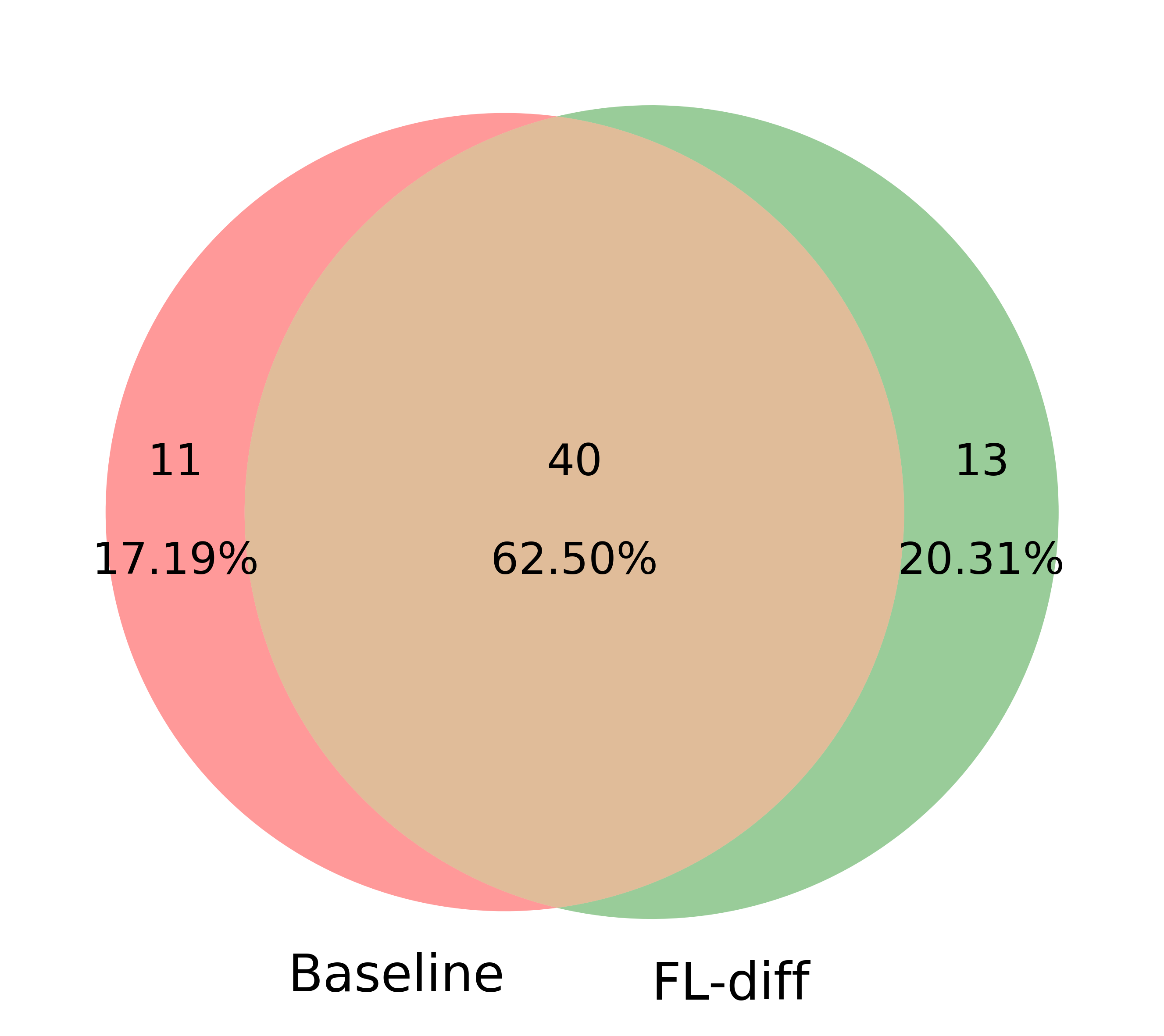} 
        \caption{FL-diff fixes 13 more bugs compared to the baseline.}
        \label{venn-diagram-DeepSeek-Coder-V2-Lite-Instruct-16B-Defects4J:g}
    \end{subfigure}
    \caption{DeepSeek-Coder-V2-Lite-Instruct-16B on Defects4J. Venn diagrams comparing the number of bugs fixed by the baseline (red) and the seven individual HAFix heuristics (green), with the overlapping region (brown) indicating bugs fixed by both the baseline and the heuristic. Numbers and percentages within each region denote the count and proportion of bugs fixed.}
    \label{fig:venn-diagram-DeepSeek-Coder-V2-Lite-Instruct-16B-Defects4J}
\end{figure}

\clearpage
\section{Additional RQ3 Results}\label{appendix:AdditionalRQ3Results}

\begin{figure}[!htbp]
  \centering
  \begin{subfigure}[b]{0.48\textwidth}
    \includegraphics[width=\textwidth]{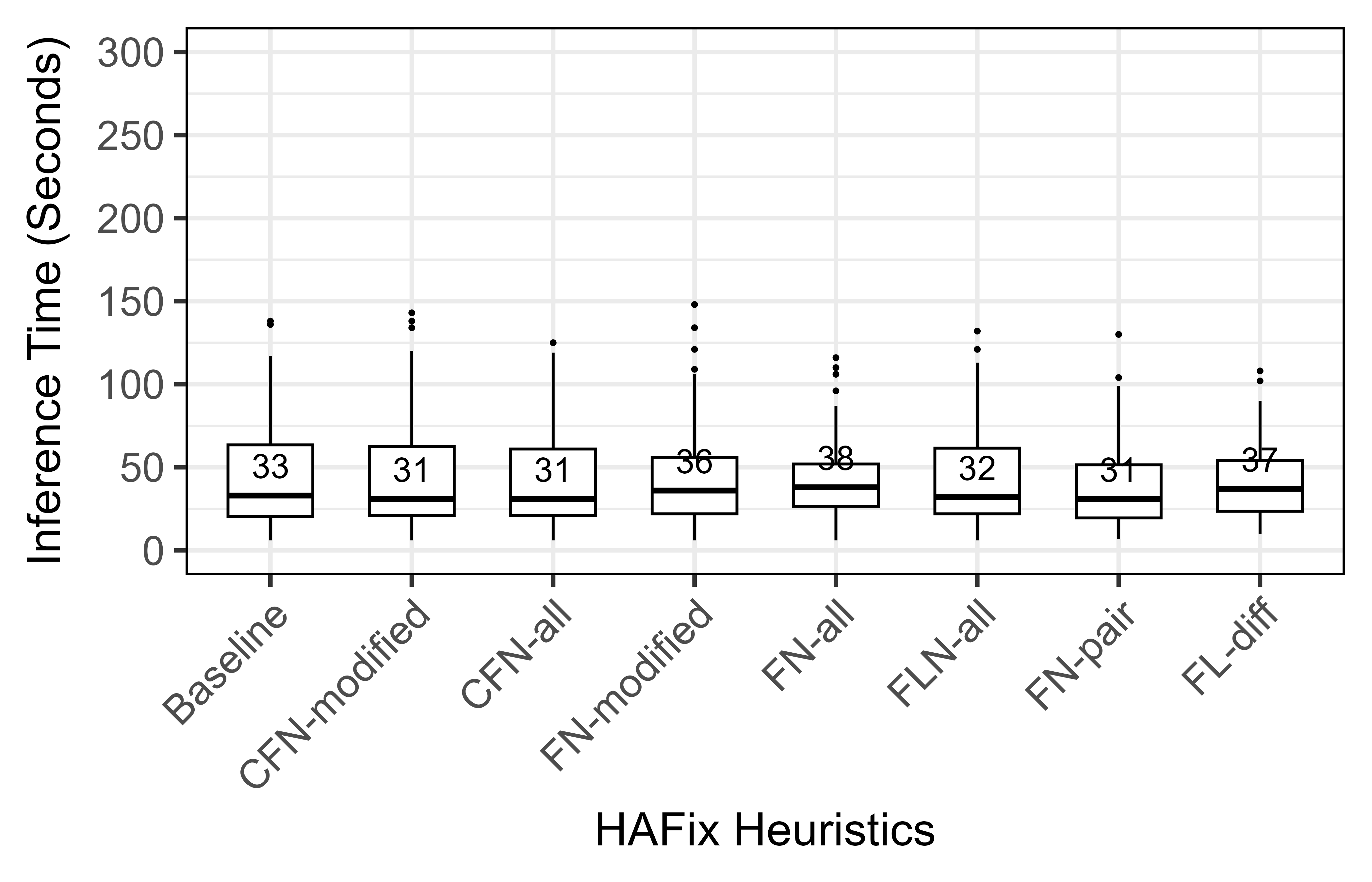}
    \caption{CodeLlama-Instruct-7B on BugsInPy}
    \label{rq3_box_plot_time_exhaustive_2_datasets_3_models:a}
  \end{subfigure}
  \hfill
  \begin{subfigure}[b]{0.48\textwidth}
    \includegraphics[width=\textwidth]{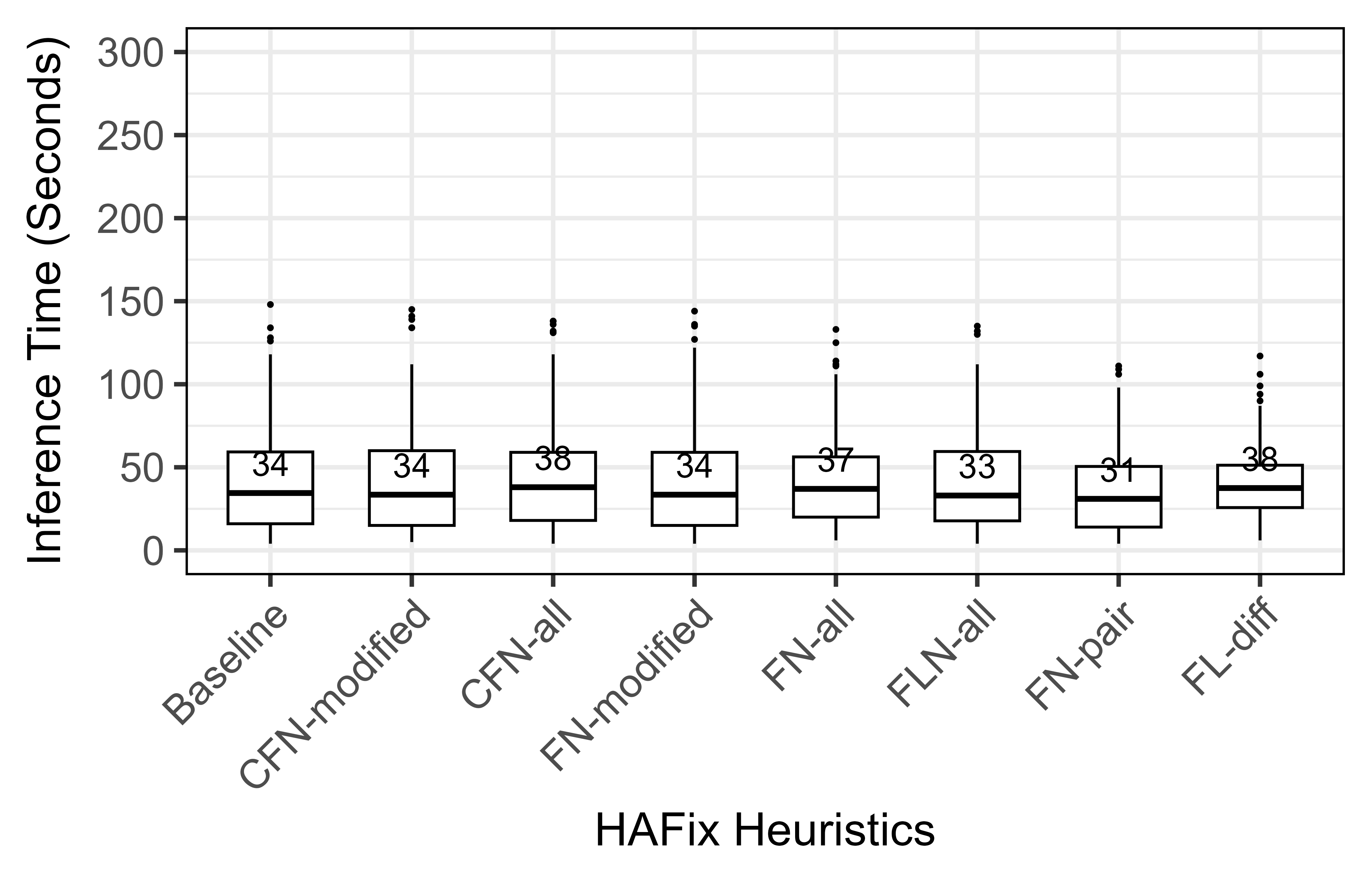}
    \caption{CodeLlama-Instruct-7B on Defects4J}
    \label{rq3_box_plot_time_exhaustive_2_datasets_3_models:b}
  \end{subfigure}

  \vspace{1em} 
    
  \begin{subfigure}[b]{0.48\textwidth}
    \includegraphics[width=\textwidth]{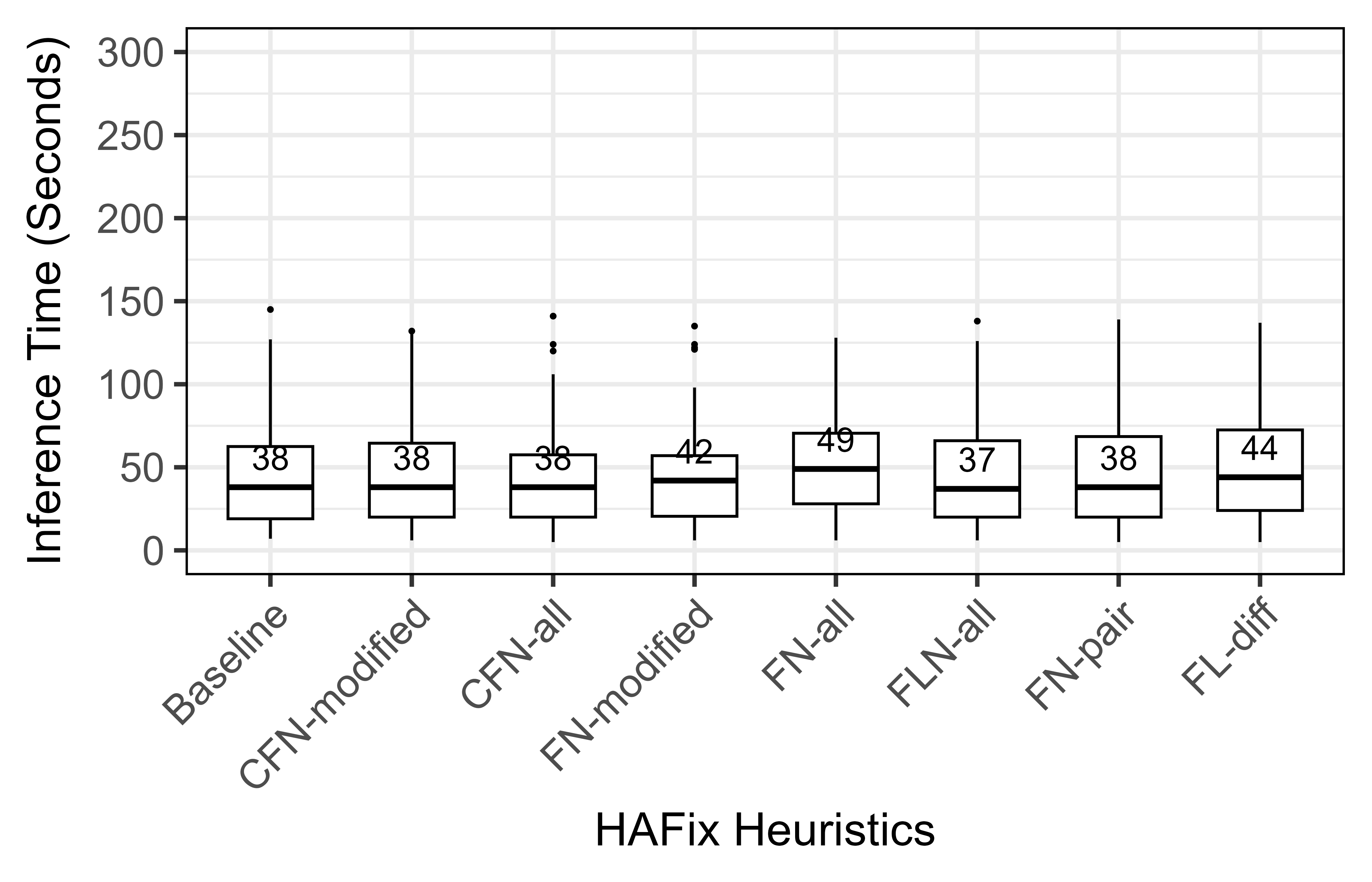}
    \caption{DeepSeek-Coder-Instruct-6.7B on \\BugsInPy}
    \label{rq3_box_plot_time_exhaustive_2_datasets_3_models:c}
  \end{subfigure}
  \hfill
  \begin{subfigure}[b]{0.48\textwidth}
    \includegraphics[width=\textwidth]{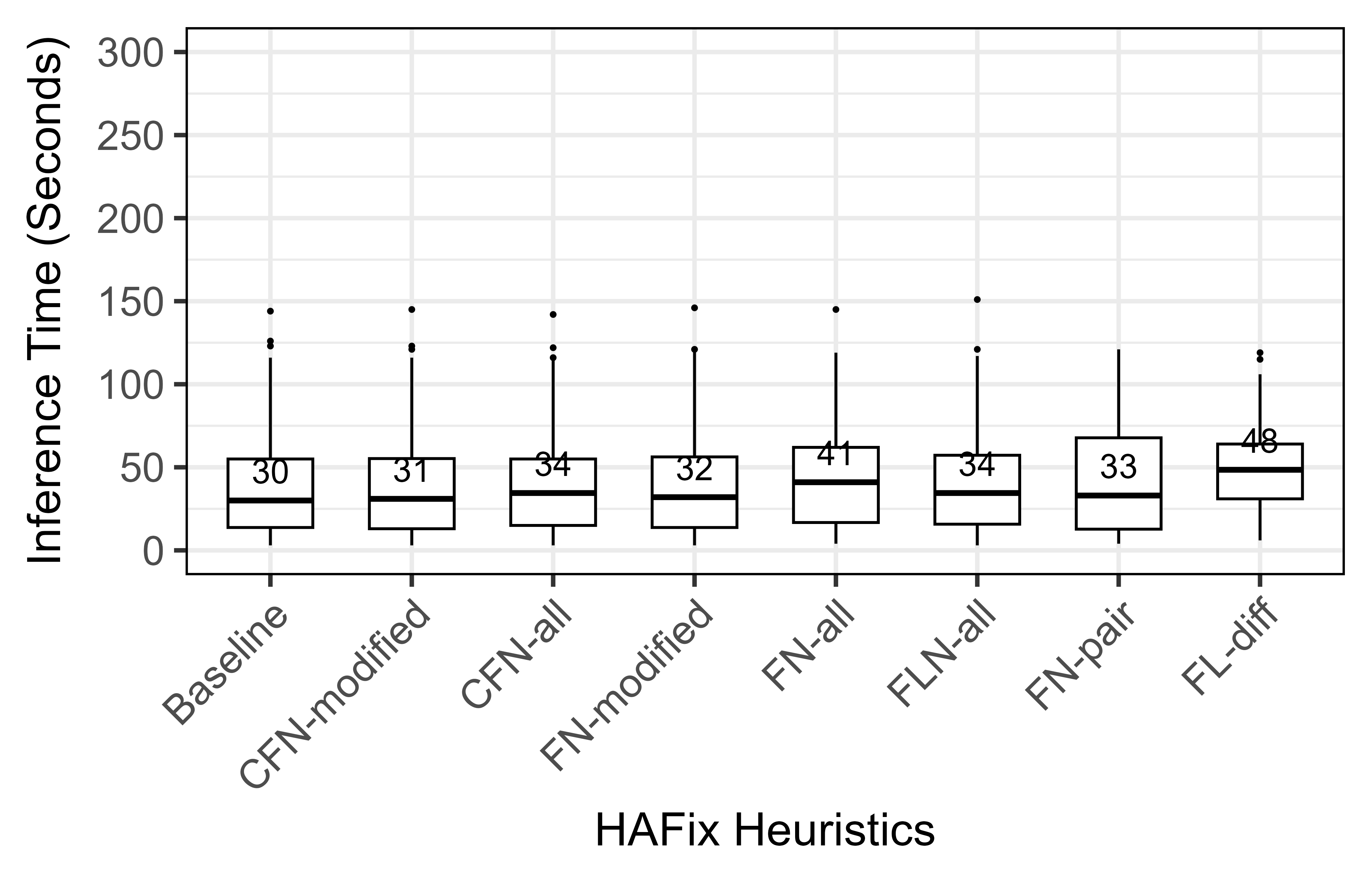}
    \caption{DeepSeek-Coder-Instruct-6.7B on \\Defects4J}
    \label{rq3_box_plot_time_exhaustive_2_datasets_3_models:d}
  \end{subfigure}

  \vspace{1em} 
    
  \begin{subfigure}[b]{0.48\textwidth}
    \includegraphics[width=\textwidth]{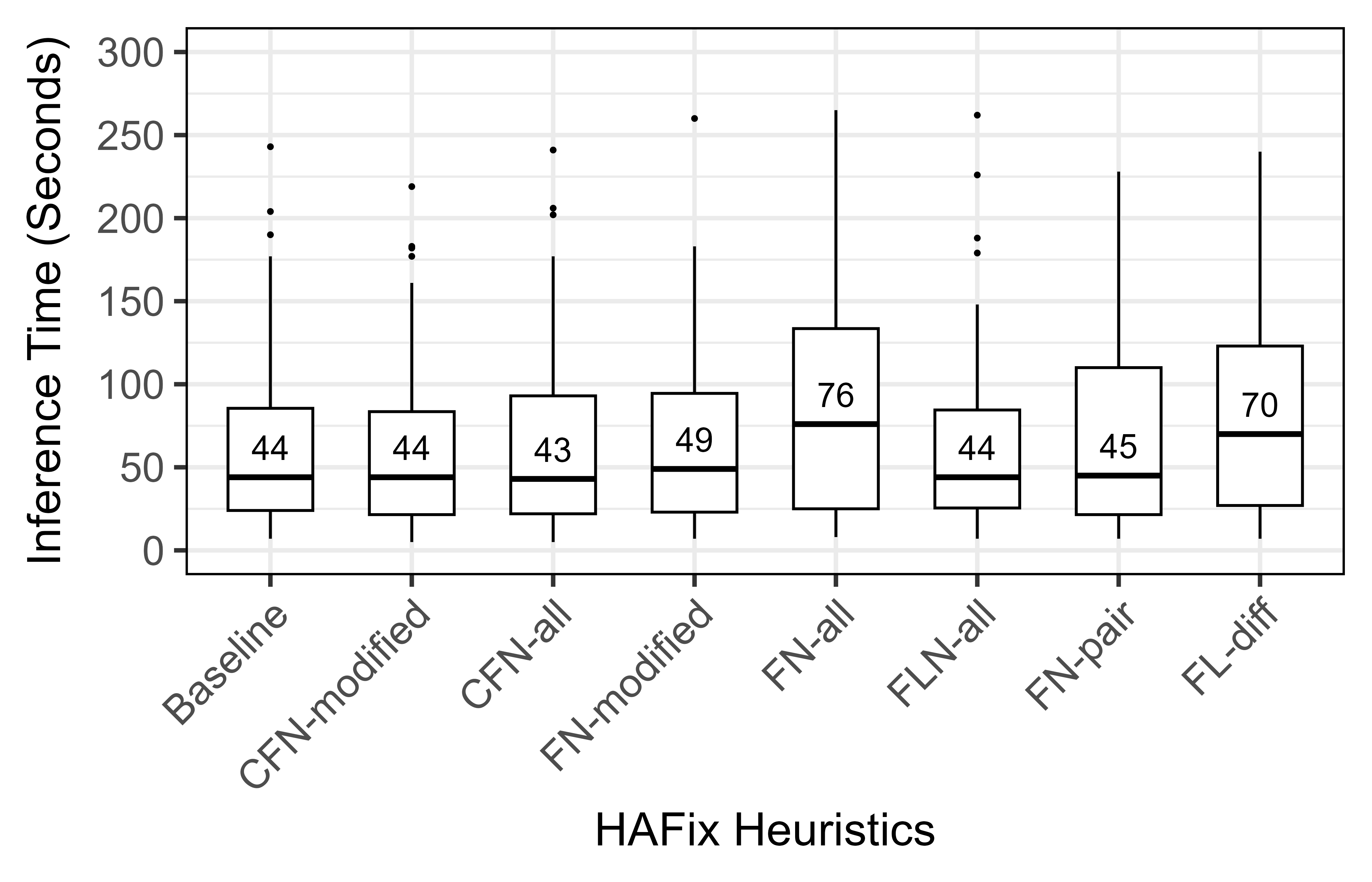}
    \caption{DeepSeek-Coder-V2-Lite-Instruct-16B \\on BugsInPy}
    \label{rq3_box_plot_time_exhaustive_2_datasets_3_models:e}
  \end{subfigure}
  \hfill
  \begin{subfigure}[b]{0.48\textwidth}
    \includegraphics[width=\textwidth]{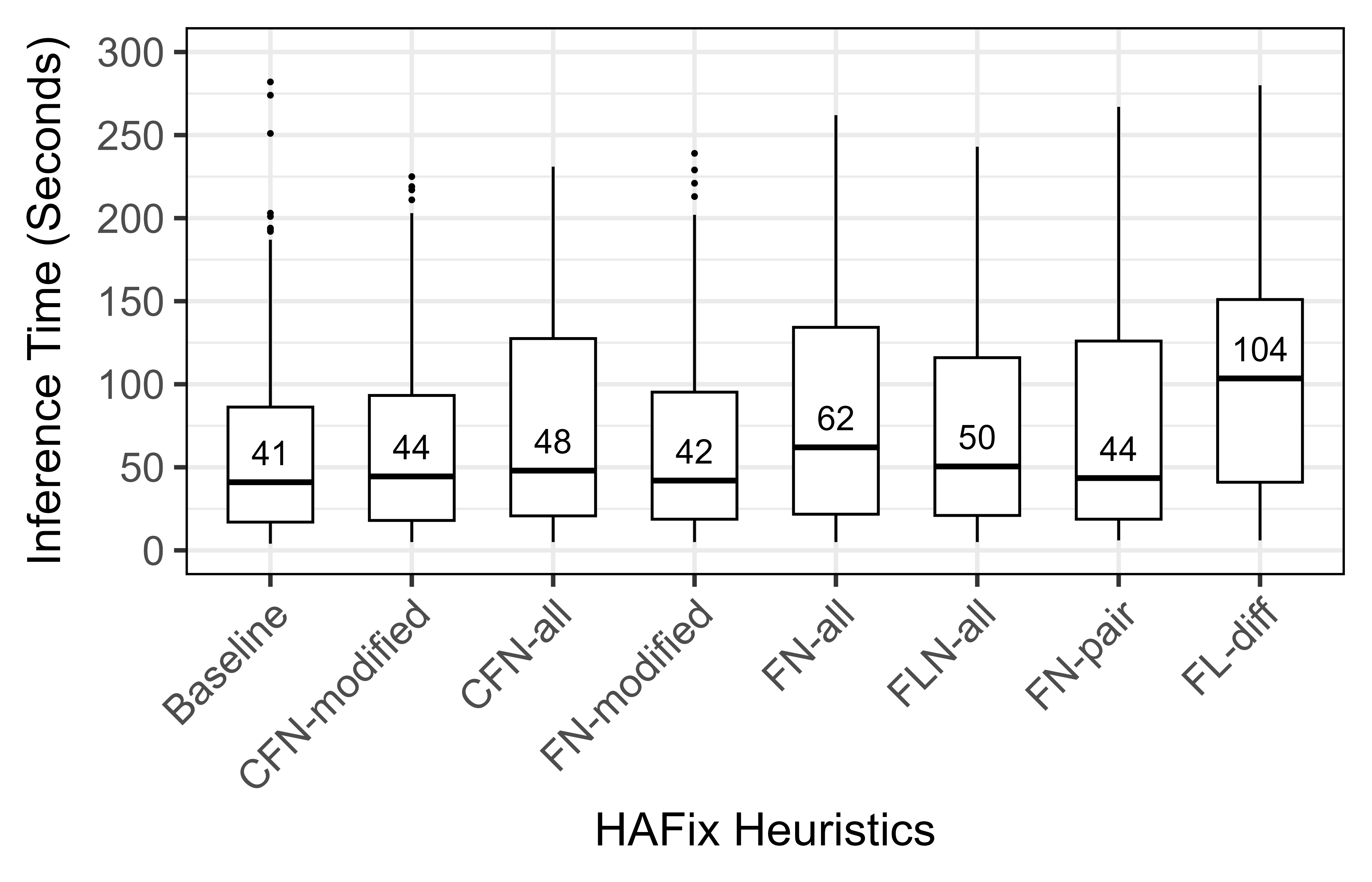}
    \caption{DeepSeek-Coder-V2-Lite-Instruct-16B \\on Defects4J}
    \label{rq3_box_plot_time_exhaustive_2_datasets_3_models:f}
  \end{subfigure}
  \caption{Distribution of inference time for each bug across different heuristics in the Exhaustive scenario, evaluated on two datasets and three models.}
  \label{fig:rq3_box_plot_time_exhaustive_2_datasets_3_models}
\end{figure}

\begin{table}[!htbp]
    \centering
    \caption{Pairwise comparisons of inference time across heuristics using the Wilcoxon signed-rank test over three models and two datasets. Values shown are $p$-values. A Bonferroni-corrected significance threshold of $\alpha = 0.0018$ ($0.05/28$) is applied for pairwise comparisons. Entries where the row heuristic is significantly faster than the column heuristic are highlighted in bold and marked with \textcolor{green}{$\uparrow$}, while those where it is significantly slower are highlighted in bold and marked with \textcolor{red}{$\downarrow$}. Non‑significant differences are left unmarked.}
    \label{tab:wilcox_rq3_box_plot_time_exhaustive_2_datasets_3_models}
    \begin{subtable}[t]{\textwidth}
        \centering
        \caption{CodeLlama-Instruct-7B on BugsInPy}
        \label{wilcox_rq3_box_plot_time_exhaustive_2_datasets_3_models:a}
        \resizebox{\textwidth}{!}{
        \begin{tabular}{cccccccc}
        \toprule
              & Baseline & CFN-modified & CFN-all & FN-modified & FN-all & FLN-all & FN-pair \\ \midrule
    CFN-modified & 0.2528    & -          & -          & -          & -          & -          & -          \\
    CFN-all      & 0.3396    & 0.6578     & -          & -          & -          & -          & -          \\
    FN-modified  & 0.1598    & 0.9325     & 0.8791     & -          & -          & -          & -          \\
    FN-all       & 0.1133    & 0.3843     & 0.1604     & 0.3518     & -          & -          & -          \\
    FLN-all      & 0.9684    & 0.4686     & 0.1591     & 0.4093     & 0.1978     & -          & -          \\
    FN-pair      & 0.0566    & 0.0028     & 0.0274     & 0.0086     & 0.0251     & 0.0343     & -          \\
    FL-diff      & 0.8833    & 0.5019     & 0.6300     & 0.6187     & 0.4551     & 0.6978     & 0.0124     \\ \bottomrule
        \end{tabular}
        }
    \end{subtable}
    \vspace{0.1cm}
    
    \begin{subtable}[t]{\textwidth}
        \centering
        \caption{CodeLlama-Instruct-7B on Defects4J}
        \label{wilcox_rq3_box_plot_time_exhaustive_2_datasets_3_models:b}
        \resizebox{\textwidth}{!}{
        \begin{tabular}{cccccccc}
        \toprule
              & Baseline & CFN-modified & CFN-all & FN-modified & FN-all & FLN-all & FN-pair \\ \midrule
CFN-modified & 0.2804   & -          & -          & -          & -          & -          & -          \\
CFN-all      & \textbf{\boldmath $1.7 \times 10^{-5}$}\,\textcolor{red}{$\downarrow$}  & \textbf{0.0001}\,\textcolor{red}{$\downarrow$}     & -          & -          & -          & -          & -          \\
FN-modified  & 0.0901  & 0.3235     & 0.0048     & -          & -          & -          & -          \\
FN-all       & \textbf{0.0004}\,\textcolor{red}{$\downarrow$}   & \textbf{0.0011}\,\textcolor{red}{$\downarrow$}     & 0.5880     & 0.0170     & -          & -          & -          \\
FLN-all      & 0.0656   & 0.3958     & 0.0101     & 0.9656     & 0.0368     & -          & -          \\
FN-pair      & 0.0020   & \textbf{0.0004}\,\textcolor{green}{$\uparrow$}     & \textbf{\boldmath $1.4 \times 10^{-8}$}\,\textcolor{green}{$\uparrow$}    & \textbf{\boldmath $3.3 \times 10^{-5}$}\,\textcolor{green}{$\uparrow$}    & \textbf{\boldmath $3.4 \times 10^{-9}$}\,\textcolor{green}{$\uparrow$}    & \textbf{\boldmath $1.5 \times 10^{-5}$}\,\textcolor{green}{$\uparrow$}
    & -          \\
FL-diff      & 0.1064   & 0.1321     & 0.5491     & 0.1027     & 0.9907     & 0.3822     & \textbf{\boldmath $5.4 \times 10^{-6}$}\,\textcolor{red}{$\downarrow$}    \\ \bottomrule
        \end{tabular}
        }
    \end{subtable}
    \vspace{0.1cm}

    \begin{subtable}[t]{\textwidth}
        \centering
        \caption{DeepSeek-Coder-Instruct-6.7B on BugsInPy}
        \label{wilcox_rq3_box_plot_time_exhaustive_2_datasets_3_models:c}
        \resizebox{\textwidth}{!}{
        \begin{tabular}{cccccccc}
        \toprule
              & Baseline & CFN-modified & CFN-all & FN-modified & FN-all & FLN-all & FN-pair \\ \midrule
CFN-modified & 0.1447   & -          & -          & -          & -          & -          & -          \\
CFN-all      & 0.8600   & 0.4756     & -          & -          & -          & -          & -          \\
FN-modified  & 0.2068   & 0.5482     & 0.4465     & -          & -          & -          & -          \\
FN-all       & \textbf{0.0018}\,\textcolor{red}{$\downarrow$}   & 0.0096     & \textbf{0.0008}\,\textcolor{red}{$\downarrow$}     & 0.0133     & -          & -          & -          \\
FLN-all      & 0.8547   & 0.2879     & 0.6376     & 0.1663     & \textbf{0.0008}\,\textcolor{green}{$\uparrow$}     & -          & -          \\
FN-pair      & 0.4240   & 0.9324     & 0.6153     & 0.4724     & 0.0152     & 0.3492     & -          \\
FL-diff      & 0.0036   & 0.0198     & 0.0046     & 0.0910     & 0.3534     & 0.0032     & 0.1551     \\ \bottomrule
        \end{tabular}
        }
    \end{subtable}
    \vspace{0.1cm}

    \begin{subtable}[t]{\textwidth}
        \centering
        \caption{DeepSeek-Coder-Instruct-6.7B on Defects4J}
        \label{wilcox_rq3_box_plot_time_exhaustive_2_datasets_3_models:d}
        \resizebox{\textwidth}{!}{
        \begin{tabular}{cccccccc}
        \toprule
              & Baseline & CFN-modified & CFN-all & FN-modified & FN-all & FLN-all & FN-pair \\ \midrule
CFN-modified & 0.3306   & -          & -          & -          & -          & -          & -          \\
CFN-all      & 0.0507   & 0.3368     & -          & -          & -          & -          & -          \\
FN-modified  & 0.0073   & 0.0302     & 0.7120     & -          & -          & -          & -          \\
FN-all       & \textbf{0.0006}\,\textcolor{red}{$\downarrow$}   & \textbf{0.0015}\,\textcolor{red}{$\downarrow$}     & 0.0025     & 0.0123     & -          & -          & -          \\
FLN-all      & \textbf{0.0014}\,\textcolor{red}{$\downarrow$}   & 0.0522     & 0.1909     & 0.1572     & 0.0800     & -          & -          \\
FN-pair      & 0.0067   & 0.0166     & 0.3105     & 0.0287     & 0.5829     & 0.1370     & -          \\
FL-diff      & \textbf{\boldmath $1.8 \times 10^{-6}$}\,\textcolor{red}{$\downarrow$}  & \textbf{\boldmath $1.7 \times 10^{-6}$}\,\textcolor{red}{$\downarrow$}  & \textbf{\boldmath $3.2 \times 10^{-6}$}\,\textcolor{red}{$\downarrow$}  & \textbf{\boldmath $2.7 \times 10^{-8}$}\,\textcolor{red}{$\downarrow$}
    & \textbf{0.0001}\,\textcolor{red}{$\downarrow$}     & \textbf{\boldmath $1.1 \times 10^{-5}$}\,\textcolor{red}{$\downarrow$}
    & \textbf{0.0004}\,\textcolor{red}{$\downarrow$}     \\ \bottomrule
        \end{tabular}
        }
    \end{subtable}
    \vspace{0.1cm}

    \begin{subtable}[t]{\textwidth}
        \centering
        \caption{DeepSeek-Coder-V2-Lite-Instruct-16B on BugsInPy}
        \label{wilcox_rq3_box_plot_time_exhaustive_2_datasets_3_models:e}
        \resizebox{\textwidth}{!}{
        \begin{tabular}{cccccccc}
        \toprule
              & Baseline & CFN-modified & CFN-all & FN-modified & FN-all & FLN-all & FN-pair \\ \midrule
CFN-modified & 0.9439   & -          & -          & -          & -          & -          & -          \\
CFN-all      & 0.3117   & 0.0765     & -          & -          & -          & -          & -          \\
FN-modified  & 0.1420   & 0.1258     & 0.5381     & -          & -          & -          & -          \\
FN-all       & \textbf{\boldmath $8.3 \times 10^{-6}$}\,\textcolor{red}{$\downarrow$}  & \textbf{\boldmath $5.2 \times 10^{-6}$}\,\textcolor{red}{$\downarrow$}  & \textbf{\boldmath $4.4 \times 10^{-6}$}\,\textcolor{red}{$\downarrow$}  & \textbf{\boldmath $4.7 \times 10^{-6}$}\,\textcolor{red}{$\downarrow$}
    & -          & -          & -          \\
FLN-all      & 0.1789   & 0.4768     & 0.6288     & 0.5595     & \textbf{\boldmath $2.9 \times 10^{-5}$}\,\textcolor{green}{$\uparrow$}
    & -          & -          \\
FN-pair      & 0.0454   & 0.0503     & 0.2866     & 0.2795     & 0.0032     & 0.1352     & -          \\
FL-diff      & \textbf{0.0003}\,\textcolor{red}{$\downarrow$}   & \textbf{0.0001}\,\textcolor{red}{$\downarrow$}     & 0.0020     & 0.0028     & 0.1482     & 0.0023     & 0.0298     \\ \bottomrule
        \end{tabular}
        }
    \end{subtable}
    \vspace{0.1cm}
    
    \begin{subtable}[t]{\textwidth}
        \centering
        \caption{DeepSeek-Coder-V2-Lite-Instruct-16B on Defects4J}
        \label{wilcox_rq3_box_plot_time_exhaustive_2_datasets_3_models:f}
        \resizebox{\textwidth}{!}{
        \begin{tabular}{cccccccc}
        \toprule
              & Baseline & CFN-modified & CFN-all & FN-modified & FN-all & FLN-all & FN-pair \\ \midrule
CFN-modified & 0.0161   & -          & -          & -          & -          & -          & -          \\
CFN-all      & \textbf{0.0001}\,\textcolor{red}{$\downarrow$}   & 0.0020     & -          & -          & -          & -          & -          \\
FN-modified  & \textbf{0.0011}\,\textcolor{red}{$\downarrow$}   & 0.0293     & 0.2412     & -          & -          & -          & -          \\
FN-all       & \textbf{\boldmath $6.4 \times 10^{-10}$}\,\textcolor{red}{$\downarrow$}  & \textbf{\boldmath $2.0 \times 10^{-8}$}\,\textcolor{red}{$\downarrow$}  & \textbf{\boldmath $4.4 \times 10^{-5}$}\,\textcolor{red}{$\downarrow$}  & \textbf{\boldmath $9.6 \times 10^{-9}$}\,\textcolor{red}{$\downarrow$}
    & -          & -          & -          \\
FLN-all      & 0.0061   & 0.3732     & 0.3682     & 0.7393     & \textbf{0.0004}\,\textcolor{green}{$\uparrow$}     & -          & -          \\
FN-pair      & 0.0256   & 0.2713     & 0.3181     & 0.2730     & \textbf{\boldmath $3.5 \times 10^{-5}$}\,\textcolor{green}{$\uparrow$}
    & 0.7172     & -          \\
FL-diff      & \textbf{\boldmath $9.7 \times 10^{-13}$}\,\textcolor{red}{$\downarrow$}  & \textbf{\boldmath $1.8 \times 10^{-12}$}\,\textcolor{red}{$\downarrow$}  & \textbf{\boldmath $5.9 \times 10^{-9}$}\,\textcolor{red}{$\downarrow$}  & \textbf{\boldmath $3.5 \times 10^{-12}$}\,\textcolor{red}{$\downarrow$}  & \textbf{\boldmath $2.0 \times 10^{-5}$}\,\textcolor{red}{$\downarrow$}  & \textbf{\boldmath $1.3 \times 10^{-9}$}\,\textcolor{red}{$\downarrow$}  & \textbf{\boldmath $5.2 \times 10^{-11}$}\,\textcolor{red}{$\downarrow$}
    \\ \bottomrule
        \end{tabular}
        }
    \end{subtable}
\end{table}


\begin{table}[!htbp]
  \centering
  \caption{Pairwise comparisons of inference tokens of bugs using Wilcoxon signed-rank test over three models and two datasets. E, ES, ES-A, and ES-U correspond to the Exhaustive, EarlyStop, ES-AccSorted, and ES-UniSorted cost scenarios, respectively. Each cell reports the $p$-value and the corresponding effect size $r_{\mathrm{rb}}$ (Rank-Biserial Correlation). A Bonferroni-corrected significance threshold of $\alpha = 0.0083$ ($0.05/6$) is applied for pairwise comparisons.}
  \label{tab:wilcox_rq3_box_plot_token_2_datasets_3_models}

  \begin{subtable}[t]{0.46\textwidth}
    \centering
    \caption{CodeLlama-Instruct-7B on BugsInPy}
    \label{wilcox_rq3_box_plot_token_2_datasets_3_models:a}
    \begin{tabular}{lccc}
      \toprule
      & ES & E & ES-A \\
      \midrule
      E & \makecell{$2.6 \times 10^{-7}$ \\ $r_{\mathrm{rb}} = 0.05$} & - & - \\
      \midrule
      ES-A & \makecell{0.0320 \\ $r_{\mathrm{rb}} = 0.76$} & \makecell{$2.6 \times 10^{-7}$ \\ $r_{\mathrm{rb}} = 0.05$} & - \\
      \midrule
      ES-U & \makecell{0.8551 \\ $r_{\mathrm{rb}} = 0.99$} & \makecell{$2.6 \times 10^{-7}$ \\ $r_{\mathrm{rb}} = 0.05$} & \makecell{0.0564 \\ $r_{\mathrm{rb}} = 0.35$} \\
      \bottomrule
    \end{tabular}
  \end{subtable}
  \hfill
  \begin{subtable}[t]{0.46\textwidth}
    \centering
    \caption{CodeLlama-Instruct-7B on Defects4J}
    \label{wilcox_rq3_box_plot_token_2_datasets_3_models:b}
    \begin{tabular}{lccc}
      \toprule
      & ES & E & ES-A \\
      \midrule
      E & \makecell{$3.6 \times 10^{-13}$ \\ $r_{\mathrm{rb}} = 0.27$} & - & - \\
      \midrule
      ES-A & \makecell{0.0111 \\ $r_{\mathrm{rb}} = 0.78$} & \makecell{$3.6 \times 10^{-13}$ \\ $r_{\mathrm{rb}} = 0.27$} & - \\
      \midrule
      ES-U & \makecell{0.5360 \\ $r_{\mathrm{rb}} = 0.95$} & \makecell{$2.5 \times 10^{-13}$ \\ $r_{\mathrm{rb}} = 0.25$} & \makecell{0.0613 \\ $r_{\mathrm{rb}} = 0.51$} \\
      \bottomrule
    \end{tabular}
  \end{subtable}

\vspace{2em}

  \begin{subtable}[t]{0.46\textwidth}
    \centering
    \caption{DeepSeek-Coder-Instruct-6.7B on \\BugsInPy}
    \label{wilcox_rq3_box_plot_token_2_datasets_3_models:c}
    \begin{tabular}{lccc}
      \toprule
      & ES & E & ES-A \\
      \midrule
      E & \makecell{$8.3 \times 10^{-7}$ \\ $r_{\mathrm{rb}} = 0.20$} & - & - \\
      \midrule
      ES-A & \makecell{0.6295 \\ $r_{\mathrm{rb}} = 0.62$} & \makecell{$5.6 \times 10^{-7}$ \\ $r_{\mathrm{rb}} = 0.15$} & - \\
      \midrule
      ES-U & \makecell{0.0831 \\ $r_{\mathrm{rb}} = 0.93$} & \makecell{$5.6 \times 10^{-7}$ \\ $r_{\mathrm{rb}} = 0.15$} & \makecell{0.0006 \\ $r_{\mathrm{rb}} = 0.29$} \\
      \bottomrule
    \end{tabular}
  \end{subtable}
  \hfill
  \begin{subtable}[t]{0.46\textwidth}
    \centering
    \caption{DeepSeek-Coder-Instruct-6.7B on \\Defects4J}
    \label{wilcox_rq3_box_plot_token_2_datasets_3_models:d}
    \begin{tabular}{lccc}
      \toprule
      & ES & E & ES-A \\
      \midrule
      E & \makecell{$3.6 \times 10^{-12}$ \\ $r_{\mathrm{rb}} = 0.39$} & - & - \\
      \midrule
      ES-A & \makecell{0.4585 \\ $r_{\mathrm{rb}} = 0.68$} & \makecell{$2.5 \times 10^{-12}$ \\ $r_{\mathrm{rb}} = 0.37$} & - \\
      \midrule
      ES-U & \makecell{0.2020 \\ $r_{\mathrm{rb}} = 0.95$} & \makecell{$2.5 \times 10^{-12}$ \\ $r_{\mathrm{rb}} = 0.37$} & \makecell{0.2666 \\ $r_{\mathrm{rb}} = 0.63$} \\
      \bottomrule
    \end{tabular}
  \end{subtable}

\vspace{2em}

  \begin{subtable}[t]{0.46\textwidth}
    \centering
    \caption{DeepSeek-Coder-V2-Lite-Instruct-16B \\on BugsInPy}
    \label{wilcox_rq3_box_plot_token_2_datasets_3_models:e}
    \begin{tabular}{lccc}
      \toprule
      & ES & E & ES-A \\
      \midrule
      E & \makecell{$4.6 \times 10^{-6}$ \\ $r_{\mathrm{rb}} = 0.39$} & - & - \\
      \midrule
      ES-A & \makecell{0.0593 \\ $r_{\mathrm{rb}} = 0.83$} & \makecell{$4 \times 10^{-6}$ \\ $r_{\mathrm{rb}} = 0.39$} & - \\
      \midrule
      ES-U & \makecell{0.8127 \\ $r_{\mathrm{rb}} = 0.96$} & \makecell{$4 \times 10^{-6}$ \\ $r_{\mathrm{rb}} = 0.39$} & \makecell{0.1702 \\ $r_{\mathrm{rb}} = 0.65$} \\
      \bottomrule
    \end{tabular}
  \end{subtable}
  \hfill
  \begin{subtable}[t]{0.46\textwidth}
    \centering
    \caption{DeepSeek-Coder-V2-Lite-Instruct-16B \\on Defects4J}
    \label{wilcox_rq3_box_plot_token_2_datasets_3_models:f}
    \begin{tabular}{lccc}
      \toprule
      & ES & E & ES-A \\
      \midrule
      E & \makecell{$7.8 \times 10^{-13}$ \\ $r_{\mathrm{rb}} = 0.31$} & - & - \\
      \midrule
      ES-A & \makecell{0.0004 \\ $r_{\mathrm{rb}} = 0.82$} & \makecell{$5.3 \times 10^{-13}$ \\ $r_{\mathrm{rb}} = 0.29$} & - \\
      \midrule
      ES-U & \makecell{0.1169 \\ $r_{\mathrm{rb}} = 0.99$} & \makecell{$5.3 \times 10^{-13}$ \\ $r_{\mathrm{rb}} = 0.29$} &                                              \makecell{0.0003 \\ $r_{\mathrm{rb}} = 0.47`$} \\
      \bottomrule
    \end{tabular}
  \end{subtable}
\end{table}

\begin{figure}[!htbp]
  \centering
  \begin{subfigure}[b]{0.48\textwidth}
    \includegraphics[width=\textwidth]{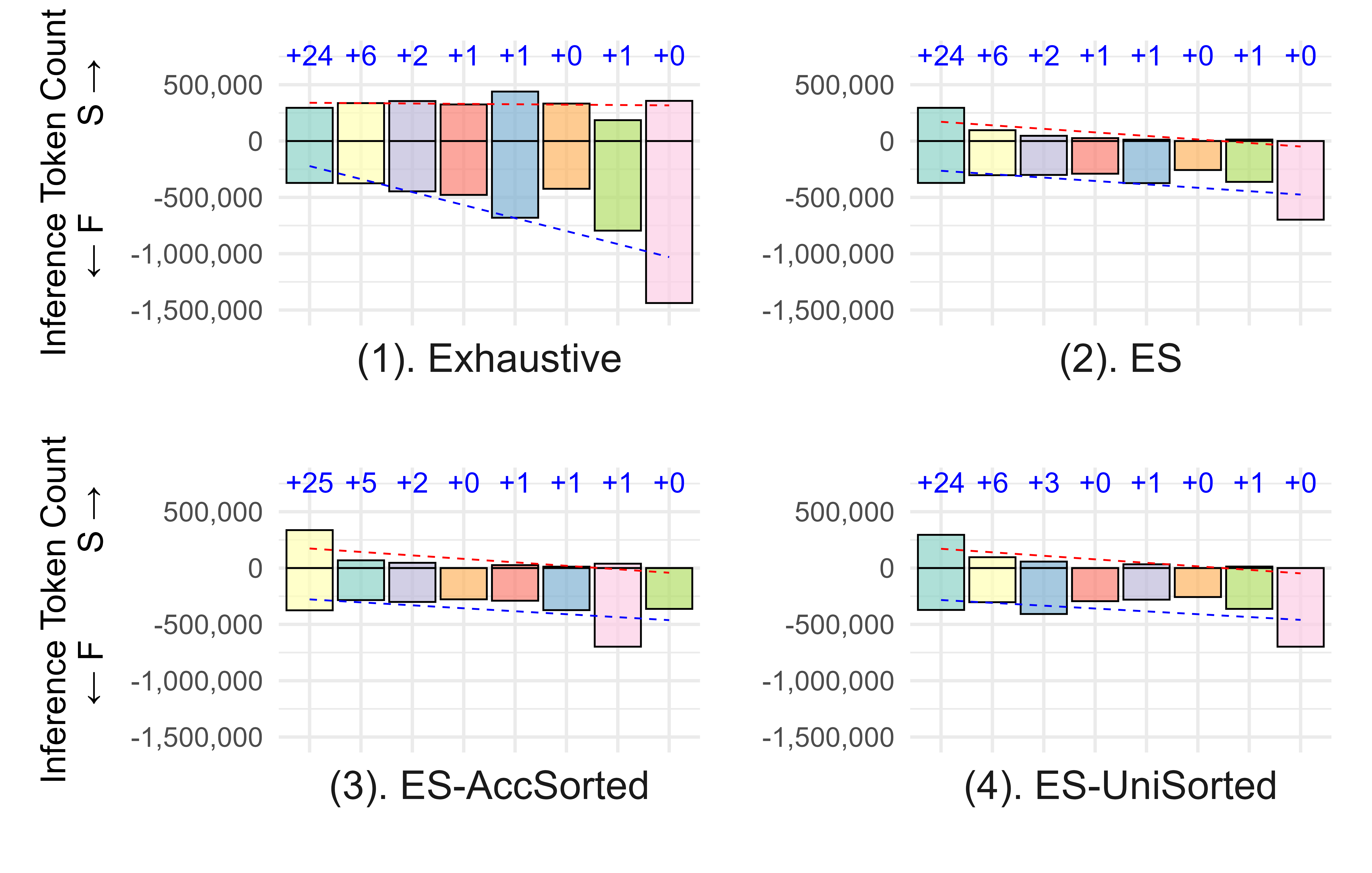}
    \caption{CodeLlama-Instruct-7B on BugsInPy}
    \label{rq3_symmetry_bar_2_datasets_3_models:a}
  \end{subfigure}
  \hfill
  \begin{subfigure}[b]{0.48\textwidth}
    \includegraphics[width=\textwidth]{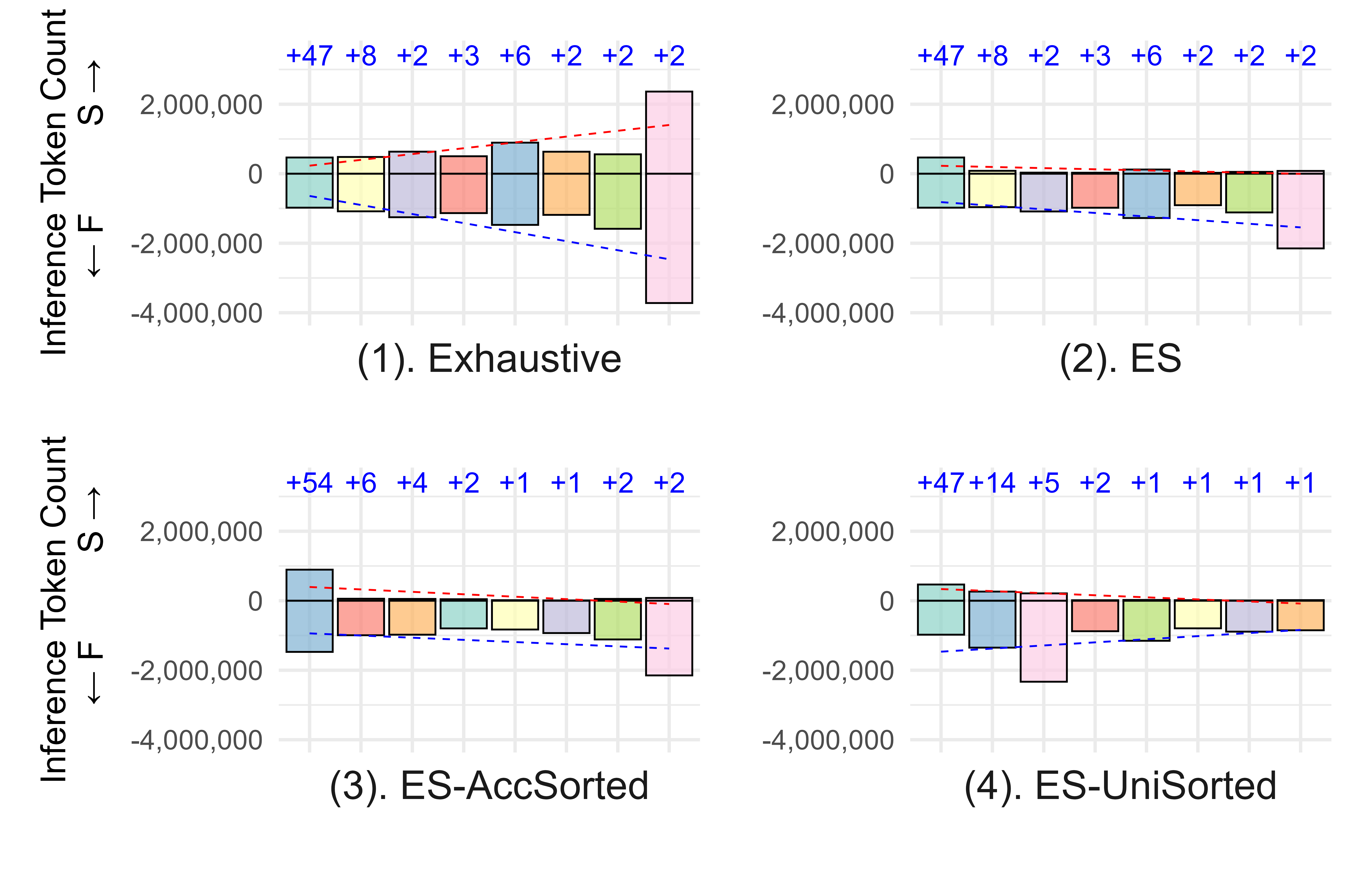}
    \caption{CodeLlama-Instruct-7B on Defects4J}
    \label{rq3_symmetry_bar_2_datasets_3_models:b}
  \end{subfigure}

  \vspace{1em} 
    
  \begin{subfigure}[b]{0.48\textwidth}
    \includegraphics[width=\textwidth]{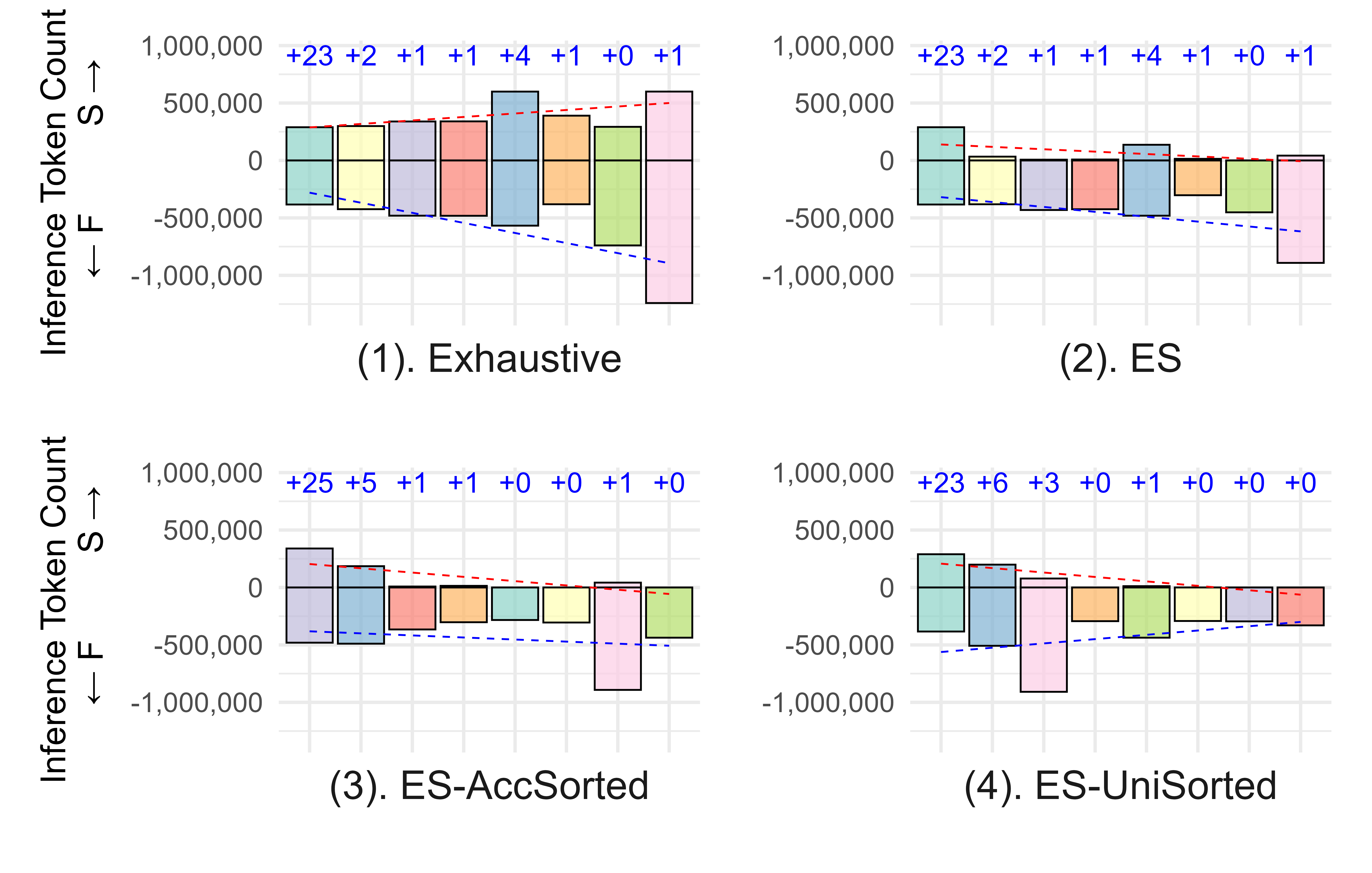}
    \caption{DeepSeek-Coder-Instruct-6.7B on \\BugsInPy}
    \label{rq3_symmetry_bar_2_datasets_3_models:c}
  \end{subfigure}
  \hfill
  \begin{subfigure}[b]{0.48\textwidth}
    \includegraphics[width=\textwidth]{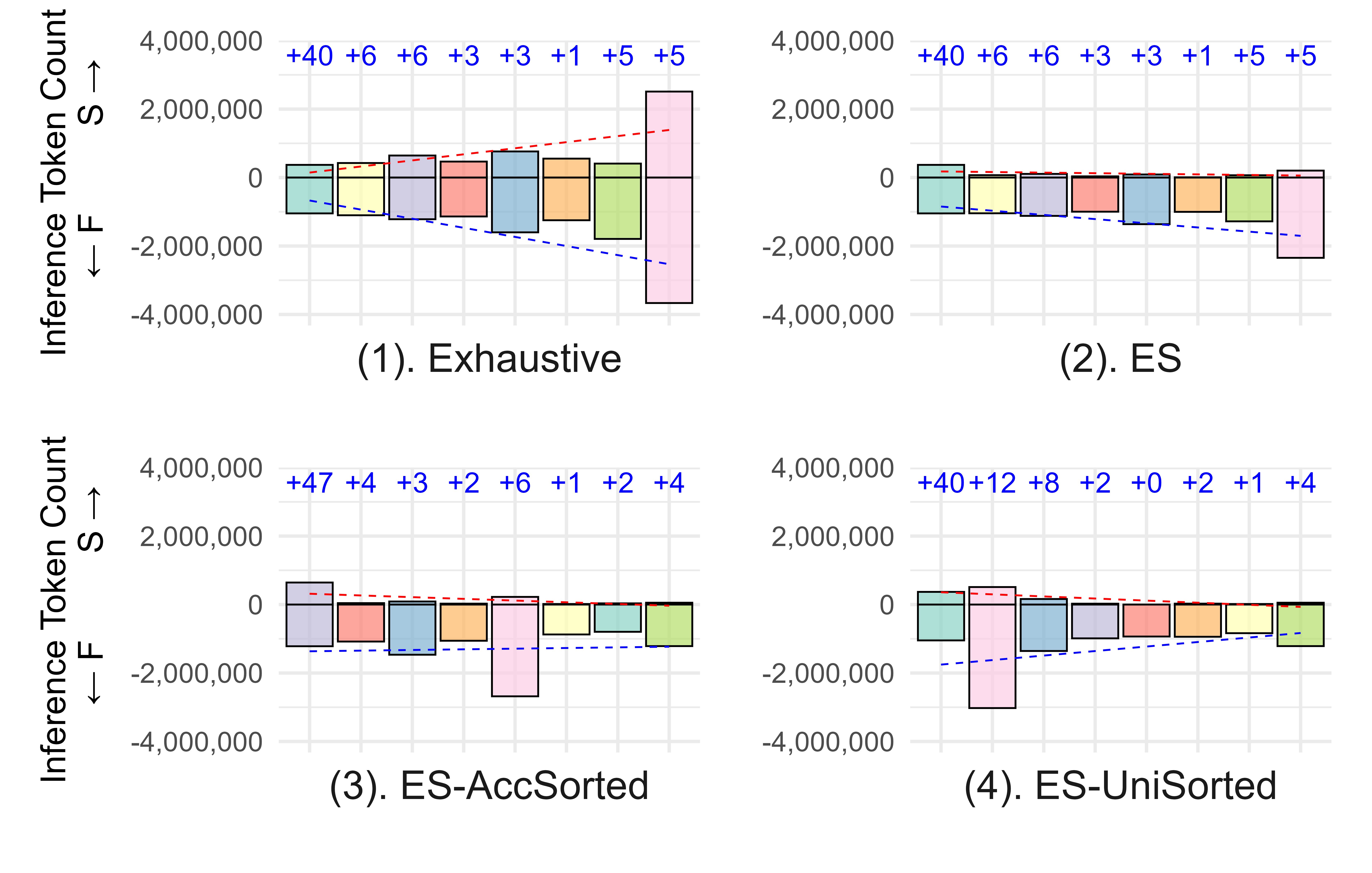}
    \caption{DeepSeek-Coder-Instruct-6.7B on \\Defects4J}
    \label{rq3_symmetry_bar_2_datasets_3_models:d}
  \end{subfigure}

  \vspace{1em} 
    
  \begin{subfigure}[b]{0.48\textwidth}
    \includegraphics[width=\textwidth]{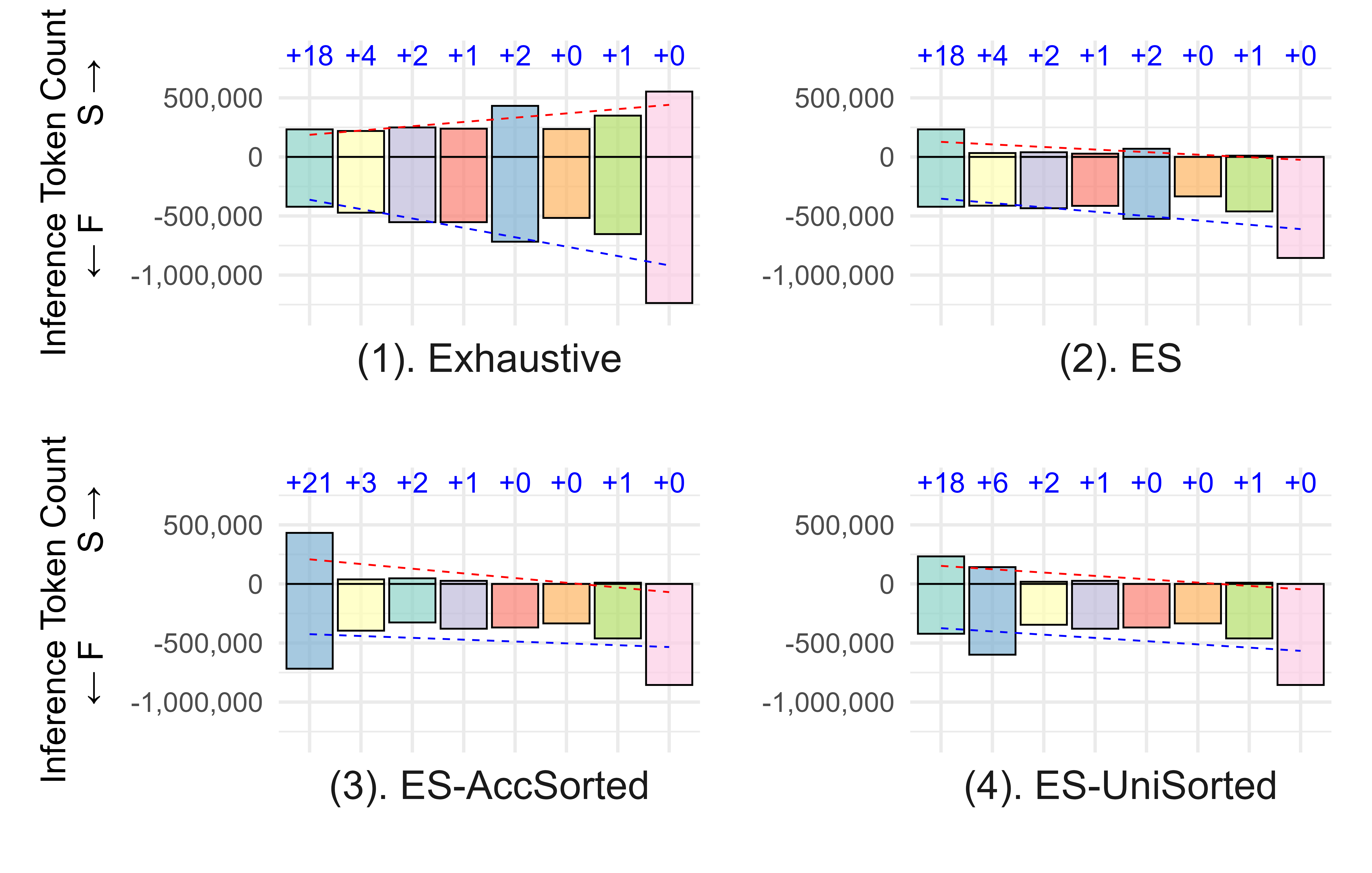}
    \caption{DeepSeek-Coder-V2-Lite-Instruct-16B \\on BugsInPy}
    \label{rq3_symmetry_bar_2_datasets_3_models:e}
  \end{subfigure}
  \hfill
  \begin{subfigure}[b]{0.48\textwidth}
    \includegraphics[width=\textwidth]{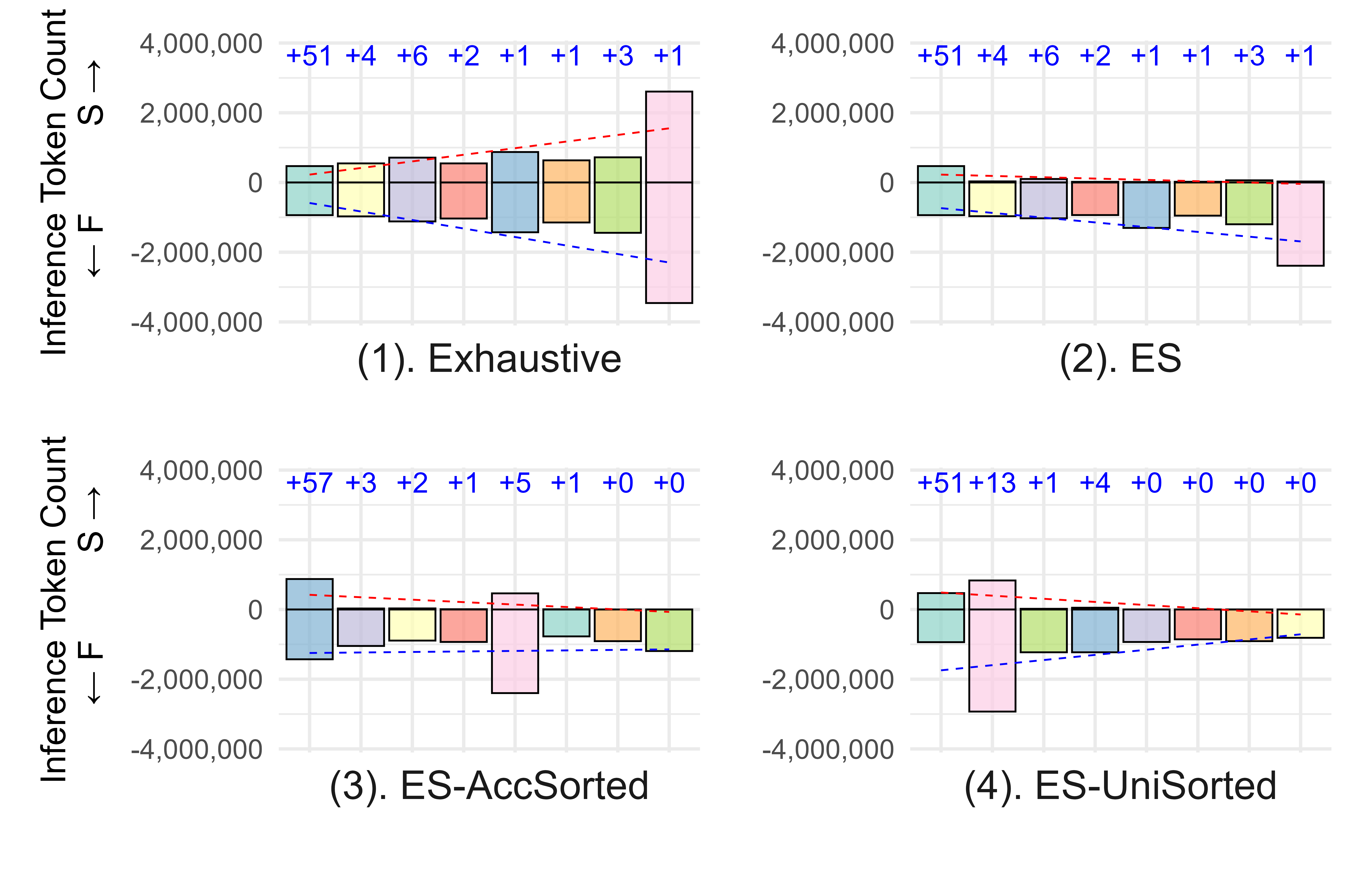}
    \caption{DeepSeek-Coder-V2-Lite-Instruct-16B \\on Defects4J}
    \label{rq3_symmetry_bar_2_datasets_3_models:f}
  \end{subfigure}

  \vspace{1em}
  \begin{adjustbox}{width=0.4\textwidth}
    \includegraphics{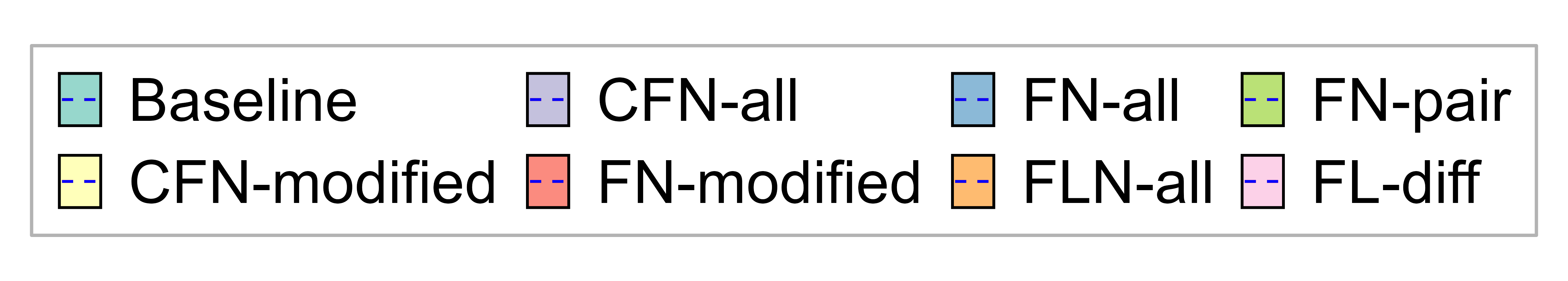}
  \end{adjustbox}
  
  \caption{The inference tokens consumption trends of different heuristics across the four cost scenarios for HAFix-Agg, evaluated on the two datasets and three models. The positive y-axis represents the inference price for successfully fixed bugs (S →), while the negative y-axis corresponds to the inference price for attempted but failed fixes (← F). Numerical values above each bar indicate the number of bugs successfully fixed on top of the preceding heuristic. The red and blue lines depict the changing trends in the number of successfully fixed bugs and failed fixes, respectively. Bars are displayed in the order of heuristic execution within each cost scenario, from left to right.}
  \label{fig:rq3_symmetry_bar_2_datasets_3_models}
\end{figure}

\end{appendices}

\clearpage
\section{Declarations}
\noindent \textbf{Data Availability Statement:}
The experiment code and datasets from this study are available in the replication package \citep{HAFix_Replication}.

\noindent \textbf{Conflict of Interest:}
All authors certify that they have no affiliations with or involvement in any organization or entity with any financial interest or non-financial interest in the subject matter or materials discussed in this manuscript.

\noindent \textbf{Funding:} This study was funded by NSERC.

\noindent \textbf{Ethical Approval:} This article does not contain any studies with human participants or animals performed by any of the authors.

\noindent \textbf{Informed Consent:} Not applicable.

\clearpage
\bibliographystyle{spbasic}      
\bibliography{bibliography}   

\end{document}